\documentclass[journal,10pt,comsoc]{IEEEtran}
\ifCLASSINFOpdf
\else
\fi
\usepackage{amsmath}
\usepackage[cmintegrals]{newtxmath}
\allowdisplaybreaks

\usepackage{graphicx}
\usepackage{subfig}
\usepackage{color}

\usepackage[hyphens]{url}
\usepackage{cite}
\usepackage{array}
\usepackage{multirow}
\usepackage[table,xcdraw]{xcolor}
\usepackage{tabularx}
\usepackage{array}
\usepackage{ragged2e}
\newcolumntype{P}[1]{>{\RaggedRight\hspace{0pt}}p{#1}}

\DeclareGraphicsExtensions{.pdf,.png,.jpg,.fig,-eps-converted-to.pdf,.jpeg}

\def\endthebibliography{%
	\def\@noitemerr{\@latex@warning{Empty `thebibliography' environment}}%
	\endlist
}
\begin{document}
%
\title{Next-generation Wireless Solutions for the Smart Factory, Smart Vehicles, the Smart Grid and Smart Cities}
%
%
%

\author{Tai Manh Ho,
	Thinh Duy Tran,
	Ti Ti Nguyen,
	S. M. Ahsan Kazmi,
	Long Bao Le,~\IEEEmembership{Senior Member,~IEEE,} 
	Choong Seon Hong,~\IEEEmembership{Senior Member,~IEEE,}
	Lajos Hanzo,~\IEEEmembership{Fellow,~IEEE,}
\thanks{T. M. Ho, T. D. Tran, T. T. Nguyen, and L. B. Le are with INRS-EMT, Université du Québec, Montréal, Québec, Canada. (e-mail: \{manhtai.ho,thinh.tran,tit.nguyen,le\}@emt.inrs.ca).}
\thanks{S. M. A. Kazmi is with Institute of Information Security and Cyberphysical Systems, Innopolis University, Kazan, Russia. (e-mail: a.kazmi@innopolis.ru).}
\thanks{C. S. Hong is with Department of Computer Science and Engineering, Kyung Hee University, South Korea. (e-mail: cshong@khu.ac.kr).}
\thanks{L. Hanzo is with School of Electronics and Computer	Science, University of Southampton, Southampton SO17 1BJ, U.K. (e-mails: lh@ecs.soton.ac.uk).}
}
\maketitle

\begin{abstract}
5G wireless systems will extend mobile communication services beyond mobile telephony, mobile broadband, and massive machine-type communication into new application domains, namely the so-called vertical domains including the smart factory, smart vehicles, smart grid, smart city, etc. Supporting these vertical domains comes with demanding requirements: high-availability, high-reliability, low-latency, and in some cases, high-accuracy positioning. In this survey, we first identify the potential key performance requirements of 5G communication in support of automation in the vertical domains and highlight the 5G enabling technologies conceived for meeting these requirements. We then discuss the key challenges faced both by industry and academia which have to be addressed in order to support automation in the vertical domains. We also provide a survey of the related research dedicated to automation in the vertical domains. Finally, our vision of 6G wireless systems is discussed briefly.
\end{abstract}

\begin{IEEEkeywords}
5G Wireless Technology, Vertical Automation, Smart Factory, Smart Vehicle, Smart Grid, Smart City.
\end{IEEEkeywords}

%
\IEEEpeerreviewmaketitle

\section{Introduction} \label{sec:intro}


The fifth-generation (5G) wireless systems have incorporated radical recent technological advancements for significantly enhancing the wireless capacity and adaptability.
In addition to supporting mobile broadband services, they will also allow to connect a massive number of different devices having diverse quality of service (QoS) requirements.

Apparently, there is no single standardization body to satisfy all aspects of the 5G wireless standard. Different stakeholders including industry, academia and governments have invested tremendous efforts into developing the 5G standards.
Among these standardization bodies, the 3rd Generation Partnership Project (3GPP) is probably the most well-known organization in defining and standardizing the 5G technology. 3GPP is the international standards body that covers mobile communications network technologies. The 5G radio and system architecture studies (5G Study Release 14) in 3GPP commenced in December 2015. 

While the first phase of 5G (5G Phase I), which corresponds to 3GPP Release 15, is meant to provide operators with a minimum set of features that enable 5G capabilities targeting the enhanced mobile broadband (eMBB) as well as ultra-reliable low-latency communications (uRLLC), and also the massive machine-type communication (mMTC) use cases. Since Release 15 has been completed by Q4 2018, the detailed design of the commercial 5G wireless system in Phase I has also been done by then. The second phase (5G Phase II) targets features that will allow the 3GPP 5G Radio to comply with stringent IMT-2020 requirements. This second phase will be developed in Release 16 and is expected to be ready by December 2019—in time for IMT-2020 submission \cite{9.2}.

\begin{figure}[!t]
	\centering
	\includegraphics[width=3.5in]{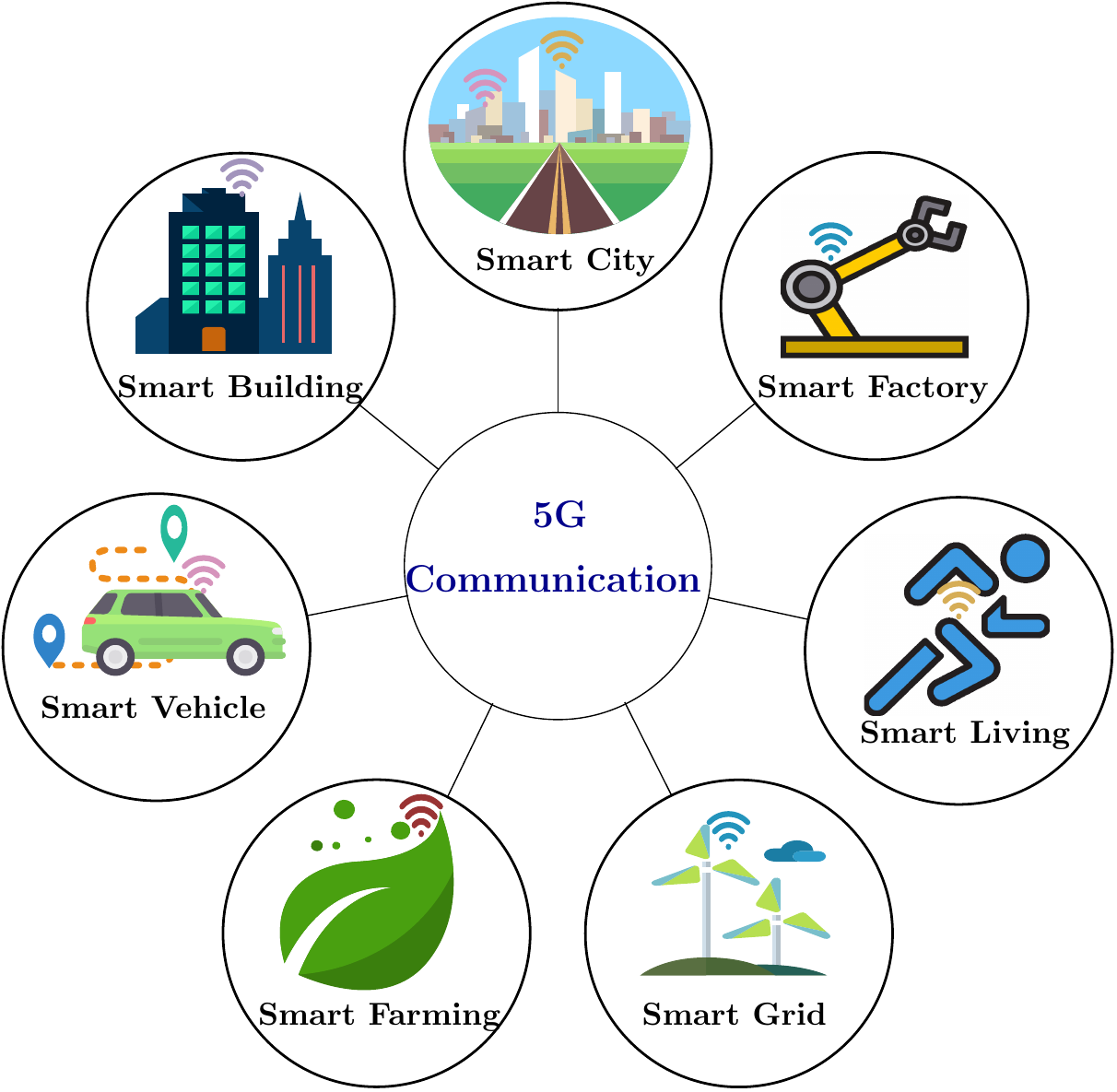}  
	\centering
	\caption{Automation in the vertical domains relying on 5G communication.}
	\label{fig:Fig1}
\end{figure} 
\begin{figure*}[!t]
	\centering
	\includegraphics[width=5in]{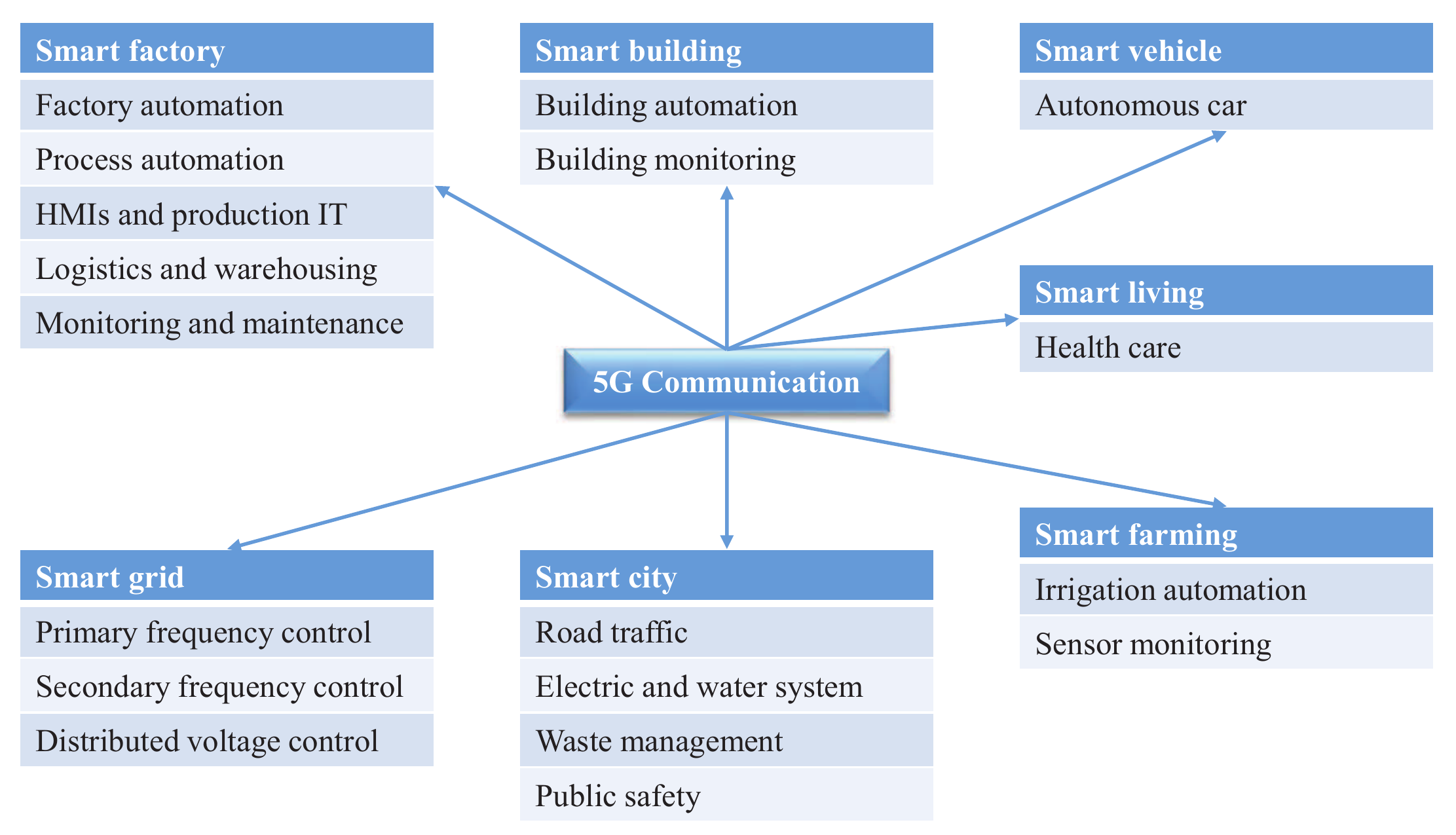}
	\caption{Vertical domains and some application areas categorized by 3GPP Release 16 \cite{1.1}.}
	\label{fig:Fig2}
\end{figure*}

Several mobile network operators (MNOs) have announced plans to deploy 5G. In the United States, AT\&T has launched the 5G mobile network in several U.S. states. Verizon is proceeding with its Fixed Wireless Access (FWA) deployment using 5G equipment provided by Ericsson, Nokia, and Samsung has launched new services. Meanwhile, in South Korea, KT Corporation (KT) and SK Telecom (SKT) have stated their intention to deploy 5G during 2019; both companies have recently launched the new network in March 2019 \cite{9.1}. 

In like with the standardization, the 5G chipset race is becoming interesting. Qualcomm was the first to announce its X50 5G modem in October 2017. Intel followed with the announcement of its XMM8060 5G modem in November 2017. Meanwhile, Samsung is reportedly working on Exynos 9820, which includes a 5G modem \cite{9.1}.

\begin{figure*}[!t]
	\centering
	\includegraphics[width=6.5in]{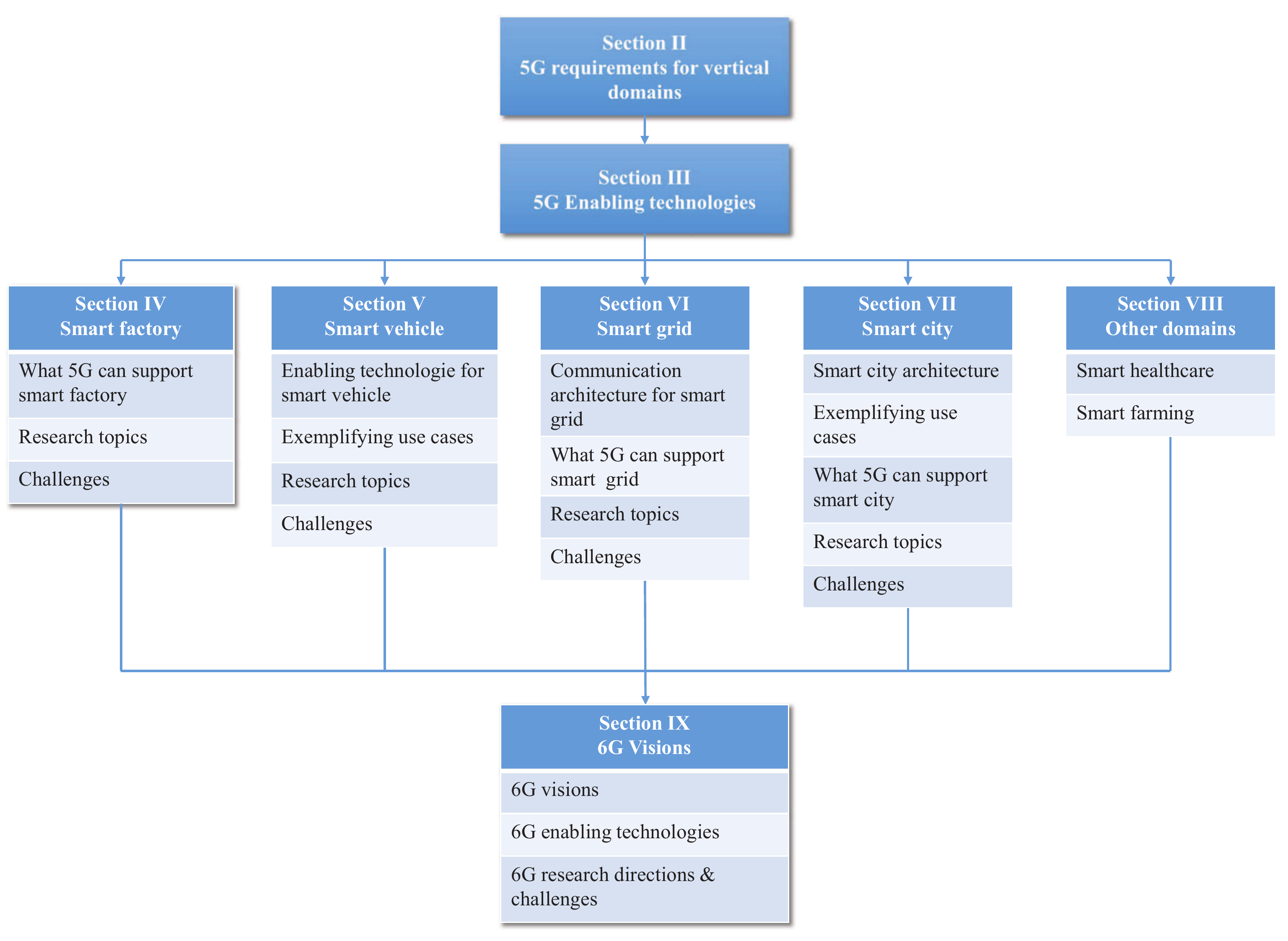}
	\caption{Survey roadmap.}
	\label{fig:Roadmaps}
\end{figure*}
In the 5G era, the high-rate, low-latency, high-reliability services enable multiple vertical domains to operate both smarter and more efficiently \cite{1.1}.  Many industrial areas were previously managed manually, but at the time of writing it is possible to perform a partial or complete automation, leading for example to smart vehicles, smart factories, smart buildings, etc.

This survey focuses on how the 5th generation wireless network can support automation in multiple vertical domains. In general, 5G facilitates low-latency, high-reliability and high service availability. Communication conceived for automation in vertical domains has to support, for instance, industrial automation, energy automation, and transportation automation (Fig. \ref{fig:Fig1}).

One of the main differences between 5G and the previous generations of cellular networks lies in 5G’s strong focus on machine-type communication and on the Internet of Things (IoT). The capabilities of 5G thus extend far beyond mobile broadband eMBB having ever-increasing data rates. 
In particular, 5G supports new services such as the radically new uRLLC, and also mMTC, especially massive IoT connectivity.
These key 5G performance features offer numerous new use cases and applications in many different vertical domains including smart factory, smart vehicle, smart grid, smart city, and more  as shown in Fig. \ref{fig:Fig2}.

\begin{table*}[tb]
	\centering
	\caption{List of the acronyms.}
	\label{table:TableAcr}
	{\renewcommand{\arraystretch}{1.4}
		\begin{tabular}{ll|ll}
			\hline
			\textbf{Acronyms} & \textbf{Meaning} & \textbf{Acronyms} & \textbf{Meaning} \\ \hline
			3GPP & 3rd Generation Partnership Project & LPWAN & Low Power Wide Area networks  \\ \hline
			5G-ACIA & 5G Alliance for Connected Industries and Automation & LTE & Long Term Evolution  \\ \hline
			5G-CAV & 5G Communications for Automation in Vertical Domains & M2M & Machine to Machine \\ \hline
			5GCAR & 5G Communication Automotive Research and innovation & MAC & Multiple access protocols  \\ \hline
			5G PPP & 5G Infrastructure Public Private Partnership & MCS & Modulation and Coding Scheme \\ \hline
			5GAA & 5G Automotive Association &  MEC & Multi-access Edge Computing \\ \hline
			 AI & Artificial Intelligence & MIMO & Multiple-Input and Multiple-Output\\ \hline
			 AR & Augmented Reality  &  ML  & Machine Learning \\ \hline
			 BS & Base Station & mMTC & Massive Machine-Type Communication \\ \hline
			CNN & Convolutional Neural Network  & NOMA & Non-Orthogonal Multiple Access \\ \hline
			 CoMP & Coordinated Multi-Point  & NFV & Network Functions Virtualization \\ \hline
			 C-V2X & Cellular Vehicle to Everything  & NR & New Radio \\ \hline
			 D2D & Device to Device & OFDM & Orthogonal Frequency-Division Multiplexing \\ \hline
			 DRL & Deep Reinforcement Learning & RAN & Radio Access Network \\ \hline
			 eMBB & Enhanced Mobile Broadband & RAT & Radio Access Technology \\ \hline
			 FAA & Factory Automation Application & SDN & Software Defined Network \\ \hline
			 HD & High Definition & UDN & Ultra-dense Network \\ \hline
			 I-IoT & Industrial Internet of Things & uRLLC & Ultra-reliable low-latency communications \\ \hline
			 ITS & Intelligent Transport System V2V & VR & Virtual Reality \\ \hline
			 IWN & Industrial Wireless Network  & V2G & Vehicle to Grid \\ \hline
			 KPI & Key Performance Indicator  &  V2V & Vehicle to Vehicle \\ \hline  
			 LoRa & Long Range  &  VLC & Visible Light Communication \\ \hline
		\end{tabular}
	}
\end{table*}
These automation aided vertical domains are being considered in Release 16  “Study on Communication for Automation in Vertical Domains” (CAV), 3GPP TR 22.804 V16.2.0 \cite{1.1}. 
There are numerous activities addressing these vertical domains worldwide. For example, in the industrial manufacturing domain, the 5G Alliance for Connected Industries and Automation (5G-ACIA) \cite{2.8} is a group made up of industrial companies, technology vendors and operators that are working on aligning the requirements of different sectors and then to communicating their priorities to 5G technology developers. 
The integration of 5G technologies in the manufacturing process has substantial potential to accelerate the transformation of the manufacturing industry and to make smart factories more efficient and productive.
Moreover, in the autonomous vehicle domain, in June 2017, the 5GCAR project \cite{3.93} brought together a strong consortium from the automotive industry and the mobile communications industry including Ericsson, Bosch, Huawei, Nokia, Orange, Viscoda, Volvo Cars, which aims to develop innovation at the intersection of those industrial sectors in order to support a fast, and successful path towards safer and more efficient future driving.
Meanwhile, in the smart city domain, commencing at the same time as the 5GCAR project, the 5GCity project \cite{5.12} is a research and innovation project forming part of the Horizon 2020 program that will design, develop, deploy and demonstrate in practical operational conditions, a distributed cloud and radio 5G-platform having a versatile hosting capability for municipalities and infrastructure owners. Both the 5GCAR and 5GCity projects are funded by the European Commission from the European Union’s Horizon 2020 research and innovation program. 
The consortium responsible for researching, developing and standardizing a direct communication technology for automated driving termed as the C2C-CC (Car-2-Car Communication Consortium) was established in Europe in 2015 \cite{3.67}. Fifteen vehicle manufacturers, over thirty suppliers and more than forty research institutions have been working together over the past few years and have developed all the required building blocks to enable vehicles to exchange information with each other.
On the other hand, in December 2018 in South Korea, in a state-sponsored project conceived for speeding up the establishment of smart factories, SK Telecom (SKT) joined faces with 18 other companies and organizations including Samsung Electronics, Microsoft Korea, Ericsson-LG, Siemens Korea to launch an alliance known as The 5G Smart Factory Alliance (5G-SFA) aiming for creating a compatible and universal solution for smart factories by unifying segmented technologies and standards.

Connectivity is a crucial requirement for the fourth Industrial Revolution, termed as “Industry 4.0”, which requires powerful and pervasive interactivity between machines, people and objects. 
5G technologies which provide seamless and ubiquitous connectivity with massive bandwidth, low-latency, and high-reliability, will support the network characteristics essential for manufacturing and critical applications. 
These requirements are  satisfied by fixed-line/wired networks in operational manufacturing processes.
However, the innovative features of 5G technology not only fulfills these requirements despite using wireless infrastructure, but also allows for higher flexibility, lower operational cost and shorter deployment time for factory floor production reconfiguration, layout changes, and other radical alterations.

Many cities around the world are realizing their own specific vision of advancing the smart cities concept by accelerating the integration of 5G technologies into their infrastructures. The innovative 5G technologies, which offer faster connections relying on their tremendous capacity, extreme reliability, and massive connectivity, will enable cities to better integrate their infrastructure, devices and their citizens. Moreover, compared to the operational 4G standard, 5G will play an important role in enabling additional smart city capabilities including various high-bandwidth and low-latency applications. For example, 5G will support the large-scale communications of connected vehicles and the traffic signal infrastructure to optimize traffic flow, as well as the deployment of a massive number of sensors for real time monitoring of the city infrastructure such as, water pipes, highways and buildings.


To elaborate a little further, 5G will revolutionize the transportation system supporting both the individual vehicles and their transportation infrastructure.
Vehicle-to-vehicle (V2V) and vehicle-to-infrastructure (V2I) communications will allow vehicles to run more efficiently while improving road safety and fuel-economy, hence promoting environmental friendliness.
New services and business models can be supported by embedding sensors into roads, railways, and airfields to communicate with smart vehicles.

Energy is the life-blood of smart factories, smart cities and smart vehicles.
The smart grid represents a state-of-the-art paradigm shift for the energy sector, which combines the traditional grid with the added benefit of communication and information control technologies for enhancing its efficiency, cleanliness, security, and privacy. 
Since renewable resources, such as solar- and wind-power are intermittent, the smart grid will require integrated monitoring and control, as well as integration with substation automation in order to control heterogeneous energy flows and plan for 'just-sufficient' standby capacity to supplement intermittent generation.
The integration of 5G into the smart grid will enable innovative solutions to facilitate the production, transmission, distribution, and 'just-in-time' exploitation of various energy types.

\begin{table*}[tb]
	\centering
	\caption{Comparison of related works on vertical domains surveys.}
	\label{table:RelatedSurvey}
	{\renewcommand{\arraystretch}{1.4}
		\begin{tabular}{llllllll}
			
			& \begin{tabular}[c]{@{}l@{}}Wang \textit{et al.}\\ \cite{3.1}-2018\end{tabular} & \begin{tabular}[c]{@{}l@{}}Islam \textit{et al.}\\ \cite{2.57}-2012\end{tabular} & \begin{tabular}[c]{@{}l@{}}Fang \textit{et al.} \\ \cite{fang2012smart}-2012\end{tabular} & \begin{tabular}[c]{@{}l@{}}Gharaibeh \textit{et al.} \\ \cite{5.13}-2017\end{tabular} & \begin{tabular}[c]{@{}l@{}}Djahel \textit{et al.} \\ \cite{5.14}-2015\end{tabular} & \begin{tabular}[c]{@{}l@{}}Eckhoff \textit{et al.} \\ \cite{5.15}-2018\end{tabular} & \begin{tabular}[c]{@{}l@{}}Sookhak \textit{et al.} \\ \cite{5.16}-2018\end{tabular} \\ \hline
			Smart vehicles & \textbf{YES} &  &  &  & \textbf{YES} &  &  \\ \hline
			Smart factories &  & \textbf{YES} &  &  &  &  &  \\ \hline
			Smart grids &  &  & \textbf{YES} &  &  &  &  \\ \hline
			Smart cities &  &  &  & \textbf{YES} & \textbf{YES} & \textbf{YES} & \textbf{YES} \\ \hline
			Security \& privacy & \textbf{YES} & \textbf{YES} & \textbf{YES} & \textbf{YES} & \textbf{YES} & \textbf{YES} & \textbf{YES} \\ \hline
			Communications & \textbf{YES} & \textbf{YES} & \textbf{YES} &  & \textbf{YES} &  &  \\ \hline
			5G technologies & \textbf{YES} &  &  &  &  &  &  \\ \hline
		\end{tabular}
	}
\end{table*}
There are several valuable survey papers that cover different aspects of diverse vertical domains as shown in Table~\ref{table:RelatedSurvey}. 
For example, the survey by Wang \textit{et al.} \cite{3.1} covers the main communication enabling technologies in autonomous vehicles.
In \cite{2.57} Islam \textit{et al.} investigate the reliability and security challenges of wireless sensor networks (WSNs) and survey their practicality in factory automation.
In \cite{fang2012smart}, Fang \textit{et al.} survey the enabling techniques including wireless communication, other than 5G for the smart grid. Gharaibeh \textit{et al.} \cite{5.13} identify techniques suitable for data security and privacy enhancement and discuss the networking and computing techniques supporting smart cities. 
Djahel \textit{et al.} \cite{5.14} review up-to-date techniques used in the different phases involved in the traffic management systems of smart cities.
Smart vehicles and social media are also considered potential facilitators of efficient traffic congestion detection and mitigation solutions. 
 Eckhoff \textit{et al.} \cite{5.15} investigate the privacy aspects of the smart city, including the application areas, enabling techniques, privacy types, attackers and the sources of attacks, constructing a specific information infrastructure for a smart city.
Sookhak \textit{et al.} \cite{5.16} present a comprehensive survey of the security and privacy issues of smart cities, highlighting the security requirements of designing a secure smart city, identifying the existing security as well as privacy solutions, and presenting the  open research issues as well as challenges of security and privacy in smart cities.

This survey is intrinsically different from the aforementioned surveys, given its communication point of view in multiple vertical domains and 5G services. 
Our objective is to provide a comprehensive survey of the enabling techniques which foster automation in multiple vertical domains and 5G communication services. We also discuss the potential challenges in incorporating 5G communication into multiple vertical domains. Specifically, the contributions of this survey can be delineated as follows:
\begin{enumerate}
	\item We illustrate the key 5G performance requirements for representative use cases selected from various application domains including the smart factory, smart vehicle, smart grids and smart city. Specifically, we categorize the requirements of representative use cases according to the three types of 5G services, namely eMBB, uRLLC, and mMTC. 
	\item  We provide an in-depth review of the enabling techniques supporting 5G communication services in order to support automation in multiple vertical domains. 
	\item We discuss objective targeted and the potential challenges, when incorporating 5G communication into individual vertical domains.
	\item We summarize the state-of-the-art 5G techniques that enable automation in multiple vertical domains.
	\item We briefly review the visions of the sixth generation (6G)  wireless system  which is expected to circumvent the limitations of the ongoing 5G deployment.
\end{enumerate}

The rest of the survey is organized as follows.  In Section \ref{sec:5G_CAV_requirements}, we discuss three types of 5G services and the key performance requirements of each type of service for automation in the vertical domains. In Section III, we portray the enabling techniques for each type of 5G services and the corresponding vertical domains. In Sections IV–VII we discuss how 5G communication can support the smart factory, smart vehicle, smart grid, and smart city concepts, respectively. In these sections, we also discuss the challenges in realizing automation in individual vertical domains. In Section VIII, we provide a brief overview of the concepts of other vertical domains such as smart farming and of smart living in term of smart healthcare. Furthermore, we briefly review the 6G visions of automation domains in Section IX. Finally, we conclude this study in Section X. A list of the acronyms used in the survey is presented in Table \ref{table:TableAcr}. For convenience, Figure \ref{fig:Roadmaps} provides a detailed structure of the survey.

\section{5G Requirements for the Automation of Vertical Domains} \label{sec:5G_CAV_requirements}

\begin{figure}[!t]
	\centering
	\includegraphics[width=3.5in]{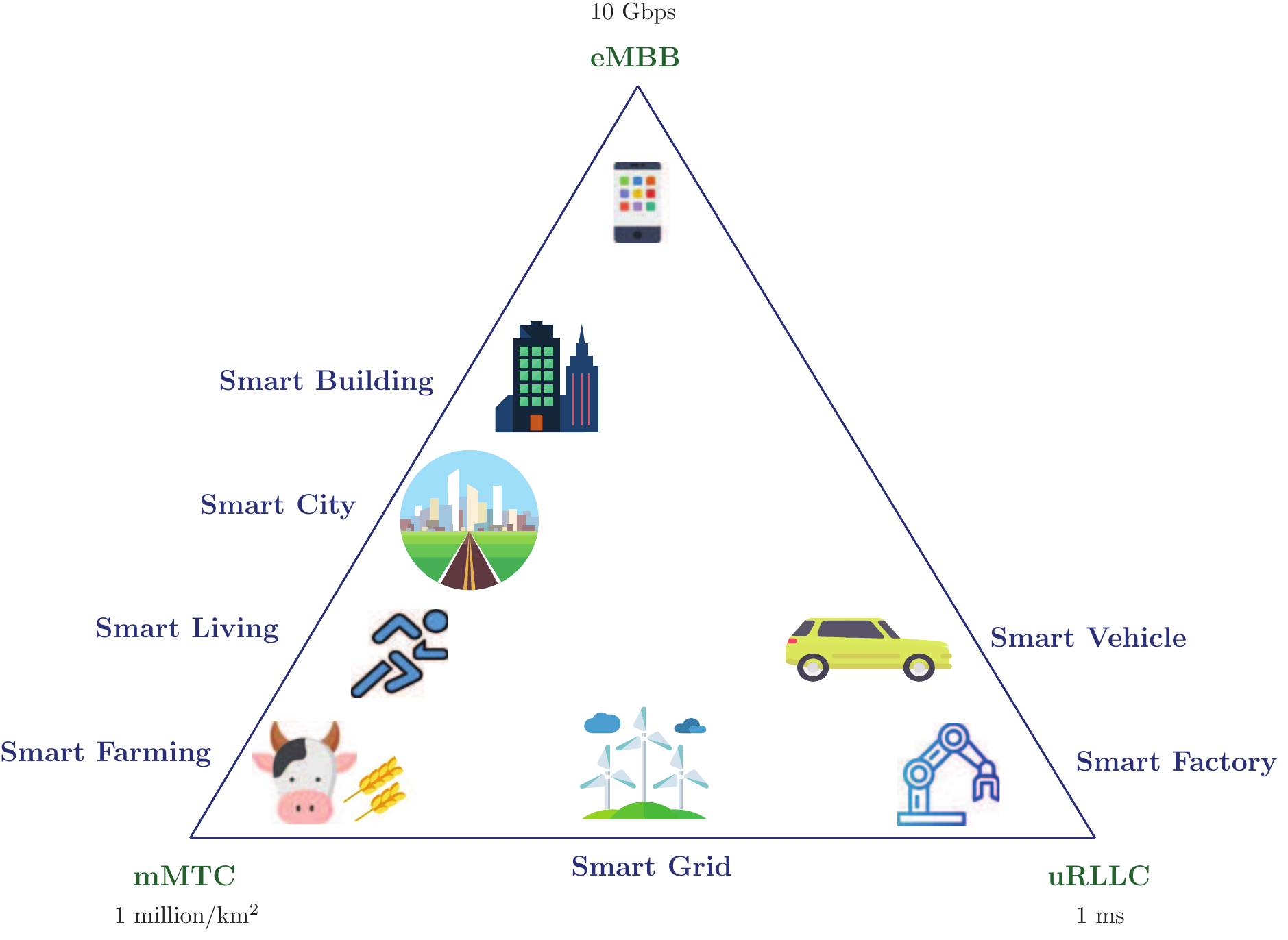}  
	\centering
	\caption{Vertical domains and arrangement in 5G service types.}
	\label{fig:Fig3}
\end{figure}

Again, according to the International Telecommunication Union (ITU) \cite{1.6}, 5G mobile network services are classified into three categories, namely eMBB, uRLLC and mMTC, to meet the requirements of automation in multiple vertical domains.
These three types of services support many aspects of different use cases in different vertical domains. 
eMBB aims for fulfilling the ever-increasing mobile traffic demand imposed by bandwidth-hungry services, such as high definition (HD) videos, virtual reality (VR) and augmented reality (AR). 
Meanwhile, the time-sensitive networking and mission-critical services such as assisted and automated driving, robot-motion control, and remote management, are satisfied by the uRLLC mode.
Furthermore,  mMTC is aimed at supporting dense connections of various device types (e.g, mobile devices, IoT devices and sensors) in crowded areas, such as smart cities and smart farming.
Fig.~\ref{fig:Fig3} depicts the arrangement of vertical domains according the above three 5G service types.

Specifically, eMBB provides extremely high data rates (several Gbps) and expands network coverage, well beyond that of 4G. 
By contrast, mMTC is designed to provide deep indoor penetration and wide-area coverage for ubiquitous connectivity of million of low-cost and battery-limited IoT devices  per square kilometer.
Finally, URLLC can facilitate mission-critical applications having stringent requirements in terms of end-to-end (E2E) latency (e.g., 1 millisecond latency), reliability and availability. This includes, for example, high-performance connectivity for applications in industrial automation and control.

\begin{table*}[tb]
	\centering
	\caption{5G performance requirement of IMT 2020 \cite{1.34}.}
	\label{table:Table1}
	{\renewcommand{\arraystretch}{1.4}
		\begin{tabular}{l|ll}
			\hline 
			\multicolumn{2}{c}{\textbf{5G Performance Index}} & \multicolumn{1}{c}{\textbf{Value}} \\ \hline
			\multicolumn{2}{l}{Peak data rate [Gbps]} & 20 Gbps (DL), 10 Gbps (UL) \\ \hline
			\multicolumn{2}{l}{Peak spectral efficiency [bps/Hz]} & 30 bps/Hz (DL), 15 bps/Hz (UL) \\ \hline
			\multicolumn{2}{l}{User experienced data rate [Mbps]} & 100 Mbps (DL), 50 Mbps (UL) \\ \hline
			& Indoor hot spot & 0.3 bps/Hz (DL), 0.21 bps/Hz (UL) for eMBB \\ \cline{2-3} 
			& Dense urban & 0.225 bps/Hz (DL), 0.15 bps/Hz (UL) for eMBB \\ \cline{2-3} 
			\multirow{-3}{*}{5th percentile user spectral efficiency} & Rural & 0.12 bps/Hz (DL), 0.045 bps/Hz (UL) for eMBB \\ \hline
			& Indoor hot spot & 9.0 bps/Hz (DL), 6.75 bps/Hz (UL) for eMBB \\ \cline{2-3} 
			& Dense urban & 7.8 bps/Hz (DL), 5.4 bps/Hz (UL) for eMBB \\ \cline{2-3} 
			\multirow{-3}{*}{Average spectral efficiency [bps/Hz]} & Rural & 3.3 bps/Hz (DL), 2.1 bps/Hz (UL) for eMBB \\ \hline
			Area traffic capacity [Mbps/m$^2$] &  & 10 Mbps/m$^2$ for indoor hotspot - eMBB \\ \hline
			& User plane & 4 ms for eMBB, 1ms for uRLLC \\ \cline{2-3} 
			\multirow{-2}{*}{Latency [ms]} & Control plane & TBD for eMBB and uRLLC \\ \hline
			\multicolumn{2}{l}{Connection density [Device/Km$^2$]} & 106 for mMTC \\ \hline
			\multicolumn{2}{l}{Energy efficiency} & TBD for eMBB \\ \hline
			\multicolumn{2}{l}{Reliability} & $1-10^{-5}$ for uRLLC \\ \hline
			& Stationary & 0 Km/h for indoor hotspot - eMBB \\ \cline{2-3} 
			& Pedestrian & 3 Km/h for Dense urban - eMBB \\ \cline{2-3} 
			& Vehicular & 120 Km/h for Rural - eMBB \\ \cline{2-3} 
			\multirow{-4}{*}{Mobility [Km/h]} & High-speed vehicular & $\sim$500 Km/h for Rural - eMBB, High-speed train \\ \hline
			\multicolumn{2}{l}{Mobility interruption time [ms]} & 0 ms \\ \hline
			\multicolumn{2}{l}{Bandwidth [MHz]} & 100 MHz $\sim$ 1 GHz \\ \hline
		\end{tabular}
	}
\end{table*}

Table \ref{table:Table1} summarizes the key system parameters along with the relevant use cases, which were approved by the ITU in November 2017 for International Mobile Telecommunications-2020 (IMT-2020) systems \cite{1.34}.
Motivated readers are referred to \cite{1.34} for their detailed definitions.
The following subsections will provide an overview of uRLLC, eMBB, and mMTC service types, which are fundamental for bringing pervasive automation of the vertical domain to fruition.

\subsection{Latency $\&$ Reliability for Ultra-Reliable Low-Latency Communication}

The uRLLC mode is a principal service type that supports mission-critical applications, such as autonomous vehicles, industrial automation and augmented reality (AR) 
\cite{1.4,1.5,1.6,1.7,1.8,1.9,1.10,1.11,1.13,1.14,1.15,1.16,1.18,1.19,1.20,1.22,1.36,1.59,1.60,1.61,1.62,1.66,1.67,1.68,1.69,1.70,1.72,1.77,1.78,1.79}.

For the uRLLC type, the first release of 5G (3GPP Release 15) already specified a latency requirement of 1 ms with a reliability of $99.999\%$ for transmission over the 5G radio interface. This permits reliable transmission of small data packets (with a size of only a few bytes) over the air within a specified time limit, as required for closed-loop control applications, for example. Low-latency communication is facilitated by the introduction of short transmission slots, allowing faster uplink and downlink transmissions. By reducing the transmission duration and interval by flexible adjustments, both the time over the air and the delay introduced by the transmitter while waiting for the next transmission opportunity are reduced \cite{2.8}.
\begin{figure}[!t]
	\centering
	\includegraphics[width=3.5in]{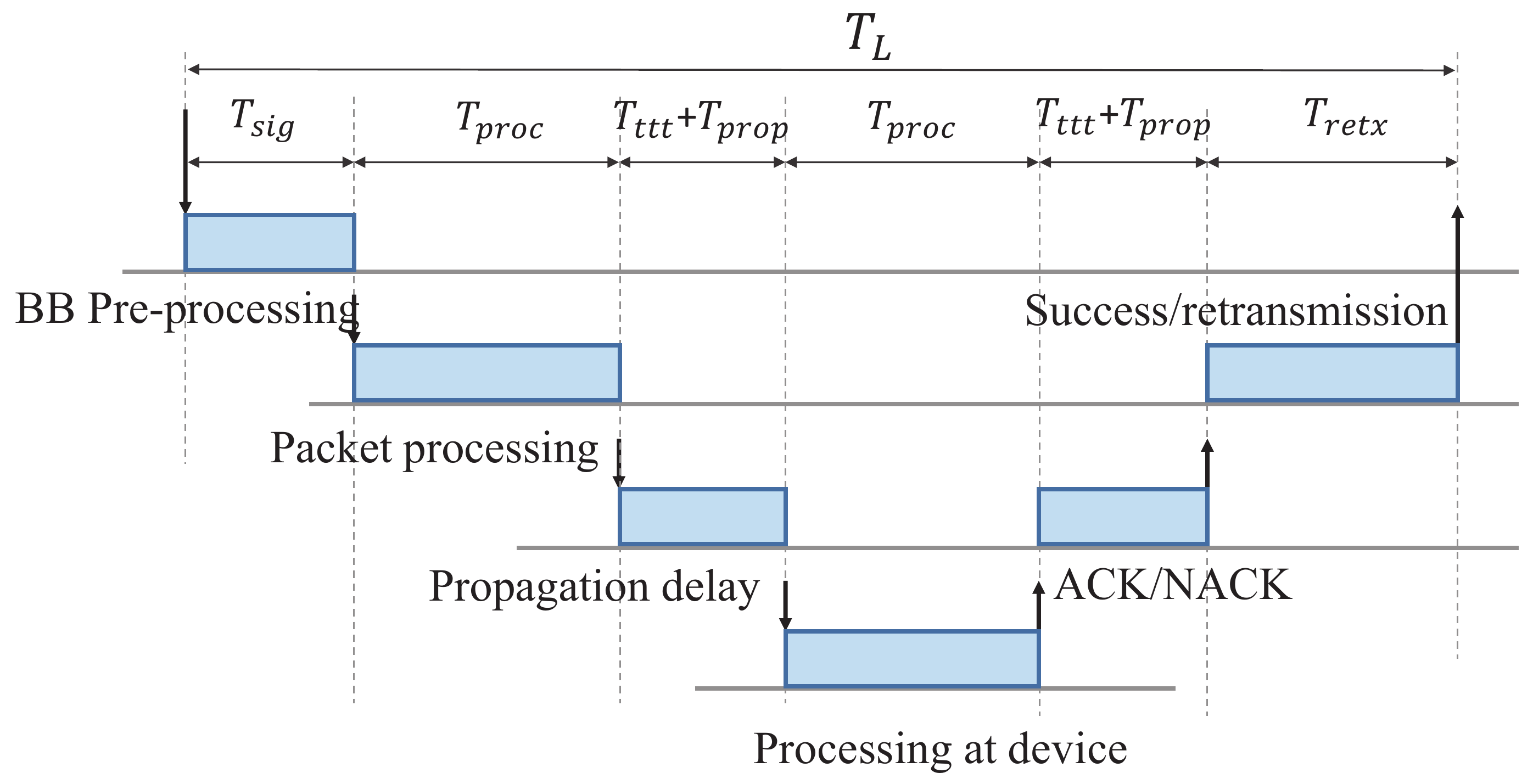}  
	\centering
	\caption{Latency components in 5G physical layer \cite{1.36}.}
	\label{fig:URLLCframe}
\end{figure} 

The latency requirement to be satisfied by the 5G physical layer for fulfilling the uRLLC is described in the 3GPP technical report \cite{1.88}. Specifically, the total communication latency can be decomposed into multiple components \cite{1.36} as shown in Fig. \ref{fig:URLLCframe}. In \cite{1.13} and \cite{1.36} Ji \textit{et al.} have studied the physical-layer design of uRLLC in 5G. 
The authors commence by describing the key requirements of the uRLLC service type.
Several physical-layer issues and enabling techniques, including multiplexing schemes, the packet and frame structure, and reliability improvement techniques are discussed as well. The associated numerologies\footnote{{In 3GPP-TR38.802 \cite{1.39}, a numerology is defined by sub-carrier spacing and cyclic prefix overhead. Multiple subcarrier spacings can be derived by scaling a basic subcarrier spacing by an integer $N$. The numerology used can be selected independently of the frequency band although it is assumed not to use a very low subcarrier spacing at very high carrier frequencies.}} and frame structure of NR are also studied in \cite{1.154} where the subframe length of New Radio (NR) is 1 ms composed of 14 OFDM symbols using 15 kHz subcarrier spacing and the cyclic prefix (CP).

Higher reliability can be accomplished for instance by using robust modulation and coding schemes (MCS) and diversity techniques. 
Powerful channel coding schemes, such as Turbo codes or low density parity check (LDPC) codes are used for the data channels in 3GPP Release 15, while tail-biting convolutional or Reed-Müller codes or Polar codes are employed for the control channels. 
Further improvements are expected to be introduced for satisfying the requirements of smart factories. 
Diversity gains can be provided by  spatial-, frequency- or time-domain diversity. 
Multi-connectivity relaying on multiple radio access technology (RAT) can be arranged via multi-carrier techniques or multiple transmission points, where  the device is connected to the radio network via multiple frequency carriers. 
Recently, applying these features for improving the transmission reliability has attracted more attention both from academia and industry \cite{2.8}.
In \cite{1.90} and \cite{6.5}, the authors describe in more detail the latency requirements of URLLC-based applications.

\subsection{Data Rate for eMBB}

The eMBB service type supports bandwidth-hungry applications such as HD video streaming, VR, and AR \cite{1.4,1.5,1.6,1.7,1.8,1.11,1.13,1.15,1.29,1.36}.
According to \cite{1.34}, the 5G eMBB service type is capable of achieving peak rates of 20 Gbps and 10 Gbps in the downlink and uplink, respectively. 
Such high data rates are mainly facilitated by having wide system bandwidth (up to 400 MHz) as well as massive MIMO and 3D MIMO schemes using a large number of antenna elements \cite{1.153}, and high modulation orders, such as 256 QAM or even higher in future 3GPP releases. 
5G is expected to operate over diverse carrier frequencies spanning from 1 to 86 GHz, and in both the licensed and unlicensed spectrum. 
At high carrier frequencies above 6 GHz, large chunks of the spectrum are still available, but they suffer from severe signal attenuation \cite{1.13,1.36,1.38,1.43,1.44,1.45,1.46,1.86}.

5G wireless networks must be able to satisfy the following three key attributes for realizing eMBB service type use cases:
\begin{itemize}
	\item High network throughput both in dense indoor and outdoor areas, such as office buildings, conference centers, downtown areas, and in stadiums.
	\item Ubiquitous connectivity to provide a seamless user experience.
	\item Extensive mobility to support various vehicular velocities of cars, buses, trains, and planes.
\end{itemize}

The eMBB service type use cases impose heterogeneous requirements.
For instance, a ‘hotspot’ area, e.g., a sporting or music event with massively connected devices, will demand high network capacity at low mobility requirement.
Conversely, passengers in a high-speed train will require a high degree of mobility but have to tolerate lower traffic capacity.
Table \ref{table:Table1} illustrates detailed specifications of various eMBB service types.

\subsection{Low Hardware Complexity, Long Battery Life, High Coverage and Device Density for mMTC}
The mMTC service type aims for supporting a massive number of connections from machine-type devices, e.g. IoT devices.
Several mMTC-based services designed for the smart city and smart agriculture rely on sensing and monitoring, while accurate metering is essential for the smart grid
\cite{1.4,1.5,1.6,1.7,1.9,1.11,1.12,1.17,1.21,1.23,1.24,1.25,1.26,1.27,1.28,1.30,1.31,1.32,1.49,1.52,1.53,1.54,1.55,1.56,1.57,1.58}.


In mMTC applications, 5G will provide connection densities far exceeding the requirement of 1,000,000 devices per km$^2$ \cite{1.34}, 20 dB coverage improvements (resulting in a coupling loss of 164 dB), and battery lifetimes exceeding 10 years (see Chapter 3 in \cite{1.35}). The support of a large device population per square km is achieved by efficient signaling. The coverage extension (20 dB better than 4G) is predominately attributable to the use of repetition of the transmitted information and owing to the reduction of active frequency bandwidth \cite{2.8}. 

The devices having a low hardware complexity and low manufacturing cost of less than a few dollars, such as IoT devices and sensors, tend to have a limited transmission bandwidth to 1 MHz or less, and a few hundreds of kbps as well as a limited output power (20 dBm). In addition, half-duplex transmission is used to avoid duplex filters. The long battery lifetime (5 - 10 years) is achieved by allowing extended discontinuous reception (eDRX) for supporting a sleep mode for a device. Finally, having a limited number of MCS and a limited  number of transmit modes is capable of reducing the complexity \cite{2.8}. These requirements are essential for the mMTC service type.
Table \ref{table:Table2} summarizes some of the most important requirements of different Communication for Automation in Vertical (CAV) use cases. Detailed explanations of these requirements can be found in \cite{1.1,1.2,1.3,1.148,1.149,3.66,1.119,1.90,1.102,2.8,3.24}.

\begin{table*}[tb]
	\centering
	\caption{The summary of 5G requirements for CAV use cases \cite{1.1,1.2,1.3,1.148,1.149,3.66,1.119,1.90,1.102,2.8,3.24}.}
	\label{table:Table2}
	{\renewcommand{\arraystretch}{1.5}
	\begin{tabularx}{\textwidth}{l|l P{1cm} X X X X X X}
		\hline
		\multicolumn{2}{l}{\textbf{Use case}} & \begin{tabular}[c]{@{}l@{}}\textbf{Latency} \\ (ms)\end{tabular} & \textbf{Reliability} & \begin{tabular}[c]{@{}l@{}}\textbf{Availability}\\ (\%)\end{tabular} & \begin{tabular}[c]{@{}l@{}}\textbf{Device}\\ \textbf{density}\end{tabular} & \begin{tabular}[c]{@{}l@{}}\textbf{Traffic} \\ \textbf{density}\end{tabular} & \begin{tabular}[c]{@{}l@{}}\textbf{User}\\ \textbf{throughput}\end{tabular} & \begin{tabular}[c]{@{}l@{}}\textbf{Mobility} \\ (km/h)\end{tabular} \\ \hline
		\multirow{15}{*}{\rotatebox[origin=c]{90}{\textbf{Smart Factory} \cite{1.1,2.8}}} & Manufacturing cell \cite{1.1}& 5 & $10^{-9}$ & $>99.9999$ & 0.33-3/m$^2$ & N/S\footnote{Not specified} & N/S & \textless{}30 \\ \cline{2-9} 
		& Machine tools \cite{1.1}& 0.5 & $10^{-9}$& $>99.9999$ & 0.33-3/m$^2$ & N/S & N/S & \textless{}30 \\ \cline{2-9} 
		& Printing machines \cite{1.1} & 2 & $10^{-9}$ & $>99.9999$ & 0.33-3/m$^2$ & N/S & N/S & \textless{}30 \\ \cline{2-9} 
		& Packaging machines \cite{1.1} & 1 & $10^{-9}$ & $>99.9999$ & 0.33-3/m$^2$ & N/S & N/S & \textless{}30 \\ \cline{2-9} 
		& \begin{tabular}[c]{@{}l@{}}Cooperative \\  motion control \cite{2.8}\end{tabular}  & 1 & $10^{-9}$ & $>99.9999$ & 0.33-3/m$^2$ & N/S & N/S & \textless{}30 \\ \cline{2-9} 
		&\begin{tabular}[c]{@{}l@{}}Video-operated \\ remote control \cite{2.8}\end{tabular} & 10-100 & $10^{-9}$ & $>99.9999$ & 0.33-3/m$^2$ & N/S & N/S & \textless{}30 \\ \cline{2-9} 
		& \begin{tabular}[c]{@{}l@{}}Assembly robots \\ or milling machines  \cite{2.8}\end{tabular}  & 4-8 & $10^{-9}$ & $>99.9999$ & 0.33-3/m$^2$ & N/S & N/S & \textless{}30 \\ \cline{2-9} 
		& Mobile cranes \cite{2.8}& 12 & $10^{-9}$ & $>99.9999$ & 0.33-3/m$^2$ & N/S & N/S & \textless{}30 \\ \cline{2-9} 
		& \begin{tabular}[c]{@{}l@{}}Process automation \\ - Monitoring \cite{1.3}\end{tabular} & 50 & $10^{-3}$ & 99.9 & 10000/plant & 10 Gbps/km$^2$ & 1 Mbps & \textless{}5 \\ \cline{2-9} 
		& \begin{tabular}[c]{@{}l@{}}Process automation \\ - Remote control \cite{1.3}\end{tabular} & 50 & $10^{-5}$ & 99.999 & 1000/km$^2$ & 100 Gbps/km$^2$ & \textless{}100 Mbps & \textless{}5 \\ \hline
		\multirow{2}{*}{\rotatebox[origin=c]{90}{\textbf{Smart Grids}}} & \begin{tabular}[c]{@{}l@{}}Electricity distribution \\ - Medium Voltage \cite{1.3}\end{tabular} & 25 & $10^{-3}$ & 99.9 & 1000/km$^2$ & 10 Gbps/km$^2$ & 10 Mbps & 0 \\ \cline{2-9} 
		& \begin{tabular}[c]{@{}l@{}}Electricity distribution \\ - High Voltage \cite{1.3}\end{tabular} & 5 & $10^{-6}$ & 99.9999 & 1000/km$^2$ & 100 Gbps/km$^2$ & 10 Mbps & 0 \\ \hline
		\multirow{4}{*}{\rotatebox[origin=c]{90}{\textbf{Smart Vehicle} \cite{1.3}}} & Autonomous driving \cite{3.66} & 5 & $10^{-5}$ & 99.999 & 500-3000/km$^2$ & N/S & 0.1-29 Mbps & urban $<100$, highway$<500$ \\ \cline{2-9} 
		& Collision warning \cite{3.24} & 10 & $10^{-3} - 10^{-5}$ & 99.999 & 500-3000/km$^2$ & N/S & 0.1-29 Mbps & urban $<100$, highway$<500$ \\ \cline{2-9} 
		& High-speed train \cite{1.3} & 10 & N/S& N/S & 1000/train & 12.5-25 Gbps/train & 25-50 Mbps & \textless{}500 \\ \hline
		\multirow{5}{*}{\rotatebox[origin=c]{90}{\textbf{ITS} \cite{1.90}}} & Road safety urban \cite{1.90}& 10-100 & $10^{-3} - 10^{-5}$ & 99.999 & 3000/km$^2$ & 10 Gbps/km$^2$ & 10 Mbps & \textless{}100 \\ \cline{2-9} 
		& Road safety highway \cite{1.90}& 10-100 & $10^{-3} - 10^{-5}$  & 99.999 & 500/km$^2$ & 10 Gbps/km$^2$ & 10 Mbps & \textless{}500 \\ \cline{2-9} 
		& Urban intersection \cite{1.90}& \textless{}100 & $10^{-5}$ & 99.999 & 3000/km$^2$ & 10 Gbps/km$^2$ & 10 Mbps & \textless{}50 \\ \cline{2-9} 
		& Traffic efficiency \cite{1.90}& \textless{}100 & $10^{-3}$ & 99.9 & 3000/km$^2$ & 10 Gbps/km$^2$ & 10 Mbps & \textless{}500 \\ \cline{2-9} 
		& Traffic jam \cite{1.119}& 8 & N/S & 95.0 & N/S & 480 Gbps/km$^2$ & 20-100 Mbps & N/S \\ \hline
		\multirow{5}{*}{\rotatebox[origin=c]{90}{\textbf{Smart City} \cite{1.119}}} & Large outdoor event \cite{1.119} & N/S & $10^{-2}$ & 99.0 & 4/m$^2$ & 900 Gbps/km$^2$ & 30 Mbps & N/S \\ \cline{2-9} 
		& Shopping mall \cite{1.119}& N/S & $10^{-2}$ & 99.0 & N/S & N/S & 60-300 Mbps & N/S \\ \cline{2-9} 
		& Stadium \cite{1.119} & N/S & $10^{-2}$ & 99.0 & 4/m$^2$ & 0.1-10 Mbps/m$^2$ & 0.3-20 Mbps & N/S \\ \cline{2-9} 
		& Dense urban \cite{1.119}& N/S & N/S & N/S & 200000/km$^2$ & 700 Gbps/km$^2$ & 60-300 Mbps & N/S \\ \cline{2-9} 
		& Media on demand \cite{1.119}& 200-5000 & TBC & 95.0 & 4000/km$^2$ & 60 Gbps/km$^2$ & 15 Mbps & N/S \\ \hline
	\end{tabularx}
	}
\end{table*}

%
\section{Enabling Techniques for 5G Communication in Automation-Aided Vertical Domains} \label{sec:EnlabingTechnologies}
\subsection{Enabling Techniques for URLLC-based Domains}

\subsubsection{Multi-connectivity, Multi-RAT}
Arising from the Coordinated Multi-Point (CoMP) transmission techniques of 4G LTE \cite{marsch2011coordinated}, multi-connectivity is the ultimate solution for reducing the transmission delay and increasing the transmission reliability \cite{1.36}. A mobile user chooses a RAT among 4G LTE, 5G NR, WiFi and other IEEE 802.x standards providing a reduced transmission latency. Furthermore, each user can also associate itself with two or more RATs at the same time to create backup communication links, thus, improving the transmission reliability. In 3GPP Release 16 \cite{1.89}, dual-connectivity is an important concept under investigation that is likely to improve the transmission reliability. D2D communications \cite{7.1} is another potential latency-reduction approach in the 5G standards.

Numerous solutions related to multi-RAT/multi-connectivity/dual-connectivity for 5G URLLC can be found in \cite{1.16,1.77,1.92,1.93,1.94,1.95,1.96,1.97}.
In \cite{1.16}, the authors improve the successful connectivity probability of 5G NR by proposing an admission control mechanism and by adjusting the proportion of multi-connectivity-mode users. Their aim is to improve the network's reliability subject to a maximum tolerable latency compared to that of a single-connectivity scenario.
In \cite{1.92}, the authors derive the capacity upper bound of dual-connectivity communications relying on mmWave communications. Similarly, the authors of \cite{1.93} study the performance evaluation methodology of dynamic multi-connectivity in mmWave ultra-dense networks (UDNs). They combine powerful methods from queuing theory, stochastic geometry, as well as system-level simulations to study the performance of the network. A thorough comparison of alternative multi-connectivity strategies is conducted.
The authors of \cite{1.94} propose an SDN-based 5G architecture supporting the uRLLC service type, where a prototype hardware implementation is used for investigating the practical effects of supporting mission-critical data in a softwarized 5G core network.
Furthermore, the extensive investigations in \cite{1.95} show that the multi-connectivity technique outperforms the CoMP philosophy in 5G UDN and in low-user density scenarios in terms of the average throughput. Moreover, the authors of \cite{1.96} and \cite{1.97} investigate the impact of multi-connectivity on the 5G URLLC service type.

\subsubsection{5G V2X Communications} \label{sec:V2X_tech}

Vehicle-to-everything (V2X) communication refers to information transmission to/from a vehicle to any entity in a vehicular communication system that incorporates several specific types of communication, such as V2I (vehicle-to-infrastructure), V2N (vehicle-to-network), V2V (vehicle-to-vehicle), V2P (vehicle-to-pedestrian), V2D (vehicle-to-device) and V2G (vehicle-to-grid) communications.

The motivation of applying 5G V2X to multiple vertical domains, especially in the automotive and smart city domain is to reduce latency to improve reliability \cite{1.36}, based on a promising business model \cite{3.56}, and to enable safe transportation \cite{2.13},\cite{3.52}. In \cite{3.56}, 5G-PPP provides a study of 5G V2X deployment including both the business models, and the stakeholders, describing how to calculate the costs and profits in such a system. The overall conclusion of this study is that a promising business case can be expected if Connected and Automated Driving (CAD) services will be provided over the network infrastructure and the user penetration rate grows slowly over time. A compelling benefit of the V2X communications technology is to support safe transportation, which is foreseen to be soon available in some countries. By sharing data, such as vehicular position and speed to surrounding vehicles and infrastructures through V2X communications, connected vehicles can enhance their awareness of potential accidents and significantly improve collision avoidance.

D2D communication together with multi-RAT is considered to be an attractive technique of reducing the latency and improving reliability \cite{1.36}. V2X communication is an ultimate evolution of D2D communication in 5G networks, which requires low-latency and high-reliability for automation \cite{3.2,3.15,3.17}. 3GPP Release 14 stated the specification of LTE-based V2X communications (Cellular-V2X) \cite{3.3}. Furthermore, V2X communications are studied in 5G networks in 3GPP Release 15 \cite{3.4}, \cite{3.5} and in the context of 5G NR in 3GPP Release 16 \cite{3.6,3.7}. 
%

Furthermore, the latency and reliability requirements of different types of V2X communications are discussed in 3GPP Release 14 \cite{3.3,3.8,3.24}. 
The requirements of latency-critical use cases in V2X can be found in Table \ref{table:Table2}.  

\subsection{Enabling Techniques for eMBB-based Application Domains}

The 4G LTE system primarily utilizes the sub-3 GHz range for expanding the network coverage, but it suffers from its limited bandwidth.
To address this issue, 5G NR has been designed for natively supporting all spectrum types (licensed, unlicensed, shared) and spectrum bands (low, mid, high). 
Early 5G deployments in 2019 targeted only the licensed spectrum following the 3GPP Release 15 specifications. 
However, future deployments based on 3GPP Release 16 and beyond are expected to also support unlicensed/shared spectrum and attractive new services beyond eMBB.

From the radio access point of view, there are three fundamental problems that have to be addressed in order to satisfy the increasing data demand: 1) How to densify networks? 2) How to deliver increased spectral efficiency? and 3) How to gain access to wider spectrum? \cite{1.101}. 
To answer these questions, several techniques have been proposed and integrated into 5G NR such as small cells for UDN scenarios \cite{1.12,1.98,1.99,1.100,7.2,7.3,7.4}, massive MIMO schemes for increasing the spectral and energy efficiency \cite{1.13,1.36,1.37,1.47,1.48,1.86}, as well ass mmWave \cite{1.13,1.36,1.38,1.43,1.44,1.45,1.46,1.86} and full-duplex based communications \cite{1.73,1.74,1.75,1.76,1.81,1.82}.

\begin{figure}[!t]
	\centering
	\includegraphics[width=3.5in]{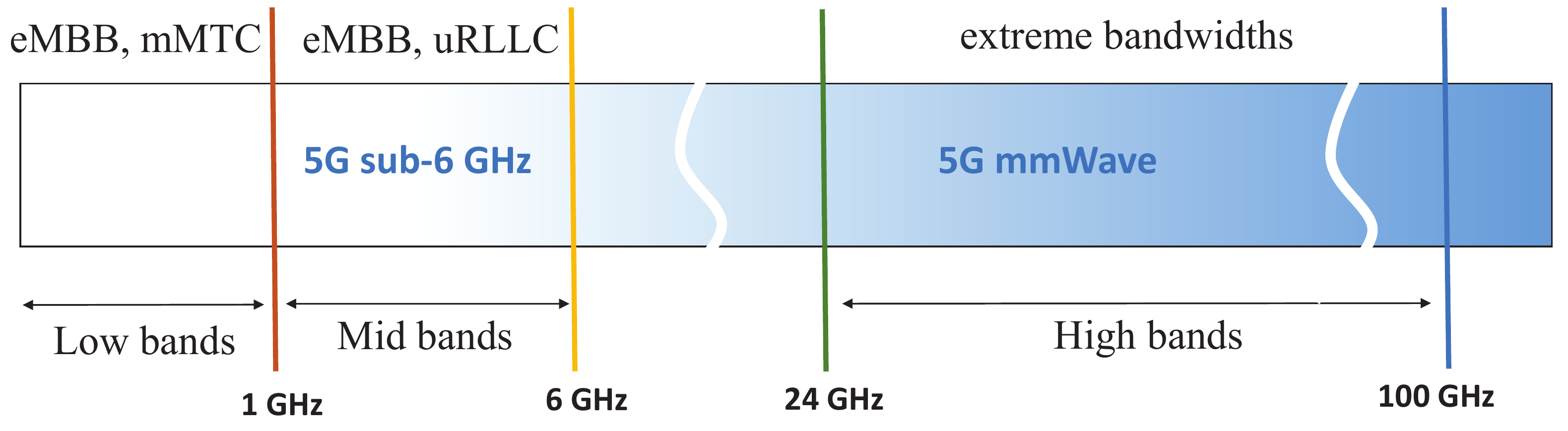}  
	\centering
	\caption{5G NR will natively support all spectrum types and bands \cite{1.101}.}
	\label{fig:5GSpectrum}
\end{figure}

However, one of the most plausible techniques is to exploit more wide spectrum bands that have not been previously  utilized for mobile applications \cite{1.101}. 
In the early 5G deployment phase, MNOs will continue to expand the coverage with the aid of LTE developments in the sub-3-GHz band, as well as involving new 5G NR deployments in selected markets, such as the 600 MHz and 700 MHz bands in the U.S. and Europe, respectively \cite{1.101}. 
By benefiting from the multi-RAT technology, devices can gain access to a wider coverage and to a Gigabit data rates by simultaneously connecting to both LTE lower bands and to higher 5G NR bands.
In order to provide multi-Gigabit connectivity for devices, 5G NR will exploit higher spectral bands in two typical ways as shown in Figs. \ref{fig:5GSpectrum}:

\begin{itemize}
	\item Massive MIMO for mid-band spectrum in the 3 to 5 GHz range. Massive MIMO and 3D MIMO schemes constitute key enabling techniques in 5G NR \cite{1.153} for achieving increased spectral efficiency and network coverage, especially at higher frequency bands (from 32 up to 256 antennas at both 3.5 and 6 GHz frequencies in a 200 MHz bandwidth \cite{1.153}).
	
	This performance gain is predominantly attributable to powerful beamforming techniques in which the transmit power is assigned to narrow beams pointing to each individual downlink user, thus imposing reduced interference and attaining spatial multiplexing for multiple users within a sector. 
	For instance, massive MIMO beamforming relying on 64 transmit antennas can deliver up to 5-fold cell capacity gain in comparison to traditional 2-antenna BSs \cite{1.101}.
	\item mmWave mode relying on advanced antenna design and sophisticated RF processing techniques is expected to deliver extremely high data rates of up to 20 Gbps \cite{1.101} when relying on a contiguous bandwidth of up to 800 MHz. 
	Given these extreme data rates, 5G NR mmWave enhances the user experience for use cases such as HD video (4k and 8k), immersive VR, interactive AR, industrial/enterprise video surveillance, and of other compelling applications.

\end{itemize}

\begin{table*}[!ht]
	\centering
	\caption{Industry standards for mMTC \cite{1.13,1.21,1.36}.}
	\label{table:mMTCStandards}
	{\renewcommand{\arraystretch}{1.4}
		\begin{tabular}{llllllll}
			\hline
			\begin{tabular}[c]{@{}l@{}}\textbf{3-GPP standards}\\ \textbf{for mMTC}\end{tabular} & \textbf{Bandwidth} & \textbf{Throughput} & \textbf{Coverage} & \textbf{Battery life} & \textbf{Mobility} & \textbf{Deployment} & \textbf{Security} \\ \hline
			Cat-M1 & 1.4 MHz & 0.8/1 Mbps Full-duplex & 160 dB (+15 dB) & 10+ years. & Connected \& idle mode & SW & Yes \\ \hline
			NB-IoT & 200 kHz & 227/250 kbps DL & 164 dB (+20 dB) & 10+ years. & Idle mode & SW & Yes \\ \hline
			EC-GSM-IoT & 600 kHz & 473/473 kbps & 164 dB (+20 dB) & 10+ years. & Idle mode & SW & Yes \\ \hline
		\end{tabular}
	}
\end{table*}
\subsection{Enabling Techniques for mMTC-based Application Domains}
\subsubsection{Enabling techniques in the physical, MAC, and network layers for mMTC}
For mMTC-based applications, the most important aspect is to support the substantial number of devices connected in a specific service area with a high connection availability. To this end, several techniques have been studied in the literature.

Sparse Code Multiple Access (SCMA) which is a non-orthogonal multiple access scheme in the code domain that is capable of supporting massive connectivity \cite{1.49,1.50,1.51,1.52,1.53,1.111}. 
Compressive sensing based multi-user detection (CS-MUD) techniques are proposed in \cite{1.54,1.55,1.56,1.57,1.58,1.109} for reducing the control signaling overhead and for reducing the complexity of data processing per device.
The CS-MUD techniques improve user detection reliability, thus reliably handling massive scale access.

Multiple access protocols (MAC protocols) designed for the mMTC service type are investigated in several works\cite{1.23,1.24,1.25,1.26,1.28,1.30,1.105,1.108,1.112,1.116,1.117}.
Due to the massive connectivity supported by the mMTC service type, MAC protocols are devised for improving the  random access capability \cite{1.23} and for authentication-overhead reduction \cite{1.24}.
A MAC protocol was designed for new NOMA waveform and for D2D-based communication in mMTC scenarios in \cite{1.25} and \cite{1.112}, respectively.
To further reduce the communication overhead, hint-based \cite{1.28}, group-paging-based \cite{1.30}, and connectionless-signaling-based communications \cite{1.105,1.108} constitute promising solutions for random access, when supporting the mMTC service type.
Moreover, the authors of \cite{1.116} analyze recent family of MAC protocols conceived for massive random access in LTE/LTE-A networks in terms of the access delay, access success rate, power efficiency, QoS guarantees, and the impact on human-type communication.
The unslotted ALOHA random access protocol considering spreading factor allocation in LoRa (Long Range) systems is analyzed in \cite{1.117}.
Other contributions on physical radio frame design and mmWave based architectures supporting massive IoT 5G networks are proposed in \cite{1.104} and \cite{1.107}, respectively.

Another aspect to be considered in supporting mMTC-based devices is the waveform design \cite{1.106,1.113,1.114}.
A robust waveform synchronization technique \cite{1.106} and the fundamental modulation waveform designed for an OFDMA-based and cognitive code division multiple access (cognitive-CDMA) \cite{1.113} network system are also studied.
The in-phase and quadrature imbalance (IQI) of NB-IoT systems relying on OFDM-IM (index modulation) are mitigated in \cite{1.114}.

Radio resource management is also of crucial importance in mMTC services in the face of a massive number of connected devices, limited radio resources and low transmission power.
Several impressive contributions have improved the resource efficiency of mMTC services \cite{1.110,1.115,1.118}.
An interference-aware radio resource allocation strategy was proposed for NB-IoT systems to maximize the throughput by considering the control channel overhead, time offset and repetition factor \cite{1.110}.
To improve the total throughput of connected devices, a data aggregation scheme was proposed relying on the cooperation of a fixed data aggregator and multiple mobile data aggregators in support of various mMTC devices  stringent QoS requirements.
Energy-efficient resource allocation was conceived in \cite{1.118} for massive MIMO decode-and-forward relay-based IoT networks.

\subsubsection{Industry Standards for mMTC}
Several industry standards have been defined for mMTC services by different international organizations and  which have been also widely deployed.
Specifically, they can be categorized as follows:
\begin{itemize}
	\item Non-3GPP standards: LoRA, Sigfox, Ingenu, Weightless, WI-SUN, 802.11ah, 802.15.4 \cite{1.13,1.21,1.36}.
	\item 3-GPP standards (protocol stacks): Cat-M1, NB-IoT, EC-GSM-IoT \cite{1.13,1.21,1.36}. The specifications of these standards are summarized in Table \ref{table:mMTCStandards}.
	\item IETF (Internet Engineering Task Force): Constrained Application protocol, HTTP/2, QUIC transport protocol, IPv6 over LPWAN working group defining encapsulation and header compression mechanisms \cite{1.31}, \cite{1.63}, \cite{1.64}, \cite{1.65}.
\end{itemize}

\subsection{New Waveforms, Numerologies and Frame Structure for 5G Radio Access}

One of the key goals of the 5G RAT is to support diverse use cases, which may be characterized by utilizing a combination of requirements of the three general services: eMBB, mMTC and uRLLC.
Addition, all the new 5G air interface should be designed in a manner that can accommodate diverse use cases having heterogeneous requirements and the large variety of deployment scenarios, including V2X communication.
Therefore, supporting pervasive heterogeneity requires a high degree of flexibility, which rules out the traditional approach of any one-size-fits-all solutions. 
Key enablers for a flexible 5G air interface are new waveforms, scalable numerologies, and a versatile frame structure \cite{1.11}.

\subsubsection{New Waveforms for 5G NR}
Even though orthogonal-subcarrier based waveforms have been widely used in 4G/TD-LTE/LTE-Advanced systems, they impose high out-of-band emission (OOBE) \cite{1.158,1.160}. 
There have been considerable arguments on whether new waveform candidates can be applied for 5G NR as an alternative to conventional orthogonal frequency-division multiplexing (OFDM) \cite{1.11,1.39,1.40,1.41,1.42,1.77,1.154,1.157,1.158,1.160}.  
Several new 5G waveform candidates such as filtered OFDM (F-OFDM), universal filtered OFDM (UF-OFDM), filter bank multicarrier (FBMC), universal filtered multicarrier (UFMC), and generalized frequency division multiplexing (GFDM) are discussed in \cite{1.158,1.160} and in the references therein.
Among them, filter-based waveforms were strong candidates for 5G networks, since they can support asynchronous transmissions by reducing the OOBE relying on innovative filter designs \cite{1.160}, which leads to increased
bandwidth efficiency, relaxed synchronization requirements, and reduced inter-user interference \cite{1.154}.
The pros and cons of several new waveforms conceived for the 5G NR air interface are discussed and summarized in Section VI of \cite{1.11}.

\begin{figure}[t]
	\centering
	\includegraphics[width=3.5in]{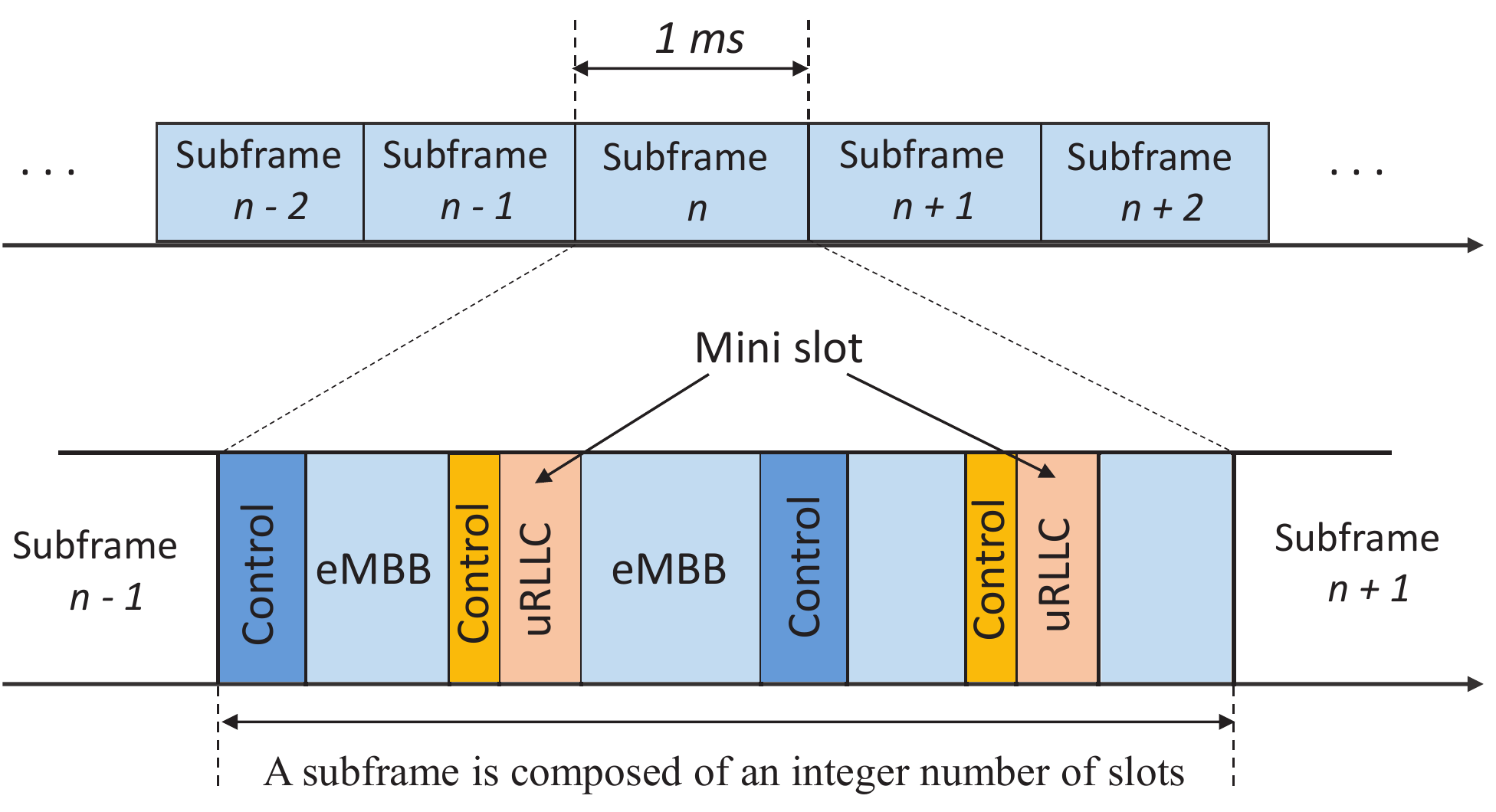}  
	\centering
	\caption{5G NR Frame structure \cite{1.154}.}
	\label{fig:5GSubframe}
\end{figure}

\begin{table}[t]
	\centering
	\caption{Numerologies for 5G NR\cite{1.101}.}
	\label{table:Numerologies}
	{\renewcommand{\arraystretch}{1.4}
		\begin{tabular}{llll}
			\hline
			\textbf{Numerologies} & \begin{tabular}[c]{@{}l@{}} \textbf{Low bands}\\ \textless 1 GHz\end{tabular} & \begin{tabular}[c]{@{}l@{}}\textbf{Mid bands}\\ 1-6 GHz\end{tabular} & \begin{tabular}[c]{@{}l@{}}\textbf{High bands}\\ \textgreater 24 GHz\end{tabular} \\ \hline
			\begin{tabular}[c]{@{}l@{}}Maximum bandwidth\\ (MHz)\end{tabular} & {50} & {100} & {400} \\ \hline
			\begin{tabular}[c]{@{}l@{}}Subcarrier spacing (kHz)\end{tabular} & 15/30 & 15/30/60 & 60/120 \\ \hline
			\begin{tabular}[c]{@{}l@{}}Scheduling interval (ms)\end{tabular} & 0.5/0.25 & 0.5/0.25/0.125 & {0.125} \\ \hline
		\end{tabular}
	}
\end{table}

\begin{table*}[t]
	\centering
	\caption{The summary of the enabling technologies for URLLC, eMBB, and mMTC.}
	\label{fig:Fig21}
	{\renewcommand{\arraystretch}{1.4}
		\begin{tabular}{l |P{9cm}l}
			\hline
			\begin{tabular}[c]{@{}l@{}}\textbf{5G service}\\ \textbf{types}\end{tabular}  & \textbf{Enabling techniques} & \textbf{References} \\ \hline
			& Ultra-dense small-cell (UDN) & \cite{1.22,1.98,1.99,1.100,7.5} \\ \cline{2-3} 
			& Full-dimension 3D and massive MIMO & \cite{1.13,1.36,1.37,1.47,1.48,1.86,1.153} \\ \cline{2-3} 
			& Millimeter-wave communication & \cite{1.13,1.36,1.38,1.43,1.44,1.45,1.46,1.86}  \\ \cline{2-3} 
			& New waveforms & \cite{1.11,1.13,1.36,1.39,1.40,1.41,1.42,1.77,1.154,1.157,1.158}  \\ \cline{2-3} 
			& Full-duplex based communications & \cite{1.73,1.74,1.75,1.76,1.81,1.82}  \\ \cline{2-3} 
			& NOMA waveform (5G NR) & \cite{1.76,1.31,1.80,1.83,1.84,1.85,1.86,1.87}  \\ \cline{2-3} 
			\multirow{-7}{*}{\textbf{eMBB}} & Un-licensed specrtum & \cite{1.101,1.103}  \\ \hline
			& Numerologies, frame structure, and packet structure & \cite{1.13,1.36,1.61,1.62,1.104,1.154,1.155,1.156,1.159,8.35} \\ \cline{2-3} 
			& Grant-Free transmission & \cite{1.66,1.67,1.68,1.69,1.70}  \\ \cline{2-3} 
			& Latency-sensitive scheduling schemes: instant scheduling, reservation-based scheduling & \cite{1.8,1.71} \\ \cline{2-3} 
			& Multi-connectivity (Multi-RAT)- Stemmed from the CoMP transmission technology, among different RAT: 4G LTE, 5G NR, WiFi, cell-free can be considered as a type of multi-connectivity & \cite{1.16,1.36,1.89,1.92,1.93,1.94,1.95,1.96,1.97} \\ \cline{2-3} 
			\multirow{-7}{*}{\textbf{uRLLC}} & V2X communications & \cite{3.1,3.2,3.3,3.4,3.5,3.6,3.7,3.8,3.9,3.10,3.11,3.12,3.13,3.14,3.15,3.16,3.17,3.18,3.19,3.20,3.21,3.22,3.23,3.24,3.25,3.26,3.27,3.28,3.29,3.30,3.31,3.32} \\ \hline
			& Sparse code multiple access (SCMA) & \cite{1.49,1.50,1.51,1.52,1.53} \\ \cline{2-3} 
			& Compressive sensing-based multi-user detection (CS-MUD) & \cite{1.49,1.54,1.55,1.56,1.57,1.58} \\ \cline{2-3} 
			& Non-3GPP standards: LoRA, Sigfox, Ingenu, Weightless, WI-SUN, 802.11ah, 802.15.4 & \cite{1.21,1.13,1.36} \\ \cline{2-3} 
			& 3-GPP standards (protocol stacks): Cat-M1, NB-IoT, EC-GSM & \cite{1.21,1.13,1.36} \\ \cline{2-3} 
			& Scheduling algorithms for 5G NR & \cite{1.22,1.28}  \\ \cline{2-3} 
			& IETF (Internet Engineering Task Force): Constrained Application protocol, QUIC transport protocol, IPv6 over LPWAN working group defining encapsulation and header compression mechanisms & \cite{1.31,1.63,1.64,1.65} \\ \cline{2-3} 
			& Multiple access protocols (MACs): message-efficient oriented protocol, integrated authentication protocol, clustering \& routing, massive V2X regime, Grouping-page (GP) based techniques & \cite{1.23,1.24,1.25,1.26,1.27,1.30,3.14,1.116} \\ \cline{2-3} 
			\multirow{-12}{*}{\textbf{mMTC}} & MAC for NOMA waveform & \cite{1.106,1.113,1.114} \\ \hline
		\end{tabular}
	}
\end{table*}

\subsubsection{Numerologies, and Frame Structure for 5G NR}
Adoption of different numerologies in the frame structure is considered as a key requirement for the new 5G waveform \cite{1.11,1.77,1.154,1.155,1.156,1.159}. 
Similar to LTE, 5G NR is based on OFDM transmission with the possibility of discrete Fourier transform (DFT) precoding for higher power amplifier efficiency in the uplink direction \cite{1.154}. 
A fixed OFDM numerology cannot fulfill the stringent requirements of a large variety of deployment scenarios, ranging from large cells operating in the sub-1 GHz frequency band to mmWave transmission having very wide spectrum allocations.
Thus the physical layer numerology should be delicately designed by carefully considering the hardware impairments, high propagation losses, and multipath propagation in the high frequency band (i.e., mmWave) \cite{1.159}.
Therefore, 5G NR requires flexible and scalable numerology with subcarrier spacings ranging from 15 kHz up to 240 kHz with a proportional change in cyclic prefix (CP) duration \cite{1.154}.
A summary of the numerologies for 5G can be found in \cite{1.11} and \cite{1.159}.

The frame structure of 5G NR is discussed in detail in \cite{1.154,1.155,1.156,1.159} in which the subframe length of 1 ms is composed of 14 OFDM symbols using 15 kHz subcarrier spacing and normal CP.
The design of the 5G NR new frame structure allows uRLLC transmission over a fraction of a slot, which is referred to as “mini-slot” transmission. 
Such transmissions can also preempt an ongoing slot-based transmission to another device, allowing for prompt transmission of data requiring very low latency \cite{1.155,1.156}. 
Each mini-slot is also able to carry control signals/channels at the beginning and/or end OFDM symbol(s). A mini-slot is the shortest unit of resource allocation/scheduling \cite{1.154}.
The new 5G NR frame structure is shown in Fig. \ref{fig:5GSubframe}.

The summary of the uRLLC, eMBB, and mMTC enabling techniques is given in Table \ref{fig:Fig21}.

\begin{table}[!t]
	\centering
	\caption{A taxonomy of the applications of deep reinforcement learning for communication and networking \cite{1.46}.}
	\label{table:DRLApplications}
	{\renewcommand{\arraystretch}{1.4}
		\begin{tabular}{llll}
			\hline
			\textbf{\begin{tabular}[c]{@{}l@{}}Network access \\ \& Rate control\end{tabular}} & \textbf{\begin{tabular}[c]{@{}l@{}}Caching \&\\ Offloading\end{tabular}} & \textbf{\begin{tabular}[c]{@{}l@{}}Security \& \\ Connectivity \\ preservation\end{tabular}} & \textbf{\begin{tabular}[c]{@{}l@{}}Miscellaneous\\ issues\end{tabular}} \\ \hline
			\begin{tabular}[c]{@{}l@{}}- Network \\ access\\ - Adaptive \\ rate control\end{tabular} & \begin{tabular}[c]{@{}l@{}}- Proactive \\ caching\\ - Data \\ offloading\end{tabular} & \begin{tabular}[c]{@{}l@{}}- Network \\ security\\ - Connectivity \\ preservation\end{tabular} & \begin{tabular}[c]{@{}l@{}}- Traffic\\ routing\\ - Resource \\ scheduling\\ - Data \\ collection\end{tabular} \\ \hline
		\end{tabular}
	}
\end{table}

\subsection{AI for 5G} \label{sec:AI}
\subsubsection{Machine Learning Categories for 5G Communications}
A pair of popular machine learning (ML) techniques may be involved for enhancing the performance of a 5G network: deep neural networks (DNN)\cite{1.145} and deep reinforcement learning (DRL)\cite{1.146}. 
Specifically, some particular types of DNN that can be employed to design and optimize various funtionalities and algorithms of the 5G wireless system are feed-forward, convolutional and recurrent neural networks.
In \cite{1.145}, the authors provide a tutorial on DNNs and their application in the 5G network.
Moreover, the DRL technique involved for 5G networking is reviewed in \cite{1.146}. Table \ref{table:DRLApplications} gives an overview of the compelling applications of DRL techniques in communications.

AI techniques can be integrated into 5G networks for enhancing their adaptability to variations of the surrounding environment by learning the network parameters.
For instance, abnormalities in network traffic, user demand, resource utilization, and other possible threats can be sensed in real time by using AI techniques \cite{1.122}, thus making 5G similar to a backbone communication network which my be involved for diverse automation-based vertical domains such as smart manufacturing, autonomous vehicles, intelligent transport systems, and the smart city concept \cite{1.120}.

\begin{figure}[!t]
	\centering
	\includegraphics[width=3.5in]{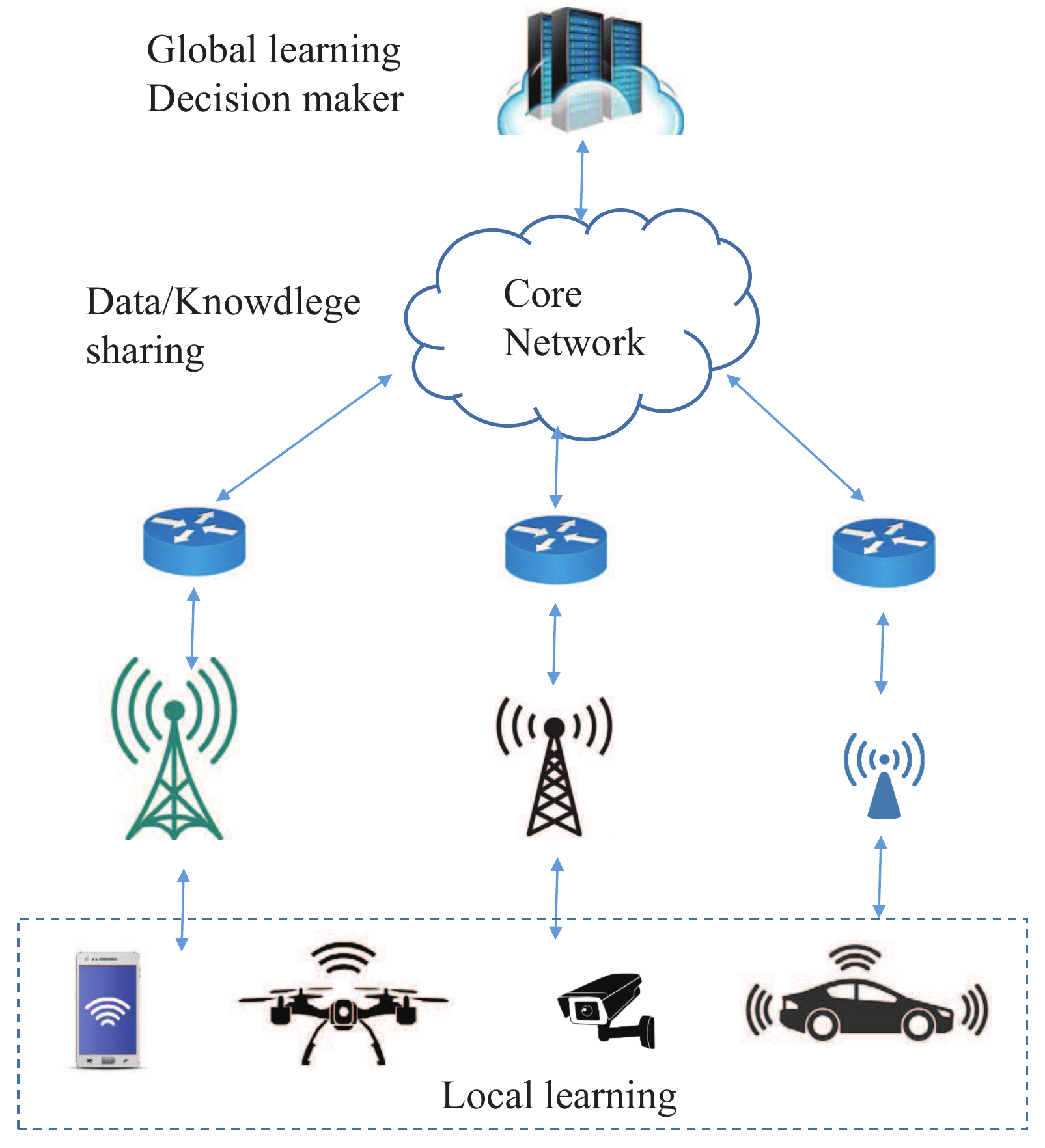}  
	\centering
	\caption{Local and global learning and decision making in large distributed networks \cite{1.120}}
	\label{fig:AI5G}
\end{figure}

Fig. \ref{fig:AI5G} depicts a distributed 5G architecture in which AI is embedded into multiple network layers. 
In such an intelligent architecture, data is distributively processed across multiple-layers for different purposes.
For example, time-series and contextual data are first locally processed by learning algorithms/mechanisms at each network node where it was originally created to construct models of the local behavior.
Then the processed data will be centrally consolidated by a decision maker at a higher network layer for learning beneficial global knowledge \cite{1.120}.

Various applications of machine learning techniques are introduced in \cite{1.147} and involved for cross-layer network design such as modulation and coding schemes for the physical layer, access control and resource allocation for the MAC and network layers.
Moreover, machine learning can also be utilized for enhancing other network aspects, such as security, privacy, and resource as well ass network management.
In the following subsections, we briefly review the state-of-the-art in AI and machine learning techniques suitable for 5G architectural design, mainly focusing on the physical layer and resource allocation.

\subsubsection{ML for the 5G Physical Layer}
The PHY layer of a wireless system has to deal with physical stochastic factors such as white noise (i.e., AWGN channel) and interference. These uncertainties in a wireless channel substantially affect the KPIs of a wireless communications system such as spectrum efficiency and transmission reliability. 
Traditionally, deterministic techniques involved for designing different PHY layer components, such as channel coding and signal detection, result in low adaptability to network dynamics.

The aforementioned challenging problem can be overcome by applying ML techniques. 
In \cite{1.124} and \cite{1.125}, O'Shea \textit{et al.} propose to interpret a communication system as an autoencoder carrying out a high-accuracy end-to-end signal reconstruction task with the help of a convolutional neural network (CNN) and unsupervised machine learning involved for modulation classification based on raw IQ samples.
ML-based signal compression \cite{1.150} and channel encoding \cite{1.151,1.152} problems were also investigated by this research group.
The generative adversarial network (GAN) concept \cite{1.132} has also been applied for synthesizing new modulation as well as coding schemes and for accurately learning the channel statistics in \cite{1.130} and \cite{1.129}, respectively.
Meanwhile, the CNN tool has also been adopted for modulation recognition by improving synchronization and equalization\cite{1.126,1.134}.
Moreover, both signal identification and classification can be performed by semi-supervised and deep learning techniques \cite{1.127,1.131}.
DRL has also been used for wireless radio control and signal detection \cite{1.135}.
A communications system composed of software-defined radio and deep learning based frame synchronization was implemented and described in \cite{1.133}, which achieve a reduction of the block error rate in comparison to a conventional differential quadrature phase-shift keying (DQPSK) transceiver.

ML-based techniques have also been involved for MIMO wireless systems \cite{1.128,1.136,1.13,8.37}.
Specifically, an unsupervised deep learning based autoencoder was used for modeling the impairments inflicted by a MIMO Rayleigh fading channel \cite{1.128,1.136} and by interference \cite{1.137}.

Signal detection and CSI estimation may also be carried out by ML techniques in physical layer design \cite{1.138,1.139,1.140}.
Neural network based regression approximation can be readily relied upon for accurately estimating the CSI \cite{1.138}.
Meanwhile, CNNs were also used for radio signal detection, localization, and identification in dispersive wide-band scenarios \cite{1.139,1.140}.

Finally, applying ML techniques to tackle resource allocation problems in wireless networks has also attracted substantial attention \cite{1.143}. Problems such as capacity maximization in different channel conditions and dynamic spectrum access can be efficiently solved by appropriate ML techniques embedded into a cognitive radio engine. Based on the knowledge gained from the learning process, the cognitive radio engine can efficiently address these problems. The authors of \cite{1.144} offered a comprehensive review for various learning problems in the context of cognitive radio networks, and emphasized the potentials of ML techniques in terms of improving the level of cognition in wireless systems.


\begin{figure*}[!t]
	\centering
	\includegraphics[width=6in]{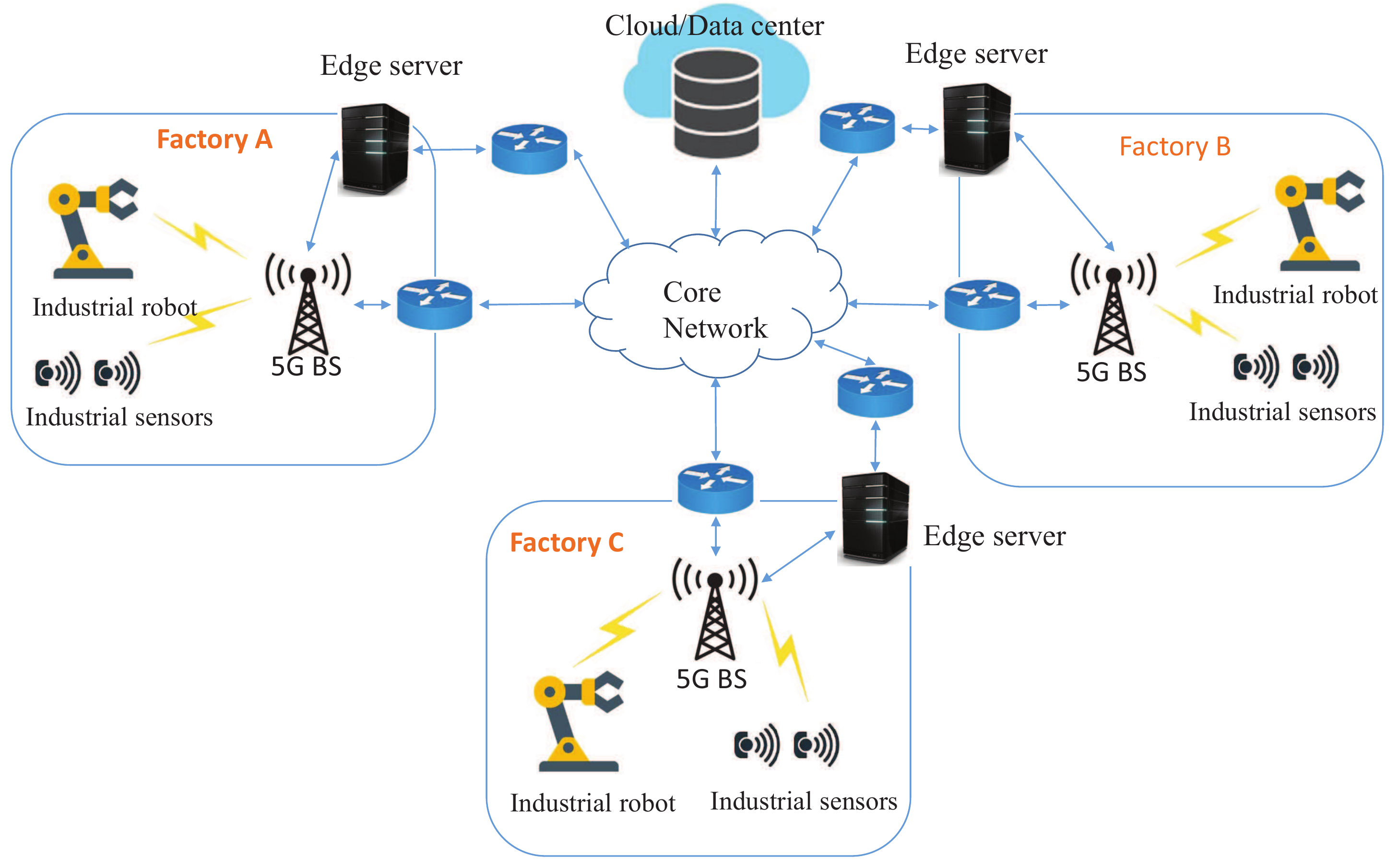}  
	\centering
	\caption{5G-enabled smart factory scenario \cite{1.1,2.8}.}
	\label{fig:Fig27}
\end{figure*}

\section{Smart Factories}
\subsection{Overview}

The "Fourth Industrial Revolution" or simply "Industry 4.0" is the inevitable trend of the manufacturing industry thanks to the innovations of 5G wireless communications, automation technologies, and artificial intelligence.
This revolution aims for achieving higher flexibility, versatility, resource efficiency, cost efficiency, worker supports, and improved quality of industrial production and logistics \cite{2.8}. 
Technically, the key advantage of 5G wireless networks is that of providing improved reliability, lower latency, seamless and ubiquitous connectivity at an extremely high throughput both for human and machine-type connections.
Having efficient wireless communication and localization services are crucial for shifting from traditional static sequential production systems to flexible and modular ones.
There are several promising application areas ranging from supply and inventory management, sensory process monitoring, to robotic control, where the 5G network can bring about benefits for a smart factory \cite{2.8}. 
%

3GPP has already defined and analyzed several concrete use cases for the ``Factory of the Future'' in the technical report TR 22.804 \cite{1.1} and in the 5G-ACIA white paper \cite{2.8}.
Specifically, smart factory based use cases in conjunction with with their specific performance requirements are classified according to the basic 5G service types eMBB, mMTC, and uRLLC.
For instance, motion control of mobile robots requiring an extremely high reliability and ultra-low latency connection is categorized as a uRLLC-based application.
Meanwhile, AR and VR applications demand high data rates, thus being considered as eMBB-based applications \cite{2.8}.

The 5G system architecture of a smart factory scenario is exemplified in Fig. \ref{fig:Fig27}. 
It illustrates that 5G is capable of supporting communications both within the factory and with other factories.
In a typical smart factory, a 5G network provides wireless connectivity between sensors, robots and humans via 5G access nodes, while inter-factory connections rely on the backbone core network.


The smart factory domain is diverse and heterogeneous in terms of use cases, applications and requirements.
Several typical smart factory use cases' requirements are stated in Table  \ref{table:Table2} (see also 3GPP TR 22.804 \cite{1.1} for further information). 
It can be observed that smart factory use cases require the most stringent connection availability, latency and reliability.
Note that in practice, these values may vary widely for different types of devices (i.e., sensors, actuators, and robots) in a smart factory.
The 5G system is envisioned as being capable of meeting or even exceeding the industrial availability/ reliability requirements of today’s production lines.

\subsection{5G Support for Smart Factories}
\subsubsection{Reach Peak Productivity with the aid of 5G}

To investigate how 5G can support multiple vertical domains, Ericsson has carried out a comprehensive survey of more than 650 decision-makers from eight key industries \cite{2.13}.
Interestingly, 5G technologies will bring huge benefits to the vast majority of smart factories.
In Table \ref{table:5G_use_cases_priority}, manufacturing executives expect 5G to expand video surveillance/streaming of manufacturing assets ($78\%$), develop better machine-to-machine sensors ($72\%$), and improve remote site safety and security ($68\%$) \cite{2.13}.


\begin{table}[tb]
	\centering
	\caption{Key 5G use cases for the high-tech manufacturing industry. Source: Ericsson survey \cite{2.13}.}
	\label{table:5G_use_cases_priority}
	{\renewcommand{\arraystretch}{1.2}
	\begin{tabular}{p{6.5cm}p{1.5cm}}
		\hline
		\textbf{Priority of 5G-based industrial applications} & \textbf{Percentage} (\%) \\ \hline
		Video surveillance/streaming of manufacturing assets and processes & 78 \\ \hline
		Machine-to-machine sensors & 72 \\ \hline
		Remote site safety/security & 68 \\ \hline
		Tracking and remote monitoring of manufacturing assets & 67 \\ \hline
		Real-time remote control of robotics & 56 \\ \hline
		Leverage remote experts via video for complex repairs in the field & 50 \\ \hline
	\end{tabular}
	}
\end{table}

\subsubsection{Massive Industrial IoT}

Several mega-factories are equipped with a massive number of devices.
As a benefit of the seamless ubiquitous connectivity provided by 5G technology of the mMTC service type, all these devices can be connected to a single network.
This gives them improved insight into the supply chain and forms a massive industrial IoT network satisfying stringent management requirements.
For example, these connected devices may help improve predictive maintenance and operational efficiency on the factory floor as well as prevent theft and  monitor quality within logistic channels.

\subsubsection{Robotics}

5G provides high-rate, ultra-low latency and high-reliability connections in support of wireless robots with increased agility for a smart factory.
This facilitates a huge productivity boost by allowing these robots to perform complex and/or dangerous tasks.
Moreover, the 5G network is capable of supporting the synchronization and management task of robots by offloading their computational workload to the edge server.
This allows a huge improvement in the number of robots in a smart factory, thus improving productivity, without increasing the processing cost.

\subsubsection{Augmented Reality and Virtual Reality}

AR and VR are popularly employed in manufacturing processes for training, machine design and maintenance as well as data visualization.
These technologies, however, require enormous data rate and low latency.
5G networks not only fulfill these requirements but also free employees from a tethered network connection, thus allowing them to work anywhere on the factory floor.

\subsubsection{AI and Machine Learning}

AI supports the 5G-based massive industrial IoT network in collecting data in real time from sensors, robots and connected devices.
With the advent of machine learning, AI can help the factory executives by providing warnings and predictions over the abnormalities in the manufacturing process.
Furthermore, AI automatically makes decisions in abnormal situations in a tactile manner with the support of high-reliability and ultra-low latency connections.

\subsection{Recent Research in the Smart Factory Domain}

\begin{table}[t]
	\centering
	\caption{Summary of recent works on LTE and 5G NR for factory automation applications (FAAs).}
	\label{table:Factory_topic2}
	{\renewcommand{\arraystretch}{1.5}
		\begin{tabular}{ll}
			\hline
			\textbf{Research topics} & \textbf{References} \\ \hline
			\begin{tabular}[c]{@{}l@{}}Use cases, architecture, \\ requirements and challenges\end{tabular} & \cite{2.1,2.2,2.15,2.22,2.56,2.3,2.14,2.25,2.32,2.35,2.54} \\ \hline
			System evaluation and testbeds & \cite{2.16,2.19,2.17,2.23,2.42,2.49,2.50} \\ \hline
			\begin{tabular}[c]{@{}l@{}}System design for uRLLC for \end{tabular} & \cite{2.4,2.23,2.25,2.26,2.21,2.39} \\ \hline
			\begin{tabular}[c]{@{}l@{}}PHY and MAC design \end{tabular} & \cite{2.5,2.7,2.44,2.43} \\ \hline
			Resource allocation & \cite{2.20,2.28,2.30,2.31,2.33,2.36,2.41,2.52,2.55} \\ \hline
			uRLLC service type for robotics & \cite{2.7,2.24,2.29,2.53} \\ \hline
			Industrial wireless networks & \cite{2.18,2.40,2.44,2.45} \\ \hline
			Industrial IoT wireless networks & \begin{tabular}[c]{@{}l@{}}\cite{2.48,2.47,2.31,2.36,2.50},\\ \cite{2.41,2.52,2.55}\end{tabular} \\ \hline
			\begin{tabular}[c]{@{}l@{}}SDN, NFV, and network  \\ virtualization for FAAs\end{tabular} & \begin{tabular}[c]{@{}l@{}}\cite{2.51,2.47,2.42,2.49,2.52}, \\ \cite{2.35}\end{tabular} \\ \hline
		\end{tabular}
	}
\end{table}

\begin{table*}[htbp]
	\centering
	\caption{Summary of benefits for representative factory automation use cases provided by 5G technologies.}
	\label{table:Factory_topic}
	{\renewcommand{\arraystretch}{1.5}
	\begin{tabularx}{\textwidth}{P{1.7cm}|P{1cm} P{2.5cm} P{2.5cm} P{3cm} X }
		\hline
		\textbf{Topic} & \textbf{Ref} & \textbf{Use case} & \textbf{Problem} & \textbf{5G Technologies, Standards, Protocols} & \textbf{Contribution}  \\ \hline
		\multirow{5}{\hsize}{Architecture, requirements, and challenges} & \noindent\cite{2.56} & Candy packaging line & System architecture design & IWSN, software defined industrial network, D2D,  edge computing & Proposing a hierarchical architecture of the smart factory, issues and challenges analytics.  \\ \cline{2-6} 
		& \cite{2.22} & Mobile robots, mobile control panels, process monitoring, & Integration of industrial Ethernet TSN with 5G mobile networks & IEEE 802.1, Link Layer Discovery Protocol (LLDP), and Precision Time Protocol (PTP) & Proposing different scenarios to combine 5G and Industrial Ethernet to a hybrid topology \\ \hline
		System evaluation and testbeds & \cite{2.16} & Printing machines, packaging machines, production line & System simulation and evaluation & Dynamic scheduling, numerology & Evaluating the performance of LTE and new 5G radio-interface design in a realistic factory. \\ \hline
		\multirow{5}{\hsize}{System design for uRLLC } & \cite{2.4} & Closed loop industrial control & D2D protocol design for uRLLC for FAAs & D2D, Multi-antenna beamforming & Proposes a novel two-phase transmission protocol for URLLC D2D in FAAs \\ \cline{2-6} 
		& \cite{2.21} & Remote control, closed loop industrial control & URLLC scheme design & Channel estimation, SISO, MISO,multi-BS variable-rate & Proposing pilot-assisted, variable-rate URLLC scheme for DL in FA network  \\ \hline
		\multirow{6}{\hsize}{PHY and MAC design and analysis} & \cite{2.5} & Closed loop industrial control, printing machine & PHY/MAC design for NR/LTE URLLC in 5G & Shortened TTI, HARQ, RTT, wideband allocation, MIMO, scalable numerology & Providing the fundamental PHY/MAC systems design for 5G URLLC with particular applications to the tactile internet.  \\ \cline{2-6} 
		& \cite{2.7} & Industrial robots & Design PHY and MAC schemes uRLLC transmission & M2M, radio channel characterization & Wideband channel measurement  to characterize and parameterize the radio environment. \\ \hline				
		\multirow{10}{\hsize}{Resource allocation} & \cite{2.28} & Steel rolling production process monitoring & RB allocation for delay-sensitive applications & Sequential Learning, SC-FDMA, predictive pre-allocation & A predictive pre-allocation framework for low-latency uplink access scheduling in industrial process automation   \\ \cline{2-6} 
		& \cite{2.30} & Sensors and actuators control & Energy-efficient resource allocation for industrial IoT & Full-duplex communications, industrial IoT, SDN & Proposing a 5G communication framework  for industrial cyber-physical IoT systems with full-duplex communication   \\ \cline{2-6} 
		& \cite{2.33} & Hot rolling process in metallurgical industry & Transmission-estimation codesign framework & Edge computing, multi-antenna beamforming. & Proposing a hierarchical transmission estimation codesign for high-accurate state estimation of I-IoT systems   \\ \cline{2-6} 
		& \cite{2.55} & Remote control, laminar cooling stage of hot rolling process & Resource allocation for wireless control system (WCS) & Control driven hybrid cooperative transmission (CHCT), hybrid decode amplify forward (HDAF) & A framework of cooperative transmission and control for multiloop WCS   \\ \hline
		Industrial wireless networks & \cite{2.18} & Light industry and heavy industry & Assessment of channel propagation at 28 and 60 GHz frequencies & 28 and 60 GHz mmWave, licensed- and unlicensed-band & Analysis of mmWave channel propagation in FA environments in both licensed (28 GHz) and unlicensed (60 GHz) bands   \\ \hline
		SDN, NFV and network slicing & \cite{2.42} & Sensor measurements & Network slicing  trial at Hamburg Seaport & Network slicing, NFV, multi-RAT & Proving that network slicing  in a real, large-scale industrial environment is technically feasible  \\ \hline
	\end{tabularx}
}
\end{table*}

\subsubsection{Use Cases, Architectures, Challenges and Requirements for Smart Factory Network}

\begin{figure*}[!t]
	\centering
	\includegraphics[width=4.5in]{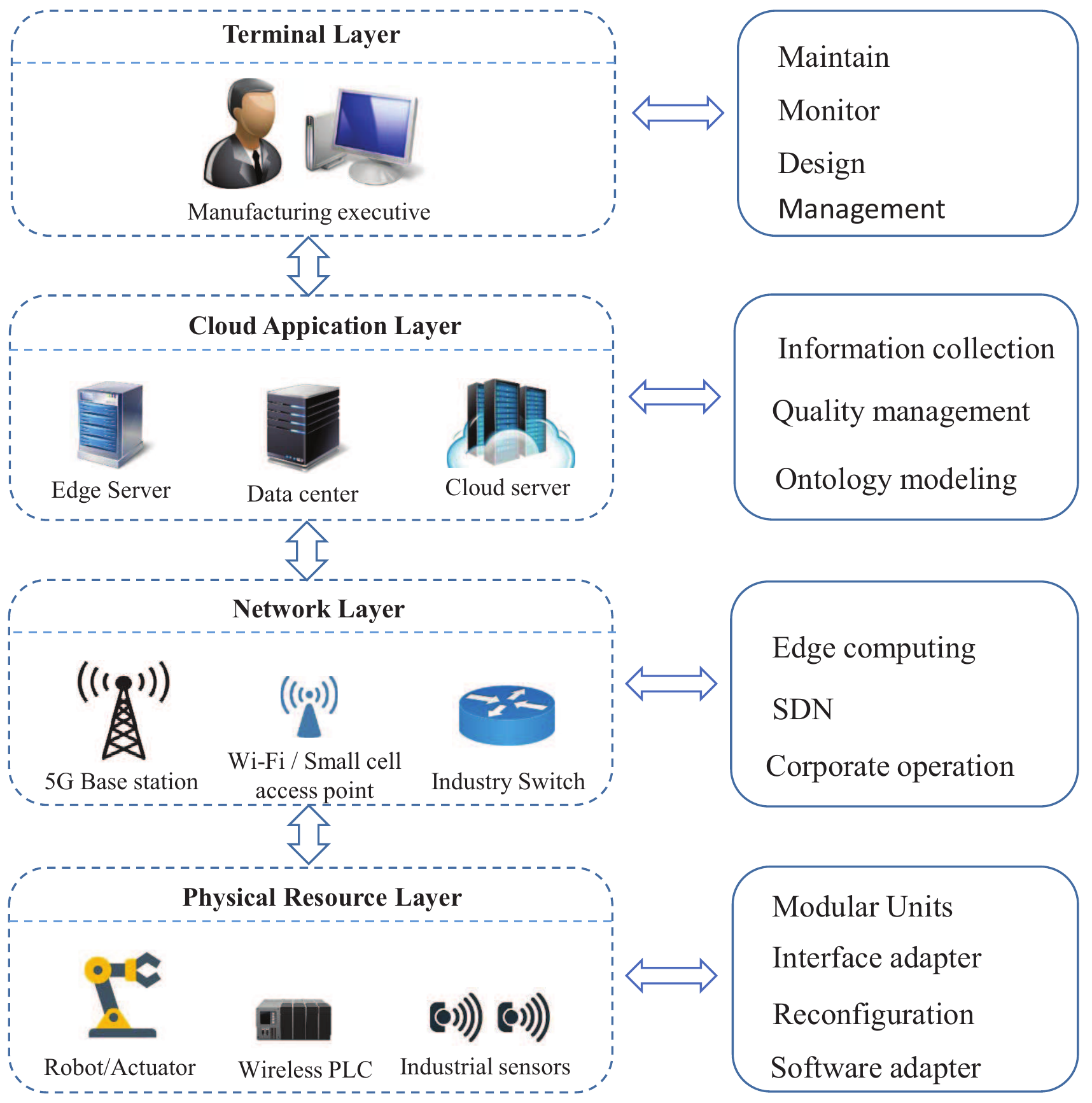}  
	\centering
	\caption{Hierarchical architecture for smart factory  \cite{2.56,2.58}.}
	\label{fig:SmartFactoryArchitecture}
\end{figure*}

Several contributions have studied 5G network deployments designed for the smart factory, including the system architecture \cite{2.1,2.56}, the requirements and challenges \cite{2.2,2.3,2.14,2.15,2.35,2.54,2.56}. The
“5Gang” consortium \cite{2.1}, which is a collaboration between industrial and academic partners, has defined a 5G system architecture suitable for the smart factory, while a hierarchical architecture is proposed in \cite{2.56}. As shown in Fig. \ref{fig:SmartFactoryArchitecture}, the architecture designed for a smart factory in \cite{2.56} includes four layers, namely the physical layer, network layer, data application layer and terminal layer.

The authors of \cite{2.3}, \cite{2.14,2.54} investigate the requirements and potential of both the operational LTE and 5G air interfaces designed for the smart factory, respectively.
The key design challenges of 5G deployments conceived for a smart factory are analyzed in \cite{2.2} and \cite{2.25}, where $99.9999\%$ service availability and less than 1 ms latency are mandatory for closed-loop motion control and for other mission-critical industrial applications.

Further use cases of 5G in the smart factory are discussed by the authors of \cite{2.1,2.2,2.14,2.15,2.22}.
The TACNET4.0 project \cite{2.15} examines representative use cases of 5G in a smart factory.
A use case considering the coexistence of the 5G network and of the Industrial Ethernet is proposed in \cite{2.22}.
For a comprehensive overview of the beneficial impact of 5G on the smart factory as well as the requirements and challenges of this technology deployment, the motivated readers might like to read \cite{2.32,2.35,2.54}.

\subsubsection{System Evaluation and Testbeds}

To investigate the benefits of 5G in the smart factory, several system level evaluations have been disseminated in \cite{2.14,2.16,2.19,2.50,2.23,2.42,2.49}.
The authors of \cite{2.14} and \cite{2.16} evaluate the system-level performance of both the LTE and of the OFDM-based 5G radio interface design concepts in realistic factory scenarios.
Moreover, the trade-off between the coverage and the attainable capacity of 5G mission-critical MTC solutions encountered in realistic factory automation applications is analyzed in \cite{2.19}.
As a further development relying on a simulator-based evaluation, the authors of \cite{2.50} conceived diverse efficiency improvements for homogeneous Industrial Wireless Sensor and Actuator Network (IWSAN).

In \cite{2.23} a testbed was constructed for evaluating the performance of D2D communications both in LTE and 5G NR in FAAs, 
while in \cite{2.42} a network-slicing testbed was conceived for large-scale 5G industrial networks supporting multi-connectivity.
Furthermore, both network-slicing and a SDN based testbed are investigated in the context of a Wind Power plant in Brande, Denmark \cite{2.49}.
These contributions demonstrate that network-slicing and SDN techniques constitute key 5G components in FAAs.
Finally, the techno-economic aspects are analyzed in an industrial M2M communication scenario in order to identify the key factors in configuring 5G for different smart factory use cases \cite{2.17}.

\subsubsection{uRLLC-based System Design for FAAs}

For FAAs, the most important KPIs are related to the 5G uRLLC service type, thus designing a wireless system for the FAAs that support the uRLLC service type is crucial.
Recently, several contributions have studied uRLLC-based system designs conceived for FAAs \cite{2.4,2.21,2.23,2.26,2.39}.
The authors of \cite{2.39} prove the feasibility of designing an OFDM-based 5G air interface capable of meeting all uRLLC service requirements.
A two-phase transmission protocol satisfying the uRLLC requirements of a multi-antenna BS aided D2D network is designed in \cite{2.4} for factory automation scenarios.
Cooperative D2D communications operating in a coexisting of LTE and 5G NR environment is evaluated in terms of both reliability and timeliness constraints in a representative factory hall based on a Software Defined Radio (SDR) testbed \cite{2.23}.
To meet the 1-ms transmission latency specification, a cooperative Automatic Repeat Request (ARQ)-based data transmission scheme exploiting the spatial diversity and a pilot-assisted rate control scheme is proposed in \cite{2.26} and \cite{2.21}, respectively.

\subsubsection{PHY and MAC Design for Smart Factory}

In order to construct an efficient wireless system meeting the uRLLC requirements for employment in FAAs, the physical and MAC layer design constitute a particularly critical step.
In this context, the authors of \cite{2.7} and \cite{2.43} study the radio propagation characteristics of several representative factory scenarios ranging from a factory automation cell provided for robots to both an open production space and to a dense factory cluster.
The authors of \cite{2.7} investigate a joint PHY and MAC design operating at a 5.85 GHz carrier frequency, while the authors of \cite{2.43} consider the 2.3 and 5.7 GHz bands.
Furthermore, the joint PHY and MAC designs for both LTE and 5G NR considering uRLLC, conceived for safety- and privacy-critical FAAs are studied in \cite{2.5}. Cross-layer optimization problems solved under mission-critical constraints are tackled in \cite{2.44}.

\subsubsection{Resource Allocation}

Resource allocation is considered as one of the pivotal techniques of satisfying the uRLLC requirements in the context of FAAs \cite{2.20,2.28,2.30,2.31,2.33,2.36,2.41,2.52,2.55}.
The resource allocation problems of wireless communications have been studied for a long time; however, resource allocation for such high-specification mission-critical machine-type communication (C-MTC) requires further research attention, especially for FAAs. Most recent studies focus on supporting uRLLC based FAAs by proposing  efficient radio resource allocation schemes \cite{2.20}, as exemplified by a steel-rolling production process in \cite{2.28}.
Resource allocation is also studied in the context of the  Industrial IoT (I-IoT) \cite{2.33} and Cyber-physical Internet of things systems (CPIoTS) \cite{2.30} whilst considering the limited availability of communication resources and in the context of an energy-efficiency-aware full-duplex mode, respectively. 
Moreover, caching aided resource allocation conceived for I-IoT is considered under specific QoS constraints in \cite{2.31}. 
The NOMA scheme termed as interleave-grid multiple access (IGMA) is proposed in \cite{2.36} which mitigates the inter-cell interference and improves the spectrum efficiency in dense I-IoT scenarios.
A resource allocation scheme is designed for an MEC based I-IoT network in \cite{2.52}.
Resource allocation is designed for interference avoidance in Cognitive I-IoTs \cite{2.41}, while in \cite{2.55} wireless control systems are investigated. 
The authors of \cite{2.41} also consider the interference between the primary and secondary users, while in \cite{2.55} the mutual interference among multiple control-loops is investigated.

\subsubsection{uRLLC Services for Robotics}

Meeting stringent uRLLC requirements is mandatory for successfully supporting automated robots in a smart factory.
For example, robots working in an automation process require $10^{-9}$ reliability and $99.9999\%$ availability with less than $10$ ms latency \cite{1.1,2.8}. Therefore, combining 5G and robotic applications becomes a promising research area in the context of Industry 4.0 \cite{2.7,2.24,2.29,2.53}.
Leveraging uRLLC capabilities of the 5G network, the authors of \cite{2.24} propose an offloading scheme from a robot to the cloud for reducing the load imposed on the robot by time-critical and computationally exhaustive operations. 
Several robot use cases supported in the automated factory ranging from picking and placing robots \cite{2.29} to UAVs and AGVs \cite{2.53} are proposed considering their specific requirements in uRLLC.

\subsubsection{Industrial Wireless Networks}

An industrial wireless network (IWN) connects numerous industrial components \cite{2.18,2.40,2.45}, where several challenges have to be overcome for successfully exploiting 5G.
The authors of \cite{2.18} study the mmWave channel properties in the 28 and 60 GHz licensed and unlicensed bands in light and heavy industrial environments classified in term of the density and the size of the industrial components involved. 
The authors of \cite{2.40} analyze several modulation candidates advocated for IWN communicating over highly dispersive multipath channels. 
Finally, the re-transmission based diversity gain is investigated in the context of IWNs as a means of improving the reliability and reducing latency \cite{2.45}.

\subsubsection{Industrial IoT Wireless Networks}

In the I-IoT wireless networks diverse industrial components with heterogeneous requirements communicate with each other via a 5G platform. 
Integrating heterogeneous devices into an IWN is a challenge, since all network components should meet their diverse requirements.
Interference aware resource allocation \cite{2.41,2.55}, caching aided resource allocation \cite{2.31} and deadlock avoidance assisted resource provision \cite{2.52} have been proposed to deal with the problem of having limited resources in the I-IoT wireless network.
The SDN and NFV concepts are considered as promising techniques for I-IoT \cite{2.47,2.52}, since these techniques are capable for reducing the network overhead, thus reducing latency and improving reliability. A comprehensive survey of the state-of-the-art I-IoT networks is provided in \cite{2.48}.

\subsubsection{SDN, NFV, Network Slicing and Network Virtualization for FAAs}

The sophisticated concepts of SDN, NFV, network slicing \cite{3.76}, and network virtualization are expected to support FAAs in industrial environments facilitating an overall latency reduction \cite{2.51,2.52}. 
The applicability of SDN and NFV in I-IoT networks is studied in \cite{2.47}, while network slicing was investigated using a testbed in \cite{2.42,2.49}.

The aforementioned advances in developing 5G for the smart factory are summarized in Table \ref{table:Factory_topic2} and Table \ref{table:Factory_topic}.

\subsection{Data-Driven Smart Factory}
The operation of a smart factory critically hinges on data. Since 5G technologies will allow capturing a huge amount of data in real-time, it is necessary to have an efficient data model suitable for emulating a smart factory to boost production.

\subsubsection{Data Processing Architecture}
Typically smart factories, machines, smart sensors, and robotic platforms generate a tremendous amount of data in support of the monitoring, maintenance and management of the production line. In \cite{2.56}, an open maintenance architecture based on manufacturing-related big data is proposed, which relies on machine learning and statistical analysis. Similarly, in \cite{2.58}, a cloud-assisted system architecture is designed for the collection of manufacturing-related big data for industrial environments, which contributes to preventive maintenance. Data processing including real-time active maintenance and offline analysis as well as prediction in the cloud is also discussed in \cite{2.58}. A hybrid cloud and edge architecture are considered as a suitable model for big data processing in manufacturing-related automation, where the cloud provides big data storage, and powerful yet cost-efficient computing for offline big data analysis by relying on shared cloud resources from multiple providers. On the other hand, edge computing processes the data at the network edge close to manufacturing points or data sources, which provides critical services that meet key requirements of a smart factory for high-reliability low-latency connectivity, real-time processing, and privacy protection \cite{2.59}. The family of hybrid cloud and edge architectures is characterized in Fig. \ref{fig:Fig27} and Fig. \ref{fig:SmartFactoryArchitecture}.

\subsubsection{Data Protection \& Security Challenges in the Smart Factory}
In a smart factory, the manufacturing processes will rely on cloud-based data storage and analytics for facilitating real-time decision making. The challenge is to develop secure and reliable data access and protection mechanisms across the data collection, storage, and processing system. For example, the sensitive data found in the workspaces should be stored on-site the manufacturing organization, rather than in the cloud, which is vulnerable to cyber attacks and can lead to substantial damage to the factory \cite{2.60}. 


\subsection{Smart Factory Challenges}

However, in order to realize the full potential of 5G in the context of the smart factory, we still have to tackle  major challenges. Some of these challenges are outlined in more detail below. 

\subsubsection{Compatibility improvement of 5G for the manufacturing industry} The development of 5G standards and technical solutions has to take into account the specific requirements of the manufacturing industry \cite{1.1,1.3,2.8}, with particular emphasize on the heterogeneous requirements posed by different industries.  
The first step towards considering the use cases and requirements of the industrial domain has already been taken by 3GPP through the study items on “Communication for Automation in Vertical domains” (3GPP TR 22.804) \cite{1.1}, and “Feasibility Study on LAN Support in 5G” (3GPP TR 22.821) \cite{1.161}. Given these specifications, both academia and industry should work together for aiming at carefully tailored industrial 5G standardization. The resultant standards will play a vital role in the  efficient support of the vertical industries by 5G.   

\subsubsection{Spectrum exploitation} 
To fulfill the stringent latency and reliability requirements posed by the massive number of heterogeneous I-IoT devices and IWNs including industrial wireless sensor network (IWSN), new licensed and/or unlicensed spectral band have to be made available \cite{1.101}. In the manufacturing workspace where all industrial devices and components may communicate through 5G, the mmWave is eminently suitable for providing a high-rate, low-latency and high-reliability. However, designing the enabling techniques to exploit mmWave communication in a smart factory is challenging and it is an open direction for future research. 

\subsubsection{Safety and security}
By integrating 5G technologies into the manufacturing process, many components in the smart factory become vulnerable to cyber-attacks. Thus, providing adequate level of cyber-security becomes an issue of vital importance that has to be addressed. Since the smart factory critically depends on connecting industrial components via the 5G platform, important data pertaining to the production in industry requires uncompromised protection. Moreover, a cyber-attack that might compromise the operation of a production line or of a complete factory may inflict enormous damage that cannot be tolerated by the industry thus reliable protection is required. Automatic techniques of detecting intrusions and cyber-attacks against vulnerable factory components connected to the 5G network must be  studied \cite{2.59}.

\section{Smart Vehicles}

In addition to supporting V2X connectivity, the 5G technology also facilitates safety improvements autonomous vehicles.
This technology thus inspires cooperation between the telecom and automotive industries, which has led to the conception of the 5G Automotive Association (5GAA).
Specifically, 5G will provide a low-latency medium for feeding the real-time transportation infrastructure conditions updates to millions of connected vehicles.
Several potential wireless communication technologies have been approved for autonomous vehicles, as shown in Table \ref{table:EnableTech_SmartVehicle}.
In this work, we focus on attention on 5G enabling techniques conceived for smart vehicles, as detailed in the following sections.

\begin{figure}[!t]
	\centering
	\includegraphics[width=3.5in]{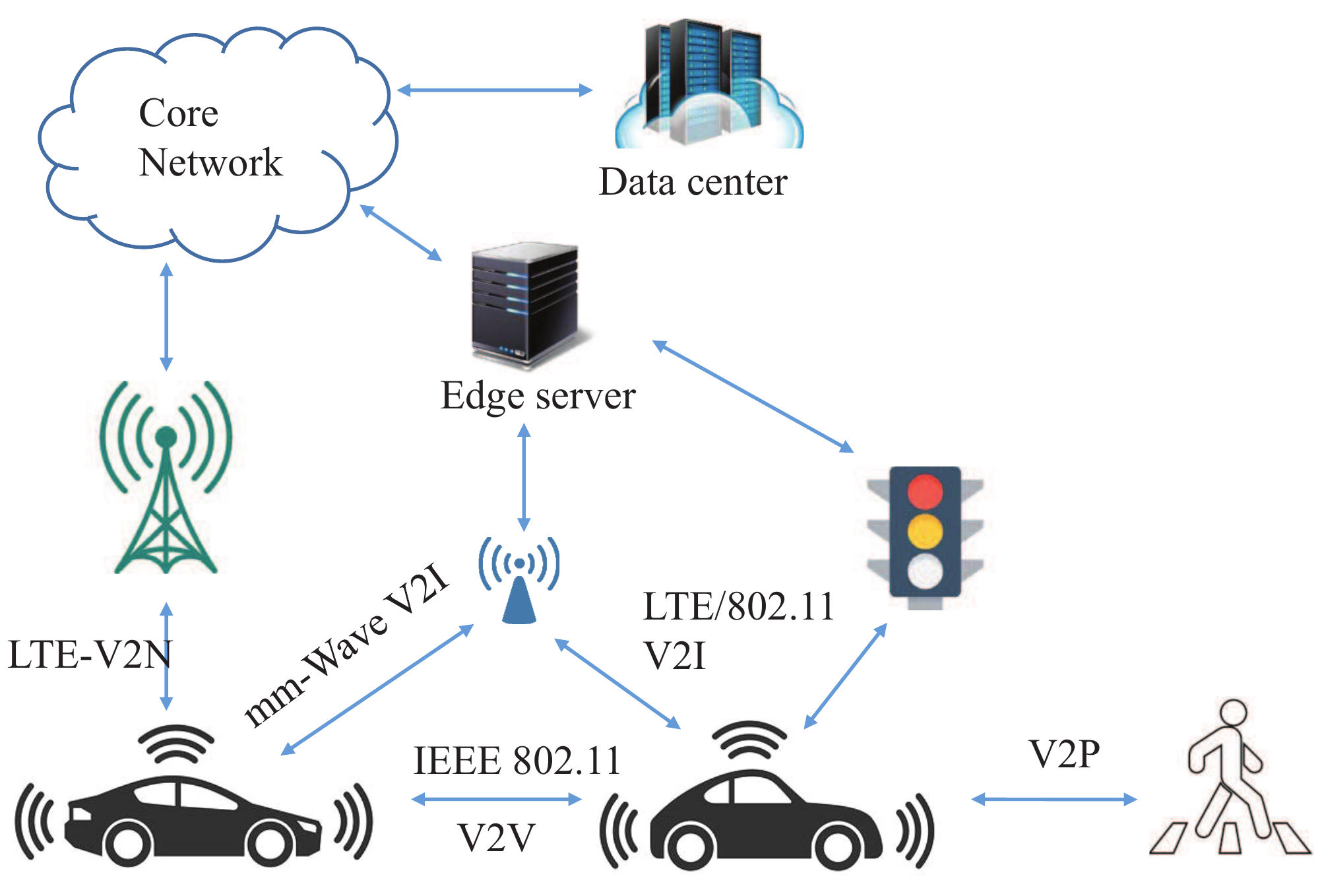}  
	\centering
	\caption{V2X Communications.}
	\label{fig:Fig33}
\end{figure}

\begin{table*}[tb]
	\centering
	\caption{Comparison potential wireless communication technologies for autonomous driving  \cite{3.1}}
	\label{table:EnableTech_SmartVehicle}
	{\renewcommand{\arraystretch}{1.2}
		\begin{tabular}{lllllll}
			\hline
			\textbf{Technology} & \textbf{Spectrum} & \textbf{Standard} & \textbf{Modulation} & \textbf{Data rate} & \textbf{Latency} & \textbf{Range} \\ \hline
			DRSC & 5.850-5.925 GHz & IEEE 802.11p & OFDM & \textless 54 Mbps & 100 ms & \textless 1 km \\ \hline
			LTE-V & N/A & LTE-V & \begin{tabular}[c]{@{}l@{}}MIMO, OFDMA,\\     SC-FDMA\end{tabular} & 1 Gbps & 50 ms & \textless 2 km \\ \hline
			5G & \begin{tabular}[c]{@{}l@{}}600 MHz - 6 GHz,\\     24-86 GHz\end{tabular} & N/A & \begin{tabular}[c]{@{}l@{}}Massive MIMO,\\     NOMA\end{tabular} & 10 Gbps & 1 ms & \textless 2 km \\ \hline
		\end{tabular}
	}
\end{table*}

\begin{figure*}[!t]
	\centering
	\includegraphics[width=5.5in]{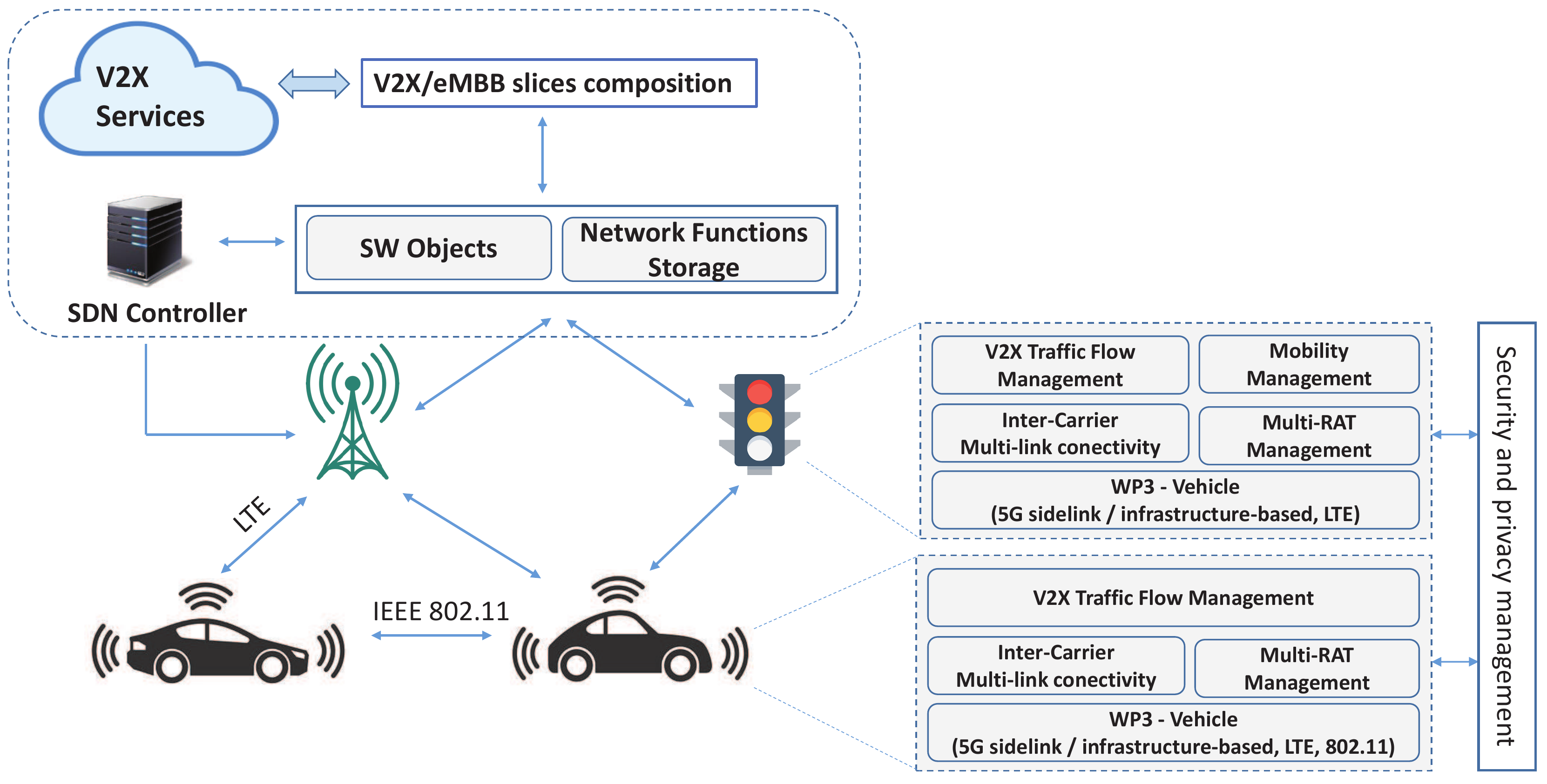}  
	\centering
	\caption{V2X architecture, source: 5GCAR \cite{3.93}.}
	\label{fig:V2XArchitecture}
\end{figure*}

\subsection{5G Enabling Techniques for Smart Vehicles}

There are numerous obstacles and traffic incidents occurring on roads that cannot be captured by on-car cameras and radars.
To this end, a potential 5G technology based on peer-to-peer communication termed as V2X, which connects any vehicle to its surroundings for exchanging road condition-related information brings about substantial benefits.
This eventually improves the autonomous capability of these smart vehicles.
Therefore, 5G V2X may be considered as the backbone communication link of smart vehicles.

\subsubsection{Cellular-V2X Technology}

The development and deployment of C-V2X has been high on the agenda of several technology corporations, such as Ericsson, Huawei, Nokia, and Qualcomm since 2016.
Collision avoidance assisted by C-V2X has been demonstrated to be capable of warning  about oncoming pedestrians, vehicles, and obstacles.
ITS and smart vehicle use cases can be readily supported by the C-V2X technology, which offers extreme adaptability for diverse transportation scenarios.
In particular, 3GPP focuses on standardizing the fundamental functions of C-V2X from its Release 15 onwards \cite{3.68}.
Technical details of V2X communications are discussed in Section \ref{sec:V2X_tech}.

\subsubsection{Edge Cloud Assisted V2X}

An advanced feature making 5G suitable for V2X communications is the Edge Cloud Computing concept \cite{3.77}.
This feature enables connected vehicles to offload their heavy computational tasks to edge cloud servers for reducing power consumption at a guaranteed latency.
Moreover, connected vehicles and other transportation infrastructure components can send their information to the edge cloud for centralized big-data processing, thus providing an up-to-date global view on the traffic conditions, and augmented decisions for ITS.
In Fig. \ref{fig:Fig33}, we illustrate the whole picture of V2X relying on multi-RAT communications assisted by an edge cloud.

\subsubsection{SDN and NFV Architectures for V2X}
The open programmability and logically centralized control features of the software defined networking (SDN) and network function virtualization (NFV) paradigms offer attractive control and management techniques\cite{3.71,3.72}, supporting both resource allocation \cite{3.73} and handover management \cite{3.74} in V2X communication.
An SDN-based multi-RAT system can be designed to facilitate for the vehicles to simultaneously receive data  from more than one network (e.g., from a cellular base station, roadside unit, and other infrastructure elements) \cite{3.73}. Moreover, SDN based packet classification is capable of reducing both the packet delay and frequent packet drop events, which in turn supports mission-critical services in V2X communication \cite{3.75}.

As illustrated in Fig. \ref{fig:V2XArchitecture}, the V2X system architecture proposed by the 5GCAR project \cite{3.93} consists of several components such as the management of traffic flows, multi-RAT connectivity and inter-operator communications. Moreover, this architecture also supports the management of links via SDN as well as the placement of virtualized network functions, and network slicing.

\subsubsection{Network Slicing Techniques for V2X}
Network Slicing is defined by 3GPP \cite{3.76} and it is controlled by SDN. The associated NFV is considered a key feature of 5G in the context of V2X communication \cite{3.91,3.92}. More explicitly, network slicing is defined as the concept of creating multiple logical networks relying on a shared physical infrastructure \cite{3.76}.
An example of network slicing conceived for V2X communication is constituted by a common physical infrastructure that can be sliced into an autonomous driving slice exchanging  safety-related messages between vehicles, an infotainment slice providing, for example, a video streaming service \cite{3.90}. 
A tele-operated driving slice as well as remote vehicle diagnostics and management slice \cite{3.92}.

\subsubsection{Machine Learning for V2X}
AI based V2X communication relying on machine learning is capable of acquiring information from diverse sources and hence predict and avoid potential traffic accidents as well as congestion, thus enhancing the comfort, safety and efficiency of smart vehicles and transportation systems. Machine learning algorithms can be implemented distributively by involving each connected vehicle in learning the local paraphernalia and making a decision for the specific situation of each vehicle. 
Machine learning exploiting situational awareness for enhancing the directionality of mmWave V2X communications \cite{3.84,3.31,3.83}  constitutes a plausible example of learning the particular environment of each connected vehicle.

The V2X systems may also be supported by Edge or Cloud computing combined with a powerful centralized computing server, which involves learning algorithms for global decision making with the aid of collecting data from connected vehicles and devices. Fig. \ref{fig:AI5G} illustrates both the local and global learning as well as decision making in a large distributed system including the V2X network. A reinforcement learning-based traffic signal control regime is proposed by constructing a cooperative mechanism among the distributed infrastructure elements and connected vehicles \cite{3.88}. A survey of AI-aided solutions conceived for V2X was provided by Tong \textit{et al.} in \cite{3.87}. Several AI techniques and applications designed for V2X communications can be found in \cite{3.87,3.89} and the references therein.

\begin{table*}[htbp]
	\centering
	\caption{Table: 5GCAR use case requirements and KPIs. Source 5G-PPP \cite{3.55}}
	\label{table:Fig34}
	{\renewcommand{\arraystretch}{1.5}
		\begin{tabular}{l|p{2.0cm}p{2.5cm}p{2.25cm}p{2.65cm}p{2.5cm}p{2.5cm}}
			\hline
			\multicolumn{2}{l}{\multirow{2}{*}{\textbf{Requirements}}} & \multicolumn{5}{|c}{\textbf{5GCAR Use Case} \cite{3.55}} \\ \cline{3-7} 
			\multicolumn{2}{p{2.0cm}}{} & \multicolumn{1}{|p{2.5cm}}{\textbf{Lane merge}} & \multicolumn{1}{p{2.25cm}}{\textbf{See-through}} & \multicolumn{1}{p{2.65cm}}{\textbf{Network assisted \newline vulnerable pedestrian \newline protection}} & \multicolumn{1}{p{2.5cm}}{\textbf{High definition \newline local map \newline acquisition}} & \multicolumn{1}{p{2.5cm}}{\textbf{Remote driving \newline for automated \newline parking}} \\ \hline
			\multirow{7}{*}{\textbf{\rotatebox[origin=c]{90}{Automotive}}} & Intersection \newline crossing time & Not applicable & Not applicable & 7 seconds & Not applicable & 1 to 6 seconds \\ \cline{2-7} 
			& Localization & 1 to 4 meters & 10 meters & 10 to 50 cm & 5 to 50 cm & 5 to 50 cm \\ \cline{2-7} 
			& Maneuver \newline completion time & 4 seconds & 4 seconds & Not applicable & Not applicable & Not applicable \\ \cline{2-7} 
			& Minimum \newline car distance & 0.9 to 2 seconds & 0.9 seconds & Not applicable & 0.9 to 2 seconds & 2 seconds \\ \cline{2-7} 
			& Mobility & 0 to 150 km/h & 0 to 30 km/h & 0 to 100 km/h & 0 to 250 km/h & 30 to 50 km/h \\ \cline{2-7} 
			& Relevance area & 250 to 350 meters & 300 to 500 meters & 40 to 70 meters & \textgreater{}250 meters & 1000 meters \\ \cline{2-7} 
			& Take over time & 10 seconds & 4 seconds & 10 seconds & 10 seconds & 10 seconds \\ \hline
			\multirow{6}{*}{\textbf{\rotatebox[origin=c]{90}{Network}}} & Availability & V2I/V2N 99\% and for V2V 99.9\% & 99\% & 99\% to 99.99\% & 99\% to 99.99\% & 100.00\% \\ \cline{2-7} 
			& Communication range & \textgreater 350 meters & 50 to 100 meters & \textgreater{}70 meters & \textgreater{}1 km & Several kms \\ \cline{2-7} 
			& Data rate & 0.350 to 6.4 Mbps & 15 to 29 Mbps & 128 kbps & 960 to 1920 kbps & 6.4 to 29 Mbps \\ \cline{2-7} 
			& Latency & \textless 30 ms & 50 ms & \textless 60 ms & \textless{}30 ms & 5 to 30 ms \\ \cline{2-7} 
			& Reliability & 99.90\% & 99\% & 99\% to 99.99\% & 99\% to 99.99\% & 100.00\% \\ \cline{2-7} 
			& Service data \newline unit size & 1200 to 16000 bytes per frame & 41700 bytes per frame & 1600 bytes per frame & 60 bytes per frame & 16000 up to 41700 bytes per frame \\ \hline
			\multirow{5}{*}{\textbf{\rotatebox[origin=c]{90}{Qualitative}}} & Cost & Medium & Medium & Medium to High & Medium to High & High \\ \cline{2-7} 
			& Power \newline consumption & Low & Low & Low & Medium to High & Low \\ \cline{2-7} 
			& Security & Privacy: High \newline Confidentiality: Low \newline Integrity: High \newline Authentication: High & Privacy: Medium  \newline Confidentiality: Low & Privacy: High  \newline Confidentiality: Low \newline Integrity: High \newline Authentication: High & Privacy: High \newline  Confidentiality: High \newline Integrity: High \newline Authentication: High & Privacy: Medium  \newline Confidentiality: Low \newline Integrity: High \newline Authentication: High \\ \hline
		\end{tabular}
	}
\end{table*}
\subsection{Use Cases and Performance Requirements}

\subsubsection{V2X Use Cases from Organizations}

3GPP defined the first V2X use cases and requirements in Release 14, which is specified in the Technical Specification (TS) 22.185 \cite{3.59} and in the Technical Report (TR) 22.885 \cite{3.58}.
In the next 3GPP Release 15 TS 22.186 \cite{3.61} and TR 22.886 \cite{3.60}, autonomous driving and the associated new requirements were well defined for multiple V2X use cases.
Meanwhile, the International Telecommunications Union Recommendation (ITU-R) M.2083-0 \cite{3.63} and ITU-R M.1890 \cite{3.64} defined several use cases for ITS envisioned for the year 2020 and beyond.
The European Telecommunications Standards Institute (ETSI) released the TR 102 638 BSA \cite{3.62} and TC ITS \cite{3.65} documents defining a set of application-oriented standards for vehicular communications in ITS.
Finally, the Next Generation Mobile Networks (NGMN) organization published a white paper identifying eight classes of use case categorized into a pair of specific domains including “Mobile broadband in vehicles” and “Airplanes connectivity”.

Several use cases and the corresponding requirements specified for smart vehicles and ITS are summarized in Table \ref{table:Table2}.
In what follows, we present a brief overview of specific use cases in the realms of  smart vehicle based on the findings of the 5GCAR project.

\subsubsection{5GCAR Project Use Case and KPIs}

In \cite{3.67}, the 5G PPP organization discussed the applicability of 5G to the tight control of future smart vehicles and to the related critical services.
Several use cases arising from the smart vehicle domain have been identified, which can be used for specifying the 5G requirements.
Specifically, having a high mobility and heterogeneous connection density constitute a pair of representative features of smart vehicles.
In this survey, we summarize the most common V2X use cases in Table \ref{table:Fig34}, which constitute the 5GCAR project's use cases \cite{3.55}.
We refer the readers to Section B of  the Appendix in \cite{3.55} for the derivation of these use case requirements.

\begin{table}[!t]
	\centering
	\caption{Summary of recent contributions on V2X communication for Smart Vehicles.}
	\label{table:V2X_topic}
	{\renewcommand{\arraystretch}{1.5}
		\begin{tabular}{ll}
			\hline
			\textbf{Research topics} & \textbf{References} \\ \hline
			\begin{tabular}[c]{@{}l@{}}Frame structure for V2X\end{tabular} & \cite{3.8,3.9,3.17} \\ \hline
			Virtual cell for V2X & \cite{3.11,3.14,3.26,5.55} \\ \hline
			\begin{tabular}[c]{@{}l@{}}Protocol and architectures \end{tabular} & \begin{tabular}[c]{@{}l@{}}\cite{3.10,3.13,3.17,3.19,3.24},\\ \cite{3.28,3.29} \end{tabular}\\ \hline
			\begin{tabular}[c]{@{}l@{}}Resource Allocation \end{tabular} & \cite{3.16,3.18,3.20,3.23,3.25,3.31} \\ \hline
			Multi-RAT for V2X & \cite{3.15,3.27,3.30,3.73} \\ \hline
			\begin{tabular}[c]{@{}l@{}}SDN, NFV architectures \end{tabular} & \cite{3.71,3.72,3.73,3.74,3.75,3.78} \\ \hline
			\begin{tabular}[c]{@{}l@{}}Network slicing for V2X\end{tabular} & \cite{3.90,3.91,3.92} \\ \hline
			\begin{tabular}[c]{@{}l@{}}Machine learning \end{tabular} & \cite{3.31,3.83,3.84,3.86,3.87,3.88} \\ \hline
			\begin{tabular}[c]{@{}l@{}}mmWave for V2X\end{tabular} & \begin{tabular}[c]{@{}l@{}}\cite{3.13,3.31,3.78,3.79,3.80,3.81,3.82},\\ \cite{3.83,3.84,3.85} \end{tabular}\\ \hline
		\end{tabular}
	}
\end{table}

\begin{table*}[htbp]
	\centering
	\caption{Summary of benefits for representative V2X use cases supported by 5G technologies.}
	\label{table:V2X_topic1}
	{\renewcommand{\arraystretch}{1.5}
		\begin{tabularx}{\textwidth}{P{1.7cm}|P{1cm} P{2.5cm} P{2.5cm} P{3cm} X }
			\hline
			\textbf{Topic} & \textbf{Ref} & \textbf{Use case} & \textbf{Problem} & \textbf{5G Technologies, Standards, Protocols} & \textbf{Contribution} \\ \hline
			Virtual cell for V2X & \cite{3.14} & High mobility V2X & Improving V2X throughput in high mobility use case & Virtual antenna arrays, Shared UE-side Distributed Antenna System, MIMO, unlicensed mm-Wave & Vehicular Shared UE-side Distributed Antenna System (Vehicular SUDAS) which enables high throughput wireless communication in high mobility V2X scenarios. \\ \hline
			Protocol and architectures & \cite{3.13} & Massive V2X use cases & Initial access collisions due to massive connection attempts & mmWave & Improvements to the reliability of the initial access procedure for 5G mmWave cellular in massive V2X communications scenarios \\ \hline
			\multirow{2}{\hsize}{Resource allocation} & \cite{3.16} & Safety information V2X  broadcasting & Scheduling and resource allocation for for V2X broadcasting & NOMA-based mixed centralized/distributed (NOMA-MCD) & Proposing a novel NOMA-MCD scheme for the V2X broadcasting system. The access latency can be reduced and the reliability can be improved by NOMA in a dense network \\ \cline{2-6} 
			& \cite{3.18} & Heavy road traffic & Multichannel conflict-free TDMA link scheduling problem for LTE V2X & LTE V2X D2D, TDMA link scheduling & The proposed heuristic MUCS substantially improves the frequency usage efficiency considering the prime use cases anticipated for LTE V2X \\ \hline
			Multi-RAT for V2X & \cite{3.27} & Dense urban vehicular network & Coexistence problem of cellular V2X users and VANET users over the unlicensed spectrum & Unlicensed spectrum & Proposing an energy sensing based spectrum sharing scheme for cellular V2X users to share the unlicensed spectrum fairly with VANET users. \\ \hline
			\multirow{2}{\hsize}{SDN, NFV for V2X} & \cite{3.71} & Autonomous driving, cooperative driving (e.g., lane-merging assistance, platooning) & Resource management in software-defined vehicular networks & Software-defined vehicular networks & Proof of concept experiment where SDN/OpenFlow programmability allows new degrees of wireless  resource management in dynamic vehicular environments \\ \cline{2-6} 
			& \cite{3.73} & Car streaming video to an operation center for traffic monitoring & Resource allocation for heterogeneous V2X network & SDN, Multi-RAT, LTE, Wi-Fi & A software-defined networking-based application-layer scheme to exploit the available bandwidth from the LTE and Wi-Fi networks in V2I communication \\ \hline
			Network slicing & \cite{3.92} & Forward collision warning,  Vehicular infotainment & Network slices customized for V2X & Network slicing, NFV, SDN & Elaborating on the role of network slicing to enable the isolated treatment and guaranteed performance of V2X \\ \hline
			\multirow{2}{\hsize}{Machine learning} & \cite{3.31} & Blockage between vehicles and BS & Beam selection with environmentawareness in mmWave vehicular systems & mmWave, beam selection, online learning & Providing the first contextual online learning algorithm for beam selection in mmWave base stations. \\ \cline{2-6} 
			& \cite{3.88} & Traffic Signal Control & Network clustering for V2X & Cooperative reinforcement learning & Proposing reinforcement learning based intelligent traffic control \\ \hline
			mmWave for V2X & \cite{3.85} & Blockage transmission & Transmission interruption issue caused by blockage & mmWave, beam tracking, beamforming & An energy-angle domain access and transmission frame structure for mmWave V2X communication scenario \\ \hline
		\end{tabularx}
	}
\end{table*}

\subsection{Recent Research Topics in the Smart Vehicle Domain}
Recently, the V2X paradigm has attracted a tremendous amount of interest from both academia and industry.

\subsubsection{Frame Structure for V2X Communications}
There are works that authors of \cite{3.8,3.9,3.17} propose frame structure improvements for V2X communications.
Specifically, the authors of \cite{3.8} conceived various LTE frame structure enhancements for V2X communications based on LTE D2D communications. 
The so-called LTE-V frame structure was designed for short-range V2X 5G communications in \cite{3.9}. 
As a further development, short-range V2X communication was designed based on the IEEE 802.11p standard in \cite{3.17}.

\subsubsection{Virtual Cell for V2X Communications}
A range of techniques was conceived for enhancing 5G cellular V2X (C-V2X) in \cite{3.11,3.14,3.26}.
Specifically, the authors of \cite{3.11} devised a new cell virtualization.
The proposed virtual cell (VC) concept is user-oriented, where multiple transmission points collaboratively serve each and every user in the network.
Here, each VC is considered as a hotspot that moves along the users' motion trajectory.
To realize a VC, several enabling techniques are relied upon, including the C-RAN concept, distributed antenna systems (DAS) and the coordinated multipoint (CoMP) philosophy.
The challenges in realizing VCs include the mitigation of inter-VC and intra-VC interference as well as admission control.
In \cite{3.14},  a shared UE-side distributed antenna system (SUDAS) is proposed. This system supports a high data rate in a high-mobility V2X communication scenario. 
Both the more conventional sub-6GHz and unlicensed mmWave bands are considered. A vehicular SUDAS channel model taking into account the Doppler spread is considered and a substantial MIMO gain is demonstrated.
Moreover, to reduce the relative frequency of cellular hand-offs, the authors of  \cite{3.26} propose a Multi-hop Moving Zone clustering scheme, which combines the 3GPP 5G cellular technology with IEEE 802.11p, for achieving high packet delivery rate and low latency requirements. In this scheme, vehicles are clustered up-to three hops using V2V communications based on IEEE 802.11p, while the cluster heads are selected by cellular-V2X (C-V2X) on the basis of multi-metrics, i.e. relative speed, distance and link life time. The multi-hop property reduces the number
of clusters, which further decreases the cost of hand-overs significantly with eNodeBs \cite{3.26}.

\subsubsection{Protocols and Architectures for V2X Communications}
Several innovative protocols and architectures are designed for V2X communications in \cite{3.10,3.17,3.28,3.19,3.13,3.24,3.29}.
Pilot optimization and interference-free pilot design are proposed for 5G V2X communications in \cite{3.10,3.28}.
Direct D2D, mmWave, and MIMO-OFDM based communications designed for 5G massive V2X networks are shown to improve the system performance in \cite{3.17,3.13} and \cite{3.24}, respectively.
A fast packet classification protocol is proposed in \cite{3.29} for 5G V2X communications.

\subsubsection{Resource Allocation for V2X Communications}
Resource allocation schemes designed for V2X communication can be found in  \cite{3.16,3.18,3.20,3.23,3.25,3.31,3.73}.
The typical idea of using NOMA waveforms for 5G V2X communication is studied in \cite{3.25}. As a further development, NOMA-based time-frequency resources were jointly optimized with  power allocation for a dense 5G V2X network in \cite{3.16}.
A heuristic vehicle under ad hoc routing algorithm is proposed for seamless link scheduling in \cite{3.18}.
In  \cite{3.31}, the mmWave communication, while in \cite{3.23} the 5G NR was adopted for achieving low latency and high reliability.
In \cite{3.20}, a novel MAC scheduling scheme termed as Segmentation MAC (SMAC) compatible with the IEEE 802.11p standard was proposed.
Additionally, the evaluation of 5G C-V2X testbeds was performed in \cite{3.33,3.12,3.22}.

\subsubsection{Multi-RAT for V2X Communications}
Multi-RAT V2X is studied in several recent contributions \cite{3.15,3.27,3.30,3.73}.
5G V2X communications achieve a high diversity gain by simultaneously transmitting data packets on the LTE-Uu and PC5 interfaces  \cite{3.15}, and meet stringent latency and reliability requirements \cite{3.30}.
The C-V2X users can access to the network via both C-V2X and by using the unlicensed-band based VANET V2X of \cite{3.27}. The aim is to reduce the collisions between the C-V2X users and the VANET V2X users operating in the unlicensed band, while maximizing the total number of active users. In \cite{3.73} a resource allocation scheme was put forward for heterogeneous multi-RAT aided SDN.

\subsubsection{mmWave for V2X Communications}
Millimeter-Wave bands have become popular candidates for 5G V2X communications, since smart vehicular systems require high-rate links in the Gbps range to acquire the necessary sensory information for autonomous driving. Recently, mmWave-style 5G V2X communications have inspired numerous studies \cite{3.13,3.31,3.78,3.79,3.80,3.81,3.82,3.83,3.84,3.85}. However, the transmission range of mmWave carriers remains limited due to severe signal attenuation. 
Consequently, mmWave carriers impose challenges, when aiming for seamless services, and typically rely on high-gain directional transmission/reception using beamforming. Therefore, these studies mostly focus on achieving high-gain directionality to improve the performance of mmWave transmission in high-mobility V2X communication.
In \cite{3.81} Lien \textit{et al.} propose a beam alignment strategy where the transmitter and receiver arrange for aligning their beam directions toward each other at the same time. The effects of mmWave propagation on V2X systems in the light of their directionality and the effect of various beamwidths are studied in \cite{3.79}. Furthermore, machine learning based approaches relying on situational awareness for achieving high-accuracy beam pairing \cite{3.84} and beam selection \cite{3.31,3.83} were also investigated in mmWave V2X communications.

\subsection{Data-Driven Smart Vehicle}
Since smart vehicles are equipped with a large number of sophisticated sensors, actuators and communication equipment, the data exchanged in smart vehicular V2X networks exhibit heterogeneous characteristics. These data could be safety-related messages exchanged between vehicles, road traffic observation data, collected sensor data or entertainment sources downloaded from the core network. These data can be exploited for supporting autonomous driving systems and for improving the driver’s experience. However, smart vehicles require an efficient data processing system in which data can be analyzed in real-time for enhancing the safety functions and autonomous driving features. Fundamentally, the data processing architecture has to be faster and smarter in order to cope with the large volume of data and with the complexity of the smart vehicular system. 

Again, the smart vehicular data processing architecture has to take advantage of cloud and edge computing for reducing the burden of on-board computation by offloading data processing and computing to the mobile edge or cloud. 
By amalgamating cloud computing with SDN and NFV, the resultant V2X system can optimize the cooperation among the smart vehicles and the infrastructure.

\begin{figure} [!t]
	\centering	
	\includegraphics[width=3.5in]{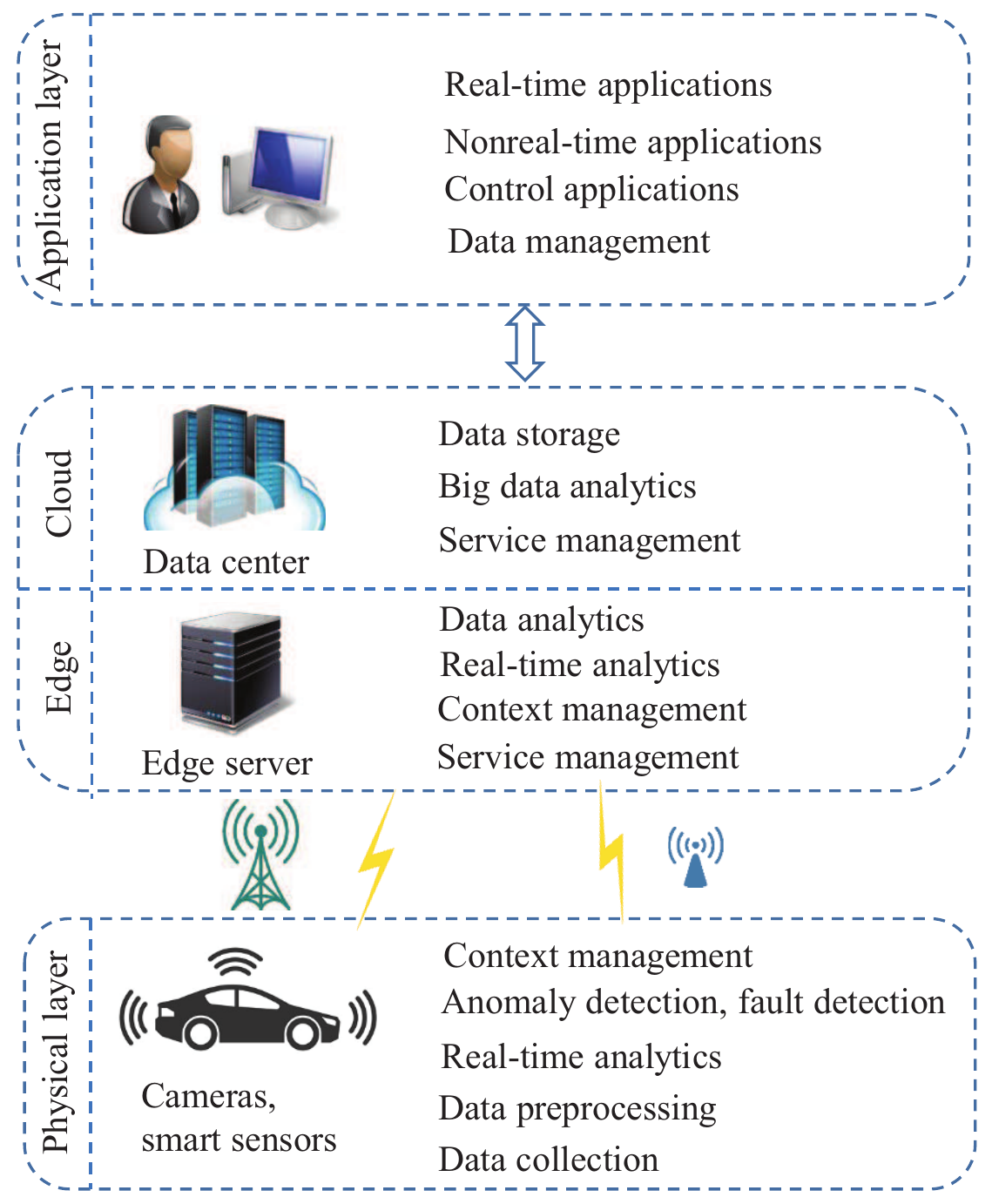}
	\caption{Data processing architecture for smart vehicle system.}
	\label{fig:V2XData}
\end{figure}
\subsubsection{Data Processing Architectures for Smart Vehicles}
Typically, the data processing architecture designed for the automation systems and smart vehicular systems consist of three layers: a physical layer or object layer, edge and cloud layer as well as application layer, as illustrated in Fig. \ref{fig:V2XData}. This three-layer architecture can also be extended to an architecture having four \cite{3.94} or five layers \cite{3.95}. The description of these layers is given as follows:
\begin{itemize}
	\item \textit{Physical layer}: this layer comprises  smart physical devices such as cameras, sensors, and actuators which are responsible for data sensing, data transmission to the upper layers and for executing control tasks. These devices communicate	with other devices using different communication technologies, including Bluetooth, WiFi, Zigbee, LAN and LTE as well as 5G \cite{3.94}.
	\item \textit{Edge cloud computing layer}: this layer is responsible for the storage and computation of data. Different services can harness it to make carefully considered decisions by using for example  machine learning techniques. The data processing can be executed at the edge (i.e., for real-time analytics) rather than in the cloud to reduce latency. 
	For example, an edge server may assist an autonomous car in performing complex video processing of the clips gleaned from the surrounding cameras to avoid collisions between vehicles at obstructed intersections due to obstacles, or to defect if a pedestrian is about to appear from a side-street \cite{3.95}.
	
	Moreover, given the high mobility of connected vehicles, the edge computing implementation faces grave challenges. Smart vehicles may lose their connection with the edge server while the associated computations are being carried out at the edge, when moving out of the service area of the edge, hence potentially losing the data arriving from the edge. A promising solution for edge computing deployment in V2X networks is to utilize multiple levels of edge servers. In the urban deployment scenario where the velocity of a smart vehicle is not too high, the edge server may rely on small cells, on road-side units, or on the transportation infrastructure. By contrast, in high-velocity highway deployment scenarios, the edge server can be implemented with the aid of macrocells to avoid the interruption of computation. 
	\item \textit{Application layer}: This layer provides drivers with the applications, services and functionalities of the edge cloud computing layer for improving the driving experiences. These applications can be categorized as real-time and	non-real-time applications. The autonomous vehicles can adapt to the prevalent traffic situations by receiving control information from applications in this layer.  Moreover, the data collected from different vehicles and infrastructure components can help the administrators to monitor and manage the system by relying on the applications in this layer.
\end{itemize}

\subsubsection{Data Analytics for Smart Vehicles}
The data collected by cameras and other smart sensors can be locally processed by the vehicles to support real-time tasks such as anomaly detection, 
detection of operational faults in smart vehicles,  where the local processing outcomes can be sent to the edge and cloud infrastructure for further analysis. The associated data analytics can be carried out by exploiting the raw power of machine learning techniques (e.g., deep learning) deployed locally on-board the smart vehicles for specific situations (e.g., context-aware data \cite{3.95}) or may be globally deployed at the edge or cloud for the global monitoring and managing of the transportation system (e.g., smart traffic lights, sign boards, etc). 
The applications in the application layer can be harnessed for performing complex processing tasks such as processing video signals received from multiple sources.
Moreover, data collected by each vehicle can be broadcast to nearby vehicles for safety-related cooperation among a group of vehicles.


\subsection{Smart Vehicle Challenges}
For the sake of achieving the seamless integration of numerous technologies in a smart vehicular system, such as sensing, communications and AI technologies, a number of challenges have to be tackled for improving the system performance, in the interest of reliable automation, and improving passenger experience in smart vehicles.

\subsubsection{Safety, Security and Privacy} 
Safety is the most important feature of a smart vehicle. Integration and deployment of technologies in the smart vehicle and V2X network have to take into account all safety-related aspects. 
Specifically, the security and privacy of the smart vehicle connected to the V2X network is of paramount important. The operations of a fully autonomous vehicle are controlled by on-board programs or by the control servers. These programs are vulnerable to malicious attack through the 5G network. Thus, it is challenging to design a sufficiently reliable security mechanism to prevent any potential attempts to take control by hostile agents.
In \cite{3.57}, important objectives are identified for improving the security and privacy aspect of 5G communications in support of smart vehicles. For instance, vehicle authentication, remote service control access, misconduct detection and encryption of the messages are among the most important objectives to be addressed. Thus, further investigations are required for developing schemes that offer security and privacy for smart vehicles.

\subsubsection{Real-time Data Acquisition and Data Processing}
Data captured in the V2X system typically must be processed in real time. Therefore, real-time data acquisition and processing becomes challenging for the smart vehicle and for the V2X network.
Moreover, every time a huge amount of data is collected from the various components and from the vehicles, which are processed to provide accurate decisions in a real-time manner. This requires efficient communications, computation, cooperation and control of the V2X system. Moreover, the rich but still limited resources of the car have to be efficiently scheduled to meet the associated stringent real-time data processing requirements.

\subsubsection{Network Design and Resource Allocation} 
Again, given the high-velocity mobility of the smart vehicle, it is important to design an efficient network architecture for the constantly changing topology of the V2X network.
Moreover, since the smart vehicle is capable of simultaneously relying on multiple communication techniques to communicate with the surrounding vehicles (i.e., IEEE 802.11 DRSC) and V2X infrastructure (LTE-V or 5G), flow optimization for multi-RAT connections and context-aware resource allocation subject to high-rate, low-latency and high-reliability requirements constitute a pair of critical challenges in realizing 5G V2X communications for improving the smart vehicle's performance. Thus, efficient resource allocation techniques have to be conceived for meeting the stringent V2X networking requirements.

\section{Smart Grid}
\subsection{Overview}
The electricity grid has evolved beyond recognitions stimulated by the increasing penetration of renewable energy resources and active interactions between the supply and  demand side. 
In this context, smart grid relies on the sophisticated integration of the distributed energy grid and of communication networks connecting the grid and the consumers \cite{fang2012smart}.
With the emergence of 5G technology, the smart grid is expected to become capable of generating, transmitting and distributing the electricity in a more efficient, sustainable, reliable, stable,  flexible and secure manner. 

\begin{figure} [!t]
	\centering	
	\includegraphics[width=3.5in]{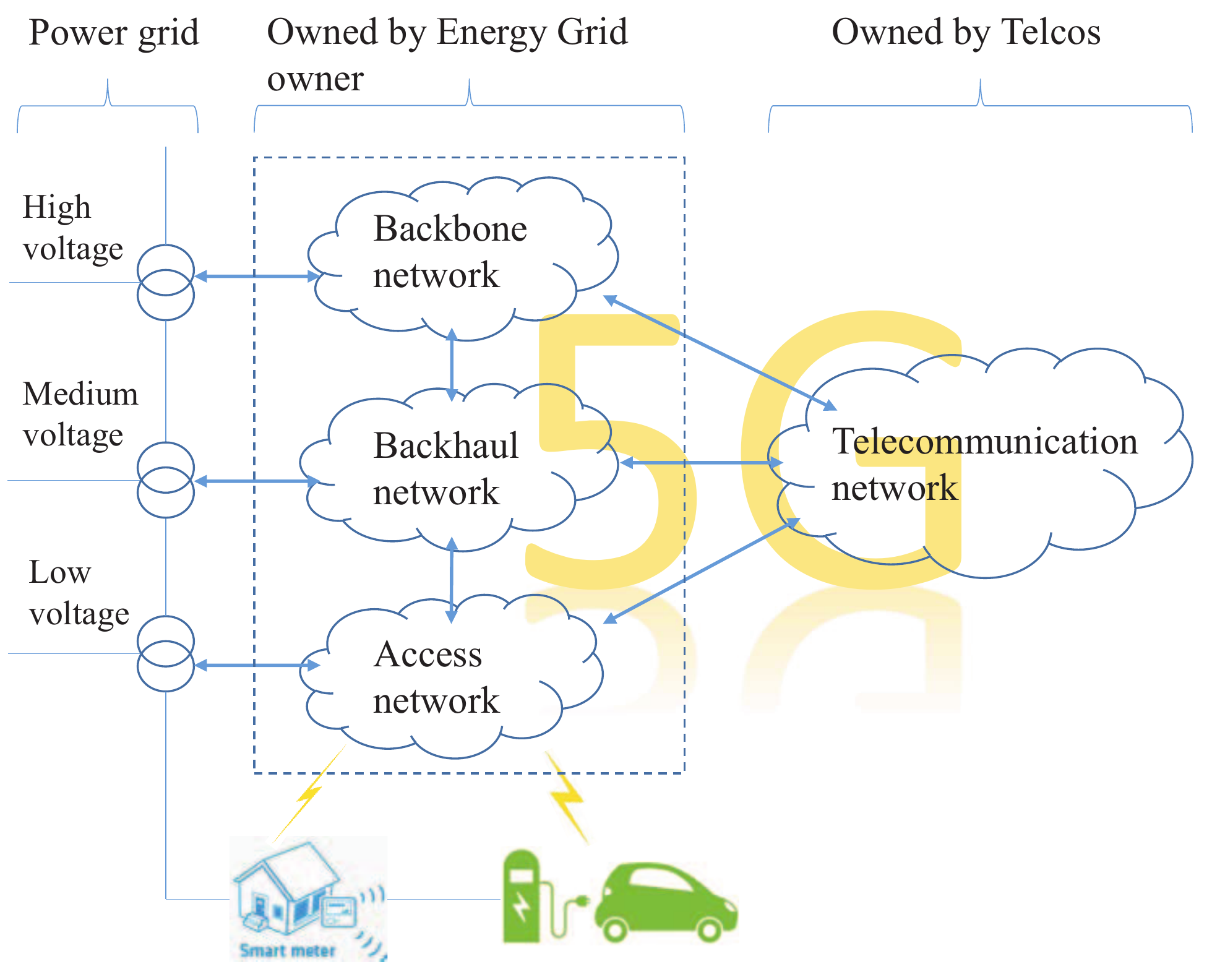}
	\caption{5G smart grid domains \cite{5gpppGrid}.}
	\label{SG_fig2}
\end{figure}

Furthermore, the user side's energy usage patterns have recently changed to adapt to the more sophisticated scenarios such as the appearance of electrified roads\footnote{{An electrified road or eroad is a road which supplies electric power to vehicles travelling on it, either through trolley wires above the road or through conductor rails embedded in its surface. The world's first electric road, which can charge commercial and passenger vehicles while on the move, has opened in Sweden \cite{thelocalse}.}} and ultra-fast charging stations for high-speed electric vehicle charging, large-demand/storage mobile units, and new massive wireless sensors, actuators as well as IoT devices \cite{5gpppGrid}. 
In Table~\ref{t:SG_tech_change}, we summarize key technical changes of the smart grid with reference to recent reports. 

\begin{table*}[htbp]
	\centering
	\caption{Key technical changes of smart grid}
	\label{t:SG_tech_change}
	{\renewcommand{\arraystretch}{1.4}
		\begin{tabularx}{\textwidth}{lX}
			\hline
			\textbf{Aspects} & \textbf{Description} \\ 
			\hline
			Demand response \cite{5gpppGrid, Gridsaxena2017efficient, zhang2018real} 
			& Prosumers, who not only consume electricity but  produce and probably share their production, will increase.  Their consumption can be optimized based on the market information. Thus, bidirectional communications between energy providers and end users becomes more sophisticated in future Smart grid for an intelligent utilization of electricity such as dynamic pricing mechanism, power consumption prediction. \\
			\hline
			Privacy \& security \cite{Gridchin2017energy} 			
			& The concern on privacy \& security will more increase when the degree of automation in Smart grid strongly increases $\implies$ requires highly secure communication system. \\
			\hline
			Distributed energy resources (DERs) \cite{GridSiemenDER1}
			& The growth in exploiting of renewable generators from large electricity production companies to small utilities and individual families require smarter infrastructure system where advanced distributed  information metering and monitoring systems are required to manage the stochastic in electricity production as well as to achieve a fast integration and accommodation of clean energy to power	distribution networks.\\
			\hline
			Nomadic and connected devices \cite{cao2017toward}
			& There will be the upsurge in electric powered nomadic and connected devices from home applications to industry, transportation, etc. Thus, hierarchical and distributed architectures will play important role to reduce the communication delay time and improve the system performance.\\
			\hline
			IoT \& big data \cite{Gridwang2017wireless} & 
			Enormous benefits can be achieved from the comprehensive understanding of smart grid data such as efficient resource allocation, quality of service improvement, monitoring and protection, etc. Big data mining is a strong tool for the analysis and processing of smart grid data. However, most smart grid data will be achieved from wireless sensors, IoT devices. Therefore, wireless communications will be key technology for data collection and transmission.\\
			\hline
	\end{tabularx} }
\end{table*}

\begin{table*}[htbp]
	\centering
	\caption{Smart grid communication network requirements\cite{5gpppGrid}.}
	\label{t:SG_tech_change2}
	{\renewcommand{\arraystretch}{1.4}
			\begin{tabular}{llll}
				\hline
				\textbf{Requirements} & \textbf{Grid access} & \textbf{Grid backhaul} & \textbf{Grid backbone} \\ \hline
				Data rate & 1 kbps & Several Mbps & \begin{tabular}[c]{@{}l@{}}Up to several Gbps\end{tabular} \\ \hline
				E2E Latency & \textless 1s & \textless 50ms & \textless 5ms \\ \hline
				Packet loss & N/S & \textless 10$^{-6}$ & \textless 10$^{-9}$ \\ \hline
				Availability & \begin{tabular}[c]{@{}l@{}}99\% equal to 9h  downtime p.a.\end{tabular} & \begin{tabular}[c]{@{}l@{}}99.99\% equal to 50 min downtime p.a.\end{tabular} & \begin{tabular}[c]{@{}l@{}}99.999\% equal to 5 min downtime p.a.\end{tabular} \\ \hline
				\begin{tabular}[c]{@{}l@{}}Failure convergence  time\end{tabular} & \textless 1s & \textless 1s & in few ms \\ \hline
				\begin{tabular}[c]{@{}l@{}}Handling of  crisis situations\end{tabular} & Not required & Mandatory & Mandatory \\ \hline
				\begin{tabular}[c]{@{}l@{}}Area coverage  radius\end{tabular} & \textless 10km & \textless 100km & \textless 1000km \\ \hline
			\end{tabular}
		}
\end{table*}

The authors of \cite{5gpppGrid, GridReserveD58} show that given its ability to support mMTC and URLLC, 5G will provide a reliable infrastructure for the smooth,  real-time information flows among myriads of devices in the smart grid relying on the recent advances in artificial intelligence (AI), they will be the key tools for driving the smart grid revolution  \cite{HuaweiGrid}. 
In the following, we will discuss in more detail the communication aspects of the smart grid, including its performance requirements, and main challenges.

\subsection{Hierarchical Architecture in Smart Grid Communications and Network Performance Requirements}

The smart grid control requirements can be grouped into three main categories, including primary frequency control, secondary frequency control and distributed voltage control \cite{1.1}. 
The primary frequency control ensures a swift response to excessive frequency variations, while secondary frequency control ensures an accurate and lasting response to frequency variations. 
The distributed voltage control analyses the impedance values for ensuring that the additional energy can be harmoniously injected into the grid, otherwise, electric inverters may throttle the energy to be added by power plants or storage systems. 
The communication network is potentially required to support the tele-traffic of a large group of up to 100,000 users and payload sizes of approximately 100 bytes with voltage measurement intervals in the order of 50 ms for primary frequency control and 200 ms for distributed voltage control \cite{1.1}.

An important issue is to appropriately design the communication network conceived for serving the hierarchical grid voltage and geographic area, as shown in Fig.~\ref{SG_fig2} \cite{5gpppGrid}.
Specifically, this hierarchical network consists of several layers including the backbone, backhaul and access networks. 
For the smart grid's communications, 5G is immediately expected to play a significant role in the access networks, which connect various grid elements in the low-voltage power grid, while the backhaul networks connect different elements operating in the medium-voltage power grid. 
Among them, the backbone network connecting the elements in the high-voltage power grid has the most stringent delay and reliability requirements \cite{5gpppGrid}. 

The communication requirements of these networks are summarized in Table.~\ref{t:SG_tech_change2}.
The authors of \cite{4.4} propose a 5G-based hierarchical architecture combing multiple eMBB, uRLLC and mMTC 5G service types at different layers of an advanced smart grid.

\subsection{5G Support for the Smart Grid}

Some interesting use cases that 5G is capable of supporting for the sophisticated smart grid of the near future can be summarized as follows.
\subsubsection{Fast Self-healing}
One of the key features of the smart grid is the prompt detection of grid disturbances and quick reaction to sudden failures to isolate them, the ability to automatically reconfigure and restore normal operation of the grid \cite{Gridelgenedy2015smart}.	
In addition to switchgear technologies, fast response to faulty in active networks requires ultra-reliable and low latency communications for monitoring and protecting devices \cite{5gpppGrid}. 
Based on 5G communications, the authors of \cite{GridSliceNetD21} study three scenarios: protection coordination, automatic reconfiguration and differential protection, with the goal of isolating faults and reconfiguring the network topology. 
To elaborate a little further on the detection and isolation of faults, the event-driven communications protocols of IEC 61850 GOOSE \cite{GridSliceNetD21} require an E2E latency of less than 5 ms with a data rate of 30/3 Mbps for the downlink and uplink, respectively.
Similarly, the synchrophasor measurements following IEC 61850 SV communications \cite{GridSliceNetD21} require the E2E latency to be less than 3 ms in conjunction with the data rate of 3.2/1 Mbps for the downlink and uplink, respectively. 
In \cite{3GPPTR22804}, the authors assume that the communication network is expected to support synchronicity between a	communication group of up to 100 UEs at a delay, which is on the order of 1 $\mu$s or below with an availability exceeding 99.9999\%. 
At the same time they also aim for supporting the differential protection with an E2E latency of 0.8 ms.
	
\subsubsection{Precise Forecast} Forecasting power generation and consumption can help distribution system operators plan for potential imbalanced situations in advance.
Owing to the employment of wireless sensors, actuators, and smart meters, the data collected in the smart grid, such as short and long-term power generation, statistics weather conditions, equipment health and consumption profiles, has grown exponentially in volume, velocity and variety. 
The obvious advantage of this is that the forecast of the weather, user energy demand and energy production will become a more accurate when more data is available \cite{zhang2018synergy}. 
Therefore, the detrimental impact of uncertainty in the smart grid can be mitigated more efficiently. 
The authors of \cite{rana2016microgrid} explore IoT and wireless sensor network based 5G communications to sense, transmit and estimate the microgrid's states.
	
\subsubsection{Real-time Remote Monitoring and Control}	
Given the recent spreading of smart meters, remote energy monitoring is not a new feature. However, with the introduction of 5G services and the IoTs, real-time remote monitoring and control have become a commercial reality. 
As discussed in \cite{garau20175g}, the 5G network supports the provision of M2M connections and multicast services, which result in a significant improvement in the performance of  network management by reducing the latency of time-critical operation during faults in the power network. 
Meanwhile, the authors of \cite{gheisarnejad2019future} consider the applicability of 5G to the secondary load frequency control problem of maritime micro-grids, where the control errors and frequency deviations are measured by PMUs each 0.01s and transmitted to their respective control center via the shared 5G infrastructure.

\subsubsection{Energy Efficiency}
The integration of renewable energy (e.g., solar and wind energy) generated at the local community level enables customer participation in the electricity market, which forms micro-grids \cite{rana2016microgrid} associated with their own specific management and operation \cite{utkarsh2018distributed}.
The authors of \cite{Gridsaxena2017efficient} explore the emerging evolved Multimedia Broadcast and Multicast Services (eMBMS) mode of the 5G system for designing an efficient wireless communication framework between the aggregator and the energy customers to minimize the energy cost for demand-response users.
Meanwhile, an energy-aware smart metering system based on small-cell networks is proposed in \cite{4.1}.
By exploiting the powerful SDN and NFV techniques, the authors of \cite{chekired2018decentralized} introduce a real-time dynamic pricing model for EV charging and for building energy management.

\subsubsection{Wireless Charging}
Plug-in charging for mobile devices and for electric vehicles typically has a limited quality of experience. 
Hence substantial research efforts have been invested in wireless or inductive charging solutions for the in-motion charging of EVs \cite{manshadi2018wireless}. 
The solution in \cite{ai2018smart} facilitate the shortening of the charging time in 5G wireless rechargeable sensor networks. 
As a further development, the authors of \cite{yin2018autonomous} consider a Stackelberg game to manage the dynamically fluctuating energy demand in multiple-receiver based wireless charging systems.

\begin{table}[!t]
	\centering
	\caption{Summary of recent contributions in the smart grid domains.}
	\label{table:Grid_topic}
	{\renewcommand{\arraystretch}{1.5}
		\begin{tabular}{ll}
			\hline
			\textbf{Research topics} & \textbf{References} \\ \hline
			\begin{tabular}[c]{@{}l@{}}Network slicing for smart grid\end{tabular} & \cite{HuaweiGrid,rehmani2018software,zhang2016sdn, lin2018self, qu2018enabling, chen2017sdn, guo2016achieving, chaudhary2018sdn} \\ \hline
			\begin{tabular}[c]{@{}l@{}}Cloud and fog computing \\for smart grid\end{tabular} & \cite{cosovic20175g, tao2017foud, chekired2017smart,rabiefog, munir2019edge, yaghmaee2018performance} \\ \hline
			\begin{tabular}[c]{@{}l@{}}IoT for smart grid\end{tabular} & \cite{munir2019edge,li2018smart,collier2017emerging,chiu2017optimized,rana2015kalman, 4.1,4.2,4.5} \\ \hline
			\begin{tabular}[c]{@{}l@{}}UAVs for smart grid\end{tabular} & \cite{zhou2018energy, nguyen2019intelligent, lim2018multi} \\ \hline
			\begin{tabular}[c]{@{}l@{}}V2G\end{tabular} & \begin{tabular}[c]{@{}l@{}}\cite{cao2017toward, manshadi2018wireless, tao2017foud, chekired2017smart},\\ \cite{hoang2017charging, wang2017distributed}  \end{tabular},  \\ \hline
			\begin{tabular}[c]{@{}l@{}}Data mining, AI and machine \\learning for smart grid \end{tabular} & \cite{Gridwang2017wireless, zhu2018big, wang2017robust, munir2019edge, wang2018deep, ni2019multistage, nguyen2019intelligent, haddad2018smart, li2017weather} \\ \hline
			\begin{tabular}[c]{@{}l@{}}Communication testbeds\\ for smart grid\end{tabular} & \cite{zhang2016sdn, kurtz2016empirical, you2018cognitive} \\ \hline
		\end{tabular}
	}
\end{table}

\begin{table*}[htbp]
	\centering
	\caption{Summary of benefits for representative smart grid use cases supported by 5G technologies.}
	\label{table:Grid_topic1}
	{\renewcommand{\arraystretch}{1.5}
		\begin{tabularx}{\textwidth}{P{1.5cm}|P{1cm} P{2cm} P{3cm} P{3cm} X }
			\hline
			\textbf{Topic} & \textbf{Ref} & \textbf{Use case} & \textbf{Problem} & \textbf{5G Technologies, Standards, Protocols} & \textbf{Contribution} \\ \hline
			\multirow{4}{\hsize}{Architecture, requirements, and challenges} & \cite{GridSliceNetD21} & Self-healing & Integration of smart grids with 5G mobile networks & URLLC, mMTC, IEC 61850 GOOSE, MEC & Exploiting 5G to provide adequate architectures for power system protection and control device peer-to-peer communications. \\ \cline{2-6} 
			& \cite{HuaweiGrid} & Smart grid's multi-slice architecture & Slice design, deployment, and enabling on the smart grid & Network slicing & Presenting four typical smart grid application scenarios enaled by 5G network slicing \\ \hline
			\multirow{5}{\hsize}{Monitoring and control} & \cite{rana2016microgrid} & Power system monitoring & Microgrid state estimation & IoT & Proposing awireless sensor network based 5G communication network to sense, transmit and estimate the microgrid states \\ \cline{2-6} 
			& \cite{cosovic20175g} & Power system monitoring & Distributed state estimation & URLLC, mMTC, IoT, MEC, SDN & Discussing the latency and reliability requirements that distributed state estimation imposes on 5G communication networks \\ \cline{2-6} 
			& \cite{gheisarnejad2019future} & Secondary load frequency control (LFC) & The application of 5G standards for the & IoT,  type-2 fractional order fuzzy PD/fuzzy PI & Providing  an overall shipboard multi microgrid scheme over 5G network and study the  fuzzy controller under the non-negligible  time delay and packet loss due to communication process. \\ \hline
			\multirow{5}{\hsize}{Resource allocation} & \cite{Gridsaxena2017efficient} & Demand response & Multicast scheduling and radio resource management for minimizing the energy cost for demand response customers & 5G HetNet, IoT, Multimedia Broadcast and Multicast Services & Providing  dynamic programming-based and greedy heuristics-based frameworks for efficient planning of 5G small cells \\ \cline{2-6} 
			& \cite{chekired2018decentralized} & Demand response & EVs charging and discharging scheduling and building energy management in SG & Dynamic pricing, SDN, NFC, cloud computing & Proposing a pricing modelbased on decentralized cloud-SDN architecture to manage the system energy \\ \hline
			V2G & \cite{tao2017foud} & V2G services and applications & Fog-based and cloud-based hybrid computing modeling & MEC & Discussing in potential V2G services and applications enabled by 5G-based hydrid fog-cloud computing \\ \hline
			Data mining, AI, and machine learning & \cite{munir2019edge} & Demand response & Minimize the energy consumption of microgrid-enabled MEC networks & MEC, model-based DRL (MDRL) & Providing a MDRL-based framework for allocating the energy generation through microgrid to MEC servers. \\ \hline
			System evaluation and testbeds for smart grid & \cite{kurtz2016empirical} & Software defined infrastructure & System simulation and evaluation & SCADA edge-cloud & Empirical comparison of fully virtualized and hardware based Bare-Metal switching for SDN-based 5G communication in critical infrastructures \\ \hline
		\end{tabularx}
	}
\end{table*}

\subsection{Recent Research Contributions in the Smart Grids Domain}
\subsubsection{Network Slicing} 
The integration of SDN into the smart grid is discussed in \cite{rehmani2018software}.
Several smart grid application scenarios such as distributed power supplies, intelligent distributed feeder automation, information acquisition in low voltage distribution systems and millisecond-resolution precise load control may be supported by 5G network slicing \cite{HuaweiGrid}, which is expected to improve both the power supply reliability and the users' QoS, while reducing the operational costs and enhancing the security.
Furthermore, the Openflow-based SDN, which allows us to dynamically configure the end-to-end paths for the efficient transmission of electric power/data, is studied in \cite{zhang2016sdn} with the objective of conceiving resilient solutions for the network's expansion and fault avoidance . A number of SDN solutions designed for supporting self-healing in the smart grids can be found in \cite{lin2018self, qu2018enabling}. Furthermore, in \cite{chen2017sdn}  plug-in electric vehicle relying on the integrated smart grid, while in \cite{guo2016achieving} traffic management, and in \cite{chaudhary2018sdn} secure communication were investigated.

\subsubsection{IoT for Smart Grid} 
Smart meters allow the utilities to collect, process and monitor power consumption of grid users in real time \cite{chiu2017optimized, collier2017emerging, 4.1,4.2,4.5}. In \cite{4.5}, the authors propose to apply the emerging 5G mmWave technique for communications between smart meters and the gateways. The authors of \cite{4.2} implement an Arduino-based\footnote{{Arduino is defined as an open source electronic platform, based on flexible and easy-to-use software and hardware. Homepage https://www.arduino.cc/} } smart metering system utilizing the mobile network for communication in order to accurately estimate the average energy usage. Furthermore, 5G-based IoT solutions are employed for effectively sensing, transmitting and estimating the micro-grid states \cite{rana2016microgrid}.  Meanwhile, the authors of \cite{li2018smart} discuss how to apply narrow-band IoT solutions for smart grid communications.
Furthermore, 5G-based IoT-solutions are adopted in \cite{rana2015kalman} for overcoming the high-accuracy voltage regulation challenges in the context of renewable distributed energy resources (DER) by transmitting the information collected by DERs to a control center.

\subsubsection{Mobile Edge Computing for Smart Grid}
In \cite{cosovic20175g}, a 5G system architecture exploiting MEC is involved for distributed state estimation in smart grids. By exploiting the 5G URLLC mode in a wide area monitoring system (WAMS)\footnote{ Wide
area monitoring system (WAMS) aims to detect and counteract power grid disturbances in real time \cite{cosovic20175g}.}, the root mean square error (RMSE) of an estimated state degrades by up to three orders of magnitude. 
In \cite{tao2017foud, chekired2017smart}, cloud computing is combined with 5G V2G networks for supporting balanced power management services while providing flexible on-demand services for EVs. 
A framework exploiting fog computing is conceived for electrical load forecasting in \cite{rabiefog}.
By considering the demand supply problem, the authors of \cite{munir2019edge} investigate a microgrid-based MEC networks’ energy supply plan with the objective of managing the uncertainty associated with renewable energy generation, while the authors of \cite{yaghmaee2018performance} study the effects of communication imperfections on the performance of the cloud-based system.

\subsubsection{Preventive Aerial Maintenance of Critical Infrastructure}
In challenging environments, where the areas to be surveyed is large and the number of checkpoints to is high, swarms of drones/UAVs can be used as the most suitable solution for automatic surveillance systems \cite{GridNRG5D11}. This technical report shows that real-time video streaming from the drones to the central controller can be readily realized by exploiting network virtualization techniques relying on 5G, when the latency of the radio access link is lower than 1 ms and the transmission rate is up to 10 Gbps. 
The authors of \cite{zhou2018energy} formulate  a large-timescale energy-efficiency optimization problem for the UAVs involved for power line inspection in a smart grid. As a further advance, a deep learning aided vision-based UAV is involved for automatically detecting common faults in power line components in \cite{nguyen2019intelligent}. In \cite{lim2018multi}, the UAV locations are optimized for reliable power network damage assessment under uncertain weather conditions.

\subsubsection{V2G}
Recently, the topic of V2G communications has attracted tremendous researches attention in the context of providing cost-efficient EVs charging \cite{cao2017toward,manshadi2018wireless, tao2017foud, chekired2017smart, hoang2017charging, wang2017distributed}.  A scalable communication solution designed for information exchange between the power grid and EVs Helsinki is studied in \cite{cao2017toward} with the objective of reducing the charging time and increasing the number of EVs charged. In \cite{hoang2017charging}, the cyber-security involves of plug-in electric vehicles are investigated. 
Meanwhile, the authors of \cite{wang2017distributed} discuss a distributed framework formulated for integrating communications and a control scheme for supporting efficient energy transfer, considering the dynamic activities of both the power grid and of the EVs.

\subsubsection{AI and Machine Learning for the Smart Grid}
The efficient management of the stochastic factors affecting the smart grid by applying data mining, AI and machine learning is studied in \cite{Gridwang2017wireless, munir2019edge,nguyen2019intelligent,  zhu2018big, wang2017robust,  wang2018deep, ni2019multistage, haddad2018smart, li2017weather}. 
Specifically, the authors of \cite{wang2018deep} use a deep CNN for learning the characteristics of consumers such as the associated daily load profiles by leveraging the socio-demographic information of consumers inferred from smart metering data. Meanwhile, large scale event classification techniques using artificial neural networks (ANNs) capable of distinguishing up to 310 classes were developed in \cite{haddad2018smart} to maintain the quality and reliability of the distributed generation systems under various impairments or operating conditions.
In \cite{li2017weather}, deep learning is involved for performing electrical load forecasting.


\subsubsection{Communication Testbeds for the Smart Grid}
Recently developed testbeds have been discussed in \cite{zhang2016sdn, kurtz2016empirical, you2018cognitive}. 
In \cite{zhang2016sdn}, an OpenFlow-based SDN allowed the operators to dynamically control and monitor the entire network using the software running on the operating system of the centralized controller. 
In the Smart Grid Laboratory of Durham University in the UK, a cognitive radio aided smart grid testbed  \cite{kurtz2016empirical} was constructed to provide real-time emulation of realistic smart grid systems, where the average round trip communication latency is about 9.7ms.
In \cite{you2018cognitive}, a prototype was built to facilitate the gradual evolutionary development of LTE towards 5G in order to support mission-critical communication in the smart grid.

\begin{figure} [!t]
	\centering	
	\includegraphics[width=3.5in]{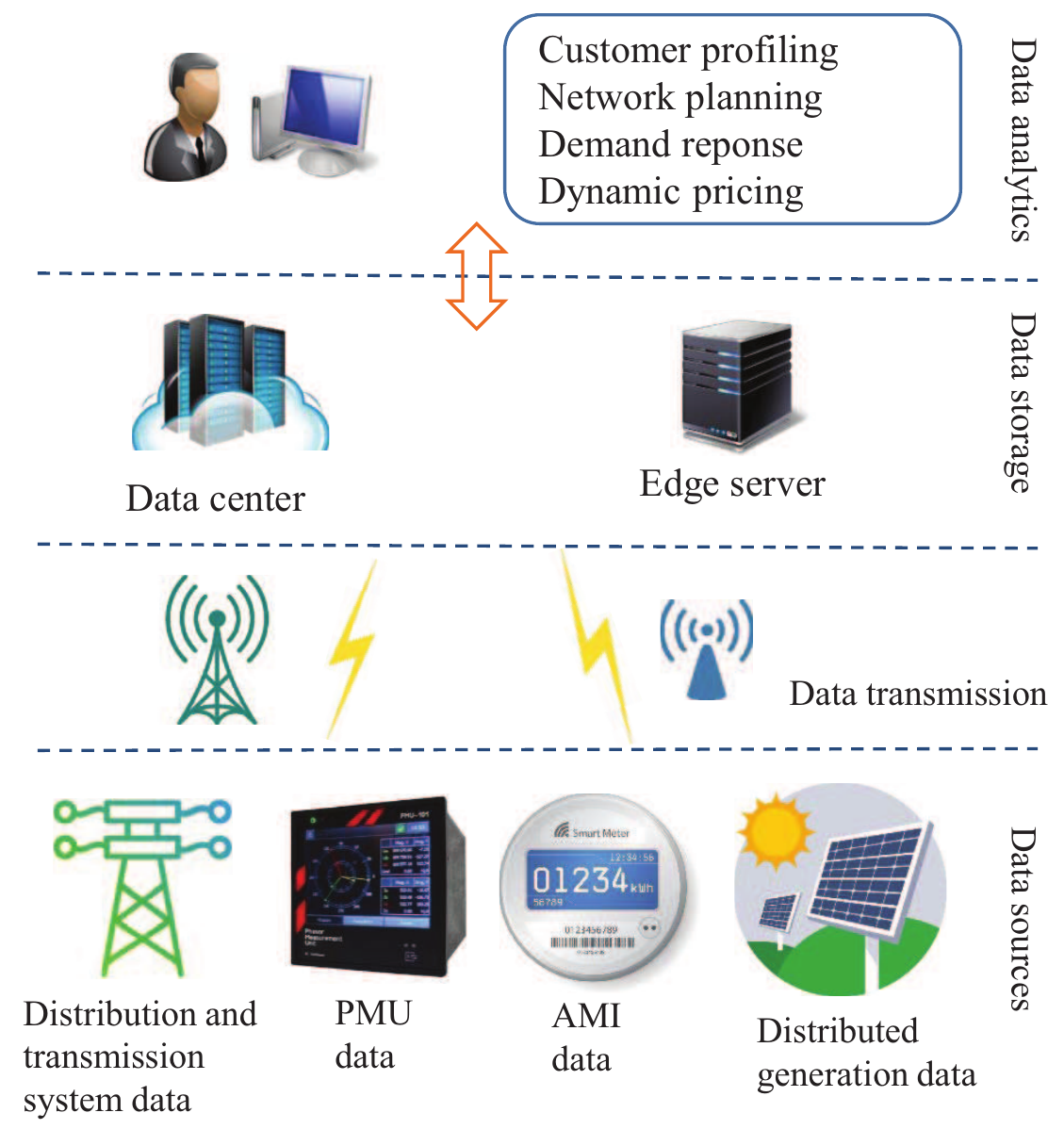}
	\caption{Data processing architecture for smart grid system \cite{Gridwang2017wireless}.}
	\label{fig:GridData}
\end{figure}

\subsection{Data-Driven Smart Grid}
In the 5G-aided smart grid, smart meters are being deployed at consumer premises to monitor their energy consumption in real time, which is securely sent back to the utility over 5G networks. These smart meters can also receive signals from the utility, conveying  information on dynamic power pricing and incentives for reducing load during peak periods \cite{4.7,4.8}. The data collected from smart meters and smart sensors are necessary for the near-real-time detection of critical situations and for their mitigation at a low latency for ensuring the grid stability. Constructing an efficient smart grid data processing system can help predict the power supply and demand equilibrium to take preemptive actions for curtailing demand by notifying consumers \cite{4.7}.

\subsubsection{Data Processing Architectures for Smart Grid}
A four-level big data computing architecture including the data sources, data transmission, data storage, and data analysis is proposed in \cite{Gridwang2017wireless} for the smart grid system which is shown in Fig. \ref{fig:GridData}.
\begin{itemize}
	\item \textit{Data source}: Smart grid data are generated by different data sources including the distribution- and transmission-related system data, phasor measurement based and advanced metering infrastructure data as well as distributed power generation data \cite{Gridwang2017wireless,4.9}. Phasor measurements constitute one of the most critical measurements in the power transmission and distribution systems, which compare the phasor to a  time reference. Another fundamental subsystem of the smart grid is the advanced metering infrastructure (AMI) which integrates multiple technologies (smart meters, communication networks, and information management systems) for intelligent communications between the energy consumers and system operators \cite{4.9}.
	\item \textit{Data storage}: Data storage is one of the most important components in the smart grid which is indispensable for supporting literally all services and functionalities. The data gleaned from diverse sources must be stored in a systematic manner for efficient contextual retrieval \cite{4.9}. Data storage can be assigned at the edge for real-time analytics and real-time applications such as grid failure detection and isolation as well as restoration, or at the data center in the cloud for global analytics and processing, such as deciding about the most appropriate demand response \cite{4.7}, prediction of the amount of energy generation and energy consumption, supporting micro-grids, and customer billing.
	\item \textit{Data analysis}: By exploiting both historical and real-time data in the smart grid, system operators becoming capable of optimizing their operation, whilst predicting and preventing potential problems  in the system before they occur \cite{4.9}. For example, a demand response strategy can utilize real-time data arriving from smart meters concerning their near-instantaneous energy consumption to dynamically adjust their load and pricing policies. Moreover, data is captured from diverse sources in the smart grid and utilized for supporting different services and applications, therefore, the specific data analysis methods differ significantly. Accordingly, they can be categorized based on their functional features, such as the associated statistical analysis methods, data mining techniques, and data visualization methods \cite{4.7}. Several existing works have considered data analytics in the smart grid. In \cite{Gridwang2017wireless}, Wang \textit{et at.} discuss  wireless communication aided big data computing architectures conceived for smart grid analytics. As further development considering the user's privacy, the authors of \cite{zhu2018big} discuss diverse defense strategies designed for protecting the consumer's energy consumption patterns. Meanwhile, in \cite{wang2017robust}, the available data is analyzed for electricity price forecasting, where Grey Correlation Analysis, Kernel functions and Principle Component Analysis, and Support Vector Machines are employed for feature-redundancy elimination, for dimensionality reduction and for classification, respectively.
	\item \textit{Data transmission}: With the benefit of recent innovation in wireless communication technologies, leading to the conception of 5G technologies, data in a smart grid can be efficiently transmitted through the integrated 5G infrastructure.
\end{itemize}

\subsubsection{Data Security in the Smart Grid}
Cybersecurity is a crucial challenge in the smart grid.  A survey of recent security advances conceived for the smart grid, relying on a data driven approach is presented in \cite{4.9}. The authors report on their comprehensive investigations dedicated to  the security vulnerabilities and solutions within the entire life-cycle of a large volume of smart grid data including four stages: data generation, data acquisition, data storage and data processing. Sophisticated solutions regarding the data generation security, data acquisition security, data storage security, data processing security, and data analytics security are discussed in \cite{4.9}.


\subsection{Smart Grid Challenges}
The major research challenges related to the smart grid are described in the following.

\subsubsection{Backward Compatibility and Evolution} 
The 5G system lends itself to seamless integration into the existing (primarily wire-bound) smart grid communication infrastructure.  

\subsubsection{Resilience, Real-time Control and Monitoring} 
The 5G system shall be able to support continuous monitoring of the current network state in real-time, to take prompt automated actions in case of problems, and to perform efficient root-cause analyses in order to avoid any undesired interruption of the production processes, which would incur substantial financial loss. More 
particularly, if a third-party network operator is involved, accurate service-level agreement (SLA) monitoring is needed for avoiding potential liability disputes in case of SLA violations. Therefore, new approaches are required for integrating 5G into the smart grid in order to fulfill the requirements of high resilience, and real time monitoring.  
	
\subsubsection{High Grade of Security and Privacy} 
Due to the bidirectional nature of communication in the smart grid, it is crucial to develop a cybersecurity mechanism for preventing cyber attacks from outside the power grid through the 5G wireless network. For example, the authors of \cite{4.3} propose an intrusion detection system so that the smart grid becomes capable of detecting price integrity and load alteration attacks. In \cite{zhang2017efficient}, an efficient and privacy-aware power injection scheme suitable for advanced metering infrastructure and 5G smart grid network slicing is proposed, which allows the utility company to harness the total amount of collected power and to prevent any attacker from reading any individual power injection bid.

\section{Smart City}
\begin{figure*}[htb!]
	\centering
	\includegraphics[width=6in]{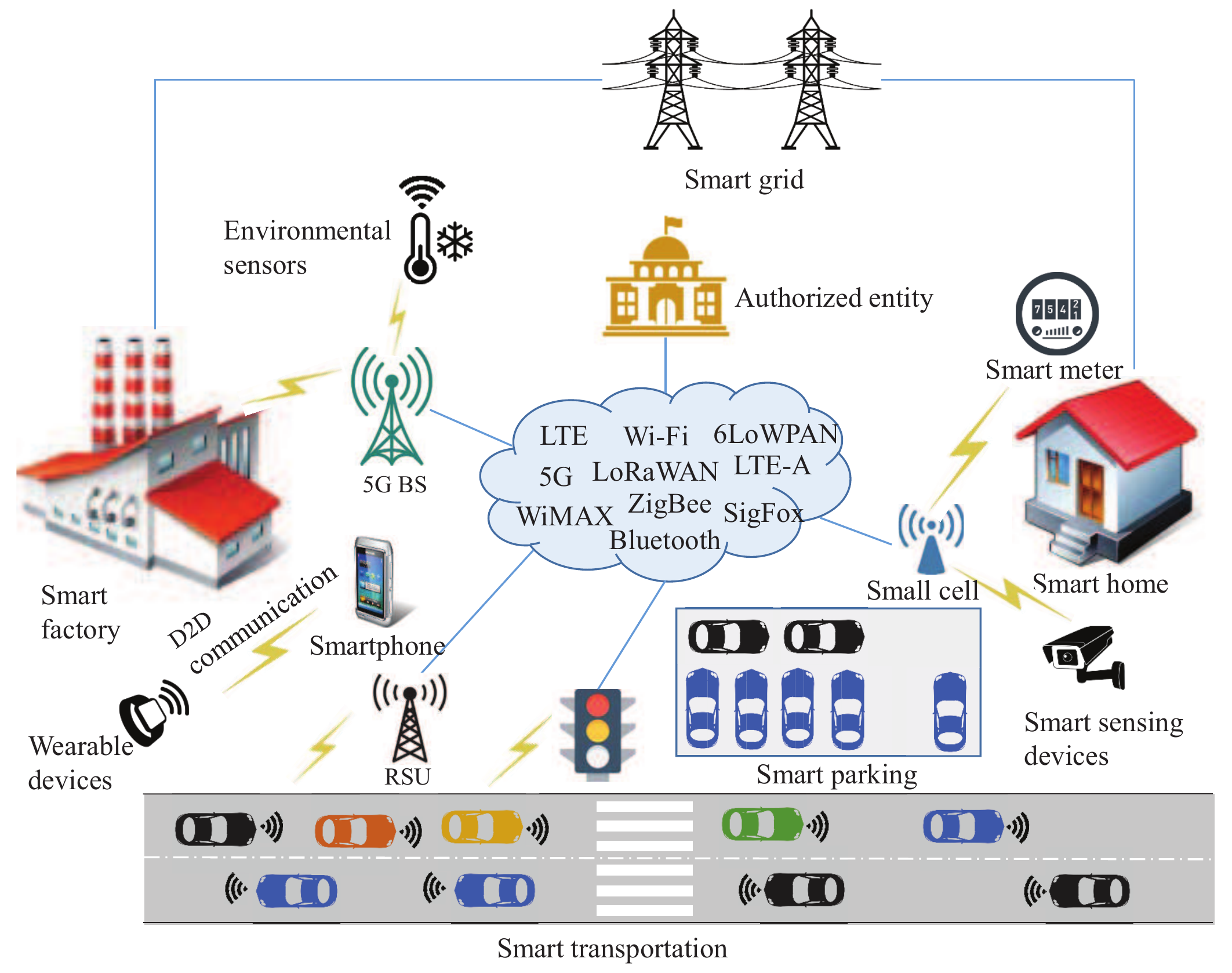}  
	\centering
	\caption{Smart city ecosystem \cite{5.3}.}
	\label{fig:SmartCityEcosystem}
\end{figure*}
The trend is that cities are striving to become “smart”, to become communities at the forefront of using data, sensors and connected devices for improving the government services and the residents’ quality of life through data analytics and automation, which rely on connectivity.
Smart cities promise to deliver remarkable advances, such as traffic congestion reduction with the aid of intelligent transportation system, energy efficiency improvements by using smart street lighting and charging, more effective health and safety inspections with predictive analytics, and many more.
Thanks to the 5G technology's innovations such as wireless sensor networks, machine-to-machine (M2M) and V2X communications, all the aforementioned advantages of a smart city can indeed be realized \cite{5.1}.

\begin{figure*}[htb!]
	\centering
	\includegraphics[width=6in]{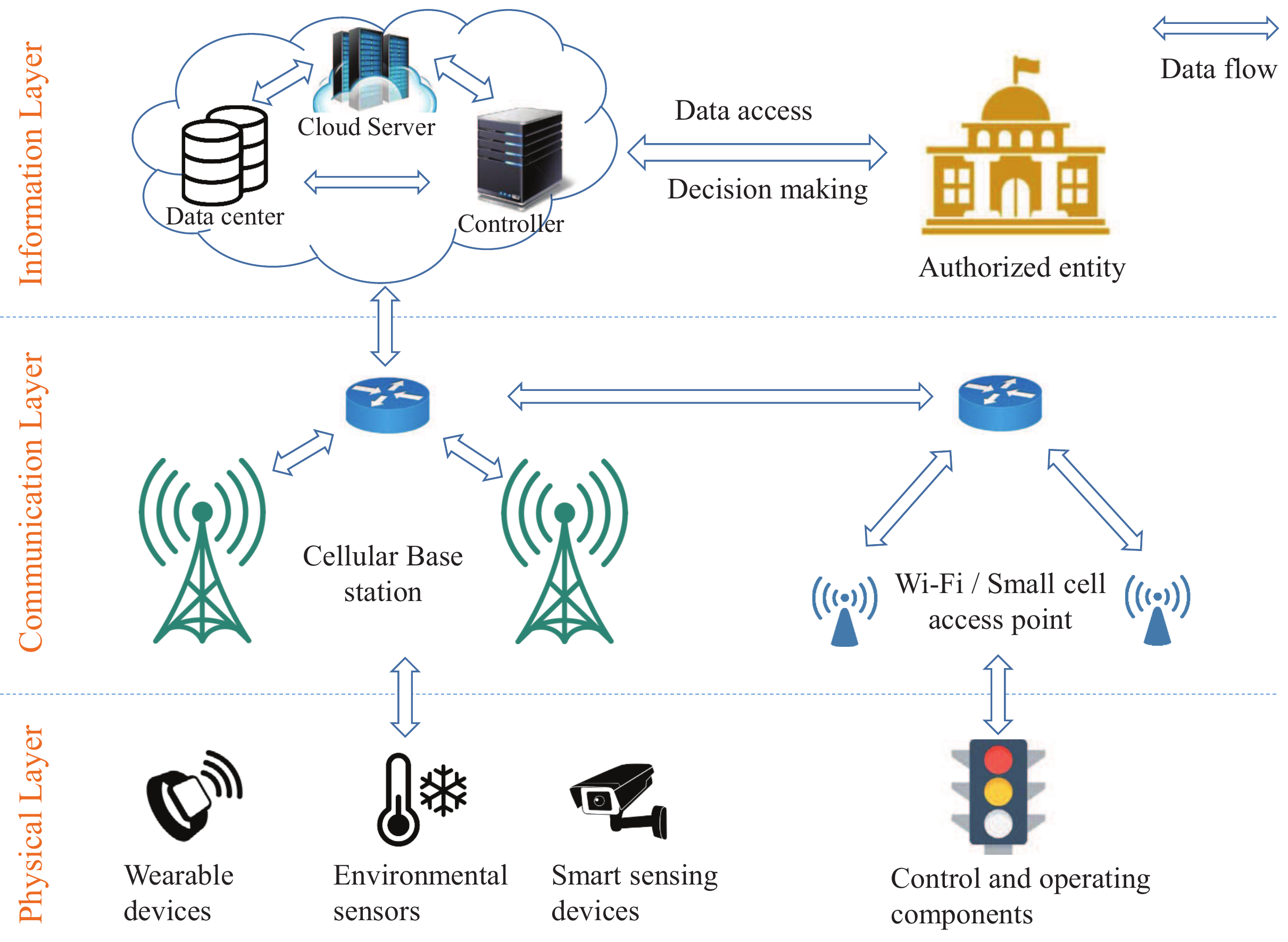}  
	\centering
	\caption{Architecture-oriented view on enabling technologies for smart cities \cite{5.3}}
	\label{fig:5Gcity}
\end{figure*}

\subsection{Smart City Architecture and Enabling Technologies}

\subsubsection{Smart City Architecture}
A smart city is a complex ecosystem characterized by the intensive use of information and communications technologies. In a smart city ecosystem, the IoT will play a fundamental role, which relies on a large-scale heterogeneous infrastructure \cite{5.62}. These heterogeneous infrastructures can be deployed at multiple layers in the smart cities architecture. The authors of \cite{5.3} propose a hierarchical architecture for smart cities based on 5G communications, which includes three main layers: the physical, communication, and information layers. Data is collected from the physical world by smart sensors and wearable devices. The communication layer typically relies on heterogeneous networks including a cellular network, small cells and Wi-Fi networks, the data is sent to the processing unit forming part of the information layer for analytics and decision feedback to the control and operating components, as shown in Fig. \ref{fig:5Gcity}.


\subsubsection{Enabling Techniques in the Smart City}
Table \ref{table:CityTaxonomyTechnologies} presents the emerging technologies of the smart city \cite{5.1,5.62}.
Specifically, these emerging technologies include 5G, software defined wireless networking (SDWN), network function virtualization (NFV), visible light communication (VLC), and others.
Modern  communication enabling technologies for the smart city include WLANs (WiFi), WiMAX, LTE, LTE-A, Bluetooth, Zigbee, Z-Wave, and LoRaWAN.
Finally, several wireless standards such as IEEE 802.11, IEEE 802.15.x, and IEEE 802.16 have been defined by the IEEE, which are eminently suitable for the smart city.

\begin{table}[!t]
	\centering
	\caption{Taxonomy of enabling communication and networking technologies for smart cities \cite{5.1,5.62}. }
	\label{table:CityTaxonomyTechnologies}
	{\renewcommand{\arraystretch}{1.4}
		\begin{tabular}{ll}
			\hline
			\textbf{\begin{tabular}[c]{@{}l@{}}Emerging \\technologies\end{tabular}} & \textbf{\begin{tabular}[c]{@{}l@{}}Communication\\ technologies\end{tabular}}  \\ \hline
			\begin{tabular}[c]{@{}l@{}}Edge, cloud computing\\ SDN/NFV/Network slicing\\ AI and machine learning\\  Visible light communication (VLC)\\ Cognitive radio networks (CRN)\\ 5G wireless technologies\end{tabular} & \begin{tabular}[c]{@{}l@{}}Wi-Fi\\ SigFox, WiMAX\\ LTE, LTE-A\\ Bluetooth\\ ZigBee\\ 6LoWPAN, LoRaWAN\\ 802.11, 802.15.x, 802.16\end{tabular} \\ \hline
		\end{tabular}
	}
\end{table}

\subsection{Exemplifying Use Cases and Performance Requirements}

The smart city use cases can be categorized according to the 5GCity project \cite{5.12} as follows:
\begin{itemize}
	\item \textit{UC1-Unauthorized waste dumping prevention:} Monitor urban areas under the risk of environmental abuse, to prevent unauthorized dumping of waste.
	\item \textit{UC2-Neutral host:} A network virtualization platform enabling flexible end-to-end network slicing concerning both computing and radio resources for different third-party network operators.
	\item \textit{UC3-Video acquisition and production:} Acquire high-quality videos via mobile applications at events and stream them through the 5G network.
	\item \textit{UC4-UHD video distribution \& immersive services:} Develop an application relying on mixed reality, solutions for user movement tracking and computer vision algorithms to create an augmented tourist guide for both indoors and outdoors.
	\item \textit{UC5: Cooperative, connected and automated mobility:} Use 5G, NFV and MEC for V2X/CCAM to significantly improve road safety, driving comfort  and smarter coordination among of connected autonomous cars, the road infrastructure and cloud services. 
	The MEC application is envisioned to operate in street cabinets to implement an always-on connection to the central cloud service. 
	Vehicles can then subscribe data alerts relevant for their surrounding area (tuneable parameter). 
	The cloud is also capable of providing the proximity caching services  for the vehicles, which may in turn stream this cached data immediately and publish any relevant data collected by the car's own intelligence system.
\end{itemize}

The requirements specified by the 5GCity project use cases are summarized in Table \ref{table:SGCityUC} \cite{5.12}.
\begin{table*}[htbp]
	\centering
	\caption{5GCity project use cases' requirements \cite{5.12}. }
	\label{table:SGCityUC}
	{\renewcommand{\arraystretch}{1.4}
		\begin{tabular}{lllllll}
			\hline
			\textbf{Requirements} & \textbf{UC1} & \textbf{UC2} & \textbf{UC3} & \textbf{UC4} & \textbf{UC5} \\ \hline
			\begin{tabular}[c]{@{}l@{}}Device Density (devices/km$^2$)\end{tabular} & \textless{}1000 & N/A & 1000-10000 & 1000-10000 & \begin{tabular}[c]{@{}l@{}}\textgreater{}= 10000 veh/km$^2$\end{tabular} \\ \hline
			Mobility (km/h) & 3-50 & 3-50 & 3-50 & 3-50 & \textgreater{}50 \\ \hline
			\begin{tabular}[c]{@{}l@{}}Infrastructure (Macro \& small cell)\end{tabular} & Medium & Medium & Medium & Medium & Medium \\ \hline
			User Data rate & \textless{}50 Mbps & \begin{tabular}[c]{@{}l@{}}0.1-1 Gbps\end{tabular} & 50-100 Mbps & \begin{tabular}[c]{@{}l@{}}0.1-1 Gbps\end{tabular} & 50-100 Mbps \\ \hline
			Latency (ms) & \textless{}=10 & \textless{}=10 & \textless{}=10 & \textless{}=10 &  \textless{}=10 \\ \hline
			\begin{tabular}[c]{@{}l@{}}Resilience and continuity\end{tabular} & 95-99\% & \textgreater{}99\% & 95-99\% & 95-99\%  & \textgreater{}99\% \\ \hline
		\end{tabular}
	}
\end{table*}

\subsection{5G Support for Smart Cities}

5G will support smart city applications by fulfilling their stringent requirements imposed by the various service types.
Specifically, as shown in Fig. \ref{fig:5Gcity}, the smart city typically requires high-throughput and high-reliability connections among sensors, network components, processing units, as well as control and other components. 
Furthermore, the 5G mMTC service types will support the ubiquitous connectivity of millions of IoT devices \cite{5.11}, such as smart sensors and wearable devices.
The main benefits of 5G for the smart city may be summarized as follows:

\begin{itemize}
	\item \textit{Citizens' experience and security enhancement:} According to \cite{2.13}, public safety and security are key factors of the smart city which maybe substantially improved by 5G technology. The priorities of public safety applications are listed in Table \ref{table:SmartCity_use_cases_priority} \cite{2.13}.
	\item \textit{Transportation system for smart cities:} Advanced 5G communication technology using UAVs/drones can be deployed for improving transportation surveillance in real time \cite{5.5}. Moreover, the 5G cellular network can be relied upon for data collection, which benefits the transportation systems and intelligent parking systems in smart cities \cite{5.6}. The 5G-based V2X communications can also be extended to  vehicular social networks for dealing with the  problem caused by road accidents, traffic congestions and other transportation issues in smart cities \cite{5.7}.
	\item  \textit{Big-data processing:} A huge amount of data can be collected by smart sensors and end-user devices. These data can be processed locally or may be offloaded to the mobile edge and cloud \cite{5.8,5.9} through 5G networks in a reliable manner. The improved centralized data processing helps executives to  make evidence-based decisions regarding the day-to-day operation of smart cities.
	\item \textit{New network paradigm:} In \cite{5.10}, the authors propose the idea of merging all individual BSs into a large virtual cell for supporting seamless operation of the network and superior traffic load management, as well as for reducing the frequency of handovers.
	\item \textit{Real-time monitoring:} Mobile crowd sensing technology relying on wearable devices and portable sensors can be applied to monitor the air quality in smart cities \cite{5.17}. The local air quality information may then be aggregated in the cloud for intelligent traffic control on order to reduce air-pollution. Another example is the use of IoT devices to monitor the energy consumption in various domains of the smart cities, thus enabling improved energy management \cite{5.18}. By relying on the mMTC service type for supporting high-density IoT devices, 5G facilitates ubiquitous IoT-based monitoring of smart cities.
\end{itemize}

\begin{table}[htbp]
	\centering
	\caption{Key 5G use cases for the public safety sector. Source: Ericsson survey \cite{2.13}}
	\label{table:SmartCity_use_cases_priority}
	{\renewcommand{\arraystretch}{1.4}
	\begin{tabular}{ll}
		\hline
		\textbf{Priority of public safety applications} & \begin{tabular}[c]{@{}l@{}} \textbf{Percentage} \%\end{tabular} \\ \hline
		Border and area security & 60 \\ \hline
		Priority and quality of aervice for public safety & 59 \\ \hline
		Enhanced GPS for faster arrival at incidents & 59 \\ \hline
		Incident area network & 57 \\ \hline
		Vehicle-to-Vehicle communications & 57 \\ \hline
		Drones for law enforcement/emergency service & 54 \\ \hline
		Video surveillance/streaming from first responders & 54 \\ \hline
		Device-to-device communication without human & 53 \\ \hline
		High-risk environment diagnostics & 50 \\ \hline
		Remote control of robots in bomb diffusing & 49 \\ \hline
		\begin{tabular}[c]{@{}l@{}}Traffic light "Green wave" for emergency vehicle \\ priority\end{tabular} & 47 \\ \hline
		Communication between agencies & 47 \\ \hline
		Visor/helmet computer with augmented/virtual reality & 46 \\ \hline
		Hazard sensors & 44 \\ \hline
	\end{tabular}
	}
\end{table}

\begin{table}[htbp]
	\centering
	\caption{Summary of recent studies on smart city domains}
	\label{table:City_topic}
	{\renewcommand{\arraystretch}{1.4}
		\begin{tabular}{ll}
			\hline
			\textbf{Research topics} & \textbf{References} \\ \hline
			\begin{tabular}[c]{@{}l@{}}Cloud and Edge computing \\for smart city\end{tabular} & \cite{5.9,5.60,5.63,5.64} \\ \hline
			\begin{tabular}[c]{@{}l@{}}IoT for smart city\end{tabular} & \cite{5.62,5.65,5.66,5.67} \\ \hline
			\begin{tabular}[c]{@{}l@{}}D2D, M2M communication \\for smart city\end{tabular} & \cite{5.68,5.69,5.70,5.71} \\ \hline
			\begin{tabular}[c]{@{}l@{}}Cell-free and cell-cooperative \\ for smart city\end{tabular} & \cite{5.10,5.54,5.55} \\ \hline
			\begin{tabular}[c]{@{}l@{}}SDN, NFV, and network slicing \\ for smart city\end{tabular} & \cite{5.42,5.43,5.44,5.45,5.46,5.47,5.48,5.49,5.50,5.51,5.52,5.53,5.72} \\ \hline
			\begin{tabular}[c]{@{}l@{}}Machine learning for smart city\end{tabular} & \cite{5.33,5.34,5.35,5.36,5.37,5.38,5.39,5.40,5.41,5.49,5.59,5.60,5.61} \\ \hline
			\begin{tabular}[c]{@{}l@{}}Security and privacy\\ for smart city\end{tabular} & \begin{tabular}[c]{@{}l@{}}\cite{5.3,5.4,5.13,3.15,5.16},\\ \cite{5.36,5.37,5.40,5.43},\\ \cite{5.60}, \cite{5.19,5.20,5.21,5.22,5.23,5.24,5.25,5.26,5.27,5.73,5.74}\end{tabular}  \\ \hline			
		\end{tabular}
	}
\end{table}

\begin{table*}[htbp]
	\centering
	\caption{Summary of benefits for representative smart city use cases supported by 5G technologies.}
	\label{table:City_topic1}
	{\renewcommand{\arraystretch}{1.5}
		\begin{tabularx}{\textwidth}{P{1.5cm}|P{1cm} P{2.5cm} P{2.5cm} P{3cm} X }
			\hline
			\textbf{Topic} & \textbf{Ref} & \textbf{Use case} & \textbf{Problem} & \textbf{5G Technologies, Standards, Protocols} & \textbf{Contribution} \\ \hline
			\multirow{5}{\hsize}{Cloud Edge computing} & \cite{5.9} & HD video streaming service, augmented reality services & MEC architecture design for video streaming in smart city & Mobile edge computing testbed & Proposing a MEC based approach to enhance users experience of video streaming in the context of smart cities. \\ \cline{2-6} 
			& \cite{5.64} & Sensor telemetry,  Physical security of Fog nodes in city of Barcelona & Management architecture  of IoT services for smart city & IoT, NFV, cloud edge, fog computing & Designing an architecture of NFV, 5G/MEC, IoT, and fog applying to a project in the city of Barcelona. \\ \hline
			IoT based smart city & \cite{5.67} & Anomaly detection of energy consumption, video surveillance in stadiums, smart city magnifier & System architecture design for IoT Smart city & Cloud, edge, fog computing & Proposing the FogFlow framework which provides a standards-based programming model for IoT services for smart cities that run over cloud and edges \\ \hline
			D2D, M2M & \cite{5.69} & Transportation video service, Environmental monitoring service & Design content delivery system for smart city & D2D & Proposing an energy-efficient content delivery system via the D2D communications for smart cities. \\ \hline
			Cell-free and cell-cooperative & \cite{5.10} & High density smart city & Design SDN-based cell-less architecture for smart city & SDN, wireless heterogeneous nework, massive MIMO, mmWave & Proposing 5G converged cell-less communication networks to support mobile terminals in smart cities considering the deployment of 5G ultra-dense wireless networks. \\ \hline
			\multirow{5}{\hsize}{SDN, NFV, and network slicing} & \cite{5.50} & City traffic congestion & Mitigating the city traffic network congestion & Software-defined vehicular networks (SDVN), MEC & Proposing a novel SDVN architecture in the smart city scenario, mitigating the traffic congestion \\ \cline{2-6} 
			& \cite{5.60} & Air quality monitoring in city of Antwerp, Belgium & Air quality monitoring & Cloud, edge, fog computing, SDN, NFV, LoRaWAN, SigFox, IEEE 802.11ah, LTE-M & An anomaly detection solution for smart city applications is presented, focusing on low-power fog computing solutions and evaluated within the scope of Antwerp City of Things testbed. \\ \hline
			Security for smart city & \cite{5.25} & Sensor real-time monitoring & Physical layer security of wireless sensor network & Multiple antenna wireless mobile sensor networks. & Investigating the secrecy performance of the wireless mobile sensor communication networks over 2-Nakagami fading channels. \\ \hline
		\end{tabularx}
	}
\end{table*}

\subsection{Recent Research on the Smart City Concept}
The smart city concept covers a wide range of research topics which have attracted attention  both from academia and industry. Given the benefit of 5G systems that rely on diverse technologies ranging from NFV, SDN, cloud and edge computing to machine learning involved for improving the system’s flexibility, scalability and robustness, the application of 5G techniques in the smart city domain have been richly documented in the literature.

\subsubsection{Cloud and Edge Computing for the Smart City}
There are numerous benefits in integrating cloud and edge computing services into smart city applications \cite{5.9,5.60,5.63,5.64}. For example, a use case of augmented reality relying on the streaming of HD video for supporting tourism through MEC in smart cities is studied in \cite{5.9}.
An anomaly detection solution designed for smart city applications and focusing on low-power edge computing is proposed in \cite{5.60}. Similarly, an anomaly detection scheme designed for air-quality monitoring using a fog computing framework based on the ETSI NFV management and orchestration architecture is investigated in \cite{5.63}. 

\subsubsection{IoT for the Smart City}
The smart city infrastructure relies on a large-scale IoT system, including smartphones, wearable devices, sensors and actuators to collect data and to execute control tasks in smart city applications. Designing IoT solution for the smart city is an attractive research topic \cite{5.62,5.65,5.66,5.67}.
Edge and cloud computing are eminently suitable for smart city IoT services \cite{5.67}. The authors of\cite{5.67} propose a fog-computing-based framework, which can be readily applied for IoT smart city platforms supporting anomaly detection in energy consumption for example. The applications, architecture and challenges of the deployment of the Internet of Vehicles (IoV) for a smart city is discussed in \cite{5.66}. The IoV architecture  proposed for the smart city in \cite{5.66} consists of several layers, including the vehicle layer, communication layer, servers and cloud layer, big data and multimedia computation layer and application layer. Moreover, a range of enabling technologies designed for IoT-based smart cities are discussed in \cite{5.62}.

\subsubsection{D2D, M2M Communication for the Smart City}
Typical smart city applications generally tend to require two different communication types, namely wide area cellular communication supporting high data rates for conveying large amount of data, and the transmission of short data packets between devices and machines, especially between IoT devices \cite{5.71}. 
In contrast to conventional cellular communication, the D2D communications between IoT devices typically exchange small data packets at high energy-efficiency and low data rates. 
To fully exploit the benefits of both of these communication types in smart city applications, numerous studies relying on 5G and D2D links have been conducted in the literature \cite{5.68,5.69,5.70,5.71}. In \cite{5.68}, the authors propose an energy-efficient resource allocation scheme for D2D communication in the smart city, while an energy-efficient multimedia content delivery scheme relying on D2D communications is proposed in \cite{5.69}. A 5G SDN based D2D communication architecture is put forward for public safety applications in the smart city by the authors of \cite{5.70}. By deploying both local and central controllers, the proposed architecture aims for maintaining communication links in even case of infrastructure damage due to a disaster \cite{5.70}.

\subsubsection{Cell-free and Cooperative Multi-cell Solutions for the Smart City}
Providing seamless and ubiquitous connectivity is a fundamental requirement in the of a smart city.
In a future wireless network, especially in  ultra-dense urban wireless access scenarios, the traditional cellular network cannot provide a sufficiently high number of connections for millions of mobile and IoT devices. In this scenario, the  cell-free concept is expected to lead to an efficient architecture of supporting massive wireless connectivity, where numerous wireless devices can be served simultaneously using multipoint transmissions and multipoint user associations \cite{8.36}. 
Utilizing high-rate, low-delay backhaul links among different BSs/APs, the overall network will appear as a cell-free distributed massive multiple-input multiple-output (massive MIMO) system from the end-device point of view \cite{8.36}. The cell-free wireless network can be implemented under the umbrella of cloud radio access networks (CRANs) or coordinated multipoint transmission (CoMP) concept by using powerful centralized processing units that manage the assignment of resources to different terminal devices \cite{5.10}. Moreover, the cell-free architecture can be realized under the massive MIMO concept which can achieve a significant performance gain over the huge traditional small-cell architecture \cite{8.32}.

\subsubsection{SDN, NFV, and Network Slicing for the Smart City}
In IoT based smart city deployments relying on a massive number of IoT devices and a complex IoT infrastructure, SDN and NFV are expected to provide an efficient solution in managing the operation of IoT devices, such as their resource allocation, data acquisition, processing and real-time decision making. The authors of \cite{5.72} propose the integration of the SDN concept with IoT devices (SDN-IoT) in support of a variety of big data applications in smart cities. An SDN based architecture is proposed for deploying IoT applications for smart cities  in \cite{5.42} in which the IoT traffic destined to multiple parties is centrally managed via a common SDN-based network infrastructure. Several SDN and NFV solutions constructed for the smart city can be found in \cite{5.42,5.43,5.44,5.45,5.46,5.47,5.48,5.49,5.50,5.51,5.52,5.53,5.72}.
Moreover, the problem of heterogeneous network traffic encountered in smart transportation systems relying on IoV applications can be efficiently handled by network slicing \cite{5.44}. In \cite{5.44} the authors propose a smart slice scheduling and road traffic management solution for vehicular fogs for accommodating the erratically fluctuating network traffic generated by vehicular applications.

\subsubsection{Machine Learning for the Smart City}
The authorities have to make numerous high-risk decisions in order to manage the operations of a smart city. Most of these decisions have to be taken in a real-time manner, which are based on the huge volume of data collected from millions of IoT devices. Thanks to AI and machine learning technologies, decision-making may be carried out in a data-driven manner to provide seamless services for smart cities. Moreover, machine learning can be involved at each end-device to improve their performance in each specific situation. In the smart city architecture of Fig. \ref{fig:5Gcity}, AI and machine learning can be implemented at each layer in support of both local and global decision making \cite{1.120}. Sophisticated machine learning solutions designed for wireless sensor networks can be found in \cite{5.33,5.34,5.35,5.36,5.37,5.38,5.39,5.40,5.41,5.49,5.59,5.60,5.61,5.73}. Support vector machine based learning mechanisms are proposed for anomaly detection in smart cities, while a wireless sensor network was involved for air-pollution detection in \cite{5.35} and for security attack detection in \cite{5.73}.


\subsubsection{Security Features for the Smart City}
Successful smart city deployments require efficient safety and security systems as well as the careful consideration of privacy. The data collected in a smart city is gleaned from various sources and the sensor devices are vulnerable to cyber-attack.
Therefore, security and privacy issues of the smart city have attracted a lot of attention in recent years. The data management, security and privacy issues of a smart city as well as the associated enabling technologies are discussed in \cite{5.3,5.4,5.13,3.15,5.16} and the references therein. Moreover, several security aspects of the smart city relying on wireless sensor network are studied in \cite{5.19,5.20,5.21,5.22,5.23,5.24,5.25,5.26,5.27,5.36,5.37,5.40,5.43,5.60,5.73,5.74}.

\begin{figure} [!t]
	\centering	
	\includegraphics[width=3.5in]{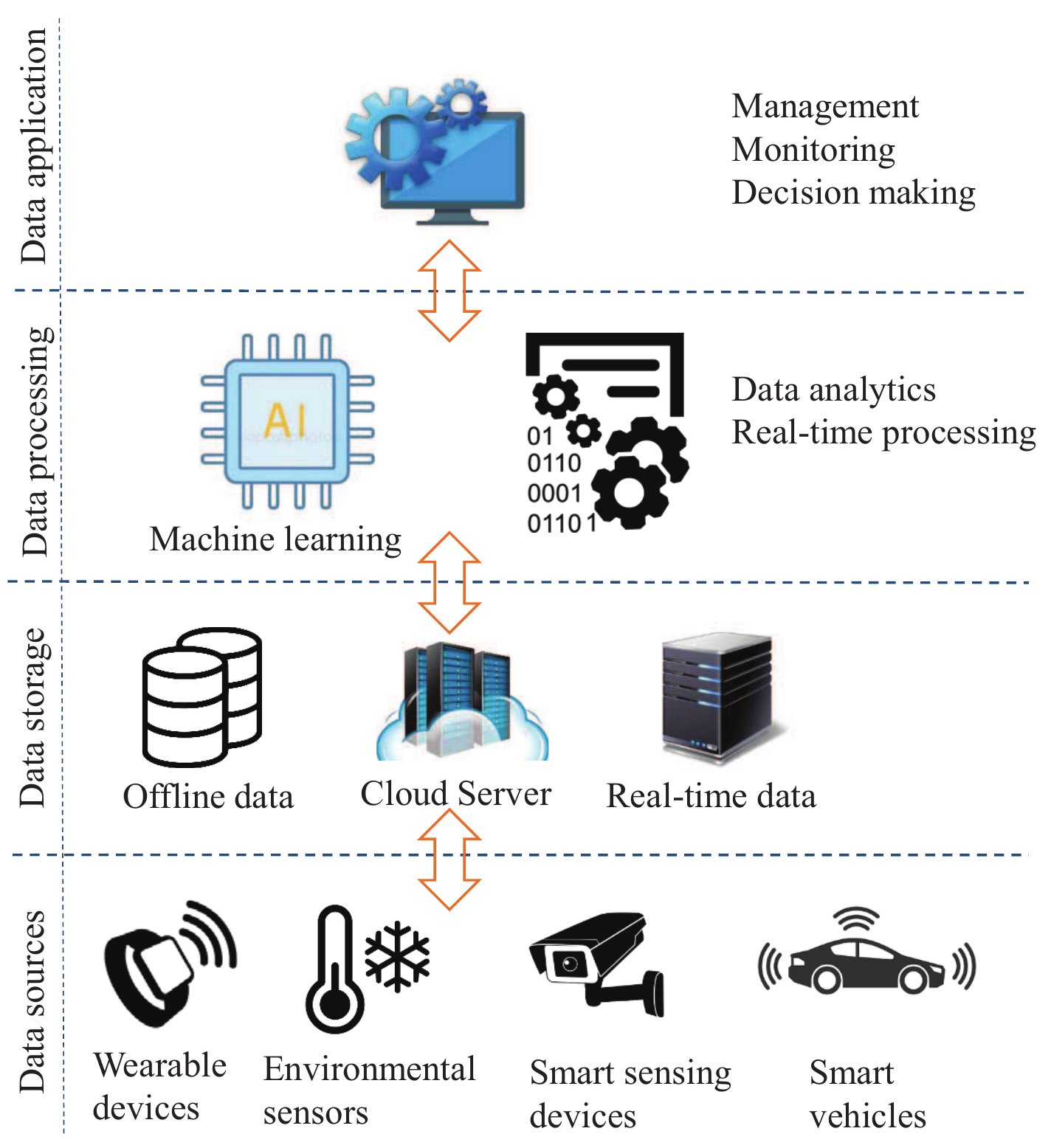}
	\caption{Data processing architecture for smart city ecosystem.}
	\label{fig:CityData}
\end{figure}
\subsection{The Data-driven Smart City}
Again, a tremendous amount of data is generated every day by the integrated IoT that can be exploited for improving safety, sustainability, efficiency and level of intelligence in a smart city.  

A novel big data processing solution conceived for extracting meaningful information from the sensor data collected in a smart city is proposed in \cite{5.75}, which relies on the data storage plane, data processing plane and data application plane.
As a further solution, a hierarchical distributed fog computing architecture was designed for big data analysis in \cite{5.76}. This hierarchical architecture relies on a sensing network layer, edge computing layer, intermediate computing layer and cloud computing layer. It was created for city-wide response and resource management in the case of a natural disaster or a large-scale service interruption \cite{5.76}. Similar four-layer architectures are also discussed in \cite{5.16} and \cite{5.24}. We might surmise the general trend that a four-layer data processing architecture including the data sources layer, data storage layer, data analytics layer and application layer is suitable for a smart city as shown in Fig. \ref{fig:CityData}.
\begin{itemize}
	\item \textit{Data sources}: Data acquisition gleans information in a smart city from multiple sources including smart sensing devices, environmental sensors, wearable devices, smart phones, smart vehicles, etc. which form part of the ubiquitous wireless sensor networks (WSNs), mobile ad hoc networks (MANETs), IoT networks, V2X networks, etc. These data sources are rather heterogeneous in term of their data volume, data format, data type, data quality and data representation standard. Therefore, to efficiently exploit these data, we need a suitable mechanism for feeding data into the decision-making process supporting the sophisticated applications and services of a smart city. Moreover, the raw data collected from heterogeneous devices is often contaminated, so it is important to have a powerful data preprocessing mechanism.
	\item \textit{Data storage}: The data collected can be stored in a data center in the cloud or at the edge close to the data sources where the data can be readily processed in a near-real-time manner in support of applications and services requiring a low-latency, such as traffic light control for example. Moreover, the data stored in the cloud can be conveniently used in support of large scale and long-term applications. For example, data acquired from environmental sensors can be stored in the cloud and used for environmental monitoring applications.
	\item \textit{Data analytics}: The data can then be analyzed to extract useful information for supporting compelling applications and services in a smart city. Data analytics can be performed  either in a near-real-time or in an offline manner, depending on the specific requirements of the applications. There are numerous techniques that can be used for analyzing data in smart city applications. In this context machine learning techniques constitute powerful tools in big data analytics in the smart city \cite{5.13,5.76}. The authors of \cite{5.13} review a number of machine learning techniques in the context of data analytics as well as data security in a smart city.
	\item \textit{Data application}: The authorized personnel can then use applications in this layer to manage and monitor the smart city. These applications rely on the information extracted from the data analytics layer and help the local authority in making a global decision. There are numerous applications in the smart city including both real-time and non-real-time applications, such as smart lighting management, smart traffic light management in support of an intelligent transport system, environmental monitoring, garbage collection and waste processing management.
\end{itemize}


\subsection{Smart City Challenges}

Although 5G can bring about substantial benefits for smart cities, several major challenges have to be addressed for improving inhabitants' experiences.

\subsubsection{Security and Privacy}
Again, ubiquitous connectivity provides numerous benefits both for a smart city's citizens and for the local authority such as improving the inhabitants' quality of life with the aid of data acquisition. However, ubiquitous connectivity in the smart city, which involves a variety of components such as communication networks, connected devices, applications and services remains prone to security and privacy threats, which requires effective countermeasures. A smart city has to collect data from extremely diverse sources to operate, monitor, and manage its services. However, designing an efficient data acquisition and data processing system capable of securing its confidential user data to guarantee user privacy is crucial, since given the ubiquitous connectivity of 5G it becomes vulnerable to unauthorized access, disclosure, disruption, modification, inspection, and annihilation.

The security and privacy issues of for the smart city are investigated in \cite{5.3,5.4,5.13,5.15,5.16,5.24}. In these studies, the authors have discussed the challenges, enabling technologies and solutions conceived for ensuring the security and privacy of data in a smart city stored in support of its applications and services. Moreover, the cybersecurity level required by key sectors of smart cities including its government, critical infrastructure, healthcare, smart buildings and transportation must be studied. Thus, designing novel approaches that can guarantee the security and privacy of both the individuals and of important service sectors of the smart city will be of crucial importance in term of elevating the smart city concept.    

\subsubsection{Data Acquisition and Data Processing} The data collected throughout smart cities are extremely diverse and voluminous. For example, data gleaned from the transportation system and environmental monitoring are very different. Moreover, these data are generally in a raw format with much useless information obfuscating the desired data. Therefore, designing selective platforms for data acquisition and processing is critical for achieving accurate centralized control, operation management, and decision making. Further investigations are required for designing efficient platforms capable of processing the deluge of heterogeneous data generated by a smart city.

\section{Other Intelligent 5G-Aided Services in the Vertical Domains}

\subsection{Smart Healthcare}

Similar to the other aforementioned automation domains, 5G is capable of providing substantial benefits in term of smart healthcare or smart living \cite{6.3} by providing suitable communication platforms.
The following representative applications of smart healthcare can be facilitated with the aid of 5G technology.

\begin{itemize}
	\item \textit{Wearable health monitoring devices:} To support healthy living, wearable clinical tracking devices known as the Internet of Medical Things (IoMT) can be used for monitoring and analyzing patients' health in a real-time manner. The collected data can then be processed at a remote medical center and fed back to the patients complemented by appropriate medical recommendations. The communication among these entities has to be performed at a high data-integrity based on 5G networks \cite{6.1,6.2,6.10,6.11}.
	
	\item \textit{Secure remote consultations:} Remote consultations, diagnosis and even surgery relying on 2D/3D scanning of images, on live HD video streaming \cite{6.4}, and on advanced VR capability, can be carried out with the aid of tele-presence services at an extremely high data rate, low latency, and high reliability using 5G connections \cite{6.5}.
	
	\item \textit{Robot-assisted and fully robotic surgery:} Surgical robots are already widely adopted at the time of writing for improving the precision of surgery and operation quality \cite{6.7}. With the aid of advanced 5G communications surgical robots can be remotely controlled by skillful doctors, thus eliminating geo-distance and regional limitations \cite{6.6,6.7}.
	
\end{itemize}

\subsection{Smart Farming}

Smart farming constitutes a revolution in the field of agriculture based on advanced communications, IoT and cloud computing \cite{5.18}.
Specifically, by harnessing IoT devices for collecting, tracking, monitoring and analyzing information such as soil moisture, weather and levels of fertilization, smart farming, brings about a tremendous productivity improvement in this vertical sector by streamlining operations, whilst supporting sustainable development.
This is accomplished with the aid of the 5G technology, which conveys the collected information through 5G wireless IoT networks \cite{6.8,6.9}  for real-time processing and decision making concerning the land, crop, livestock, logistics, and machinery.

\section{Visions for 6G Wireless Systems}

%
%

\begin{figure}[!t]
	\centering
	\includegraphics[width=3.5in]{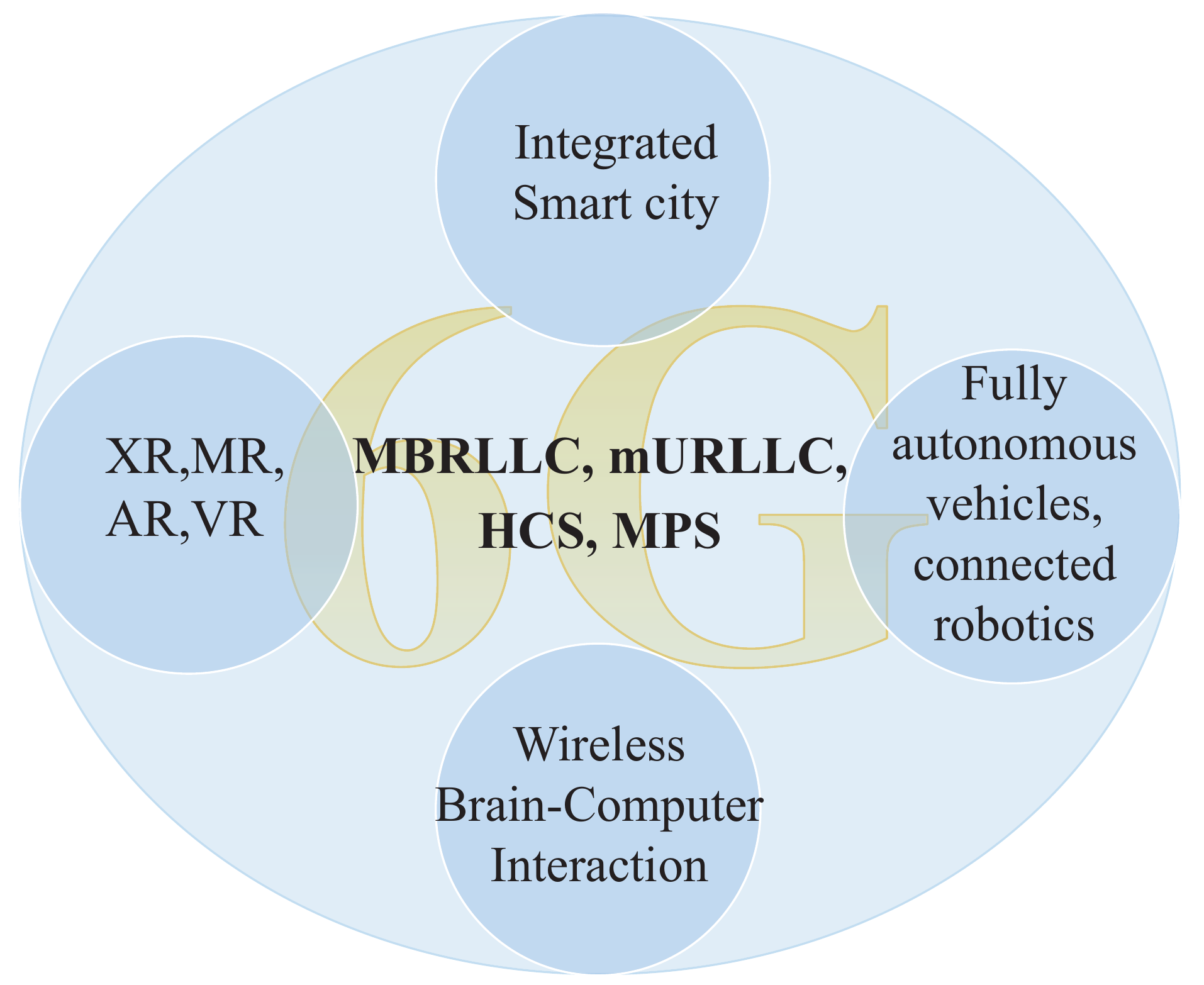}  
	\centering
	\caption{6G vision \cite{8.2}.}
	\label{fig:6G}
\end{figure}

\begin{table*}[htbp]
	\centering
	\caption{Requirements and features of 5G versus 6G \cite{8.1,8.2}. }
	\label{table:6GRequirements}
	{\renewcommand{\arraystretch}{1.4}
		\begin{tabular}{lll}
			\hline
			\textbf{Requirements} & \textbf{5G} & \textbf{6G} \\ \hline
			Service types & \begin{tabular}[c]{@{}l@{}} eMBB/mMTC/uRLLC\end{tabular} & \begin{tabular}[c]{@{}l@{}} MBRLLC/mURLLC/HCS/MPS\end{tabular} \\ \hline
			Individual data rate & 1 Gbps & 100 Gbps \\ \hline
			Peak DL data rate & 20 Gpbs & \textgreater 1 Tbps \\ \hline
			Latency & 5 ms & \textless 1 ms \\ \hline
			Mobility & 500 km/h & up to 1000 km/h \\ \hline
			Reliability & up to 99.9999 \% & up to 99.99999 \% \\ \hline
			Frequency bands & \begin{tabular}[c]{@{}l@{}}- sub-6GHz\\ - mmWave for fixed access $\leq$ 28 GHz\end{tabular} & \begin{tabular}[c]{@{}l@{}}- sub-6GHz\\ - mmWave for mobile access\\ - sub-THz band \\ - Non-RF, e.g, optical, VLC, laser...\end{tabular} \\ \hline
			Power consumption & Not defined & Ultra low \\ \hline
			Processing delay & $\leq$ 100 ns & $\leq$ 10 ns \\ \hline
			Network architecture & \begin{tabular}[c]{@{}l@{}}- Dense and ultra-dense of sub-6GHz\\ - $\leq$ 100-m mmWave cell\end{tabular} & \begin{tabular}[c]{@{}l@{}}- Cell-free smart surfaces at mmWave frequency\\ - Drone-based hotspots and tethered balloons\\ - Satellite networks\\ - Tiny THz cells\end{tabular} \\ \hline
			Security and privacy & High & Very high \\ \hline
			Network orientation & User-centric & Service-centric \\ \hline
			Wireless power transfer / Wireless charging & Not support & Support (BS to devices power transfer) \\ \hline
			Smart city components & Separate & Integrated \\ \hline
			Autonomous V2X & Partially & Fully \\ \hline
		\end{tabular}
	}
\end{table*}

\subsection{6G Vision}
Whilst the most sophisticated operating modes of the 5G systems are still being developed and deployed, their base-line modes are already operational. At the same time, the sixth generation (6G) mobile system starts attracting attentions since it is expected to overcome several challenges and limitation of 5G \cite{8.2}. Specifically, the unprecedented proliferation of new Internet of Everything (IoE) applications, such as eXtended reality (XR) services offered for connected autonomous systems including autonomous aerial vehicles (AAV) (UAV \cite{8.8,8.9,8.1}). None of the URLLC, eMBB, and mMTC 5G modes is capable of satisfying these applications, which simultaneously require  high reliability, low latency and ultrahigh data rates, both for the uplink and downlink \cite{8.2}.

The 6G system may be designed to support the aforementioned applications by satisfying either to unprecedented performance targets of microsecond latency and Terabit throughputs \cite{8.1,8.2,8.3}. These KPIs can be supported by new service classes, which may be expected to evolve from the URLLC, eMBB, and mMTC modes. The 6G system of the future may also support Mobile Broadband Reliable Low Latency Communication (MBRLLC), Massive URLLC (mURLLC), Human-Centric Services (HCS), and Multi-Purpose Communications, Computing, Control, Localization, and Sensing (3CLS) and Energy Services (MPS) \cite{8.2}. Specifically, the MBRLLC service type is expected to support stringent rate-reliability-latency requirements, energy efficiency, and rate-reliability-latency in mobile environments. The mURLLC service type supports ultra high reliability, massive connectivity and extreme reliability. The HSC service type supports demanding quality-of-physical-experience (QoPE) specifications by capturing raw wireless metrics as well as human and physical factors. The MPS service type supports stable control, low computing latency, high localization accuracy, as well as sensing and mapping accuracy \cite{8.2}. 

Building upon the 5G vision, 6G techniques relying on AI \cite{8.1,8.2,8.15} are expected to support multiple vertical domains in the smart society \cite{8.7} of the near future, smart city with its billion connected IoE devices \cite{8.1,8.2}, smart factory and Industry 4.0 \cite{8.1,8.13}, telemedicine or smart healthcare, connected autonomous systems \cite {8.2} including fully autonomous vehicles \cite{8.1}.

Several contributions have speculated on what 6G will be like, including new use cases \cite{8.1,8.5}, new service classes and applications \cite{8.2}, the performance requirements \cite{8.1,8.2,8.3}, and enabling technologies \cite{8.1,8.2,8.7}. In the following, we briefly introduce several enabling techniques that might find their way into the 6G system which can be potentially applied to CAV.

\subsection{6G Enabling Technologies}

To support the new 6G services and to guarantee their performance requirements, a range of promising solutions are proposed in \cite{8.1,8.2}. 

\subsubsection{AI Aided Holistic Systems Optimization}
Recently, numerous studies have involved AI for wireless communications, ranging from physical layer designs to resource allocation  \cite{8.1,8.2,8.15} (please see Section \ref{sec:AI} for details). Therefore, it is reasonable to argue that AI will play an important role in the optimization of the 6G system designed for example for fully autonomous (aerial) vehicles and for autonomous robots in the smart factory and in surgery.

\subsubsection{Terahertz (THz) Spectrum}
To satisfy the explosive growth of mobile data demands and bandwidth-hungry flawless multimedia applications in 6G, the terahertz (THz) band may be explored for short-range scenarios \cite{8.2,8.6,8.10,8.11,8.14,8.22,8.23,8.24}, where attocells  having a small radius will have to be used \cite{8.2}. 

\subsubsection{Visible Light Communication (VLC) - LiFi}
Visible light communication (VLC) uses light emitting diodes (LEDs) to transmit data wirelessly \cite{8.1,8.2,8.38}, which is also known as light-fidelity (LiFi) \cite{8.17,8.25}. These Li-Fi techniques are capable of achieving a downlink rate in excess of 3 Gbps from a single micro-LED \cite {8.17}, thus becoming promising candidates to meet the data rate requirement of 6G, especially when the LED-bandwidth is improved. VLC will also be useful in V2V communications, and in scenarios in which traditional RF communication is less efficient, such as underwater communication, in-cabin communications of airplanes, etc. \cite{8.1}.

\subsubsection{Cell-Free Networks}
Ubiquitous seamless connectivity/coverage can be obtained by cell-free techniques in which multiple BSs form a cooperative group/region to serve a group of mobile terminals. This technique can be implemented by adopting an SDN controller to support multiple BSs, which cooperatively serve a group of mobile users \cite{5.10}, or by using the cell-free massive MIMO technique of \cite{8.32}, or by employing the distributed coordinated multipoint transmission philosophy relying on non-orthogonal multiple access (NOMA) \cite{8.36,8.39,8.40}. The cell-free technique is capable of significantly improving the network throughput in comparison to the conventional cellular technique \cite{8.32}, hence it is a promising solution for the 6G system.

\subsubsection{Wireless Power Transfer and Energy Harvesting}
Even though numerous researchers have studied wireless power transfer and energy harvesting, they are insufficiently nature for 5G deployment.
However, in a 6G network, the communication distance in attocells will be much shorter than in 5G. Therefore, it becomes more realistic to rely on power transfer from 6G BSs to connected devices \cite{8.1,8.2}. 
Moreover, energy can also be harvested from ambient RF signals and renewable sources for low-power applications.

\begin{table}[!t]
	\centering
	\caption{6G enabling technologies and research topics}
	\label{table:6G_topic}
	{\renewcommand{\arraystretch}{1.5}
		\begin{tabular}{ll}
			\hline
			\textbf{Research topics} & \textbf{References} \\ \hline
			\begin{tabular}[c]{@{}l@{}}AI integrated\end{tabular} & \cite{8.1,8.2,8.15} \\ \hline
			\begin{tabular}[c]{@{}l@{}}Terahertz (THz) spectrum\end{tabular} & \cite{8.2,8.6,8.10,8.11,8.14,8.22,8.23,8.24} \\ \hline
			\begin{tabular}[c]{@{}l@{}}Visible light communication (VLC) \\ and LiFi\end{tabular} & \cite{8.1,8.2,8.17,8.25} \\ \hline
			\begin{tabular}[c]{@{}l@{}}Cell-free networks\end{tabular} & \cite{8.32,8.36} \\ \hline
			\begin{tabular}[c]{@{}l@{}}Wireless power transfer \\and energy harvesting\end{tabular} & \cite{8.1,8.2} \\ \hline
			\begin{tabular}[c]{@{}l@{}}Physical layer and waveform design\end{tabular} & \cite{8.16,8.33,8.34} \\ \hline
			\begin{tabular}[c]{@{}l@{}}Coexistence of RF and \\Li-Fi communications\end{tabular} & \cite{8.1,8.2} \\ \hline	
		\end{tabular}
	}
\end{table}

\subsection{6G Research Directions and Challenges}

While numerous studies have been dedicated to the physical layer design of the 5G system (see Section \ref{sec:EnlabingTechnologies}), physical layer design for the 6G system is considered to be key challenge \cite{8.4}.  For example, for THz transmission, new transceiver architectures are required to overcome the excessive  path-loss at THz frequencies. Another challenge in the physical design is the coexistence of heterogeneous THz, mmWave and microwave cells \cite{8.2}.
Moreover, radical new waveform design is required \cite{8.16,8.33,8.34}, which is the main component of any air interface.
The 6G system may also  witness the co-existence of RF and Li-Fi communication links \cite{8.1,8.2}.
In order to perform critical tasks in fully autonomous vehicles and robotic surgery relying on the 6G system, low-latency, high-reliability, and low-complexity AI algorithms have to be found by future \cite{8.1,8.2}.

Apart from the above evolutionary ideas of improving the the
performance of the 5G systems in the quest for defining a more
powerful 6G system, in \cite{9.3,9.4} the vision of integrating the
conventional terrestrial cellular system with both satellite systems
as well as with 'mobile base-stations' constituted by airplanes in the
sky was proposed.  Naturally, this ambitious concept requires
substantial further research both in terms of the physical-layer
enabling techniques as well as in terms of networking solutions.
Air-to-air communications may rely either on mm-wave solutions, since
there are substantial unused spectral bands in this frequency range or
on Free-space-optical (FSO) solutions \cite{9.5}.  The satellite-to-satellite
links may also rely on FSO communications, which facilitates
high-bandwidth information exchange. The optimization of this system
poses substantial challenges, because instead of simply minimizing
a single metric of the system, such as the BER, the delay, the power
consumption or the complexity, multi-component Pareto-optimization
will gradually become the norm, as predicted in \cite{9.3}.

\subsubsection{VLC Systems}
In the physical layer of VLC systems conventionally asymmetrically clipped optical OFDM (ACO-OFDM) has been advocated OFDM for Optical Communications \cite{9.6}.
However, Layered asymmetrically clipped optical orthogonal frequency division multiplexing (LACO-OFDM) \cite{9.7} is capable of improving the spectral efficiency of ACO-OFDM, where
multiple orthogonal frequency domain layers are sequentially
superimposed to form LACO-OFDM, with each superimposed layer filling
the empty subcarriers left by the previous layer. In \cite{9.8} the closed-form bit error ratio (BER) considering the effect of thermal noise, clipping distortion, inter-layer interference and the
bit rate difference between layers is analysed. Given that the BER
performance of LACO-OFDM is closely related to its peak-to-average
power ratio (PAPR) distribution, the authors also provided the
analytical expression of the PAPR. Furthermore, in \cite{9.9,9.10} sophisticated near-capacity coding schemes were designed for VLC systems. The associated networking aspects were investigated in \cite{9.11,9.12}. 
As another sophisticated VLC solution, Hierarchical
Colour-Shift-Keying Aided Layered Video Streaming was also conceived
for the Visible Light Downlink in \cite{9.13}. As another powerful solution, near-capacity
three-stage concatenated coding was conceived in \cite{9.14}. As an attractive application of VLC, video streaming was proposed in \cite{9.15}.

\subsubsection{Index Modulation}
In classic OFDM the widely known transmission philosophy is that we
transmit for example 1 bit/subcarrier using BPSK in a subblock of 4
subcarriers, which hence results in a total of 4 bits/subblock. By
contrast, in index-modulation-aided OFDM we only activate for example
one of the four subcarriers and the resultant index of the specific
activation pattern allows us to transmit 2 extra bits, yielding a
total of 3 bits/subcarrier block. In order to match the 4 bit/subblock
throughput of the conventional OFDM, we have to use 2 bit/subcarrier
QPSK on the activated subcarrier. Fortunately, we can indeed afford to
use the extra power we saved by only activating 1 out of 4
subcarriers.
Although index modulation has a 50-year history \cite{9.16}, as detailed in this tutorial/survey paper, it has only gain popularity in recent years. This is partly, because intially it was
shown to have a somewhat limited gain in \cite{9.17}
for low-order modulation schemes, such as BPSK. However, given that
only a fraction of the OFDM subcarriers is activated in a subblock,
we arrive at sparsely populated subblocks, which hence lend themselves
to compressed sensing, as detailed in \cite{9.18},  which results in very substantial performance benefits.

\subsubsection{Spatial Modulation}
The spatial modulation philosophy is quite similar to the
above-mentioned IM regime, but these principles are applied to the
antennas, rather than to the OFDM subcarriers. More explicitly, we can
activate one out of four MIMO elements and implicitly convey two extra
bits by detecting, which specific antennas were activated. However,
they exhibit an extra benefit over IM, because in the above example
we would only need a single RF chain, which is the most expensive
part of a transceiver \cite{9.19,9.20,9.21,9.22}.

\subsubsection{Full-Duplex Transceivers}
Conventional transceivers are unable to transmit and receive
simultaneously, because the received signal typically has a much lower
power than that of the transmit signal.  Since the high-power
transmitter amplifier does not have a perfectly linear amplifier
characteristic, it imposes non-linear distortion, which manifests
itself in terms of 3rd and 5th-order upper harmonics, which saturate
the receiver's automatic gain control circuit. Nonetheless,
based on the radical recent advances in signal processing, a
range of analogue and digital self-interference techniques
emerged, which bring full-duplex communications within reach
during the 6G research cycle \cite{9.23,9.24}.

\section{Conclusion}
In this survey, we have presented a research outlook concerning the potential applications of 5G technologies in multiple vertical domains from a communication perspective.
Various use cases of the smart factory, smart vehicles, smart grid, and of the smart city are expected to  benefit from the emerging high-performance of 5G technologies.
We have briefly touched upon eMBB, uRLLC and mMTC 5G modes. Subsequently, we have provided a detailed description on the associated enabling technologies behind these 5G services types. We have then surveyed the current state-of-the-art concerning these enabling technologies.
Next, we have discussed how 5G can support multiple vertical domains including the smart factory, smart vehicles, smart grid, and the smart city. We have also listed some major challenges and identified possible future research directions for each domain. Moreover, we have highlighted several benefits 5G can offer for smart healthcare and smart farming.
Finally, we have briefly introduced our vision of the 6G wireless network, which can overcome limitations of the 5G system.


%

%
%

\ifCLASSOPTIONcaptionsoff
  \newpage
\fi



%

\bibliographystyle{IEEEtran}  
\bibliography{Survey_5G}

\begin{thebibliography}{100}
\providecommand{\url}[1]{#1}
\csname url@samestyle\endcsname
\providecommand{\newblock}{\relax}
\providecommand{\bibinfo}[2]{#2}
\providecommand{\BIBentrySTDinterwordspacing}{\spaceskip=0pt\relax}
\providecommand{\BIBentryALTinterwordstretchfactor}{4}
\providecommand{\BIBentryALTinterwordspacing}{\spaceskip=\fontdimen2\font plus
\BIBentryALTinterwordstretchfactor\fontdimen3\font minus
  \fontdimen4\font\relax}
\providecommand{\BIBforeignlanguage}[2]{{%
\expandafter\ifx\csname l@#1\endcsname\relax
\typeout{** WARNING: IEEEtran.bst: No hyphenation pattern has been}%
\typeout{** loaded for the language `#1'. Using the pattern for}%
\typeout{** the default language instead.}%
\else
\language=\csname l@#1\endcsname
\fi
#2}}
\providecommand{\BIBdecl}{\relax}
\BIBdecl

\bibitem{9.2}
\BIBentryALTinterwordspacing
3GPP, ``{3GPP Release 16},'' Jul. 2018. [Online]. Available:
  \url{http://www.3gpp.org/release-16}
\BIBentrySTDinterwordspacing

\bibitem{9.1}
\BIBentryALTinterwordspacing
ABI-Research, ``{5G} is standardized. now what?'' Jun. 2018. [Online].
  Available: \url{https://www.abiresearch.com/blogs/5g-standardized-now-what/}
\BIBentrySTDinterwordspacing

\bibitem{1.1}
\BIBentryALTinterwordspacing
3GPP, ``Study on communication for automation in vertical domains (release
  16),'' in \emph{TR 22.804 V16.2.0}, Dec. 2018. [Online]. Available:
  \url{https://portal.3gpp.org/desktopmodules/Specifications/SpecificationDetails.aspx?specificationId=3187}
\BIBentrySTDinterwordspacing

\bibitem{2.8}
5G-ACIA, ``5g for connected industries and automation,'' in \emph{{5G}-ACIA
  White paper}.\hskip 1em plus 0.5em minus 0.4em\relax 5G-ACIA, Nov. 2018.

\bibitem{3.93}
\BIBentryALTinterwordspacing
5GCAR, ``{WP4: 5G} {V2X} system and architecture.'' [Online]. Available:
  \url{https://5gcar.eu/?page_id=248}
\BIBentrySTDinterwordspacing

\bibitem{5.12}
\BIBentryALTinterwordspacing
5G-PPP, ``{5GCity} project: Use cases for smart cities to improve the {5G}
  ecosystem using a strong partnership between industrial and research area,''
  in \emph{IEEE {5G} SUMMIT Trento}.\hskip 1em plus 0.5em minus 0.4em\relax
  IEEE, Mar. 2018. [Online]. Available:
  \url{https://www.5gcity.eu/wp-content/uploads/2018/04/Trento-IEEE-2018.pdf}
\BIBentrySTDinterwordspacing

\bibitem{3.67}
\BIBentryALTinterwordspacing
------, ``White paper on automotive vertical sector,'' in \emph{5G-PPP White
  Paper}.\hskip 1em plus 0.5em minus 0.4em\relax 5G-PPP, Oct. 2015. [Online].
  Available:
  \url{https://5g-ppp.eu/wp-content/uploads/2014/02/5G-PPP-White-Paper-on-Automotive-Vertical-Sectors.pdf}
\BIBentrySTDinterwordspacing

\bibitem{3.1}
J.~Wang, J.~Liu, and N.~Kato, ``Networking and communications in autonomous
  driving: A survey,'' \emph{IEEE Commun. Surveys Tuts.}, pp. 1--1,
  2018.

\bibitem{2.57}
K.~{Islam}, W.~{Shen}, and X.~{Wang}, ``Wireless sensor network reliability and
  security in factory automation: A survey,'' \emph{IEEE Trans. Systems, Man, and Cybernetics, Part C (Applications and Reviews)}, vol.~42,
  no.~6, pp. 1243--1256, Nov 2012.

\bibitem{fang2012smart}
X.~Fang, S.~Misra, G.~Xue, and D.~Yang, ``Smart grid—the new and improved
  power grid: A survey,'' \emph{IEEE Commun. Surveys \& Tuts}, vol.~14, no.~4,
  pp. 944--980, Dec. 2012.

\bibitem{5.13}
A.~{Gharaibeh}, M.~A. {Salahuddin}, S.~J. {Hussini}, A.~{Khreishah},
  I.~{Khalil}, M.~{Guizani}, and A.~{Al-Fuqaha}, ``Smart cities: A survey on
  data management, security, and enabling technologies,'' \emph{IEEE
  Commun. Surveys Tuts.}, vol.~19, no.~4, pp. 2456--2501,
  Fourthquarter 2017.

\bibitem{5.14}
S.~{Djahel}, R.~{Doolan}, G.~{Muntean}, and J.~{Murphy}, ``A
  communications-oriented perspective on traffic management systems for smart
  cities: Challenges and innovative approaches,'' \emph{IEEE Commun. Surveys Tuts.}, vol.~17, no.~1, pp. 125--151, Firstquarter 2015.

\bibitem{5.15}
D.~{Eckhoff} and I.~{Wagner}, ``Privacy in the smart city—applications,
  technologies, challenges, and solutions,'' \emph{IEEE Commun. Surveys Tuts.}, vol.~20, no.~1, pp. 489--516, Firstquarter 2018.

\bibitem{5.16}
M.~{Sookhak}, H.~{Tang}, Y.~{He}, and F.~R. {Yu}, ``Security and privacy of
  smart cities: A survey, research issues and challenges,'' \emph{IEEE
  Commun. Surveys Tuts.}, pp. 1--1, 2018.

\bibitem{1.6}
\BIBentryALTinterwordspacing
Huawei, ``{5G} network architecture white paper,'' in \emph{Huawei technologies
  co., ltd. White paper}, 2016. [Online]. Available:
  \url{https://www.huawei.com/minisite/hwmbbf16/insights/5G-Nework-Architecture-Whitepaper-en.pdf}
\BIBentrySTDinterwordspacing

\bibitem{1.34}
\BIBentryALTinterwordspacing
ITU-R, ``Minimum requirements related to technical performance for {IMT}-2020
  radio interface(s),'' in \emph{Report ITU-R M.2410-0 (11/2017)}.\hskip 1em
  plus 0.5em minus 0.4em\relax International Telecommunications Union
  Radiocommunication Sector, Nov 2017. [Online]. Available:
  \url{https://www.itu.int/pub/R-REP-M.2410-2017}
\BIBentrySTDinterwordspacing

\bibitem{1.4}
A.~Al-Fuqaha, M.~Guizani, M.~Mohammadi, M.~Aledhari, and M.~Ayyash, ``Internet
  of things: A survey on enabling technologies, protocols, and applications,''
  \emph{IEEE Commun. Surveys Tuts.}, vol.~17, no.~4, pp. 2347--2376,
  Fourthquarter 2015.

\bibitem{1.5}
M.~Agiwal, A.~Roy, and N.~Saxena, ``Next generation {5G} wireless networks: A
  comprehensive survey,'' \emph{IEEE Commun. Surveys Tuts.},
  vol.~18, no.~3, pp. 1617--1655, thirdquarter 2016.

\bibitem{1.7}
P.~Popovski, K.~F. Trillingsgaard, O.~Simeone, and G.~Durisi, ``{5G} wireless
  network slicing for {eMBB, URLLC, and mMTC}: A communication-theoretic
  view,'' \emph{IEEE Access}, vol.~6, pp. 55\,765--55\,779, 2018.

\bibitem{1.8}
A.~Anand, G.~D. Veciana, and S.~Shakkottai, ``Joint scheduling of {URLLC and
  eMBB} traffic in {5G} wireless networks,'' in \emph{IEEE INFOCOM 2018 - IEEE
  Conference on Computer Communications}, Honolulu, HI, US, April 2018, pp.
  1970--1978.

\bibitem{1.9}
S.~Lien, S.~Hung, D.~Deng, and Y.~J. Wang, ``Efficient ultra-reliable and low
  latency communications and massive machine-type communications in {5G} new
  radio,'' in \emph{GLOBECOM 2017 - 2017 IEEE Global Communications
  Conference}, Singapore, Dec 2017, pp. 1--7.

\bibitem{1.10}
P.~Popovski, J.~J. Nielsen, C.~Stefanovic, E.~d.~Carvalho, E.~Strom, K.~F.
  Trillingsgaard, A.~Bana, D.~M. Kim, R.~Kotaba, J.~Park, and R.~B. Sorensen,
  ``Wireless access for ultra-reliable low-latency communication: Principles
  and building blocks,'' \emph{IEEE Network}, vol.~32, no.~2, pp. 16--23, Mar.
  2018.

\bibitem{1.11}
M.~Shafi, A.~F. Molisch, P.~J. Smith, T.~Haustein, P.~Zhu, P.~D. Silva,
  F.~Tufvesson, A.~Benjebbour, and G.~Wunder, ``{5G}: A tutorial overview of
  standards, trials, challenges, deployment, and practice,'' \emph{IEEE Journal
  on Selected Areas in Communications}, vol.~35, no.~6, pp. 1201--1221, Jun.
  2017.

\bibitem{1.13}
H.~Ji, S.~Park, J.~Yeo, Y.~Kim, J.~Lee, and B.~Shim, ``Introduction to ultra
  reliable and low latency communications in {5G},'' \emph{Computing Research
  Repository (CoRR) abs/1704.05565}, 2017.

\bibitem{1.14}
C.-P. Li, J.~Jiang, W.~Chen, T.~Ji, and J.~Smee, ``{5G} ultra-reliable and
  low-latency systems design,'' in \emph{2017 European Conference on Networks
  and Communications (EuCNC)}, Oulu, Finland, Jun. 2017, pp. 1--5.

\bibitem{1.15}
A.~A. Esswie and K.~I. Pedersen, ``Opportunistic spatial preemptive scheduling
  for {URLLC and eMBB} coexistence in multi-user {5G} networks,'' \emph{IEEE
  Access}, vol.~6, pp. 38\,451--38\,463, 2018.

\bibitem{1.16}
N.~H. Mahmood, M.~Lopez, D.~Laselva, K.~Pedersen, and G.~Berardinelli,
  ``Reliability oriented dual connectivity for {URLLC} services in {5G} new
  radio,'' in \emph{2018 15th International Symposium on Wireless Communication
  Systems (ISWCS)}, Lisbon, Portugal, Aug. 2018, pp. 1--6.

\bibitem{1.18}
Z.~Li, M.~A. Uusitalo, H.~Shariatmadari, and B.~Singh, ``{5G} {URLLC}: Design
  challenges and system concepts,'' in \emph{2018 15th International Symposium
  on Wireless Communication Systems (ISWCS)}, Lisbon, Portugal, Aug. 2018, pp.
  1--6.

\bibitem{1.19}
B.~Chang, ``{URLLC} design for real-time control in wireless control systems,''
  in \emph{2018 IEEE {5G} World Forum {(5GWF)}}, Silicon Valley, CA, USA, Jul.
  2018, pp. 437--439.

\bibitem{1.20}
G.~Pocovi, K.~I. Pedersen, and P.~Mogensen, ``Joint link adaptation and
  scheduling for {5G} ultra-reliable low-latency communications,'' \emph{IEEE
  Access}, vol.~6, pp. 28\,912--28\,922, 2018.

\bibitem{1.22}
A.~Karimi, K.~I. Pedersen, N.~H. Mahmood, J.~Steiner, and P.~Mogensen, ``{5G}
  centralized multi-cell scheduling for {URLLC}: Algorithms and system-level
  performance,'' \emph{IEEE Access}, vol.~6, pp. 72\,253--72\,262, 2018.

\bibitem{1.36}
H.~Ji, S.~Park, J.~Yeo, Y.~Kim, J.~Lee, and B.~Shim, ``Ultra-reliable and
  low-latency communications in {5G} downlink: Physical layer aspects,''
  \emph{IEEE Wireless Communications}, vol.~25, no.~3, pp. 124--130, 2018.

\bibitem{1.59}
M.~Bennis, M.~Debbah, and H.~V. Poor., ``Ultra reliable and low-latency
  wireless communication: Tail, risk, and scale,'' \emph{Proceedings of the
  IEEE}, vol. 106, no.~10, pp. 1834--1853, 2018.

\bibitem{1.60}
G.~J. Sutton, J.~Zeng, R.~P. Liu, W.~Ni, D.~N. Nguyen, B.~A. Jayawickrama,
  X.~Huang, M.~Abolhasan, and Z.~Zhang, ``Enabling ultra-reliable and
  low-latency communications through unlicensed spectrum,'' \emph{IEEE
  Network}, vol.~32, no.~2, pp. 70--77, 2018.

\bibitem{1.61}
A.~Bana, K.~F. Trillingsgaard, P.~Popovski, and E.~de~Carvalho, ``Short packet
  structure for ultra-reliable machine-type communication: Tradeoff between
  detection and decoding,'' in \emph{2018 IEEE Inter. Conf. on
  Acoustics, Speech and Signal Processing (ICASSP)}, Calgary, AB, Canada, April
  2018, pp. 6608--6612.

\bibitem{1.62}
M.~Iwabuchi, A.~Benjebbour, Y.~Kishiyama, G.~Ren, C.~Tang, T.~Tian, L.~Gu,
  T.~Takada, and T.~Kashima, ``{5G} field experimental trials on {URLLC} using
  new frame structure,'' in \emph{2017 IEEE Globecom Workshops (GC Wkshps)},
  Singapore, Dec 2017, pp. 1--6.

\bibitem{1.66}
C.~Wang, Y.~Chen, Y.~Wu, and L.~Zhang, ``Performance evaluation of grant-free
  transmission for uplink {URLLC} services,'' in \emph{2017 IEEE 85th Vehicular
  Technology Conference (VTC Spring)}, Sydney, NSW, Australia, June 2017, pp.
  1--6.

\bibitem{1.67}
T.~Jacobsen, R.~Abreu, G.~Berardinelli, K.~Pedersen, P.~Mogensen, I.~Z. Kovacs,
  and T.~K. Madsen, ``System level analysis of uplink grant-free transmission
  for {URLLC},'' in \emph{2017 IEEE Globecom Workshops (GC Wkshps)}, Singapore,
  Dec 2017, pp. 1--6.

\bibitem{1.68}
Z.~Zhou, R.~Ratasuk, N.~Mangalvedhe, and A.~Ghosh, ``Resource allocation for
  uplink grant-free ultra-reliable and low latency communications,'' in
  \emph{2018 IEEE 87th Veh. Tech. Conference (VTC Spring)}, Porto,
  Portugal, Jun. 2018, pp. 1--5.

\bibitem{1.69}
R.~Abreu, T.~Jacobsen, G.~Berardinelli, K.~Pedersen, I.~Z. Kovács, and
  P.~Mogensen, ``Power control optimization for uplink grant-free {URLLC},'' in
  \emph{2018 IEEE Wireless Communications and Networking Conference (WCNC)},
  Barcelona, Spain, Apr. 2018, pp. 1--6.

\bibitem{1.70}
G.~Berardinelli, N.~H. Mahmood, R.~Abreu, T.~Jacobsen, K.~Pedersen, I.~Z.
  Kovács, and P.~Mogensen, ``Reliability analysis of uplink grant-free
  transmission over shared resources,'' \emph{IEEE Access}, vol.~6, pp.
  23\,602--23\,611, 2018.

\bibitem{1.72}
G.~Pocovi, H.~Shariatmadari, G.~Berardinelli, K.~Pedersen, J.~Steiner, and
  Z.~Li, ``Achieving ultra-reliable low-latency communications: Challenges and
  envisioned system enhancements,'' \emph{IEEE Network}, vol.~32, no.~2, pp.
  8--15, 2018.

\bibitem{1.77}
J.~Sachs, G.~Wikstrom, T.~Dudda, R.~Baldemair, and K.~Kittichokechai, ``{5G}
  radio network design for ultra-reliable low-latency communication,''
  \emph{IEEE Network}, vol.~32, no.~2, pp. 24--31, March 2018.

\bibitem{1.78}
C.~She, C.~Yang, and T.~Q.~S. Quek, ``Radio resource management for
  ultra-reliable and low-latency communications,'' \emph{IEEE Communications
  Magazine}, vol.~55, no.~6, pp. 72--78, June 2017.

\bibitem{1.79}
H.~Shariatmadari, S.~Iraji, R.~Jantti, P.~Popovski, Z.~Li, and M.~A. Uusitalo,
  ``Fifth-generation control channel design: Achieving ultrareliable
  low-latency communications,'' \emph{IEEE Veh. Tech. Mag.},
  vol.~13, no.~2, pp. 84--93, June 2018.

\bibitem{1.88}
\BIBentryALTinterwordspacing
G.~T.~R. 38.913, ``Study on scenarios and requirements for next generation
  access technologies (release 14),'' v14.2.0 2017. [Online]. Available:
  \url{http://www.3gpp.org/ftp//Specs/archive/38_series/38.913/38913-e20.zip}
\BIBentrySTDinterwordspacing

\bibitem{1.39}
G.~R. 14, ``Study on new radio access technology physical layer aspects
  (release 14),'' in \emph{3GPP Tech. Rep. 38.802}, 2017.

\bibitem{1.154}
S.~{Lien}, S.~{Shieh}, Y.~{Huang}, B.~{Su}, Y.~{Hsu}, and H.~{Wei}, ``{5G} new
  radio: Waveform, frame structure, multiple access, and initial access,''
  \emph{IEEE Commun. Mag.}, vol.~55, no.~6, pp. 64--71, June 2017.

\bibitem{1.90}
P.~Schulz, M.~Matthe, H.~Klessig, M.~Simsek, G.~Fettweis, J.~Ansari, S.~A.
  Ashraf, B.~Almeroth, J.~Voigt, I.~Riedel, A.~Puschmann, A.~Mitschele-Thiel,
  M.~Muller, T.~Elste, and M.~Windisch, ``Latency critical iot applications in
  {5G}: Perspective on the design of radio interface and network
  architecture,'' \emph{IEEE Commun. Mag.}, vol.~55, no.~2, pp.
  70--78, February 2017.

\bibitem{6.5}
J.~{Sachs}, L.~A.~A. {Andersson}, J.~{Araújo}, C.~{Curescu}, J.~{Lundsjö},
  G.~{Rune}, E.~{Steinbach}, and G.~{Wikström}, ``Adaptive {5G} low-latency
  communication for tactile internet services,'' \emph{Proceedings of the
  IEEE}, vol. 107, no.~2, pp. 325--349, Feb 2019.

\bibitem{1.29}
D.~Demmer, R.~Gerzaguet, J.~Dore, and D.~L. Ruyet, ``Analytical study of {5G}
  nr {eMBB} co-existence,'' in \emph{2018 25th Inter. Conf. on
  Telecommunications (ICT)}, St. Malo, France, 2018, pp. 186--190.

\bibitem{1.153}
J.~{Zhang}, Z.~{Zheng}, Y.~{Zhang}, J.~{Xi}, X.~{Zhao}, and G.~{Gui}, ``3d
  {MIMO} for {5G} {NR}: Several observations from 32 to massive 256 antennas
  based on channel measurement,'' \emph{IEEE Commun. Mag.}, vol.~56,
  no.~3, pp. 62--70, March 2018.

\bibitem{1.38}
S.~Han, C.~I, Z.~Xu, and C.~Rowell, ``Large-scale antenna systems with hybrid
  analog and digital beamforming for millimeter wave {5G},'' \emph{IEEE
  Communications Magazine}, vol.~53, no.~1, pp. 186--194, 2015.

\bibitem{1.43}
M.~Xiao, S.~Mumtaz, Y.~Huang, L.~Dai, Y.~Li, M.~Matthaiou, G.~K. Karagiannidis,
  E.~Björnson, K.~Yang, C.~I, and A.~Ghosh, ``Millimeter wave communications
  for future mobile networks,'' \emph{IEEE J.  Sel. Areas Commun.}, vol.~35, no.~9, pp. 1909--1935, 2017.

\bibitem{1.44}
S.~A. Busari, K.~M.~S. Huq, S.~Mumtaz, L.~Dai, and J.~Rodriguez,
  ``Millimeter-wave massive {MIMO} communication for future wireless systems: A
  survey,'' \emph{IEEE Commun. Surveys Tuts.}, vol.~20, no.~2, pp.
  836--869, 2018.

\bibitem{1.45}
X.~Wang, L.~Kong, F.~Kong, F.~Qiu, M.~Xia, S.~Arnon, and G.~Chen, ``Millimeter
  wave communication: A comprehensive survey,'' \emph{IEEE Commun. Surveys Tuts.}, vol.~20, no.~3, pp. 1616--1653, 2018.

\bibitem{1.46}
I.~A. Hemadeh, K.~Satyanarayana, M.~El-Hajjar, and L.~Hanzo, ``Millimeter-wave
  communications: Physical channel models, design considerations, antenna
  constructions, and link-budget,'' \emph{IEEE Commun. Surveys Tuts.}, vol.~20, no.~2, pp. 870--913, 2018.

\bibitem{1.86}
B.~Wang, L.~Dai, Z.~Wang, N.~Ge, and S.~Zhou, ``Spectrum and energy-efficient
  beamspace {MIMO}-{NOMA} for millimeter-wave communications using lens antenna
  array,'' \emph{IEEE J.  Sel. Areas Commun.}, vol.~35,
  no.~10, pp. 2370--2382, Oct 2017.

\bibitem{1.12}
N.~H. Mahmood, M.~Lauridsen, G.~Berardinelli, D.~Catania, and P.~Mogensen,
  ``Radio resource management techniques for {eMBB and mMTC} services in {5G}
  dense small cell scenarios,'' in \emph{2016 IEEE 84th Veh. Tech.
  Conference (VTC-Fall)}, Montreal, QC, Canada, Sep. 2016, pp. 1--5.

\bibitem{1.17}
B.~A. Jayawickrama, Y.~He, E.~Dutkiewicz, and M.~D. Mueck, ``Scalable spectrum
  access system for massive machine type communication,'' \emph{IEEE Network},
  vol.~32, no.~3, pp. 154--160, May 2018.

\bibitem{1.21}
C.~Bockelmann, N.~K. Pratas, G.~Wunder, S.~Saur, M.~Navarro, D.~Gregoratti,
  G.~Vivier, E.~D. Carvalho, Y.~Ji, .~Stefanović, P.~Popovski, Q.~Wang,
  M.~Schellmann, E.~Kosmatos, P.~Demestichas, M.~Raceala-Motoc, P.~Jung,
  S.~Stanczak, and A.~Dekorsy, ``Towards massive connectivity support for
  scalable {mMTC} communications in {5G} networks,'' \emph{IEEE Access},
  vol.~6, pp. 28\,969--28\,992, 2018.

\bibitem{1.23}
T.~Huang, Y.~Ren, K.~C. Lin, and Y.~Tseng, ``r-hint: A message-efficient random
  access response for {mMTC} in {5G} networks,'' in \emph{2017 IEEE 28th Annual
  International Symposium on Personal, Indoor, and Mobile Radio Communications
  (PIMRC)}, Montreal, QC, Canada, 2017, pp. 1--6.

\bibitem{1.24}
N.~K. Pratas, S.~Pattathil, .~Stefanović, and P.~Popovski, ``Massive
  machine-type communication ({mMTC}) access with integrated authentication,''
  in \emph{2017 IEEE Inter. Conf. on Communications (ICC)}, Paris,
  2017, pp. 1--6.

\bibitem{1.25}
N.~Ye, A.~Wang, X.~Li, H.~Yu, A.~Li, and H.~Jiang, ``A random non-orthogonal
  multiple access scheme for {mMTC},'' in \emph{2017 IEEE 85th Vehicular
  Technology Conference (VTC Spring)}, Sydney, NSW, Australia, 2017, pp. 1--6.

\bibitem{1.26}
Z.~Li, J.~Chen, R.~Ni, S.~Chen, X.~Li, and Q.~Zhao, ``Enabling heterogeneous
  {mMTC} by energy-efficient and connectivity-aware clustering and routing,''
  in \emph{2017 IEEE Globecom Workshops (GC Wkshps)}, Singapore, 2017, pp.
  1--6.

\bibitem{1.27}
Y.~Yang, G.~Song, W.~Zhang, X.~Ge, and C.~Wang, ``Neighbor-aware multiple
  access protocol for {5G} {mMTC} applications,'' \emph{China Communications},
  vol.~13, no. Supplement2, pp. 80--88, 2016.

\bibitem{1.28}
Y.~Xu, Y.~Ren, K.~C. Lin, and Y.~Tseng, ``A hint-based random access protocol
  for {mMTC} in {5G} mobile network,'' in \emph{2018 IEEE 15th International
  Conference on Mobile Ad Hoc and Sensor Systems (MASS)}, Chengdu, China, 2018,
  pp. 388--396.

\bibitem{1.30}
W.~Cao, A.~Dytso, G.~Feng, H.~V. Poor, and Z.~Chen, ``Differentiated
  service-aware group paging for massive machine-type communication,''
  \emph{IEEE Trans. Commun.}, vol.~66, no.~11, pp. 5444--5456,
  2018.

\bibitem{1.31}
S.~Cioni, R.~D. Gaudenzi, O.~D.~R. Herrero, and N.~Girault, ``On the satellite
  role in the era of {5G} massive machine type communications,'' \emph{IEEE
  Network}, vol.~32, no.~5, pp. 54--61, 2018.

\bibitem{1.32}
S.~Park, H.~Seo, H.~Ji, and B.~Shim, ``Joint active user detection and channel
  estimation for massive machine-type communications,'' in \emph{2017 IEEE 18th
  International Workshop on Signal Processing Advances in Wireless
  Communications (SPAWC)}, Sapporo, Japan, 2017, pp. 1--5.

\bibitem{1.49}
C.~Bockelmann, N.~Pratas, H.~Nikopour, K.~Au, T.~Svensson, C.~Stefanovic,
  P.~Popovski, and A.~Dekorsy, ``Massive machine-type communications in {5G}:
  physical and mac-layer solutions,'' \emph{IEEE Commun. Mag.},
  vol.~54, no.~9, 2016.

\bibitem{1.52}
S.~Moon, H.~Lee, and J.~Lee, ``Sara: Sparse code multiple access-applied random
  access for iot devices,'' \emph{IEEE Internet  Things J.}, vol.~5,
  no.~4, pp. 3160--3174, 2018.

\bibitem{1.53}
F.~Wei, W.~Chen, Y.~Wu, J.~Li, and Y.~Luo, ``Toward {5G} wireless interface
  technology: Enabling nonorthogonal multiple access in the sparse code
  domain,'' \emph{IEEE Veh. Tech. Mag.}, vol.~13, no.~4, pp.
  18--27, 2018.

\bibitem{1.54}
C.~Bockelmann, H.~F. Schepker, and A.~Dekorsy, ``Compressive sensing based
  multi-user detection for machine-to-machine communication,''
  \emph{Trans. Emerging Telecommunications Technologies}, vol.~24,
  no.~4, pp. 389--400, 2013.

\bibitem{1.55}
B.~K. Jeong, B.~Shim, and K.~B. Lee, ``A compressive sensing-based active user
  and symbol detection technique for massive machine-type communications,'' in
  \emph{2018 IEEE Inter. Conf. on Acoustics, Speech and Signal
  Processing (ICASSP)}, 2018, pp. 6623--6627.

\bibitem{1.56}
W.~Chen and F.~Xiao, ``Optimized pilot design for joint compressive sensing
  multi-user and channel detection in massive mtc,'' in \emph{2017 IEEE
  Globecom Workshops (GC Wkshps)}, Singapore, 2017, pp. 1--5.

\bibitem{1.57}
J.~Ahn, B.~Shim, and K.~B. Lee, ``Expectation propagation-based active user
  detection and channel estimation for massive machine-type communications,''
  in \emph{2018 IEEE Inter. Conf. on Communications Workshops (ICC
  Workshops)}, Kansas City, MO, USA, 2018, pp. 1--6.

\bibitem{1.58}
F.~Monsees, M.~Woltering, C.~Bockelmann, and A.~Dekorsy, ``Compressive sensing
  multi-user detection for multicarrier systems in sporadic machine type
  communication,'' in \emph{2015 IEEE 81st Veh. Tech. Conference (VTC
  Spring)}, Glasgow, UK, 2015, pp. 1--5.

\bibitem{1.35}
J.~F.~M. A.~Osseiran and P.~Marsch, \emph{{5G} Mobile and Wireless
  Communications Technology}.\hskip 1em plus 0.5em minus 0.4em\relax New York,
  NY, USA: Cambridge University Press, 2016.

\bibitem{1.2}
\BIBentryALTinterwordspacing
3GPP, ``Feasibility study on new services and markets technology enablers for
  critical communications; stage 1 (release 14),'' in \emph{TR 22.862 V14.1.0},
  Oct. 2016. [Online]. Available:
  \url{https://portal.3gpp.org/desktopmodules/Specifications/SpecificationDetails.aspx?specificationId=3014}
\BIBentrySTDinterwordspacing

\bibitem{1.3}
\BIBentryALTinterwordspacing
------, ``Service requirements for next generation new services and markets
  (release 16),'' in \emph{TS 22.261 V16.6.0}, Dec. 2018. [Online]. Available:
  \url{https://portal.3gpp.org/desktopmodules/Specifications/SpecificationDetails.aspx?specificationId=3107}
\BIBentrySTDinterwordspacing

\bibitem{1.148}
\BIBentryALTinterwordspacing
ICT, ``Scenarios, requirements and {KPIs} for {5G} mobile and wireless
  system,'' in \emph{ICT-317669 METIS project, Deliverable D1.1}.\hskip 1em
  plus 0.5em minus 0.4em\relax ICT, Apr. 2013. [Online]. Available:
  \url{https://www.metis2020.com/wp-content/uploads/deliverables/METIS_D1.1_v1.pdf}
\BIBentrySTDinterwordspacing

\bibitem{1.149}
\BIBentryALTinterwordspacing
------, ``Updated scenarios, requirements and {KPIs} for {5G} mobile and
  wireless system with recommendations for future investigations,'' in
  \emph{ICT-317669 METIS project, Deliverable D1.5}.\hskip 1em plus 0.5em minus
  0.4em\relax ICT, Apr. 2015. [Online]. Available:
  \url{https://www.metis2020.com/wp-content/uploads/deliverables/METIS_D1.5_v1.pdf}
\BIBentrySTDinterwordspacing

\bibitem{3.66}
\BIBentryALTinterwordspacing
NGMN, ``{NGMN 5G} white paper,'' in \emph{{NGMN 5G} White paper}.\hskip 1em
  plus 0.5em minus 0.4em\relax NGMN Alliance, Feb. 2015. [Online]. Available:
  \url{https://www.ngmn.org/fileadmin/ngmn/content/images/news/ngmn_news/NGMN_5G_White_Paper_V1_0.pdf}
\BIBentrySTDinterwordspacing

\bibitem{1.119}
\BIBentryALTinterwordspacing
Ericsson, ``{5G} systems – enabling the transformation of industry and
  society,'' in \emph{Ericsson white paper}.\hskip 1em plus 0.5em minus
  0.4em\relax Ericsson, Jan. 2017. [Online]. Available:
  \url{https://www.ericsson.com/assets/local/publications/white-papers/wp-5g-systems.pdf}
\BIBentrySTDinterwordspacing

\bibitem{1.102}
\BIBentryALTinterwordspacing
{5G}-Americas, ``{5G} services and use cases,'' Nov. 2017. [Online]. Available:
  \url{http://www.5gamericas.org/files/3215/1190/8811/5G_Services_and_Use_Cases.pdf}
\BIBentrySTDinterwordspacing

\bibitem{3.24}
Z.~Amjad, A.~Sikora, B.~Hilt, and J.~Lauffenburger, ``Low latency {V2X}
  applications and network requirements: Performance evaluation,'' in
  \emph{2018 IEEE Intelligent Vehicles Symposium (IV)}, Changshu, China, June
  2018, pp. 220--225.

\bibitem{marsch2011coordinated}
P.~Marsch and G.~P. Fettweis, \emph{Coordinated Multi-Point in Mobile
  Communications: from theory to practice}.\hskip 1em plus 0.5em minus
  0.4em\relax Cambridge University Press, 2011.

\bibitem{1.89}
\BIBentryALTinterwordspacing
3GPP, ``Dual connectivity (release 16),'' 2019. [Online]. Available:
  \url{http://www.3gpp.org/DynaReport/FeatureOrStudyItemFile-800058.htm}
\BIBentrySTDinterwordspacing

\bibitem{7.1}
S.~M.~A. {Kazmi}, N.~H. {Tran}, T.~M. {Ho}, A.~{Manzoor}, D.~{Niyato}, and
  C.~S. {Hong}, ``Coordinated device-to-device communication with
  non-orthogonal multiple access in future wireless cellular networks,''
  \emph{IEEE Access}, vol.~6, pp. 39\,860--39\,875, 2018.

\bibitem{1.92}
D.~Moltchanov, A.~Ometov, S.~Andreev, and Y.~Koucheryavy, ``Upper bound on
  capacity of {5G} mmwave cellular with multi-connectivity capabilities,''
  \emph{Electronics Letters}, vol.~54, no.~11, pp. 724--726, 2018.

\bibitem{1.93}
V.~Petrov, D.~Solomitckii, A.~Samuylov, M.~A. Lema, M.~Gapeyenko,
  D.~Moltchanov, S.~Andreev, V.~Naumov, K.~Samouylov, M.~Dohler, and
  Y.~Koucheryavy, ``Dynamic multi-connectivity performance in ultra-dense urban
  mmwave deployments,'' \emph{IEEE J.  Sel. Areas Commun.}, vol.~35, no.~9, pp. 2038--2055, Sep. 2017.

\bibitem{1.94}
V.~Petrov, M.~A. Lema, M.~Gapeyenko, K.~Antonakoglou, D.~Moltchanov, F.~Sardis,
  A.~Samuylov, S.~Andreev, Y.~Koucheryavy, and M.~Dohler, ``Achieving
  end-to-end reliability of mission-critical traffic in softwarized {5G}
  networks,'' \emph{IEEE J.  Sel. Areas Commun.}, vol.~36,
  no.~3, pp. 485--501, March 2018.

\bibitem{1.95}
X.~Ba, Y.~Wang, H.~Hai, Y.~Chen, and Z.~Liu, ``Performance comparison of
  multi-connectivity with comp in {5G} ultra-dense network,'' in \emph{2018
  IEEE 87th Veh. Tech. Conference (VTC Spring)}, Porto, Portugal,
  2018, pp. 1--5.

\bibitem{1.96}
C.~She, Z.~Chen, C.~Yang, T.~Q.~S. Quek, Y.~Li, and B.~Vucetic, ``Improving
  network availability of ultra-reliable and low-latency communications with
  multi-connectivity,'' \emph{IEEE Trans. Commun.}, vol.~66,
  no.~11, pp. 5482--5496, Nov 2018.

\bibitem{1.97}
T.~Hößler, M.~Simsek, and G.~P. Fettweis, ``Mission reliability for {URLLC}
  in wireless networks,'' \emph{IEEE Commun. Letters}, vol.~22, no.~11,
  pp. 2350--2353, Nov 2018.

\bibitem{3.56}
\BIBentryALTinterwordspacing
5GPPP, ``A study on {5G} {V2X} deployment version 1.0,'' in \emph{{5G} PPP
  Automotive White Paper}.\hskip 1em plus 0.5em minus 0.4em\relax 5GPPP, Feb.
  2018. [Online]. Available:
  \url{https://5g-ppp.eu/wp-content/uploads/2018/02/5G-PPP-Automotive-WG-White-Paper_Feb.2018.pdf}
\BIBentrySTDinterwordspacing

\bibitem{2.13}
\BIBentryALTinterwordspacing
Ericsson, ``Opportunities in {5G}: The view from eight industries,'' in
  \emph{Ericsson White Paper}.\hskip 1em plus 0.5em minus 0.4em\relax Ericsson,
  2016. [Online]. Available:
  \url{https://app-eu.clickdimensions.com/blob/ericssoncom-ar0ma/files/5g_industry_survey_report_final.pdf}
\BIBentrySTDinterwordspacing

\bibitem{3.52}
P.~Koopman and M.~Wagner, ``Autonomous vehicle safety: An interdisciplinary
  challenge,'' \emph{IEEE Intelligent Transport. Systems Mag.}, vol.~9,
  no.~1, pp. 90--96, Spring 2017.

\bibitem{3.2}
\BIBentryALTinterwordspacing
5G-PPP, ``{5G} automotive vision,'' Oct. 2015. [Online]. Available:
  \url{https://5g-ppp.eu/wp-content/uploads/2014/02/5G-PPP-White-Paper-on-Automotive-Vertical-Sectors.pdf}
\BIBentrySTDinterwordspacing

\bibitem{3.15}
J.~Lianghai, A.~Weinand, B.~Han, and H.~D. Schotten, ``Multi-rats support to
  improve {V2X} communication,'' in \emph{2018 IEEE Wireless Commun.
  Netw. Conference (WCNC)}, April 2018, pp. 1--6.

\bibitem{3.17}
J.~Contreras-Castillo, S.~Zeadally, and J.~A. Guerrero-Ibañez, ``Internet of
  vehicles: Architecture, protocols, and security,'' \emph{IEEE Internet 
  Things J.}, vol.~5, no.~5, pp. 3701--3709, Oct 2018.

\bibitem{3.3}
\BIBentryALTinterwordspacing
3GPP, ``Technical specification group radio access network; study on
  {LTE}-based {V2X} services,'' in \emph{3GPP Release 14 TR 36.885 V14.0.0},
  Jun. 2016. [Online]. Available:
  \url{http://www.3gpp.org/ftp/Specs/archive/36_series/36.885/36885-e00.zip}
\BIBentrySTDinterwordspacing

\bibitem{3.4}
\BIBentryALTinterwordspacing
------, ``Architecture enhancements for {V2X} services,'' in \emph{3GPP Release
  15 TS 23.285 (2018-12)}, Dec. 2018. [Online]. Available:
  \url{https://portal.3gpp.org/desktopmodules/Specifications/SpecificationDetails.aspx?specificationId=3078}
\BIBentrySTDinterwordspacing

\bibitem{3.5}
\BIBentryALTinterwordspacing
------, ``Evolved universal terrestrial radio access (e-utra); user equipment
  (ue) radio transmission and reception,'' in \emph{3GPP Release 15 TS 36.101},
  Oct. 2018. [Online]. Available:
  \url{https://portal.3gpp.org/desktopmodules/Specifications/SpecificationDetails.aspx?specificationId=2411}
\BIBentrySTDinterwordspacing

\bibitem{3.6}
\BIBentryALTinterwordspacing
------, ``Study on architecture enhancements for eps and {5G} system to support
  advanced {V2X} services,'' in \emph{3GPP Release TR 23.786}, Dec. 2018.
  [Online]. Available:
  \url{https://portal.3gpp.org/desktopmodules/Specifications/SpecificationDetails.aspx?specificationId=3244}
\BIBentrySTDinterwordspacing

\bibitem{3.7}
\BIBentryALTinterwordspacing
------, ``Study on {NR} vehicle-to-everything ({V2X}),'' in \emph{3GPP Release
  TS 38.885}, Nov. 2018. [Online]. Available:
  \url{https://portal.3gpp.org/desktopmodules/Specifications/SpecificationDetails.aspx?specificationId=3497}
\BIBentrySTDinterwordspacing

\bibitem{3.8}
S.~Chen, J.~Hu, Y.~Shi, Y.~Peng, J.~Fang, R.~Zhao, and L.~Zhao,
  ``Vehicle-to-everything ({V2X}) services supported by {LTE}-based systems and
  {5G},'' \emph{IEEE Commun. Standards Mag.}, vol.~1, no.~2, pp.
  70--76, 2017.

\bibitem{1.101}
\BIBentryALTinterwordspacing
Qualcomm and Nokia, ``Making {5G} a reality: Addressing the strong mobile
  broadband demand in 2019 \& beyond,'' Sept. 2017. [Online]. Available:
  \url{https://www.qualcomm.com/media/documents/files/making-5g-a-reality-addressing-the-strong-mobile-broadband-demand-in-2019-beyond.pdf}
\BIBentrySTDinterwordspacing

\bibitem{1.98}
M.~Kamel, W.~Hamouda, and A.~Youssef, ``Ultra-dense networks: A survey,''
  \emph{IEEE Commun. Surveys Tuts.}, vol.~18, no.~4, pp. 2522--2545,
  Fourthquarter 2016.

\bibitem{1.99}
F.~Al-Turjman, E.~Ever, and H.~Zahmatkesh, ``Small cells in the forthcoming
  {5G}/iot: Traffic modelling and deployment overview,'' \emph{IEEE
  Commun. Surveys Tuts.}, pp. 1--1, 2018.

\bibitem{1.100}
\BIBentryALTinterwordspacing
S.~Vahid, R.~Tafazolli, and M.~Filo, \emph{Small Cells for {5G} Mobile
  Networks}.\hskip 1em plus 0.5em minus 0.4em\relax John Wiley \& Sons, Ltd,
  2015, ch.~3, pp. 63--104. [Online]. Available:
  \url{https://onlinelibrary.wiley.com/doi/abs/10.1002/9781118867464.ch3}
\BIBentrySTDinterwordspacing

\bibitem{7.2}
T.~M. {Ho}, N.~H. {Tran}, L.~{Le}, Z.~{Han}, S.~M.~A. {Kazmi}, and C.~S.
  {Hong}, ``Network virtualization with energy efficiency optimization for
  wireless heterogeneous networks,'' \emph{IEEE Trans. Mobile
  Comput.}, pp. 1--1, 2018.

\bibitem{7.3}
T.~M. {Ho}, N.~H. {Tran}, C.~T. {Do}, S.~M.~A. {Kazmi}, E.~{Huh}, and C.~S.
  {Hong}, ``Power control for interference management and {QoS} guarantee in
  heterogeneous networks,'' \emph{IEEE Commun. Letters}, vol.~19, no.~8,
  pp. 1402--1405, Aug 2015.

\bibitem{7.4}
S.~M. {Ahsan Kazmi}, N.~H. {Tran}, W.~{Saad}, L.~B. {Le}, T.~M. {Ho}, and C.~S.
  {Hong}, ``Optimized resource management in heterogeneous wireless networks,''
  \emph{IEEE Commun. Letters}, vol.~20, no.~7, pp. 1397--1400, July
  2016.

\bibitem{1.37}
H.~Ji, Y.~Kim, J.~Lee, E.~Onggosanusi, Y.~Nam, J.~Zhang, B.~Lee, and B.~Shim,
  ``Overview of {Full-Dimension} {MIMO} in {LTE-Advanced Pro},'' \emph{IEEE
  Communications Magazine}, vol.~55, no.~2, pp. 176--184, 2017.

\bibitem{1.47}
Q.~Nadeem, A.~Kammoun, M.~Debbah, and M.~Alouini, ``Design of {5G} full
  dimension massive {MIMO} systems,'' \emph{IEEE Trans.
  Commun.}, vol.~66, no.~2, pp. 726--740, 2018.

\bibitem{1.48}
G.~Xu, Y.~Li, J.~Yuan, R.~Monroe, S.~Rajagopal, S.~Ramakrishna, Y.~H. Nam,
  J.~Seol, J.~Kim, M.~M.~U. Gul, A.~Aziz, and J.~Zhang, ``Full dimension {MIMO}
  {(FD-MIMO)}: Demonstrating commercial feasibility,'' \emph{IEEE J.
  Sel. Areas Commun.}, vol.~35, no.~8, pp. 1876--1886, 2017.

\bibitem{1.73}
\BIBentryALTinterwordspacing
D.~Bharadia, E.~McMilin, and S.~Katti, ``Full duplex radios,'' \emph{SIGCOMM
  Comput. Commun. Rev.}, vol.~43, no.~4, pp. 375--386, Aug. 2013. [Online].
  Available: \url{http://doi.acm.org/10.1145/2534169.2486033}
\BIBentrySTDinterwordspacing

\bibitem{1.74}
T.~Riihonen and R.~Wichman, \emph{Full-Duplex Protocol Design for {5G}
  Networks}.\hskip 1em plus 0.5em minus 0.4em\relax Cambridge University Press,
  2017, p. 172–187.

\bibitem{1.75}
A.~Sabharwal, P.~Schniter, D.~Guo, D.~W. Bliss, S.~Rangarajan, and R.~Wichman,
  ``In-band full-duplex wireless: Challenges and opportunities,'' \emph{IEEE
  J.  Sel. Areas Commun.}, vol.~32, no.~9, pp. 1637--1652,
  2014.

\bibitem{1.76}
Y.~Sun, D.~W.~K. Ng, Z.~Ding, and R.~Schober, ``Optimal joint power and
  subcarrier allocation for full-duplex multicarrier non-orthogonal multiple
  access systems,'' \emph{IEEE Trans. Commun.}, vol.~65, no.~3,
  pp. 1077--1091, March 2017.

\bibitem{1.81}
Y.~Sun, D.~W.~K. Ng, J.~Zhu, and R.~Schober, ``Multi-objective optimization for
  robust power efficient and secure full-duplex wireless communication
  systems,'' \emph{IEEE Trans. Wireless Commun.}, vol.~15,
  no.~8, pp. 5511--5526, Aug 2016.

\bibitem{1.82}
Z.~Zhang, K.~Long, A.~V. Vasilakos, and L.~Hanzo, ``Full-duplex wireless
  communications: Challenges, solutions, and future research directions,''
  \emph{Proceedings of the IEEE}, vol. 104, no.~7, pp. 1369--1409, July 2016.

\bibitem{1.50}
Y.~Wu, C.~Wang, Y.~Chen, and A.~Bayesteh, ``Sparse code multiple access for
  {5G} radio transmission,'' in \emph{2017 IEEE 86th Veh. Tech.
  Conference (VTC-Fall)}, Toronto, ON, Canada, 2017, pp. 1--6.

\bibitem{1.51}
M.~Moltafet, N.~Mokari, M.~R. Javan, H.~Saeedi, and H.~Pishro-Nik, ``A new
  multiple access technique for {5G}: Power domain sparse code multiple access
  {(PSMA)},'' \emph{IEEE Access}, vol.~6, pp. 747--759, 2018.

\bibitem{1.111}
J.~Kim, K.~Lee, J.~Kim, H.~Wang, M.~Na, and D.~Hong, ``A novel {SCMA} system
  for coexistence of active users and inactive users,'' \emph{IEEE
  Commun. Letters}, vol.~21, no.~12, pp. 2730--2733, Dec 2017.

\bibitem{1.109}
S.~Li, L.~D. Xu, and X.~Wang, ``Compressed sensing signal and data acquisition
  in wireless sensor networks and internet of things,'' \emph{IEEE Trans.
  Industrial Informat.}, vol.~9, no.~4, pp. 2177--2186, Nov 2013.

\bibitem{1.105}
R.~Bhatia, B.~Gupta, S.~Benno, J.~Esteban, D.~Samardzija, M.~Tavares, and T.~V.
  Lakshman, ``Massive machine type communications over {5G} using lean
  protocols and edge proxies,'' in \emph{2018 IEEE {5G} World Forum {(5GWF)}},
  July 2018, pp. 462--467.

\bibitem{1.108}
M.~Centenaro, L.~Vangelista, S.~Saur, A.~Weber, and V.~Braun, ``Comparison of
  collision-free and contention-based radio access protocols for the internet
  of things,'' \emph{IEEE Trans. Commun.}, vol.~65, no.~9, pp.
  3832--3846, Sep. 2017.

\bibitem{1.112}
J.~Lianghai, B.~Han, M.~Liu, and H.~D. Schotten, ``Applying device-to-device
  communication to enhance iot services,'' \emph{IEEE Commun. Standards
  Mag.}, vol.~1, no.~2, pp. 85--91, 2017.

\bibitem{1.116}
M.~S. Ali, E.~Hossain, and D.~I. Kim, ``{LTE/LTE-A} random access for massive
  machine-type communications in smart cities,'' \emph{IEEE Commun.
  Mag.}, vol.~55, no.~1, pp. 76--83, January 2017.

\bibitem{1.117}
J.~Lim and Y.~Han, ``Spreading factor allocation for massive connectivity in
  {LoRa} systems,'' \emph{IEEE Commun. Letters}, vol.~22, no.~4, pp.
  800--803, April 2018.

\bibitem{1.104}
A.~Ijaz, L.~Zhang, M.~Grau, A.~Mohamed, S.~Vural, A.~U. Quddus, M.~A. Imran,
  C.~H. Foh, and R.~Tafazolli, ``Enabling massive iot in {5G} and beyond
  systems: Phy radio frame design considerations,'' \emph{IEEE Access}, vol.~4,
  pp. 3322--3339, 2016.

\bibitem{1.107}
B.~P.~S. Sahoo, C.~Chou, C.~Weng, and H.~Wei, ``Enabling millimeter-wave {5G}
  networks for massive iot applications: A closer look at the issues impacting
  millimeter-waves in consumer devices under the {5G} framework,'' \emph{IEEE
  Consumer Electronics Magazine}, vol.~8, no.~1, pp. 49--54, Jan 2019.

\bibitem{1.106}
J.~Zhang, M.~Wang, M.~Hua, W.~Yang, and X.~You, ``Robust synchronization
  waveform design for massive iot,'' \emph{IEEE Trans. Wireless
  Commun.}, vol.~16, no.~11, pp. 7551--7559, Nov 2017.

\bibitem{1.113}
S.~Hu, H.~Guo, C.~Jin, Y.~Huang, B.~Yu, and S.~Li, ``Frequency-domain
  oversampling for cognitive cdma systems: Enabling robust and massive multiple
  access for internet of things,'' \emph{IEEE Access}, vol.~4, pp. 4583--4589,
  2016.

\bibitem{1.114}
A.~Tusha, S.~Doğan, and H.~Arslan, ``Iqi mitigation for narrowband iot systems
  with ofdm-im,'' \emph{IEEE Access}, vol.~6, pp. 44\,626--44\,634, 2018.

\bibitem{1.110}
H.~Malik, H.~Pervaiz, M.~M. Alam, Y.~L. Moullec, A.~Kuusik, and M.~A. Imran,
  ``Radio resource management scheme in nb-iot systems,'' \emph{IEEE Access},
  vol.~6, pp. 15\,051--15\,064, 2018.

\bibitem{1.115}
T.~Salam, W.~U. Rehman, and X.~Tao, ``Cooperative data aggregation and dynamic
  resource allocation for massive machine type communication,'' \emph{IEEE
  Access}, vol.~6, pp. 4145--4158, 2018.

\bibitem{1.118}
T.~Lv, Z.~Lin, P.~Huang, and J.~Zeng, ``Optimization of the energy-efficient
  relay-based massive {IoT} network,'' \emph{IEEE Internet Things J.},
  vol.~5, no.~4, pp. 3043--3058, Aug 2018.

\bibitem{1.63}
\BIBentryALTinterwordspacing
IETF, ``{RFC} 7252: The constrained application protocol {(CoAP)},'' Jun. 2014.
  [Online]. Available: \url{https://tools.ietf.org/html/rfc7252}
\BIBentrySTDinterwordspacing

\bibitem{1.64}
\BIBentryALTinterwordspacing
------, ``{RFC} 7540: Hypertext transfer protocol version 2 {(HTTP/2)},'' May
  2015. [Online]. Available: \url{https://tools.ietf.org/html/rfc7540}
\BIBentrySTDinterwordspacing

\bibitem{1.65}
\BIBentryALTinterwordspacing
------, ``{QUIC: A UDP-Based} multiplexed and secure transport,'' Dec. 2017.
  [Online]. Available:
  \url{https://tools.ietf.org/html/draft-ietf-quic-transport-17}
\BIBentrySTDinterwordspacing

\bibitem{1.158}
A.~{Hazareena} and B.~A. {Mustafa}, ``A survey: On the waveforms for {5G},'' in
  \emph{2018 Second Inter. Conf. on Electronics, Communication and
  Aerospace Technology (ICECA)}, March 2018, pp. 64--67.

\bibitem{1.160}
Y.~Liu, X.~Chen, Z.~Zhong, B.~Ai, D.~Miao, Z.~Zhao, J.~Sun, Y.~Teng, and
  H.~Guan, ``Waveform candidates for {5G} networks: Analysis and comparison,''
  \emph{arXiv preprint arXiv:1609.02427}, 2016.

\bibitem{1.40}
M.~Renfors, J.~Yli-Kaakinen, and M.~Valkama, ``Power amplifier effects on
  frequency localized {5G} candidate waveforms,'' in \emph{European Wireless
  2016; 22th European Wireless Conference}, Oulu, Finland, 2016, pp. 1--5.

\bibitem{1.41}
M.~Abdelaziz, L.~Anttila, M.~Renfors, and M.~Valkama, ``{PAPR} reduction and
  digital predistortion for non-contiguous waveforms with well-localized
  spectrum,'' in \emph{2016 International Symposium on Wireless Communication
  Systems (ISWCS)}, Poznan, Poland, 2016, pp. 581--585.

\bibitem{1.42}
J.~Yli-Kaakinen, T.~Levanen, S.~Valkonen, K.~Pajukoski, J.~Pirskanen,
  M.~Renfors, and M.~Valkama, ``Efficient fast-convolution-based waveform
  processing for {5G} physical layer,'' \emph{IEEE J.  Sel. Areas Commun.}, vol.~35, no.~6, pp. 1309--1326, 2017.

\bibitem{1.157}
\BIBentryALTinterwordspacing
R.~{Vannithamby} and S.~{Talwar}, \emph{New Physical‐layer Waveforms for
  {5G}}.\hskip 1em plus 0.5em minus 0.4em\relax Wiley, 2017. [Online].
  Available: \url{https://ieeexplore.ieee.org/document/8043580}
\BIBentrySTDinterwordspacing

\bibitem{7.5}
S.~M.~A. {Kazmi}, N.~H. {Tran}, T.~M. {Ho}, T.~Z. {Oo}, T.~{LeAnh}, S.~{Moon},
  and C.~S. {Hong}, ``Resource management in dense heterogeneous networks,'' in
  \emph{2015 17th Asia-Pacific Network Operations and Management Symposium
  (APNOMS)}, Aug 2015, pp. 440--443.

\bibitem{1.80}
Z.~Wei, D.~W.~K. Ng, J.~Yuan, and H.~Wang, ``Optimal resource allocation for
  power-efficient mc-noma with imperfect channel state information,''
  \emph{IEEE Trans. Commun.}, vol.~65, no.~9, pp. 3944--3961,
  Sep. 2017.

\bibitem{1.83}
S.~M.~R. Islam, N.~Avazov, O.~A. Dobre, and K.~Kwak, ``Power-domain
  non-orthogonal multiple access ({NOMA}) in {5G} systems: Potentials and
  challenges,'' \emph{IEEE Commun. Surveys Tuts.}, vol.~19, no.~2,
  pp. 721--742, Secondquarter 2017.

\bibitem{1.84}
Q.~Sun, S.~Han, C.~I, and Z.~Pan, ``On the ergodic capacity of {MIMO} {NOMA}
  systems,'' \emph{IEEE Wireless Commun. Letters}, vol.~4, no.~4, pp.
  405--408, Aug 2015.

\bibitem{1.85}
H.~Zhang, N.~Yang, K.~Long, M.~Pan, G.~K. Karagiannidis, and V.~C.~M. Leung,
  ``Secure communications in {NOMA} system: Subcarrier assignment and power
  allocation,'' \emph{IEEE J.  Sel. Areas Commun.},
  vol.~36, no.~7, pp. 1441--1452, July 2018.

\bibitem{1.87}
Q.~Wu, W.~Chen, D.~W.~K. Ng, and R.~Schober, ``Spectral and energy-efficient
  wireless powered iot networks: {NOMA} or {TDMA}?'' \emph{IEEE Trans.
  Veh. Tech.}, vol.~67, no.~7, pp. 6663--6667, July 2018.

\bibitem{1.103}
\BIBentryALTinterwordspacing
G.~T.~. V13.0.0, ``Study on licensed-assisted access to unlicensed spectrum,''
  Jun. 2015. [Online]. Available:
  \url{http://www.3gpp.org/ftp/Specs/archive/36_series/36.889/36889-d00.zip}
\BIBentrySTDinterwordspacing

\bibitem{1.155}
E.~{Dahlman} and S.~{Parkvall}, ``{NR} - the new {5G} radio-access
  technology,'' in \emph{2018 IEEE 87th Veh. Tech. Conference (VTC
  Spring)}, June 2018, pp. 1--6.

\bibitem{1.156}
S.~{Parkvall}, E.~{Dahlman}, A.~{Furuskar}, and M.~{Frenne}, ``{NR}: The new
  {5G} radio access technology,'' \emph{IEEE Communications Standards
  Magazine}, vol.~1, no.~4, pp. 24--30, Dec 2017.

\bibitem{1.159}
J.~{Vihriälä}, A.~A. {Zaidi}, V.~{Venkatasubramanian}, N.~{He}, E.~{Tiirola},
  J.~{Medbo}, E.~{Lähetkangas}, K.~{Werner}, K.~{Pajukoski}, A.~{Cedergren},
  and R.~{Baldemair}, ``Numerology and frame structure for {5G} radio access,''
  in \emph{2016 IEEE 27th Annual International Symposium on Personal, Indoor,
  and Mobile Radio Communications (PIMRC)}, Sep. 2016, pp. 1--5.

\bibitem{8.35}
Z.~E. {Ankarali}, B.~{Peköz}, and H.~{Arslan}, ``Flexible radio access beyond
  {5G}: A future projection on waveform, numerology, and frame design
  principles,'' \emph{IEEE Access}, vol.~5, pp. 18\,295--18\,309, 2017.

\bibitem{1.71}
\BIBentryALTinterwordspacing
G.~T.~S. 38.211, ``Technical specification group radio access network, {NR}
  (release 15),'' Dec. 2017. [Online]. Available:
  \url{http://www.3gpp.org/ftp//Specs/archive/38_series/38.211/38211-f00.zip}
\BIBentrySTDinterwordspacing

\bibitem{3.9}
R.~Molina-Masegosa and J.~Gozalvez, ``{LTE-V} for sidelink {5G} {V2X} vehicular
  communications: A new {5G} technology for short-range vehicle-to-everything
  communications,'' \emph{IEEE Veh. Tech. Mag.}, vol.~12, no.~4,
  pp. 30--39, Dec 2017.

\bibitem{3.10}
Y.~Yang, S.~Dang, Y.~He, and M.~Guizani, ``Markov decision-based pilot
  optimization for {5G} {V2X} vehicular communications,'' \emph{IEEE Internet
  of Things Journal}, pp. 1--1, 2018.

\bibitem{3.11}
T.~Sahin, M.~Klugel, C.~Zhou, and W.~Kellerer, ``Virtual cells for {5G} {V2X}
  communications,'' \emph{IEEE Communications Standards Magazine}, vol.~2,
  no.~1, pp. 22--28, MARCH 2018.

\bibitem{3.12}
H.~Cao, S.~Gangakhedkar, A.~R. Ali, M.~Gharba, and J.~Eichinger, ``A {5G} {V2X}
  testbed for cooperative automated driving,'' in \emph{2016 IEEE Vehicular
  Networking Conference (VNC)}, Dec 2016, pp. 1--4.

\bibitem{3.13}
A.~Orsino, O.~Galinina, S.~Andreev, O.~N.~C. Yilmaz, T.~Tirronen, J.~Torsner,
  and Y.~Koucheryavy, ``Improving initial access reliability of {5G} mmwave
  cellular in massive {V2X} communications scenarios,'' in \emph{2018 IEEE
  Inter. Conf. on Communications (ICC)}, May 2018, pp. 1--7.

\bibitem{3.14}
R.~Shrivastava, M.~Breiling, and A.~Krishnamoorthy, ``Vehicular sudas for {5G}
  high mobility {V2X} scenarios,'' in \emph{2017 IEEE Conference on Standards
  for Communications and Networking (CSCN)}, Sep. 2017, pp. 104--108.

\bibitem{3.16}
B.~Di, L.~Song, Y.~Li, and G.~Y. Li, ``Non-orthogonal multiple access for
  high-reliable and low-latency {V2X} communications in {5G} systems,''
  \emph{IEEE J.  Sel. Areas Commun.}, vol.~35, no.~10, pp.
  2383--2397, Oct 2017.

\bibitem{3.18}
Z.~Naghsh and S.~Valaee, ``Mucs: A new multichannel conflict-free link
  scheduler for cellular {V2X} systems,'' in \emph{2018 IEEE International
  Conference on Communications (ICC)}, May 2018, pp. 1--7.

\bibitem{3.19}
S.~Husain, A.~Kunz, A.~Prasad, E.~Pateromichelakis, K.~Samdanis, and J.~Song,
  ``The road to {5G} {V2X}: Ultra-high reliable communications,'' in \emph{2018
  IEEE Conference on Standards for Communications and Networking (CSCN)}, Oct
  2018, pp. 1--6.

\bibitem{3.20}
C.~Han, M.~Dianati, Y.~Cao, F.~Mccullough, and A.~Mouzakitis, ``Adaptive
  network segmentation and channel allocation in large-scale {V2X}
  communication networks,'' \emph{IEEE Trans. Commun.},
  vol.~67, no.~1, pp. 405--416, Jan 2019.

\bibitem{3.21}
C.~Tranoris, S.~Denazis, L.~Guardalben, J.~Pereira, and S.~Sargento, ``Enabling
  cyber-physical systems for {5G} networking: A case study on the automotive
  vertical domain,'' in \emph{2018 IEEE/ACM 4th International Workshop on
  Software Engineering for Smart Cyber-Physical Systems (SEsCPS)}, May 2018,
  pp. 37--40.

\bibitem{3.22}
H.~Cao, S.~Gangakhedkar, A.~R. Ali, M.~Gharba, and J.~Eichinger, ``A testbed
  for experimenting {5G}-{V2X} requiring ultra reliability and low-latency,''
  in \emph{WSA 2017; 21th International ITG Workshop on Smart Antennas}, March
  2017, pp. 1--4.

\bibitem{3.23}
T.~Soni, A.~R. Ali, K.~Ganesan, and M.~Schellmann, ``Adaptive numerology a
  solution to address the demanding {QoS} in {5G}-{V2X},'' in \emph{2018 IEEE
  Wireless Communications and Networking Conference (WCNC)}, Barcelona, Spain,
  April 2018, pp. 1--6.

\bibitem{3.25}
B.~Di, L.~Song, Y.~Li, and Z.~Han, ``{V2X} meets {NOMA}: Non-orthogonal
  multiple access for {5G}-enabled vehicular networks,'' \emph{IEEE Wireless
  Communications}, vol.~24, no.~6, pp. 14--21, Dec 2017.

\bibitem{3.26}
Z.~Khan and P.~Fan, ``A multi-hop moving zone (mmz) clustering scheme based on
  cellular-{V2X},'' \emph{China Communications}, vol.~15, no.~7, pp. 55--66,
  July 2018.

\bibitem{3.27}
P.~Wang, B.~Di, H.~Zhang, K.~Bian, and L.~Song, ``Cellular {V2X} communications
  in unlicensed spectrum: Harmonious coexistence with vanet in {5G} systems,''
  \emph{IEEE Trans. Wireless Communications}, vol.~17, no.~8, pp.
  5212--5224, Aug 2018.

\bibitem{3.28}
H.~Chen, R.~Zhang, W.~Zhai, X.~Liang, and G.~Song, ``Interference-free pilot
  design and channel estimation using zcz sequences for {MIMO-OFDM-based C-V2X}
  communications,'' \emph{China Communications}, vol.~15, no.~7, pp. 47--54,
  July 2018.

\bibitem{3.29}
W.~Pak, ``Fast packet classification for {V2X} services in {5G} networks,''
  \emph{Journal of Communications and Networks}, vol.~19, no.~3, pp. 218--226,
  2017.

\bibitem{3.30}
J.~Lianghai, A.~Weinand, B.~Han, and H.~D. Schotten, ``Applying multiradio
  access technologies for reliability enhancement in vehicle-to-everything
  communication,'' \emph{IEEE Access}, vol.~6, pp. 23\,079--23\,094, 2018.

\bibitem{3.31}
G.~H. Sim, S.~Klos, A.~Asadi, A.~Klein, and M.~Hollick, ``An online
  context-aware machine learning algorithm for {5G} mmwave vehicular
  communications,'' \emph{IEEE/ACM Trans. Netw.}, vol.~26, no.~6,
  pp. 2487--2500, Dec 2018.

\bibitem{3.32}
M.~Fallgren, M.~Dillinger, J.~Alonso-Zarate, M.~Boban, T.~Abbas, K.~Manolakis,
  T.~Mahmoodi, T.~Svensson, A.~Laya, and R.~Vilalta, ``Fifth-generation
  technologies for the connected car: Capable systems for vehicle-to-anything
  communications,'' \emph{IEEE Veh. Tech. Mag.}, vol.~13, no.~3,
  pp. 28--38, Sep. 2018.

\bibitem{1.145}
\BIBentryALTinterwordspacing
M.~Chen, U.~Challita, W.~Saad, C.~Yin, and M.~Debbah, ``Machine learning for
  wireless networks with artificial intelligence: {A} tutorial on neural
  networks,'' \emph{CoRR}, vol. abs/1710.02913, 2017. [Online]. Available:
  \url{http://arxiv.org/abs/1710.02913}
\BIBentrySTDinterwordspacing

\bibitem{1.146}
\BIBentryALTinterwordspacing
T.~T. Anh, N.~C. Luong, D.~Niyato, Y.~Liang, and D.~I. Kim, ``Deep
  reinforcement learning for time scheduling in rf-powered backscatter
  cognitive radio networks,'' \emph{CoRR}, vol. abs/1810.04520, 2018. [Online].
  Available: \url{http://arxiv.org/abs/1810.04520}
\BIBentrySTDinterwordspacing

\bibitem{1.122}
R.~Li, Z.~Zhao, X.~Zhou, G.~Ding, Y.~Chen, Z.~Wang, and H.~Zhang, ``Intelligent
  {5G}: When cellular networks meet artificial intelligence,'' \emph{IEEE
  Wireless Communications}, vol.~24, no.~5, pp. 175--183, Oct. 2017.

\bibitem{1.120}
\BIBentryALTinterwordspacing
Ericsson, ``Artificial intelligence and machine learning in next-generation
  systems,'' in \emph{Ericsson white paper GFMC-18:000260}.\hskip 1em plus
  0.5em minus 0.4em\relax Ericsson, Jun. 2018. [Online]. Available:
  \url{https://www.ericsson.com/assets/local/publications/white-papers/s30213986_wp_machineintelligence-maj18.pdf}
\BIBentrySTDinterwordspacing

\bibitem{1.147}
Q.~Mao, F.~Hu, and Q.~Hao, ``Deep learning for intelligent wireless networks: A
  comprehensive survey,'' \emph{IEEE Commun. Surveys Tuts.},
  vol.~20, no.~4, pp. 2595--2621, Fourthquarter 2018.

\bibitem{1.124}
T.~O’Shea and J.~Hoydis, ``An introduction to deep learning for the physical
  layer,'' \emph{IEEE Trans. Cognitive Commun. Netw.},
  vol.~3, no.~4, pp. 563--575, Dec 2017.

\bibitem{1.125}
T.~J. O'Shea, K.~Karra, and T.~C. Clancy, ``Learning to communicate: Channel
  auto-encoders, domain specific regularizers, and attention,'' in \emph{2016
  IEEE International Symposium on Signal Processing and Information Technology
  (ISSPIT)}, Limassol, Cyprus, Dec 2016, pp. 223--228.

\bibitem{1.150}
T.~J. {O'Shea}, J.~{Corgan}, and T.~C. {Clancy}, ``Unsupervised representation
  learning of structured radio communication signals,'' in \emph{2016 First
  International Workshop on Sensing, Processing and Learning for Intelligent
  Machines (SPLINE)}, July 2016, pp. 1--5.

\bibitem{1.151}
T.~{Gruber}, S.~{Cammerer}, J.~{Hoydis}, and S.~t.~{Brink}, ``On deep
  learning-based channel decoding,'' in \emph{2017 51st Annual Conference on
  Information Sciences and Systems (CISS)}, March 2017, pp. 1--6.

\bibitem{1.152}
\BIBentryALTinterwordspacing
S.~Cammerer, T.~Gruber, J.~Hoydis, and S.~ten Brink, ``Scaling deep
  learning-based decoding of polar codes via partitioning,'' \emph{CoRR}, vol.
  abs/1702.06901, 2017. [Online]. Available:
  \url{http://arxiv.org/abs/1702.06901}
\BIBentrySTDinterwordspacing

\bibitem{1.132}
I.~Goodfellow, J.~Pouget-Abadie, M.~Mirza, B.~Xu, D.~Warde-Farley, S.~Ozair,
  A.~Courville, and Y.~Bengio, ``Generative adversarial nets,'' in
  \emph{Advances in neural information processing systems}, 2014, pp.
  2672--2680.

\bibitem{1.130}
T.~J. O'Shea, T.~Roy, N.~West, and B.~C. Hilburn, ``Physical layer
  communications system design over-the-air using adversarial networks,''
  \emph{arXiv preprint arXiv:1803.03145}, 2018.

\bibitem{1.129}
\BIBentryALTinterwordspacing
T.~J. O'Shea, T.~Roy, and N.~West, ``Approximating the void: Learning
  stochastic channel models from observation with variational generative
  adversarial networks,'' \emph{CoRR}, vol. abs/1805.06350, 2018. [Online].
  Available: \url{http://arxiv.org/abs/1805.06350}
\BIBentrySTDinterwordspacing

\bibitem{1.126}
N.~E. West and T.~O'Shea, ``Deep architectures for modulation recognition,'' in
  \emph{2017 IEEE International Symposium on Dynamic Spectrum Access Networks
  (DySPAN)}, Piscataway, NJ, USA, Mar. 2017, pp. 1--6.

\bibitem{1.134}
T.~J. O’Shea, J.~Corgan, and T.~C. Clancy, ``Convolutional radio modulation
  recognition networks,'' in \emph{Inter. Conf. on engineering
  applications of neural networks}.\hskip 1em plus 0.5em minus 0.4em\relax
  Aberdeen, UK: Springer, 2016, pp. 213--226.

\bibitem{1.127}
T.~J. O'Shea, N.~West, M.~Vondal, and T.~C. Clancy, ``Semi-supervised radio
  signal identification,'' in \emph{2017 19th Inter. Conf. on
  Advanced Communication Technology (ICACT)}, Bongpyeong, South Korea, Feb.
  2017, pp. 33--38.

\bibitem{1.131}
T.~J. O’Shea, T.~Roy, and T.~C. Clancy, ``Over-the-air deep learning based
  radio signal classification,'' \emph{IEEE Journal of Selected Topics in
  Signal Processing}, vol.~12, no.~1, pp. 168--179, Feb. 2018.

\bibitem{1.135}
\BIBentryALTinterwordspacing
T.~J. O'Shea and T.~C. Clancy, ``Deep reinforcement learning radio control and
  signal detection with kerlym, a gym {RL} agent,'' \emph{CoRR}, vol.
  abs/1605.09221, 2016. [Online]. Available:
  \url{http://arxiv.org/abs/1605.09221}
\BIBentrySTDinterwordspacing

\bibitem{1.133}
S.~Dörner, S.~Cammerer, J.~Hoydis, and S.~t.~Brink, ``Deep learning based
  communication over the air,'' \emph{IEEE Journal of Selected Topics in Signal
  Processing}, vol.~12, no.~1, pp. 132--143, Feb 2018.

\bibitem{1.128}
T.~J. O'Shea, T.~Erpek, and T.~C. Clancy, ``Physical layer deep learning of
  encodings for the {MIMO} fading channel,'' in \emph{2017 55th Annual Allerton
  Conference on Communication, Control, and Computing (Allerton)}, Monticello,
  IL, USA, Oct. 2017, pp. 76--80.

\bibitem{1.136}
\BIBentryALTinterwordspacing
------, ``Deep learning based {{MIMO}} communications,'' \emph{CoRR}, vol.
  abs/1707.07980, 2017. [Online]. Available:
  \url{http://arxiv.org/abs/1707.07980}
\BIBentrySTDinterwordspacing

\bibitem{8.37}
C.~{Jiang}, H.~{Zhang}, Y.~{Ren}, Z.~{Han}, K.~{Chen}, and L.~{Hanzo},
  ``Machine learning paradigms for next-generation wireless networks,''
  \emph{IEEE Wireless Communications}, vol.~24, no.~2, pp. 98--105, April 2017.

\bibitem{1.137}
T.~Erpek, T.~J. O'Shea, and T.~C. Clancy, ``Learning a physical layer scheme
  for the {MIMO} interference channel,'' in \emph{2018 IEEE International
  Conference on Communications (ICC)}, Kansas City, MO, USA, May 2018, pp.
  1--5.

\bibitem{1.138}
T.~O'Shea, K.~Karra, and T.~C. Clancy, ``Learning approximate neural estimators
  for wireless channel state information,'' in \emph{2017 IEEE 27th
  International Workshop on Machine Learning for Signal Processing (MLSP)},
  Tokyo, Japan, Sep. 2017, pp. 1--7.

\bibitem{1.139}
T.~J. O'Shea, T.~Roy, and T.~Erpek, ``Spectral detection and localization of
  radio events with learned convolutional neural features,'' in \emph{2017 25th
  European Signal Processing Conference (EUSIPCO)}, Kos, Greece, Aug. 2017, pp.
  331--335.

\bibitem{1.140}
T.~O'Shea, T.~Roy, and T.~C. Clancy, ``Learning robust general radio signal
  detection using computer vision methods,'' in \emph{2017 51st Asilomar
  Conference on Signals, Systems, and Computers}, Pacific Grove, CA, USA, Oct.
  2017, pp. 829--832.

\bibitem{1.143}
C.~Clancy, J.~Hecker, E.~Stuntebeck, and T.~O'Shea, ``Applications of machine
  learning to cognitive radio networks,'' \emph{IEEE Wireless Communications},
  vol.~14, no.~4, pp. 47--52, Aug. 2007.

\bibitem{1.144}
M.~Bkassiny, Y.~Li, and S.~K. Jayaweera, ``A survey on machine-learning
  techniques in cognitive radios,'' \emph{IEEE Commun. Surveys Tuts.}, vol.~15, no.~3, pp. 1136--1159, Third 2013.

\bibitem{2.1}
S.~Ludwig, M.~Karrenbauer, A.~Fellan, H.~D. Schotten, H.~Buhr, S.~Seetaraman,
  N.~Niebert, A.~Bernardy, V.~Seelmann, V.~Stich, A.~Hoell, C.~Stimming, H.~Wu,
  S.~Wunderlich, M.~Taghouti, F.~Fitzek, C.~Pallasch, N.~Hoffmann, W.~Herfs,
  E.~Eberhardt, and T.~Schildknecht, ``A {5G} architecture for the factory of
  the future,'' in \emph{2018 IEEE 23rd Inter. Conf. on Emerging
  Technologies and Factory Automation (ETFA)}, vol.~1, Turin, Italy, Sep. 2018,
  pp. 1409--1416.

\bibitem{2.2}
S.~Gangakhedkar, H.~Cao, A.~R. Ali, K.~Ganesan, M.~Gharba, and J.~Eichinger,
  ``Use cases, requirements and challenges of {5G} communication for industrial
  automation,'' in \emph{2018 IEEE Inter. Conf. on Communications
  Workshops (ICC Workshops)}, Kansas City, MO, USA, May 2018, pp. 1--6.

\bibitem{2.15}
M.~Gundall, J.~Schneider, H.~D. Schotten, M.~Aleksy, D.~Schulz, N.~Franchi,
  N.~Schwarzenberg, C.~Markwart, R.~Halfmann, P.~Rost, D.~Wübben, A.~Neumann,
  M.~Düngen, T.~Neugebauer, R.~Blunk, M.~Kus, and J.~Grießbach, ``{5G} as
  enabler for industrie 4.0 use cases: Challenges and concepts,'' in \emph{2018
  IEEE 23rd Inter. Conf. Emerging Tech. Factory
  Automation (ETFA)}, vol.~1, Turin, Italy, Sep. 2018, pp. 1401--1408.

\bibitem{2.22}
A.~Neumann, L.~Wisniewski, R.~S. Ganesan, P.~Rost, and J.~Jasperneite,
  ``Towards integration of industrial ethernet with {5G} mobile networks,'' in
  \emph{2018 14th IEEE International Workshop Factory Commun. Systems
  (WFCS)}, Imperia, Italy, Jun. 2018, pp. 1--4.

\bibitem{2.56}
B.~Chen, J.~Wan, L.~Shu, P.~Li, M.~Mukherjee, and B.~Yin, ``Smart factory of
  industry 4.0: Key technologies, application case, and challenges,''
  \emph{IEEE Access}, vol.~6, pp. 6505--6519, Mar. 2018.

\bibitem{2.3}
B.~Holfeld, D.~Wieruch, T.~Wirth, L.~Thiele, S.~A. Ashraf, J.~Huschke,
  I.~Aktas, and J.~Ansari, ``Wireless communication for factory automation: an
  opportunity for {LTE} and {5G} systems,'' \emph{IEEE Communications
  Magazine}, vol.~54, no.~6, pp. 36--43, Jun. 2016.

\bibitem{2.14}
O.~N.~C. Yilmaz, Y.~.~E. Wang, N.~A. Johansson, N.~Brahmi, S.~A. Ashraf, and
  J.~Sachs, ``Analysis of ultra-reliable and low-latency {5G} communication for
  a factory automation use case,'' in \emph{2015 IEEE Inter. Conf.
  Commun. Workshop (ICCW)}, London, UK, Jun. 2015, pp. 1190--1195.

\bibitem{2.25}
B.~Singh, Z.~Li, O.~Tirkkonen, M.~A. Uusitalo, and P.~Mogensen,
  ``Ultra-reliable communication in a factory environment for {5G} wireless
  networks: Link level and deployment study,'' in \emph{2016 IEEE 27th Annual
  Inter. Symp. Personal, Indoor, Mobile Radio Commun.
  (PIMRC)}, Valencia, Spain, Sep. 2016, pp. 1--5.

\bibitem{2.32}
M.~Wollschlaeger, T.~Sauter, and J.~Jasperneite, ``The future of industrial
  communication: Automation networks in the era of the internet of things and
  industry 4.0,'' \emph{IEEE Indus. Electronics Mag.}, vol.~11, no.~1,
  pp. 17--27, March 2017.

\bibitem{2.35}
M.~Ehrlich, D.~Krummacker, C.~Fischer, R.~Guillaume, S.~S.~P. Olaya,
  A.~Frimpong, H.~de~Meer, M.~Wollschlaeger, H.~D. Schotten, and
  J.~Jasperneite, ``Software- defined networking as an enabler for future
  industrial network management,'' in \emph{2018 IEEE 23rd International
  Conference on Emerging Technologies and Factory Automation (ETFA)}, vol.~1,
  Turin, Italy, Sep. 2018, pp. 1109--1112.

\bibitem{2.54}
A.~Varghese and D.~Tandur, ``Wireless requirements and challenges in industry
  4.0,'' in \emph{2014 Inter. Conf. on Contemporary Computing and
  Informatics (IC3I)}, Mysore, India, Nov. 2014, pp. 634--638.

\bibitem{2.16}
S.~A. Ashraf, I.~Aktas, E.~Eriksson, K.~W. Helmersson, and J.~Ansari,
  ``Ultra-reliable and low-latency communication for wireless factory
  automation: From {LTE} to {5G},'' in \emph{2016 IEEE 21st Inter.
  Conf. Emerging Tech. Factory Automation (ETFA)}, Berlin,
  Germany, Sep. 2016, pp. 1--8.

\bibitem{2.19}
N.~Brahmi, O.~N.~C. Yilmaz, K.~W. Helmersson, S.~A. Ashraf, and J.~Torsner,
  ``Deployment strategies for ultra-reliable and low-latency communication in
  factory automation,'' in \emph{2015 IEEE Globecom Workshops (GC Wkshps)}, San
  Diego, CA, USA, Dec. 2015, pp. 1--6.

\bibitem{2.17}
J.~S. Walia, H.~Hämmäinen, and H.~Flinck, ``Future scenarios and value
  network configurations for industrial {5G},'' in \emph{2017 8th Inter.
  Conf. Network Future (NOF)}, London, UK, Nov. 2017, pp.
  79--84.

\bibitem{2.23}
H.~A. Munz and J.~Ansari, ``An empirical study on using d2d relaying in {5G}
  for factory automation,'' in \emph{2018 IEEE Wireless Commun.
  Netw. Conf. Workshops (WCNCW)}, Barcelona, Spain, Apr. 2018, pp.
  149--154.

\bibitem{2.42}
P.~Rost, M.~Breitbach, H.~Roreger, B.~Erman, C.~Mannweiler, R.~Miller, and
  I.~Viering, ``Customized industrial networks: Network slicing trial at
  hamburg seaport,'' \emph{IEEE Wireless Commun.}, vol.~25, no.~5, pp.
  48--55, Oct. 2018.

\bibitem{2.49}
V.~Theodorou, K.~V. Katsaros, A.~Roos, E.~Sakic, and V.~Kulkarni,
  ``Cross-domain network slicing for industrial applications,'' in \emph{2018
  European Conf. Netw. Commun. (EuCNC)}, Ljubljana,
  Slovenia, Slovenia, Jun. 2018, pp. 209--213.

\bibitem{2.50}
N.~Ericsson, T.~Lennvall, J.~Åkerberg, and M.~Björkman, ``Custom simulation
  of industrial wireless sensor and actuator network for improved efficiency
  during research and development,'' in \emph{2017 22nd IEEE International
  Conf. Emerging Tech. Factory Automation (ETFA)}, Limassol,
  Cyprus, Sep. 2017, pp. 1--8.

\bibitem{2.4}
L.~Liu and W.~Yu, ``A d2d-based protocol for ultra-reliable wireless
  communications for industrial automation,'' \emph{IEEE Trans.
  Wireless Commun.}, vol.~17, no.~8, pp. 5045--5058, Aug. 2018.

\bibitem{2.26}
Y.~Gao, T.~Yang, and B.~Hu, ``Improving the transmission reliability in smart
  factory through spatial diversity with arq,'' in \emph{2016 IEEE/CIC
  Inter. Conf. Commun. China (ICCC)}, Chengdu, China,
  Jul 2016, pp. 1--5.

\bibitem{2.21}
R.~Jurdi, S.~R. Khosravirad, and H.~Viswanathan, ``Variable-rate ultra-reliable
  and low-latency communication for industrial automation,'' in \emph{2018 52nd
  Annual Conf. Informat. Sciences Syst. (CISS)}, Princeton, NJ,
  USA, Mar. 2018, pp. 1--6.

\bibitem{2.39}
O.~N.~C. Yilmaz, Y.~.~E. Wang, N.~A. Johansson, N.~Brahmi, S.~A. Ashraf, and
  J.~Sachs, ``Analysis of ultra-reliable and low-latency {5G} communication for
  a factory automation use case,'' in \emph{2015 IEEE Inter. Conf.
  Commun. Workshop (ICCW)}, London, UK, Jun. 2015, pp. 1190--1195.

\bibitem{2.5}
C.~Li, C.~Li, K.~Hosseini, S.~B. Lee, J.~Jiang, W.~Chen, G.~Horn, T.~Ji, J.~E.
  Smee, and J.~Li, ``{5G}-based systems design for tactile internet,''
  \emph{Proceedings of the IEEE}, vol. 107, no.~2, pp. 307--324, Feb. 2019.

\bibitem{2.7}
B.~Holfeld, D.~Wieruch, L.~Raschkowski, T.~Wirth, C.~Pallasch, W.~Herfs, and
  C.~Brecher, ``Radio channel characterization at 5.85 ghz for wireless m2m
  communication of industrial robots,'' in \emph{2016 IEEE Wireless
  Commun. Netw. Conf.}, Doha, Qatar, Apr. 2016, pp. 1--7.

\bibitem{2.44}
K.~A. Maria, N.~Sutisna, T.~T. Nguyen, D.~K. Lam, Y.~Nagao, L.~Lanante,
  M.~Kurosaki, and H.~Ochi, ``Energy efficient industrial wireless system
  through cross layer optimization,'' in \emph{2018 IEEE Inter.
  Conf. Indus. Tech.(ICIT)}, Lyon, France, Feb. 2018, pp.
  1586--1591.

\bibitem{2.43}
D.~A. Wassie, I.~Rodriguez, G.~Berardinelli, F.~M.~L. Tavares, T.~B. Sorensen,
  and P.~Mogensen, ``Radio propagation analysis of industrial scenarios within
  the context of ultra-reliable communication,'' in \emph{2018 IEEE 87th
  Veh. Tech. Conf. (VTC Spring)}, Porto, Portugal, June 2018,
  pp. 1--6.

\bibitem{2.20}
S.~Auroux, D.~Parruca, and H.~Karl, ``Joint real-time scheduling and
  interference coordination for wireless factory automation,'' in \emph{2016
  IEEE 27th Annual International Symposium on Personal, Indoor, and Mobile
  Radio Communications (PIMRC)}, Valencia, Spain, Sep. 2016, pp. 1--6.

\bibitem{2.28}
M.~Li, X.~Guan, C.~Hua, C.~Chen, and L.~Lyu, ``Predictive pre-allocation for
  low-latency uplink access in industrial wireless networks,'' in \emph{IEEE
  INFOCOM 2018 - IEEE Conference on Computer Communications}, Honolulu, HI,
  USA, Apr. 2018, pp. 306--314.

\bibitem{2.30}
S.~Li, Q.~Ni, Y.~Sun, G.~Min, and S.~Al-Rubaye, ``Energy-efficient resource
  allocation for industrial cyber-physical {IoT} systems in {5G} era,''
  \emph{IEEE Trans. Industrial Informatics}, vol.~14, no.~6, pp.
  2618--2628, Jun. 2018.

\bibitem{2.31}
P.~Duan, Y.~Jia, L.~Liang, J.~Rodriguez, K.~M.~S. Huq, and G.~Li,
  ``Space-reserved cooperative caching in {5G} heterogeneous networks for
  industrial {IoT},'' \emph{IEEE Trans. Industrial Informatics},
  vol.~14, no.~6, pp. 2715--2724, Jun. 2018.

\bibitem{2.33}
L.~Lyu, C.~Chen, S.~Zhu, and X.~Guan, ``{5G} enabled codesign of
  energy-efficient transmission and estimation for industrial {IoT} systems,''
  \emph{IEEE Trans. Industrial Informatics}, vol.~14, no.~6, pp.
  2690--2704, Jun. 2018.

\bibitem{2.36}
S.~Hu, B.~Yu, C.~Qian, Y.~Xiao, Q.~Xiong, C.~Sun, and Y.~Gao, ``Nonorthogonal
  interleave-grid multiple access scheme for industrial internet of things in
  {5G} network,'' \emph{IEEE Trans. Industrial Informatics}, vol.~14,
  no.~12, pp. 5436--5446, Dec. 2018.

\bibitem{2.41}
T.~Zheng, Y.~Qin, H.~Zhang, and S.~Kuo, ``Adaptive power control for mutual
  interference avoidance in industrial internet-of-things,'' \emph{China
  Communications}, vol.~13, no. Supplement 1, pp. 124--131, 2016.

\bibitem{2.52}
E.~E. Ugwuanyi, S.~Ghosh, M.~Iqbal, and T.~Dagiuklas, ``Reliable resource
  provisioning using bankers’ deadlock avoidance algorithm in mec for
  industrial {IoT},'' \emph{IEEE Access}, vol.~6, pp. 43\,327--43\,335, Aug
  2018.

\bibitem{2.55}
L.~Lyu, C.~Chen, S.~Zhu, N.~Cheng, B.~Yang, and X.~Guan, ``Control performance
  aware cooperative transmission in multiloop wireless control systems for
  industrial {IoT} applications,'' \emph{IEEE Internet of Things Journal},
  vol.~5, no.~5, pp. 3954--3966, Oct 2018.

\bibitem{2.24}
F.~Voigtländer, A.~Ramadan, J.~Eichinger, C.~Lenz, D.~Pensky, and A.~Knoll,
  ``{5G} for robotics: Ultra-low latency control of distributed robotic
  systems,'' in \emph{2017 International Symposium on Computer Science and
  Intelligent Controls (ISCSIC)}, Budapest, Hungary, Oct. 2017, pp. 69--72.

\bibitem{2.29}
J.~Ansari, I.~Aktas, C.~Brecher, C.~Pallasch, N.~Hoffmann, M.~Obdenbusch,
  M.~Serror, K.~Wehrle, and J.~Gross, ``Demo: a realistic use-case for wireless
  industrial automation and control,'' in \emph{2017 Inter. Conf.
  on Networked Systems (NetSys)}, Gottingen, Germany, Mar. 2017, pp. 1--2.

\bibitem{2.53}
A.~Fellan, C.~Schellenberger, M.~Zimmermann, and H.~D. Schotten, ``Enabling
  communication technologies for automated unmanned vehicles in industry 4.0,''
  in \emph{2018 Inter. Conf. on Information and Communication
  Technology Convergence (ICTC)}, Jeju, South Korea, Oct. 2018, pp. 171--176.

\bibitem{2.18}
D.~Solomitckii, A.~Orsino, S.~Andreev, Y.~Koucheryavy, and M.~Valkama,
  ``Characterization of mmwave channel properties at 28 and 60 ghz in factory
  automation deployments,'' in \emph{2018 IEEE Wireless Communications and
  Networking Conference (WCNC)}, Barcelona, Spain, Apr. 2018, pp. 1--6.

\bibitem{2.40}
A.~Lizeaga, M.~Mendicute, P.~M. Rodríguez, and I.~Val, ``Evaluation of
  wcp-coqam, gfdm-oqam and fbmc-oqam for industrial wireless communications
  with cognitive radio,'' in \emph{2017 IEEE International Workshop of
  Electronics, Control, Measurement, Signals and their Application to
  Mechatronics (ECMSM)}, Donostia-San Sebastian, Spain, May 2017, pp. 1--6.

\bibitem{2.45}
K.~A. Maria, N.~Sutisna, Y.~Nagao, L.~Lanante, M.~Kurosaki, B.~Sai, and
  H.~Ochi, ``Channel selectivity schemes for re-transmission diversity in
  industrial wireless system,'' in \emph{2017 International Symposium on
  Electronics and Smart Devices (ISESD)}, Yogyakarta, Indonesia, Oct. 2017, pp.
  207--212.

\bibitem{2.48}
H.~Xu, W.~Yu, D.~Griffith, and N.~Golmie, ``A survey on industrial internet of
  things: A cyber-physical systems perspective,'' \emph{IEEE Access}, vol.~6,
  pp. 78\,238--78\,259, 2018.

\bibitem{2.47}
D.~Behnke, M.~Müller, P.~Bök, and J.~Bonnet, ``Intelligent network services
  enabling industrial {IoT} systems for flexible smart manufacturing,'' in
  \emph{2018 14th Inter. Conf. on Wireless and Mobile Computing,
  Networking and Communications (WiMob)}, Limassol, Cyprus, Oct. 2018, pp.
  1--4.

\bibitem{2.51}
G.~Marchetto, R.~Sisto, J.~Yusupov, and A.~Ksentinit, ``Formally verified
  latency-aware vnf placement in industrial internet of things,'' in \emph{2018
  14th IEEE International Workshop on Factory Communication Systems (WFCS)},
  Imperia, Italy, Jun. 2018, pp. 1--9.

\bibitem{2.58}
J.~{Wan}, S.~{Tang}, D.~{Li}, S.~{Wang}, C.~{Liu}, H.~{Abbas}, and A.~V.
  {Vasilakos}, ``A manufacturing big data solution for active preventive
  maintenance,'' \emph{IEEE Trans. Industrial Informatics}, vol.~13,
  no.~4, pp. 2039--2047, Aug 2017.

\bibitem{3.76}
3GPP, ``Study on architecture for next generation system (release 14),'' in
  \emph{TR 23.799 14.0.0}, Dec. 2016.

\bibitem{2.59}
\BIBentryALTinterwordspacing
BDVA, ``Big data challenges in smart manufacturing: a discussion paper for bdva
  and effra research and innovation roadmap alignment,'' 2018. [Online].
  Available:
  \url{http://www.bdva.eu/sites/default/files/BDVA_SMI_Discussion_Paper_Web_Version.pdf}
\BIBentrySTDinterwordspacing

\bibitem{2.60}
M.~H. u.~{Rehman}, E.~{Ahmed}, I.~{Yaqoob}, I.~A.~T. {Hashem}, M.~{Imran}, and
  S.~{Ahmad}, ``Big data analytics in industrial {IoT} using a concentric
  computing model,'' \emph{IEEE Commun. Mag.}, vol.~56, no.~2, pp.
  37--43, Feb 2018.

\bibitem{1.161}
\BIBentryALTinterwordspacing
3GPP, ``Feasibility study on lan support in {5G} (release 16),'' in \emph{3GPP
  TR 22.821 V16.1.0}, Jun. 2018. [Online]. Available:
  \url{https://portal.3gpp.org/desktopmodules/Specifications/SpecificationDetails.aspx?specificationId=3281}
\BIBentrySTDinterwordspacing

\bibitem{3.68}
\BIBentryALTinterwordspacing
NGMN, ``{V2X} white paper,'' in \emph{NGMN Alliance White Paper}.\hskip 1em
  plus 0.5em minus 0.4em\relax NGMN Alliance, Jun. 2018. [Online]. Available:
  \url{http://5gaa.org/wp-content/uploads/2018/08/V2X_white_paper_v1_0.pdf}
\BIBentrySTDinterwordspacing

\bibitem{3.77}
R.~{Soua}, I.~{Turcanu}, F.~{Adamsky}, D.~{Führer}, and T.~{Engel},
  ``Multi-access edge computing for vehicular networks: A position paper,'' in
  \emph{2018 IEEE Globecom Workshops (GC Wkshps)}, Dec 2018, pp. 1--6.

\bibitem{3.71}
R.~{Dos Reis Fontes}, C.~{Campolo}, C.~{Esteve Rothenberg}, and A.~{Molinaro},
  ``From theory to experimental evaluation: Resource management in
  software-defined vehicular networks,'' \emph{IEEE Access}, vol.~5, pp.
  3069--3076, 2017.

\bibitem{3.72}
R.~{Vilalta}, S.~{Vía}, F.~{Mira}, R.~{Casellas}, R.~{Muñoz},
  J.~{Alonso-Zarate}, A.~{Kousaridas}, and M.~{Dillinger}, ``Control and
  management of a connected car using sdn/nfv, fog computing and yang data
  models,'' in \emph{2018 4th IEEE Conference on Network Softwarization and
  Workshops (NetSoft)}, June 2018, pp. 378--383.

\bibitem{3.73}
W.~{Huang}, L.~{Ding}, D.~{Meng}, J.~{Hwang}, Y.~{Xu}, and W.~{Zhang},
  ``Qoe-based resource allocation for heterogeneous multi-radio communication
  in software-defined vehicle networks,'' \emph{IEEE Access}, vol.~6, pp.
  3387--3399, 2018.

\bibitem{3.74}
H.~{Rizvi} and J.~{Akram}, ``Handover management in {5G} software defined
  network based {V2X} communication,'' in \emph{2018 12th International
  Conference on Open Source Systems and Technologies (ICOSST)}, Dec 2018, pp.
  22--26.

\bibitem{3.75}
W.~{Pak}, ``Fast packet classification for {V2X} services in {5G} networks,''
  \emph{Journal of Communications and Networks}, vol.~19, no.~3, pp. 218--226,
  2017.

\bibitem{3.91}
C.~{Campolo}, A.~{Molinaro}, A.~{Iera}, R.~R. {Fontes}, and C.~E. {Rothenberg},
  ``Towards {5G} network slicing for the {V2X} ecosystem,'' in \emph{2018 4th
  IEEE Conference on Network Softwarization and Workshops (NetSoft)}, June
  2018, pp. 400--405.

\bibitem{3.92}
C.~{Campolo}, A.~{Molinaro}, A.~{Iera}, and F.~{Menichella}, ``{5G} network
  slicing for vehicle-to-everything services,'' \emph{IEEE Wireless
  Communications}, vol.~24, no.~6, pp. 38--45, Dec 2017.

\bibitem{3.90}
H.~{Khan}, P.~{Luoto}, M.~{Bennis}, and M.~{Latva-aho}, ``On the application of
  network slicing for {5G-{V2X}},'' in \emph{European Wireless 2018; 24th
  European Wireless Conference}, May 2018, pp. 1--6.

\bibitem{3.84}
Y.~{Wang}, M.~{Narasimha}, and R.~W. {Heath}, ``Towards robustness: Machine
  learning for mmwave {V2X} with situational awareness,'' in \emph{2018 52nd
  Asilomar Conference on Signals, Systems, and Computers}, Oct 2018, pp.
  1577--1581.

\bibitem{3.83}
A.~{Asadi}, S.~{Müller}, G.~H. {Sim}, A.~{Klein}, and M.~{Hollick}, ``Fml:
  Fast machine learning for {5G} mmwave vehicular communications,'' in
  \emph{IEEE INFOCOM 2018 - IEEE Conference on Computer Communications}, April
  2018, pp. 1961--1969.

\bibitem{3.88}
W.~{Liu}, G.~{Qin}, Y.~{He}, and F.~{Jiang}, ``Distributed cooperative
  reinforcement learning-based traffic signal control that integrates {V2X}
  networks’ dynamic clustering,'' \emph{IEEE Trans. Vehicular
  Technology}, vol.~66, no.~10, pp. 8667--8681, Oct 2017.

\bibitem{3.87}
W.~{Tong}, A.~{Hussain}, W.~X. {Bo}, and S.~{Maharjan}, ``Artificial
  intelligence for vehicle-to-everything: A survey,'' \emph{IEEE Access},
  vol.~7, pp. 10\,823--10\,843, 2019.

\bibitem{3.89}
H.~{Ye}, L.~{Liang}, G.~Y. {Li}, J.~{Kim}, L.~{Lu}, and M.~{Wu}, ``Machine
  learning for vehicular networks: Recent advances and application examples,''
  \emph{IEEE Veh. Tech. Mag.}, vol.~13, no.~2, pp. 94--101, June
  2018.

\bibitem{3.55}
\BIBentryALTinterwordspacing
5GPPP, ``Deliverable d2.1 {5GCAR} scenarios, use cases, requirements and
  {KPIs},'' in \emph{Version: v1.0 2017-08-31, Fifth Generation Communication
  Automotive Research and innovation}.\hskip 1em plus 0.5em minus 0.4em\relax
  5GPPP, Aug. 2017. [Online]. Available:
  \url{https://5gcar.eu/wp-content/uploads/2017/05/5GCAR_D2.1_v1.0.pdf}
\BIBentrySTDinterwordspacing

\bibitem{3.59}
\BIBentryALTinterwordspacing
3GPP, ``Service requirements for {V2X} services,'' in \emph{3GPP TS 22.185},
  Jul. 2018. [Online]. Available:
  \url{https://portal.3gpp.org/desktopmodules/Specifications/SpecificationDetails.aspx?specificationId=2989}
\BIBentrySTDinterwordspacing

\bibitem{3.58}
\BIBentryALTinterwordspacing
------, ``Study on {LTE} support for vehicle-to-everything ({V2X}) services,''
  in \emph{3GPP TR 22.885}, Dec. 2015. [Online]. Available:
  \url{https://portal.3gpp.org/desktopmodules/Specifications/SpecificationDetails.aspx?specificationId=2898}
\BIBentrySTDinterwordspacing

\bibitem{3.61}
\BIBentryALTinterwordspacing
------, ``Service requirements for enhanced {V2X} scenarios release 16,'' in
  \emph{3GPP TS 22.186}, Dec. 2018. [Online]. Available:
  \url{https://portal.3gpp.org/desktopmodules/Specifications/SpecificationDetails.aspx?specificationId=3180}
\BIBentrySTDinterwordspacing

\bibitem{3.60}
\BIBentryALTinterwordspacing
------, ``Study on enhancement of 3gpp support for {5G} {V2X} services release
  16,'' in \emph{3GPP TR 22.886}, Dec. 2018. [Online]. Available:
  \url{https://portal.3gpp.org/desktopmodules/Specifications/SpecificationDetails.aspx?specificationId=3108}
\BIBentrySTDinterwordspacing

\bibitem{3.63}
\BIBentryALTinterwordspacing
ITU-R, ``{IMT} vision - "framework and overall objectives of the future
  development of {IMT} for 2020 and beyond,'' in \emph{ITU-R M.2083-0}, Sep.
  2015. [Online]. Available:
  \url{https://www.itu.int/dms_pubrec/itu-r/rec/m/r-rec-m.2083-0-201509-i!!pdf-e.pdf}
\BIBentrySTDinterwordspacing

\bibitem{3.64}
\BIBentryALTinterwordspacing
------, ``Intelligent transport systems - guidelines and objectives,'' in
  \emph{ITU-R M.1890}, Apr. 2011. [Online]. Available:
  \url{https://www.itu.int/dms_pubrec/itu-r/rec/m/r-rec-m.2083-0-201509-i!!pdf-e.pdf}
\BIBentrySTDinterwordspacing

\bibitem{3.62}
\BIBentryALTinterwordspacing
ETSI, ``Intelligent transport systems (its); vehicular communications; basic
  set of applications; definitions,'' in \emph{ETSI TR 102 638}, Jun. 2009.
  [Online]. Available:
  \url{https://www.etsi.org/deliver/etsi_tr/102600_102699/102638/01.01.01_60/tr_102638v010101p.pdf}
\BIBentrySTDinterwordspacing

\bibitem{3.65}
\BIBentryALTinterwordspacing
------, ``Intelligent transport systems (its); cooperative its (c-its); release
  1,'' in \emph{ETSI TR 101 607}, May 2013. [Online]. Available:
  \url{https://www.etsi.org/deliver/etsi_tr/101600_101699/101607/01.01.01_60/tr_101607v010101p.pdf}
\BIBentrySTDinterwordspacing

\bibitem{5.55}
L.~{Wang}, T.~{Han}, Q.~{Li}, J.~{Yan}, X.~{Liu}, and D.~{Deng}, ``Cell-less
  communications in {5G} vehicular networks based on vehicle-installed access
  points,'' \emph{IEEE Wireless Communications}, vol.~24, no.~6, pp. 64--71,
  Dec 2017.

\bibitem{3.78}
C.~R. {Storck} and F.~{Duarte-Figueiredo}, ``{5G} {V2X} ecosystem providing
  entertainment on board using mm wave communications,'' in \emph{2018 IEEE
  10th Latin-American Conference on Communications (LATINCOM)}, Nov 2018, pp.
  1--6.

\bibitem{3.86}
W.~{Li}, F.~{Zhang}, Y.~{Zhang}, and Z.~{Feng}, ``Adaptive sample weight for
  machine learning computer vision algorithms in {V2X} systems,'' \emph{IEEE
  Access}, vol.~7, pp. 4676--4687, 2019.

\bibitem{3.79}
B.~{Antonescu}, M.~T. {Moayyed}, and S.~{Basagni}, ``mmwave channel propagation
  modeling for {V2X} communication systems,'' in \emph{2017 IEEE 28th Annual
  International Symposium on Personal, Indoor, and Mobile Radio Communications
  (PIMRC)}, Oct 2017, pp. 1--6.

\bibitem{3.80}
T.~{Shimizu}, V.~{Va}, G.~{Bansal}, and R.~W. {Heath}, ``Millimeter wave {V2X}
  communications: Use cases and design considerations of beam management,'' in
  \emph{2018 Asia-Pacific Microwave Conference (APMC)}, Nov 2018, pp. 183--185.

\bibitem{3.81}
S.~{Lien}, Y.~{Kuo}, D.~{Deng}, H.~{Tsai}, A.~{Vinel}, and A.~{Benslimane},
  ``Latency-optimal mmwave radio access for {V2X} supporting next generation
  driving use cases,'' \emph{IEEE Access}, vol.~7, pp. 6782--6795, 2019.

\bibitem{3.82}
I.~{Mavromatis}, A.~{Tassi}, R.~J. {Piechocki}, and A.~{Nix}, ``mmwave system
  for future its: A mac-layer approach for {V2X} beam steering,'' in \emph{2017
  IEEE 86th Veh. Tech. Conference (VTC-Fall)}, Sep. 2017, pp. 1--6.

\bibitem{3.85}
S.~{Huang}, Y.~{Gao}, W.~{Xu}, Y.~{Gao}, and Z.~{Feng}, ``Energy-angle domain
  initial access and beam tracking in millimeter wave {V2X} communications,''
  \emph{IEEE Access}, vol.~7, pp. 9340--9350, 2019.

\bibitem{3.33}
J.~Lianghai, A.~Weinand, B.~Han, and H.~D. Schotten, ``Feasibility study of
  enabling {V2X} communications by {LTE-Uu} radio interface,'' in \emph{2017
  IEEE/CIC Inter. Conf. on Communications in China (ICCC)},
  Qingdao, China, Oct 2017, pp. 1--6.

\bibitem{3.94}
M.~Gupta and R.~Sandhu, ``Authorization framework for secure cloud assisted
  connected cars and vehicular internet of things,'' in \emph{Proceedings of
  the 23nd ACM on Symposium on Access Control Models and Technologies}.\hskip
  1em plus 0.5em minus 0.4em\relax ACM, 2018, pp. 193--204.

\bibitem{3.95}
R.~{Iqbal}, T.~A. {Butt}, M.~O. {Shafique}, M.~W.~A. {Talib}, and T.~{Umer},
  ``Context-aware data-driven intelligent framework for fog infrastructures in
  internet of vehicles,'' \emph{IEEE Access}, vol.~6, pp. 58\,182--58\,194,
  2018.

\bibitem{3.57}
\BIBentryALTinterwordspacing
5GPPP, ``{5G-PPP} white paper on automotive vertical sector,'' in
  \emph{{5G-PPP} White Paper}.\hskip 1em plus 0.5em minus 0.4em\relax 5GPPP,
  Oct. 2015. [Online]. Available:
  \url{https://5g-ppp.eu/wp-content/uploads/2014/02/5G-PPP-White-Paper-on-Automotive-Vertical-Sectors.pdf}
\BIBentrySTDinterwordspacing

\bibitem{5gpppGrid}
5G-PPP, ``{5G} and energy,'' Tech. Rep., Sept. 2015.

\bibitem{thelocalse}
\BIBentryALTinterwordspacing
T.~Local, ``World's first electrified public road opens in sweden,'' in
  \emph{thelocalsweden}, Apr. 2018. [Online]. Available:
  \url{https://www.thelocal.se/20180413/worlds-first-electric-road-opens-in-sweden}
\BIBentrySTDinterwordspacing

\bibitem{Gridsaxena2017efficient}
N.~Saxena, A.~Roy, and H.~Kim, ``Efficient {5G} small cell planning with embms
  for optimal demand response in smart grids,'' \emph{IEEE Trans. Ind.
  Informat.}, vol.~13, no.~3, pp. 1471--1481, June 2017.

\bibitem{zhang2018real}
C.~Zhang, Q.~Wang, J.~Wang, P.~Pinson, J.~M. Morales, and J.~{\O}stergaard,
  ``Real-time procurement strategies of a proactive distribution company with
  aggregator-based demand response,'' \emph{IEEE Trans. Smart Grid}, vol.~9,
  no.~2, pp. 766--776, 2018.

\bibitem{Gridchin2017energy}
W.-L. Chin, W.~Li, and H.-H. Chen, ``Energy big data security threats in
  {IoT}-based smart grid communications,'' \emph{IEEE Commun. Mag.}, vol.~55,
  no.~10, pp. 70--75, oct. 2017.

\bibitem{GridSiemenDER1}
Siemens, ``The utility view of achieving agility in the distributed energy era
  through a holistic and flexible digitalization approach,'' Tech. Rep., 2017.

\bibitem{cao2017toward}
Y.~Cao, O.~Kaiwartya, R.~Wang, T.~Jiang, Y.~Cao, N.~Aslam, and G.~Sexton,
  ``Toward efficient, scalable, and coordinated on-the-move ev charging
  management,'' \emph{IEEE Wireless Commun.}, vol.~24, no.~2, pp. 66--73, 2017.

\bibitem{Gridwang2017wireless}
K.~Wang, Y.~Wang, X.~Hu, Y.~Sun, D.-J. Deng, A.~Vinel, and Y.~Zhang, ``Wireless
  big data computing in smart grid,'' \emph{IEEE Wireless Commun.}, vol.~24,
  no.~2, pp. 58--64, Apr. 2017.

\bibitem{GridReserveD58}
Reserve, ``D5.8: Report on validation of {ICT} concepts using live {5G}
  network, gateway and {Pan-European} infrastructure, v1,'' Tech. Rep., 2018.

\bibitem{HuaweiGrid}
{China Telecom}, {China's State grid}, and {Huawei}, ``{5G} network slicing
  enabling the smart grid,'' Tech. Rep., Jan. 2018.

\bibitem{4.4}
H.~C. {Leligou}, T.~{Zahariadis}, L.~{Sarakis}, E.~{Tsampasis}, A.~{Voulkidis},
  and T.~E. {Velivassaki}, ``Smart grid: a demanding use case for {5G}
  technologies,'' in \emph{2018 IEEE Inter. Conf. on Pervasive
  Computing and Communications Workshops (PerCom Workshops)}, March 2018, pp.
  215--220.

\bibitem{Gridelgenedy2015smart}
M.~Elgenedy, A.~Massoud, and S.~Ahmed, ``Smart grid self-healing: Functions,
  applications, and developments,'' in \emph{Workshop on Smart Grid and
  Renewable Energy}.\hskip 1em plus 0.5em minus 0.4em\relax IEEE, 2015, pp.
  1--6.

\bibitem{GridSliceNetD21}
SliceNet, ``D2.1: Vertical sector requirements analysis and use case
  definition,'' Tech. Rep., 2017.

\bibitem{3GPPTR22804}
3GPP, ``Study on communication for automation in vertical domains (release
  16),'' Tech. Rep., 2018.

\bibitem{zhang2018synergy}
N.~Zhang, P.~Yang, J.~Ren, D.~Chen, L.~Yu, and X.~Shen, ``Synergy of big data
  and {5G} wireless networks: opportunities, approaches, and challenges,''
  \emph{IEEE Wireless Commun.}, vol.~25, no.~1, pp. 12--18, 2018.

\bibitem{rana2016microgrid}
M.~M. Rana, L.~Li, and S.~Su, ``Microgrid state estimation using the {IoT} with
  {5G} technology,'' in \emph{Internet of Things ({IoT}) in {5G} Mobile
  Technologies}.\hskip 1em plus 0.5em minus 0.4em\relax Springer, 2016, pp.
  175--195.

\bibitem{garau20175g}
M.~Garau, M.~Anedda, C.~Desogus, E.~Ghiani, M.~Murroni, and G.~Celli, ``A {5G}
  cellular technology for distributed monitoring and control in smart grid,''
  in \emph{IEEE BMSB}.\hskip 1em plus 0.5em minus 0.4em\relax IEEE, 2017, pp.
  1--6.

\bibitem{gheisarnejad2019future}
M.~Gheisarnejad, M.-H. Khooban, and T.~Dragicevic, ``The future {5G} network
  based secondary load frequency control in maritime microgrids,'' \emph{IEEE
  Journal of Emerging and Selected Topics in Power Electronics}, 2019.

\bibitem{utkarsh2018distributed}
K.~Utkarsh, D.~Srinivasan, A.~Trivedi, W.~Zhang, and T.~Reindl, ``Distributed
  model-predictive real-time optimal operation of a network of smart
  microgrids,'' \emph{IEEE Trans. Smart Grid}, 2018.

\bibitem{4.1}
M.~{Zeinali}, J.~{Thompson}, C.~{Khirallah}, and N.~{Gupta}, ``Evolution of
  home energy management and smart metering communications towards {5G},'' in
  \emph{2017 8th Inter. Conf. on the Network of the Future (NOF)},
  Nov 2017, pp. 85--90.

\bibitem{chekired2018decentralized}
D.~A. Chekired, L.~Khoukhi, and H.~T. Mouftah, ``Decentralized cloud-sdn
  architecture in smart grid: A dynamic pricing model,'' \emph{IEEE
  Trans. Industrial Informatics}, vol.~14, no.~3, pp. 1220--1231,
  2018.

\bibitem{manshadi2018wireless}
S.~D. Manshadi, M.~E. Khodayar, K.~Abdelghany, and H.~{\"U}ster, ``Wireless
  charging of electric vehicles in electricity and transportation networks,''
  \emph{IEEE Trans. Smart Grid}, vol.~9, no.~5, pp. 4503--4512, 2018.

\bibitem{ai2018smart}
Z.~Ai, Y.~Liu, F.~Song, and H.~Zhang, ``A smart collaborative charging
  algorithm for mobile power distribution in {5G} networks,'' \emph{IEEE
  Access}, 2018.

\bibitem{yin2018autonomous}
H.~Yin, M.~Fu, M.~Liu, J.~Song, and C.~Ma, ``Autonomous power control in a
  reconfigurable 6.78-mhz multiple-receiver wireless charging system,''
  \emph{IEEE Trans. Ind. Electron.}, vol.~65, no.~8, pp. 6177--6187, Aug. 2018.

\bibitem{rehmani2018software}
\BIBentryALTinterwordspacing
M.~H. Rehmani, A.~Davy, B.~Jennings, and C.~Assi, ``Software defined networks
  based smart grid communication: A comprehensive survey,'' 2018. [Online].
  Available: \url{https://arxiv.org/abs/1801.04613}
\BIBentrySTDinterwordspacing

\bibitem{zhang2016sdn}
X.~Zhang, K.~Wei, L.~Guo, W.~Hou, and J.~Wu, ``Sdn-based resilience solutions
  for smart grids,'' in \emph{2016 Inter. Conf. on Software
  Networking (ICSN)}.\hskip 1em plus 0.5em minus 0.4em\relax IEEE, 2016, pp.
  1--5.

\bibitem{lin2018self}
H.~Lin, C.~Chen, J.~Wang, J.~Qi, D.~Jin, Z.~T. Kalbarczyk, and R.~K. Iyer,
  ``Self-healing attack-resilient {PMU} network for power system operation,''
  \emph{IEEE Trans. Smart Grid}, vol.~9, no.~3, pp. 1551--1565, May
  2018.

\bibitem{qu2018enabling}
Y.~Qu, X.~Liu, D.~Jin, Y.~Hong, and C.~Chen, ``Enabling a resilient and
  self-healing pmu infrastructure using centralized network control,'' in
  \emph{ACM ADN-NFV Security}.\hskip 1em plus 0.5em minus 0.4em\relax ACM,
  2018, pp. 13--18.

\bibitem{chen2017sdn}
N.~Chen, M.~Wang, N.~Zhang, X.~S. Shen, and D.~Zhao, ``{SDN}-based framework
  for the {PEV} integrated smart grid,'' \emph{IEEE Network}, vol.~31, no.~2,
  pp. 14--21, 2017.

\bibitem{guo2016achieving}
W.~Guo, V.~Mahendran, and S.~Radhakrishnan, ``Achieving throughput fairness in
  smart grid using sdn-based flow aggregation and scheduling,'' in \emph{IEEE
  WiMob}.\hskip 1em plus 0.5em minus 0.4em\relax IEEE, 2016, pp. 1--7.

\bibitem{chaudhary2018sdn}
R.~Chaudhary, G.~S. Aujla, S.~Garg, N.~Kumar, and J.~J. Rodrigues,
  ``{SDN-enabled} multi-attribute-based secure communication for smart grid in
  {IIoT} environment,'' \emph{IEEE Trans. Ind. Informat.}, vol.~14, no.~6, pp.
  2629--2640, 2018.

\bibitem{cosovic20175g}
M.~Cosovic, A.~Tsitsimelis, D.~Vukobratovic, J.~Matamoros, and C.~Anton-Haro,
  ``{5G} mobile cellular networks: Enabling distributed state estimation for
  smart grids,'' \emph{IEEE Comun. Mag.}, vol.~55, no.~10, pp. 62--69.

\bibitem{tao2017foud}
M.~Tao, K.~Ota, and M.~Dong, ``Foud: integrating fog and cloud for {5G}-enabled
  {V2G} networks,'' \emph{IEEE Network}, vol.~31, no.~2, pp. 8--13, 2017.

\bibitem{chekired2017smart}
D.~A. Chekired and L.~Khoukhi, ``Smart grid solution for charging and
  discharging services based on cloud computing scheduling,'' \emph{IEEE Trans.
  Ind. Informat.}, vol.~13, no.~6, pp. 3312--3321, Dec. 2017.

\bibitem{rabiefog}
A.~H. Rabie, S.~H. Ali, H.~A. Ali, and A.~I. Saleh, ``A fog based load
  forecasting strategy for smart grids using big electrical data,''
  \emph{Cluster Computing}, pp. 1--30.

\bibitem{munir2019edge}
M.~S. Munir, S.~F. Abedin, N.~H. Tran, and C.~S. Hong, ``When edge computing
  meets microgrid: A deep reinforcement learning approach,'' \emph{IEEE
  Internet Things J.}, vol.~PP, no.~99, 2019.

\bibitem{yaghmaee2018performance}
M.~H. Yaghmaee, A.~Leon-Garcia, and M.~Moghaddassian, ``On the performance of
  distributed and cloud-based demand response in smart grid,'' \emph{IEEE
  Trans. Smart Grid}, vol.~9, no.~5, pp. 5403--5417, 2018.

\bibitem{li2018smart}
Y.~Li, X.~Cheng, Y.~Cao, D.~Wang, and L.~Yang, ``Smart choice for the smart
  grid: Narrowband internet of things ({NB-IoT}),'' \emph{IEEE Internet Things
  J.}, vol.~5, no.~3, pp. 1505--1515, 2018.

\bibitem{collier2017emerging}
S.~E. Collier, ``The emerging enernet: Convergence of the smart grid with the
  internet of things,'' \emph{IEEE Ind. App. Mag.}, vol.~23, no.~2, pp. 12--16,
  2017.

\bibitem{chiu2017optimized}
T.-C. Chiu, Y.-Y. Shih, A.-C. Pang, and C.-W. Pai, ``Optimized day-ahead
  pricing with renewable energy demand-side management for smart grids,''
  \emph{IEEE Internet Things J.}, vol.~4, no.~2, pp. 374--383, 2017.

\bibitem{rana2015kalman}
M.~Rana, L.~Li, and S.~Su, ``Kalman filter based microgrid state estimation and
  control using the {IoT} with {5G} networks,'' in \emph{2015 IEEE PES
  Asia-Pacific Power and Energy Engineering Conference (APPEEC)}.\hskip 1em
  plus 0.5em minus 0.4em\relax IEEE, 2015, pp. 1--5.

\bibitem{4.2}
and M.~F.~{Ariska}, B.~{Siregar}, U.~{Andyani}, and F.~{Fahmi}, ``Power meter
  monitoring for home appliances based on mobile data communication,'' in
  \emph{2017 Inter. Conf. on Smart Cities, Automation Intelligent
  Computing Systems (ICON-SONICS)}, Nov 2017, pp. 116--120.

\bibitem{4.5}
W.~{Fatnassi} and Z.~{Rezki}, ``Increasing the reliability of smart metering
  system using millimeter wave technology,'' in \emph{2018 IEEE International
  Conference on Communications Workshops (ICC Workshops)}, May 2018, pp. 1--6.

\bibitem{zhou2018energy}
Z.~Zhou, C.~Zhang, C.~Xu, F.~Xiong, Y.~Zhang, and T.~Umer, ``Energy-efficient
  industrial internet of uavs for power line inspection in smart grid,''
  \emph{IEEE Trans. Ind. Informat.}, vol.~14, no.~6, pp. 2705--2714, 2018.

\bibitem{nguyen2019intelligent}
V.~N. Nguyen, R.~Jenssen, and D.~Roverso, ``Intelligent monitoring and
  inspection of power line components powered by uavs and deep learning,''
  \emph{IEEE Power Energy Tech. Syst. J.}, vol.~PP, no.~99, 2019.

\bibitem{lim2018multi}
G.~J. Lim, S.~Kim, J.~Cho, Y.~Gong, and A.~Khodaei, ``{Multi-UAV}
  pre-positioning and routing for power network damage assessment,'' \emph{IEEE
  Trans. Smart Grid}, vol.~9, no.~4, pp. 3643--3651, 2018.

\bibitem{hoang2017charging}
D.~T. Hoang, P.~Wang, D.~Niyato, and E.~Hossain, ``Charging and discharging of
  plug-in electric vehicles {(PEVs)} in vehicle-to-grid {(V2G)} systems: A
  cyber insurance-based model,'' \emph{IEEE Access}, vol.~5, pp. 732--754, jan.
  2017.

\bibitem{wang2017distributed}
K.~Wang, L.~Gu, X.~He, S.~Guo, Y.~Sun, A.~Vinel, and J.~Shen, ``Distributed
  energy management for vehicle-to-grid networks,'' \emph{IEEE network},
  vol.~31, no.~2, pp. 22--28, Apr. 2017.

\bibitem{zhu2018big}
L.~Zhu, M.~Li, Z.~Zhang, X.~Du, and M.~Guizani, ``Big data mining of users’
  energy consumption patterns in the wireless smart grid,'' \emph{IEEE Wireless
  Commun.}, vol.~25, no.~1, pp. 84--89, Feb. 2018.

\bibitem{wang2017robust}
K.~Wang, C.~Xu, Y.~Zhang, S.~Guo, and A.~Zomaya, ``Robust big data analytics
  for electricity price forecasting in the smart grid,'' \emph{IEEE Trans. Big
  Data}, vol.~5, no.~1.

\bibitem{wang2018deep}
Y.~Wang, Q.~Chen, D.~Gan, J.~Yang, D.~S. Kirschen, and C.~Kang, ``Deep
  learning-based socio-demographic information identification from smart meter
  data,'' \emph{IEEE Trans. Smart Grid}, vol.~PP, no.~99, 2018.

\bibitem{ni2019multistage}
Z.~Ni and S.~Paul, ``A multistage game in smart grid security: A reinforcement
  learning solution,'' \emph{IEEE Trans. Neural Netw. Learn. Syst.}, vol.~PP,
  no.~99, 2019.

\bibitem{haddad2018smart}
R.~J. Haddad, B.~Guha, Y.~Kalaani, and A.~El-Shahat, ``Smart distributed
  generation systems using artificial neural network-based event
  classification,'' \emph{IEEE Power Energy Tech. Syst. J.}, vol.~5, no.~2, pp.
  18--26, 2018.

\bibitem{li2017weather}
L.~Li, K.~Ota, and M.~Dong, ``When weather matters: {IoT}-based electrical load
  forecasting for smart grid,'' \emph{IEEE Commun. Mag.}, vol.~55, no.~10, pp.
  46--51, Oct. 2017.

\bibitem{kurtz2016empirical}
F.~Kurtz, N.~Dorsch, and C.~Wietfeld, ``Empirical comparison of virtualized and
  bare-metal switching for sdn-based {5G} communication in critical
  infrastructures,'' in \emph{2016 IEEE NetSoft Conference and Workshops
  (NetSoft)}.\hskip 1em plus 0.5em minus 0.4em\relax IEEE, 2016, pp. 453--458.

\bibitem{you2018cognitive}
M.~You, Q.~Liu, and H.~Sun, ``A cognitive radio enabled smart grid testbed
  based on software defined radio and real time digital simulator,'' in
  \emph{2018 IEEE Inter. Conf. on Communications Workshops (ICC
  Workshops)}.\hskip 1em plus 0.5em minus 0.4em\relax IEEE, 2018, pp. 1--6.

\bibitem{GridNRG5D11}
NRG5, ``D1.1: Use case scenarios analysis and {5G} requirements,'' Tech. Rep.,
  2017.

\bibitem{4.7}
Y.~{Simmhan}, S.~{Aman}, A.~{Kumbhare}, R.~{Liu}, S.~{Stevens}, Q.~{Zhou}, and
  V.~{Prasanna}, ``Cloud-based software platform for big data analytics in
  smart grids,'' \emph{Computing in Science Engineering}, vol.~15, no.~4, pp.
  38--47, July 2013.

\bibitem{4.8}
N.~H. Tran, C.~Pham, M.~N. Nguyen, S.~Ren, and C.~S. Hong, ``Incentivizing
  energy reduction for emergency demand response in multi-tenant mixed-use
  buildings,'' \emph{IEEE Trans. Smart Grid}, vol.~9, no.~4, pp.
  3701--3715, 2018.

\bibitem{4.9}
S.~{Tan}, D.~{De}, W.~{Song}, J.~{Yang}, and S.~K. {Das}, ``Survey of security
  advances in smart grid: A data driven approach,'' \emph{IEEE Commun. Surveys Tuts.}, vol.~19, no.~1, pp. 397--422, Firstquarter 2017.

\bibitem{4.3}
F.~B. {Saghezchi}, G.~{Mantas}, J.~{Ribeiro}, M.~{Al-Rawi}, S.~{Mumtaz}, and
  J.~{Rodriguez}, ``Towards a secure network architecture for smart grids in
  {5G} era,'' in \emph{2017 13th International Wireless Communications and
  Mobile Computing Conference (IWCMC)}, June 2017, pp. 121--126.

\bibitem{zhang2017efficient}
Y.~Zhang, J.~Zhao, and D.~Zheng, ``Efficient and privacy-aware power injection
  over ami and smart grid slice in future {5G} networks,'' \emph{Mobile
  Information Systems}, vol. 2017, 2017.

\bibitem{5.3}
K.~{Zhang}, J.~{Ni}, K.~{Yang}, X.~{Liang}, J.~{Ren}, and X.~S. {Shen},
  ``Security and privacy in smart city applications: Challenges and
  solutions,'' \emph{IEEE Commun. Mag.}, vol.~55, no.~1, pp.
  122--129, January 2017.

\bibitem{5.1}
I.~{Yaqoob}, I.~A.~T. {Hashem}, Y.~{Mehmood}, A.~{Gani}, S.~{Mokhtar}, and
  S.~{Guizani}, ``Enabling communication technologies for smart cities,''
  \emph{IEEE Commun. Mag.}, vol.~55, no.~1, pp. 112--120, Jan. 2017.

\bibitem{5.62}
Y.~{Mehmood}, F.~{Ahmad}, I.~{Yaqoob}, A.~{Adnane}, M.~{Imran}, and
  S.~{Guizani}, ``Internet-of-things-based smart cities: Recent advances and
  challenges,'' \emph{IEEE Commun. Mag.}, vol.~55, no.~9, pp.
  16--24, Sep. 2017.

\bibitem{5.11}
A.~{Zanella}, N.~{Bui}, A.~{Castellani}, L.~{Vangelista}, and M.~{Zorzi},
  ``Internet of things for smart cities,'' \emph{IEEE Internet of Things
  Journal}, vol.~1, no.~1, pp. 22--32, Feb. 2014.

\bibitem{5.5}
H.~{Menouar}, I.~{Guvenc}, K.~{Akkaya}, A.~S. {Uluagac}, A.~{Kadri}, and
  A.~{Tuncer}, ``Uav-enabled intelligent transportation systems for the smart
  city: Applications and challenges,'' \emph{IEEE Commun. Mag.},
  vol.~55, no.~3, pp. 22--28, Mar. 2017.

\bibitem{5.6}
Y.~{Liu}, X.~{Weng}, J.~{Wan}, X.~{Yue}, H.~{Song}, and A.~V. {Vasilakos},
  ``Exploring data validity in transportation systems for smart cities,''
  \emph{IEEE Commun. Mag.}, vol.~55, no.~5, pp. 26--33, May 2017.

\bibitem{5.7}
Z.~{Ning}, F.~{Xia}, N.~{Ullah}, X.~{Kong}, and X.~{Hu}, ``Vehicular social
  networks: Enabling smart mobility,'' \emph{IEEE Commun. Mag.},
  vol.~55, no.~5, pp. 16--55, May 2017.

\bibitem{5.8}
D.~{Mazza}, D.~{Tarchi}, and G.~E. {Corazza}, ``A unified urban mobile cloud
  computing offloading mechanism for smart cities,'' \emph{IEEE Communications
  Magazine}, vol.~55, no.~3, pp. 30--37, Mar. 2017.

\bibitem{5.9}
T.~{Taleb}, S.~{Dutta}, A.~{Ksentini}, M.~{Iqbal}, and H.~{Flinck}, ``Mobile
  edge computing potential in making cities smarter,'' \emph{IEEE
  Communications Magazine}, vol.~55, no.~3, pp. 38--43, Mar. 2017.

\bibitem{5.10}
T.~{Han}, X.~{Ge}, L.~{Wang}, K.~S. {Kwak}, Y.~{Han}, and X.~{Liu}, ``{5G}
  converged cell-less communications in smart cities,'' \emph{IEEE
  Communications Magazine}, vol.~55, no.~3, pp. 44--50, Mar. 2017.

\bibitem{5.17}
H.~{Vahdat-Nejad} and M.~{Asef}, ``Architecture design of the air pollution
  mapping system by mobile crowd sensing,'' \emph{IET Wireless Sensor Systems},
  vol.~8, no.~6, pp. 268--275, 2018.

\bibitem{5.18}
F.~Al-Turjman and C.~Altrjman, ``Energy consumption monitoring in {IoT}-based
  smart cities,'' \emph{Intelligence in {IoT}-enabled Smart Cities}, p.~7,
  2018.

\bibitem{5.60}
J.~{Santos}, P.~{Leroux}, T.~{Wauters}, B.~{Volckaert}, and F.~{De Turck},
  ``Anomaly detection for smart city applications over {5G} low power wide area
  networks,'' in \emph{NOMS 2018 - 2018 IEEE/IFIP Network Operations and
  Management Symposium}, April 2018, pp. 1--9.

\bibitem{5.63}
J.~Santos, T.~Wauters, B.~Volckaert, and F.~De~Turck, ``Fog computing: Enabling
  the management and orchestration of smart city applications in {5G}
  networks,'' \emph{Entropy}, vol.~20, no.~1, p.~4, 2017.

\bibitem{5.64}
F.~{van Lingen}, M.~{Yannuzzi}, A.~{Jain}, R.~{Irons-Mclean}, O.~{Lluch},
  D.~{Carrera}, J.~L. {Perez}, A.~{Gutierrez}, D.~{Montero}, J.~{Marti},
  R.~{Maso}, and a.~J.~P.~{Rodriguez}, ``The unavoidable convergence of {NFV,
  5G, and Fog}: A model-driven approach to bridge cloud and edge,'' \emph{IEEE
  Communications Magazine}, vol.~55, no.~8, pp. 28--35, Aug 2017.

\bibitem{5.65}
P.~{Yadav} and S.~{Vishwakarma}, ``Application of internet of things and big
  data towards a smart city,'' in \emph{2018 3rd Inter. Conf. On
  Internet of Things: Smart Innovation and Usages {(IoT-SIU)}}, Feb 2018, pp.
  1--5.

\bibitem{5.66}
L.~{Ang}, K.~P. {Seng}, G.~K. {Ijemaru}, and A.~M. {Zungeru}, ``Deployment of
  iov for smart cities: Applications, architecture, and challenges,''
  \emph{IEEE Access}, vol.~7, pp. 6473--6492, 2019.

\bibitem{5.67}
B.~{Cheng}, G.~{Solmaz}, F.~{Cirillo}, E.~{Kovacs}, K.~{Terasawa}, and
  A.~{Kitazawa}, ``Fogflow: Easy programming of {IoT} services over cloud and
  edges for smart cities,'' \emph{IEEE Internet of Things Journal}, vol.~5,
  no.~2, pp. 696--707, April 2018.

\bibitem{5.68}
C.~{Kai}, H.~{Li}, L.~{Xu}, Y.~{Li}, and T.~{Jiang}, ``Energy-efficient
  device-to-device communications for green smart cities,'' \emph{IEEE
  Trans. Industrial Informat.}, vol.~14, no.~4, pp. 1542--1551,
  April 2018.

\bibitem{5.69}
L.~{Zhou}, D.~{Wu}, J.~{Chen}, and Z.~{Dong}, ``Greening the smart cities:
  Energy-efficient massive content delivery via d2d communications,''
  \emph{IEEE Trans. Industrial Informatics}, vol.~14, no.~4, pp.
  1626--1634, April 2018.

\bibitem{5.70}
M.~{Usman}, A.~A. {Gebremariam}, U.~{Raza}, and F.~{Granelli}, ``A
  software-defined device-to-device communication architecture for public
  safety applications in {5G} networks,'' \emph{IEEE Access}, vol.~3, pp.
  1649--1654, 2015.

\bibitem{5.71}
\BIBentryALTinterwordspacing
H.~T. {Mouftah}, M.~{Erol-Kantarci}, and M.~H. {Rehmani}, \emph{{5G} and D2D
  Communications at the Service of Smart Cities}.\hskip 1em plus 0.5em minus
  0.4em\relax Wiley, 2019, pp. 147--169. [Online]. Available:
  \url{https://ieeexplore.ieee.org/document/8654114}
\BIBentrySTDinterwordspacing

\bibitem{5.54}
V.~{Poulkov}, ``The unified wireless smart access for smart cities in the
  context of a cyber physical system,'' in \emph{2017 Global Wireless Summit
  (GWS)}, Oct 2017, pp. 12--16.

\bibitem{5.42}
.~{Ogrodowczyk}, B.~{Belter}, and M.~{LeClerc}, ``{IoT} ecosystem over
  programmable sdn infrastructure for smart city applications,'' in \emph{2016
  Fifth European Workshop on Software-Defined Networks (EWSDN)}, Oct 2016, pp.
  49--51.

\bibitem{5.43}
A.~{Irfan}, N.~{Taj}, and S.~A. {Mahmud}, ``A novel secure {SDN/LTE} based
  architecture for smart grid security,'' in \emph{2015 IEEE International
  Conference on Computer and Information Technology; Ubiquitous Computing and
  Communications; Dependable, Autonomic and Secure Computing; Pervasive
  Intelligence and Computing}, Oct 2015, pp. 762--769.

\bibitem{5.44}
K.~{Xiong}, S.~{Leng}, J.~{Hu}, X.~{Chen}, and K.~{Yang}, ``Smart network
  slicing for vehicular {Fog-RANs},'' \emph{IEEE Trans. Vehicular
  Technology}, pp. 1--1, 2019.

\bibitem{5.45}
M.~{Dighriri}, A.~S.~D. {Alfoudi}, G.~M. {Lee}, and T.~{Baker}, ``Data traffic
  model in machine to machine communications over {5G} network slicing,'' in
  \emph{2016 9th Inter. Conf. on Developments in eSystems
  Engineering (DeSE)}, Aug 2016, pp. 239--244.

\bibitem{5.46}
M.~S. {Carmo}, S.~{Jardim}, A.~V. {Neto}, R.~{Aguiar}, and D.~{Corujo},
  ``Towards fog-based slice-defined wlan infrastructures to cope with future
  {5G} use cases,'' in \emph{2017 IEEE 16th International Symposium on Network
  Computing and Applications (NCA)}, Oct 2017, pp. 1--5.

\bibitem{5.47}
A.~{Hakiri} and A.~{Gokhale}, ``Work-in-progress: Towards real-time smart city
  communications using software defined wireless mesh networking,'' in
  \emph{2018 IEEE Real-Time Systems Symposium (RTSS)}, Dec 2018, pp. 177--180.

\bibitem{5.48}
R.~{Abhishek}, S.~{Zhao}, and D.~{Medhi}, ``Spartacus: Service priority
  adaptiveness for emergency traffic in smart cities using software-defined
  networking,'' in \emph{2016 IEEE International Smart Cities Conference
  (ISC2)}, Sep. 2016, pp. 1--4.

\bibitem{5.49}
M.~S. {Munir}, S.~F. {Abedin}, M.~G.~R. {Alam}, N.~H. {Tran}, and C.~S. {Hong},
  ``Intelligent service fulfillment for software defined networks in smart
  city,'' in \emph{2018 Inter. Conf. on Information Networking
  (ICOIN)}, Jan 2018, pp. 516--521.

\bibitem{5.50}
M.~{Li}, P.~{Si}, and Y.~{Zhang}, ``Delay-tolerant data traffic to
  software-defined vehicular networks with mobile edge computing in smart
  city,'' \emph{IEEE Trans. Veh. Tech.}, vol.~67, no.~10,
  pp. 9073--9086, Oct 2018.

\bibitem{5.51}
M.~{Mukherjee}, L.~{Shu}, T.~{Zhao}, K.~{Li}, and H.~{Wang}, ``Low control
  overhead-based sleep scheduling in software-defined wireless sensor
  networks,'' in \emph{2016 IEEE 18th Inter. Conf. on High
  Performance Computing and Communications; IEEE 14th Inter. Conf.
  on Smart City; IEEE 2nd Inter. Conf. on Data Science and Systems
  (HPCC/SmartCity/DSS)}, Dec 2016, pp. 1236--1237.

\bibitem{5.52}
M.~{Habibzadeh}, W.~{Xiong}, M.~{Zheleva}, E.~K. {Stern}, B.~H. {Nussbaum}, and
  T.~{Soyata}, ``Smart city sensing and communication sub-infrastructure,'' in
  \emph{2017 IEEE 60th International Midwest Symposium on Circuits and Systems
  (MWSCAS)}, Aug 2017, pp. 1159--1162.

\bibitem{5.53}
G.~{Merlino}, D.~{Bruneo}, F.~{Longo}, A.~{Puliafito}, and S.~{Distefano},
  ``Software defined cities: A novel paradigm for smart cities through {IoT}
  clouds,'' in \emph{2015 IEEE 12th Intl Conf on Ubiquitous Intelligence and
  Computing and 2015 IEEE 12th Intl Conf on Autonomic and Trusted Computing and
  2015 IEEE 15th Intl Conf on Scalable Computing and Communications and Its
  Associated Workshops (UIC-ATC-ScalCom)}, Aug 2015, pp. 909--916.

\bibitem{5.72}
S.~{Din}, M.~M. {Rathore}, A.~{Ahmad}, A.~{Paul}, and M.~{Khan}, ``{SDIoT}:
  Software defined internet of thing to analyze big data in smart cities,'' in
  \emph{2017 IEEE 42nd Conference on Local Computer Networks Workshops (LCN
  Workshops)}, Oct 2017, pp. 175--182.

\bibitem{5.33}
F.~{Jameel}, S.~{Wyne}, S.~J. {Nawaz}, and Z.~{Chang}, ``Propagation channels
  for mmwave vehicular communications: State-of-the-art and future research
  directions,'' \emph{IEEE Wireless Communications}, vol.~26, no.~1, pp.
  144--150, February 2019.

\bibitem{5.34}
T.~S. {Brisimi}, C.~G. {Cassandras}, C.~{Osgood}, I.~C. {Paschalidis}, and
  Y.~{Zhang}, ``Sensing and classifying roadway obstacles in smart cities: The
  street bump system,'' \emph{IEEE Access}, vol.~4, pp. 1301--1312, 2016.

\bibitem{5.35}
R.~{Jain} and H.~{Shah}, ``An anomaly detection in smart cities modeled as
  wireless sensor network,'' in \emph{2016 Inter. Conf. on Signal
  and Information Processing (IConSIP)}, Oct 2016, pp. 1--5.

\bibitem{5.36}
E.~M. d.~L.~{Pinto}, R.~{Lachowski}, M.~E. {Pellenz}, M.~C. {Penna}, and R.~D.
  {Souza}, ``A machine learning approach for detecting spoofing attacks in
  wireless sensor networks,'' in \emph{2018 IEEE 32nd Inter. Conf.
  on Advanced Information Networking and Applications (AINA)}, May 2018, pp.
  752--758.

\bibitem{5.37}
R.~{Rossini}, E.~{Ferrera}, D.~{Conzon}, and C.~{Pastrone}, ``Wsns
  self-calibration approach for smart city applications leveraging incremental
  machine learning techniques,'' in \emph{2016 8th IFIP International
  Conference on New Technologies, Mobility and Security (NTMS)}, Nov 2016, pp.
  1--7.

\bibitem{5.38}
B.~{Qolomany}, A.~{Al-Fuqaha}, D.~{Benhaddou}, and A.~{Gupta}, ``Role of deep
  lstm neural networks and wi-fi networks in support of occupancy prediction in
  smart buildings,'' in \emph{2017 IEEE 19th Inter. Conf. on High
  Performance Computing and Communications; IEEE 15th Inter. Conf.
  on Smart City; IEEE 3rd Inter. Conf. on Data Science and Systems
  (HPCC/SmartCity/DSS)}, Dec 2017, pp. 50--57.

\bibitem{5.39}
Z.~{Feng} and C.~{Hua}, ``Machine learning-based rf jamming detection in
  wireless networks,'' in \emph{2018 Third Inter. Conf. on Security
  of Smart Cities, Industrial Control System and Communications (SSIC)}, Oct
  2018, pp. 1--6.

\bibitem{5.40}
P.~{Gu}, R.~{Khatoun}, Y.~{Begriche}, and A.~{Serhrouchni}, ``Support vector
  machine (svm) based sybil attack detection in vehicular networks,'' in
  \emph{2017 IEEE Wireless Communications and Networking Conference (WCNC)},
  March 2017, pp. 1--6.

\bibitem{5.41}
Z.~{Tang}, A.~{Liu}, and C.~{Huang}, ``Social-aware data collection scheme
  through opportunistic communication in vehicular mobile networks,''
  \emph{IEEE Access}, vol.~4, pp. 6480--6502, 2016.

\bibitem{5.59}
P.~A. Lorimer, V.~M.-F. Diec, and B.~Kantarci, ``Covers-up: Collaborative
  verification of smart user profiles for social sustainability of smart
  cities,'' \emph{Sustainable Cities and Society}, vol.~38, pp. 348--358, 2018.

\bibitem{5.61}
D.~A. {Awan}, R.~L.~G. {Cavalcante}, Z.~{Utkovski}, and S.~{Stanczak},
  ``Set-theoretic learning for detection in cell-less c-ran systems,'' in
  \emph{2018 IEEE Global Conference on Signal and Information Processing
  (GlobalSIP)}, Nov 2018, pp. 589--593.

\bibitem{5.4}
R.~{Khatoun} and S.~{Zeadally}, ``Cybersecurity and privacy solutions in smart
  cities,'' \emph{IEEE Commun. Mag.}, vol.~55, no.~3, pp. 51--59,
  Mar. 2017.

\bibitem{5.19}
K.~{Kotobi} and M.~{Sartipi}, ``Efficient and secure communications in smart
  cities using edge, caching, and blockchain,'' in \emph{2018 IEEE
  International Smart Cities Conference (ISC2)}, Sep. 2018, pp. 1--6.

\bibitem{5.20}
R.~{Kumar} and S.~{Rajalakshmi}, ``Mobile sensor cloud computing: Controlling
  and securing data processing over smart environment through mobile sensor
  cloud computing (mscc),'' in \emph{2013 Inter. Conf. on Computer
  Sciences Applications}, Dec 2013, pp. 687--694.

\bibitem{5.21}
A.~{Mosenia} and N.~K. {Jha}, ``A comprehensive study of security of
  internet-of-things,'' \emph{IEEE Trans. Emerging Topics
  Computing}, vol.~5, no.~4, pp. 586--602, Oct 2017.

\bibitem{5.22}
A.~S. {Sani}, D.~{Yuan}, P.~L. {Yeoh}, W.~{Bao}, S.~{Chen}, and B.~{Vucetic},
  ``A lightweight security and privacy-enhancing key establishment for internet
  of things applications,'' in \emph{2018 IEEE Inter. Conf. on
  Communications (ICC)}, May 2018, pp. 1--6.

\bibitem{5.23}
Y.~{Liu}, Z.~{Pang}, G.~{Dán}, D.~{Lan}, and S.~{Gong}, ``A taxonomy for the
  security assessment of ip-based building automation systems: The case of
  thread,'' \emph{IEEE Trans. Industrial Informatics}, vol.~14, no.~9,
  pp. 4113--4123, Sep. 2018.

\bibitem{5.24}
L.~{Cui}, G.~{Xie}, Y.~{Qu}, L.~{Gao}, and Y.~{Yang}, ``Security and privacy in
  smart cities: Challenges and opportunities,'' \emph{IEEE Access}, vol.~6, pp.
  46\,134--46\,145, 2018.

\bibitem{5.25}
H.~{Wang}, L.~{Xu}, W.~{Lin}, P.~{Xiao}, and R.~{Wen}, ``Physical layer
  security performance of wireless mobile sensor networks in smart city,''
  \emph{IEEE Access}, vol.~7, pp. 15\,436--15\,443, 2019.

\bibitem{5.26}
R.~{Paul}, P.~{Baidya}, S.~{Sau}, K.~{Maity}, S.~{Maity}, and S.~B. {Mandal},
  ``{IoT} based secure smart city architecture using blockchain,'' in
  \emph{2018 2nd Inter. Conf. on Data Science and Business
  Analytics (ICDSBA)}, Sep. 2018, pp. 215--220.

\bibitem{5.27}
and Y.~{Bandung}, ``Design of secure {IoT} platform for smart home system,'' in
  \emph{2018 5th Inter. Conf. on Information Technology, Computer,
  and Electrical Engineering (ICITACEE)}, Sep. 2018, pp. 114--119.

\bibitem{5.73}
V.~{Garcia-Font}, C.~{Garrigues}, and H.~{Rifà-Pous}, ``An architecture for
  the analysis and detection of anomalies in smart city wsns,'' in \emph{2015
  IEEE First International Smart Cities Conference (ISC2)}, Oct 2015, pp. 1--6.

\bibitem{5.74}
S.~{Tousley} and S.~{Rhee}, ``Smart and secure cities and communities,'' in
  \emph{2018 IEEE International Science of Smart City Operations and Platforms
  Engineering in Partnership with Global City Teams Challenge (SCOPE-GCTC)},
  April 2018, pp. 7--11.

\bibitem{8.36}
Y.~Al-Eryani and E.~Hossain, ``Delta-oma (d-oma): A new method for massive
  multiple access in 6g,'' \emph{arXiv preprint arXiv:1901.07100}, 2019.

\bibitem{8.32}
H.~Q. {Ngo}, A.~{Ashikhmin}, H.~{Yang}, E.~G. {Larsson}, and T.~L. {Marzetta},
  ``Cell-free massive {MIMO} versus small cells,'' \emph{IEEE Trans.
  Wireless Communications}, vol.~16, no.~3, pp. 1834--1850, March 2017.

\bibitem{5.75}
X.~{He}, K.~{Wang}, H.~{Huang}, and B.~{Liu}, ``Qoe-driven big data
  architecture for smart city,'' \emph{IEEE Commun. Mag.}, vol.~56,
  no.~2, pp. 88--93, Feb 2018.

\bibitem{5.76}
B.~{Tang}, Z.~{Chen}, G.~{Hefferman}, S.~{Pei}, T.~{Wei}, H.~{He}, and
  Q.~{Yang}, ``Incorporating intelligence in fog computing for big data
  analysis in smart cities,'' \emph{IEEE Trans. Industrial
  Informat.}, vol.~13, no.~5, pp. 2140--2150, Oct 2017.

\bibitem{6.3}
5G-PPP, ``{5G-PPP} white paper on ehealth vertical sector,'' in \emph{5G and
  e-Health}.\hskip 1em plus 0.5em minus 0.4em\relax 5G-PPP, Sept. 2015.

\bibitem{6.1}
F.~{Al-Turjman} and S.~{Alturjman}, ``Context-sensitive access in industrial
  internet of things {(IIoT)} healthcare applications,'' \emph{IEEE
  Trans. Industrial Informatics}, vol.~14, no.~6, pp. 2736--2744, June
  2018.

\bibitem{6.2}
H.~{Karvonen}, M.~{Hämäläinen}, J.~{Iinatti}, and C.~{Pomalaza-Ráez},
  ``Coexistence of wireless technologies in medical scenarios,'' in \emph{2017
  European Conference on Networks and Communications (EuCNC)}, June 2017, pp.
  1--5.

\bibitem{6.10}
S.~M.~R. {Islam}, D.~{Kwak}, M.~H. {Kabir}, M.~{Hossain}, and K.~{Kwak}, ``The
  internet of things for health care: A comprehensive survey,'' \emph{IEEE
  Access}, vol.~3, pp. 678--708, 2015.

\bibitem{6.11}
V.~Oleshchuk and R.~Fensli, ``Remote patient monitoring within a future {5G}
  infrastructure,'' \emph{Wireless Personal Communications}, vol.~57, no.~3,
  pp. 431--439, 2011.

\bibitem{6.4}
J.~{Nightingale}, P.~{Salva-Garcia}, J.~M.~A. {Calero}, and Q.~{Wang},
  ``{5G-QoE}: Qoe modelling for ultra-hd video streaming in {5G} networks,''
  \emph{IEEE Trans. Broadcasting}, vol.~64, no.~2, pp. 621--634, June
  2018.

\bibitem{6.7}
J.~Marescaux, J.~Leroy, F.~Rubino, M.~Smith, M.~Vix, M.~Simone, and D.~Mutter,
  ``Transcontinental robot-assisted remote telesurgery: feasibility and
  potential applications,'' \emph{Annals of surgery}, vol. 235, no.~4, p. 487,
  2002.

\bibitem{6.6}
D.~{Soldani}, F.~{Fadini}, H.~{Rasanen}, J.~{Duran}, T.~{Niemela},
  D.~{Chandramouli}, T.~{Hoglund}, K.~{Doppler}, T.~{Himanen}, J.~{Laiho}, and
  N.~{Nanavaty}, ``{5G} mobile systems for healthcare,'' in \emph{2017 IEEE
  85th Veh. Tech. Conference (VTC Spring)}, June 2017, pp. 1--5.

\bibitem{6.8}
A.~Z. Abbasi, N.~Islam, Z.~A. Shaikh \emph{et~al.}, ``A review of wireless
  sensors and networks' applications in agriculture,'' \emph{Computer Standards
  \& Interfaces}, vol.~36, no.~2, pp. 263--270, 2014.

\bibitem{6.9}
T.~Ojha, S.~Misra, and N.~S. Raghuwanshi, ``Wireless sensor networks for
  agriculture: The state-of-the-art in practice and future challenges,''
  \emph{Computers and Electronics in Agriculture}, vol. 118, pp. 66--84, 2015.

\bibitem{8.2}
W.~Saad, M.~Bennis, and M.~Chen, ``A vision of {6G} wireless systems:
  Applications, trends, technologies, and open research problems,'' \emph{arXiv
  preprint arXiv:1902.10265}, 2019.

\bibitem{8.1}
F.~Tariq, M.~Khandaker, K.-K. Wong, M.~Imran, M.~Bennis, and M.~Debbah, ``A
  speculative study on {6G},'' \emph{arXiv preprint arXiv:1902.06700}, 2019.

\bibitem{8.8}
M.~{Mozaffari}, A.~{Taleb Zadeh Kasgari}, W.~{Saad}, M.~{Bennis}, and
  M.~{Debbah}, ``Beyond {5G} with {UAVs}: Foundations of a 3d wireless cellular
  network,'' \emph{IEEE Trans. Wireless Commun.}, vol.~18,
  no.~1, pp. 357--372, Jan 2019.

\bibitem{8.9}
B.~{Li}, Z.~{Fei}, and Y.~{Zhang}, ``Uav communications for {{5G}} and beyond:
  Recent advances and future trends,'' \emph{IEEE Internet Things J.},
  pp. 1--1, 2019.

\bibitem{8.3}
K.~{David} and H.~{Berndt}, ``{6G} vision and requirements: Is there any need
  for beyond {5G}?'' \emph{IEEE Veh. Tech. Mag.}, vol.~13, no.~3,
  pp. 72--80, Sep. 2018.

\bibitem{8.15}
T.~{Wang}, S.~{Wang}, and Z.~{Zhou}, ``Machine learning for {5G} and beyond:
  From model-based to data-driven mobile wireless networks,'' \emph{China
  Commun.}, vol.~16, no.~1, pp. 165--175, Jan 2019.

\bibitem{8.7}
M.~{Katz}, M.~{Matinmikko-Blue}, and M.~{Latva-Aho}, ``{6Genesis} flagship
  program: Building the bridges towards {6G}-enabled wireless smart society and
  ecosystem,'' in \emph{2018 IEEE 10th Latin-American Conf.
  Commun. (LATINCOM)}, Nov 2018, pp. 1--9.

\bibitem{8.13}
G.~{Berardinelli}, N.~H. {Mahmood}, I.~{Rodriguez}, and P.~{Mogensen}, ``Beyond
  {5G} wireless irt for industry 4.0: Design principles and spectrum aspects,''
  in \emph{2018 IEEE Globecom Workshops (GC Wkshps)}, Dec 2018, pp. 1--6.

\bibitem{8.5}
A.~{Yastrebova}, R.~{Kirichek}, Y.~{Koucheryavy}, A.~{Borodin}, and
  A.~{Koucheryavy}, ``Future networks 2030: Architecture and requirements,'' in
  \emph{2018 10th International Congress on Ultra Modern Telecommunications and
  Control Systems and Workshops (ICUMT)}, Nov 2018, pp. 1--8.

\bibitem{8.6}
Y.~{Xing} and T.~S. {Rappaport}, ``Propagation measurement system and approach
  at 140 {GHz-Moving} to {6G} and above 100 {GHz},'' in \emph{2018 IEEE Global
  Commun. Conference (GLOBECOM)}, Dec 2018, pp. 1--6.

\bibitem{8.10}
H.~{Elayan}, O.~{Amin}, R.~M. {Shubair}, and M.~{Alouini}, ``Terahertz
  communication: The opportunities of wireless technology beyond {5G},'' in
  \emph{2018 Inter. Conf. Advanced Commun. Tech. Netw. (CommNet)}, April 2018, pp. 1--5.

\bibitem{8.11}
J.~{Doré}, Y.~{Corre}, S.~{Bicais}, J.~{Palicot}, E.~{Faussurier},
  D.~{Ktenas}, and F.~{Bader}, ``Above-90ghz spectrum and single-carrier
  waveform as enablers for efficient tbit/s wireless communications,'' in
  \emph{2018 25th Inter. Conf. Telecommun.(ICT)}, June
  2018, pp. 274--278.

\bibitem{8.14}
A.~A. {Boulogeorgos}, A.~{Alexiou}, T.~{Merkle}, C.~{Schubert}, R.~{Elschner},
  A.~{Katsiotis}, P.~{Stavrianos}, D.~{Kritharidis}, P.~{Chartsias},
  J.~{Kokkoniemi}, M.~{Juntti}, J.~{Lehtomaki}, A.~{Teixeira}, and
  F.~{Rodrigues}, ``Terahertz technologies to deliver optical network quality
  of experience in wireless systems beyond {5G},'' \emph{IEEE Communications
  Magazine}, vol.~56, no.~6, pp. 144--151, June 2018.

\bibitem{8.22}
N.~{Khalid} and O.~B. {Akan}, ``Experimental throughput analysis of low-thz
  {MIMO} communication channel in {5G} wireless networks,'' \emph{IEEE Wireless
  Commun. Letters}, vol.~5, no.~6, pp. 616--619, Dec 2016.

\bibitem{8.23}
P.~T. {Dat}, A.~{Kanno}, T.~{Umezawa}, N.~{Yamamoto}, and T.~{Kawanishi},
  ``Millimeter- and terahertz-wave radio-over-fiber for {5G} and beyond,'' in
  \emph{2017 IEEE Photonics Society Summer Topical Meeting Series (SUM)}, July
  2017, pp. 165--166.

\bibitem{8.24}
A.~{Kanno}, ``Millimeter- and terahertz-wave radio systems enabled by photonics
  technology for {5G/IoT} and beyond,'' in \emph{2018 IEEE 7th Inter.
  Conf. Photonics (ICP)}, April 2018, pp. 1--3.

\bibitem{8.38}
N.~{Ishikawa}, S.~{Sugiura}, and L.~{Hanzo}, ``50 years of permutation, spatial
  and index modulation: From classic rf to visible light communications and
  data storage,'' \emph{IEEE Commun. Surveys Tuts.}, vol.~20, no.~3,
  pp. 1905--1938, thirdquarter 2018.

\bibitem{8.17}
H.~{Haas}, L.~{Yin}, Y.~{Wang}, and C.~{Chen}, ``What is lifi?'' \emph{J.
  Lightwave Tech.}, vol.~34, no.~6, pp. 1533--1544, March 2016.

\bibitem{8.25}
H.~Haas, ``Lifi is a paradigm-shifting {5G} technology,'' \emph{Reviews in
  Physics}, vol.~3, pp. 26--31, 2018.

\bibitem{8.39}
Y.~{Liu}, Z.~{Qin}, M.~{Elkashlan}, Z.~{Ding}, A.~{Nallanathan}, and
  L.~{Hanzo}, ``Nonorthogonal multiple access for {5G} and beyond,''
  \emph{Proceedings of the IEEE}, vol. 105, no.~12, pp. 2347--2381, Dec 2017.

\bibitem{8.40}
L.~{Dai}, B.~{Wang}, Z.~{Ding}, Z.~{Wang}, S.~{Chen}, and L.~{Hanzo}, ``A
  survey of non-orthogonal multiple access for {5G},'' \emph{IEEE
  Commun. Surveys Tuts.}, vol.~20, no.~3, pp. 2294--2323,
  thirdquarter 2018.

\bibitem{8.16}
``{5G} and beyond waveforms,'' in \emph{2017 24th Inter. Conf. on
  Telecommunications (ICT)}, May 2017, pp. 1--42.

\bibitem{8.33}
M.~{Elkourdi}, B.~{Peköz}, E.~{Güvenkaya}, and H.~{Arslan}, ``Waveform design
  principles for {5G} and beyond,'' in \emph{2016 IEEE 17th Annual Wireless and
  Microwave Technology Conference (WAMICON)}, April 2016, pp. 1--6.

\bibitem{8.34}
A.~F. Demir, M.~Elkourdi, M.~Ibrahim, and H.~Arslan, ``Waveform design for {5G}
  and beyond,'' \emph{arXiv preprint arXiv:1902.05999}, 2019.

\bibitem{8.4}
V.~{Raghavan} and J.~{Li}, ``Evolution of physical-layer communications
  research in the post-{5G} era,'' \emph{IEEE Access}, vol.~7, pp.
  10\,392--10\,401, 2019.

\bibitem{9.3}
J.~{Zhang}, T.~{Chen}, S.~{Zhong}, J.~{Wang}, W.~{Zhang}, X.~{Zuo}, R.~G.
  {Maunder}, and L.~{Hanzo}, ``Aeronautical $ad~hoc$ networking for the
  internet-above-the-clouds,'' \emph{Proceedings of the IEEE}, vol. 107, no.~5,
  pp. 868--911, May 2019.

\bibitem{9.4}
X.~{Huang}, J.~A. {Zhang}, Ren-PingLiu, Y.~J. {Guo}, and L.~{Hanzo},
  ``Airplane-aided integrated networking for {6G} wireless - will it work?''
  \emph{IEEE Veh.Tech. Mag.}.

\bibitem{9.5}
L.~{Kong}, W.~{Xu}, L.~{Hanzo}, H.~{Zhang}, and C.~{Zhao}, ``Performance of a
  free-space-optical relay-assisted hybrid {RF/FSO} system in generalized
  $m$-distributed channels,'' \emph{IEEE Photonics J.}, vol.~7, no.~5, pp.
  1--19, Oct 2015.

\bibitem{9.6}
J.~{Armstrong}, ``Ofdm for optical communications,'' \emph{Journal of Lightwave
  Technology}, vol.~27, no.~3, pp. 189--204, Feb 2009.

\bibitem{9.7}
Q.~Wang, C.~Qian, X.~Guo, Z.~Wang, D.~G. Cunningham, and I.~H. White, ``Layered
  {ACO-OFDM} for intensity-modulated direct-detection optical wireless
  transmission,'' \emph{Optics Express}, vol.~23, no.~9, pp. 12\,382--12\,393,
  2015.

\bibitem{9.8}
X.~{Zhang}, Q.~{Wang}, R.~{Zhang}, S.~{Chen}, and L.~{Hanzo}, ``Performance
  analysis of layered {ACO-OFDM},'' \emph{IEEE Access}, vol.~5, pp.
  18\,366--18\,381, 2017.

\bibitem{9.9}
Z.~{Babar}, X.~{Zhang}, P.~{Botsinis}, D.~{Alanis}, D.~{Chandra}, S.~X. {Ng},
  and L.~{Hanzo}, ``Near-capacity multilayered code design for
  {LACO-OFDM-Aided} optical wireless systems,'' \emph{IEEE Trans.
  Veh. Tech.}, vol.~68, no.~4, pp. 4051--4054, April 2019.

\bibitem{9.10}
X.~{Zhang}, Z.~{Babar}, R.~{Zhang}, S.~{Chen}, and L.~{Hanzo}, ``Multi-class
  coded layered asymmetrically clipped optical {OFDM},'' \emph{IEEE
  Trans. Commun.}, vol.~67, no.~1, pp. 578--589, Jan 2019.

\bibitem{9.11}
S.~{Feng}, R.~{Zhang}, W.~{Xu}, and L.~{Hanzo}, ``Multiple access design for
  ultra-dense {VLC} networks: Orthogonal vs non-orthogonal,'' \emph{IEEE
  Trans. Commun.}, vol.~67, no.~3, pp. 2218--2232, March 2019.

\bibitem{9.12}
X.~{Li}, R.~{Zhang}, and L.~{Hanzo}, ``Cooperative load balancing in hybrid
  visible light communications and {WiFi},'' \emph{IEEE Trans.
  Commun.}, vol.~63, no.~4, pp. 1319--1329, April 2015.

\bibitem{9.13}
C.~{Zhu}, Y.~{Huo}, J.~{Jiang}, H.~{Sun}, C.~{Dong}, R.~{Zhang}, and
  L.~{Hanzo}, ``Hierarchical colour-shift-keying aided layered video streaming
  for the visible light downlink,'' \emph{IEEE Access}, vol.~4, pp. 3127--3152,
  2016.

\bibitem{9.14}
J.~{Jiang}, R.~{Zhang}, and L.~{Hanzo}, ``Analysis and design of three-stage
  concatenated color-shift keying,'' \emph{IEEE Trans. Veh.
  Tech.}, vol.~64, no.~11, pp. 5126--5136, Nov 2015.

\bibitem{9.15}
J.~{Jiang}, Y.~{Huo}, F.~{Jin}, P.~{Zhang}, Z.~{Wang}, Z.~{Xu}, H.~{Haas}, and
  L.~{Hanzo}, ``Video streaming in the multiuser indoor visible light
  downlink,'' \emph{IEEE Access}, vol.~3, pp. 2959--2986, 2015.

\bibitem{9.16}
N.~{Ishikawa}, S.~{Sugiura}, and L.~{Hanzo}, ``50 years of permutation, spatial
  and index modulation: From classic {RF} to visible light communications and
  data storage,'' \emph{IEEE Commun. Surveys Tuts.}, vol.~20, no.~3,
  pp. 1905--1938, thirdquarter 2018.

\bibitem{9.17}
------, ``Subcarrier-index modulation aided {OFDM} - {Will} it work?''
  \emph{IEEE Access}, vol.~4, pp. 2580--2593, 2016.

\bibitem{9.18}
H.~{Zhang}, L.~{Yang}, and L.~{Hanzo}, ``Compressed sensing improves the
  performance of subcarrier index-modulation-assisted {OFDM},'' \emph{IEEE
  Access}, vol.~4, pp. 7859--7873, 2016.

\bibitem{9.19}
M.~{Di Renzo}, H.~{Haas}, A.~{Ghrayeb}, S.~{Sugiura}, and L.~{Hanzo}, ``Spatial
  modulation for generalized {MIMO}: Challenges, opportunities, and
  implementation,'' \emph{Proceedings of the IEEE}, vol. 102, no.~1, pp.
  56--103, Jan 2014.

\bibitem{9.20}
P.~{Yang}, M.~{Di Renzo}, Y.~{Xiao}, S.~{Li}, and L.~{Hanzo}, ``Design
  guidelines for spatial modulation,'' \emph{IEEE Commun. Surveys Tuts.}, vol.~17, no.~1, pp. 6--26, Firstquarter 2015.

\bibitem{9.21}
S.~{Sugiura}, S.~{Chen}, and L.~{Hanzo}, ``Coherent and differential space-time
  shift keying: A dispersion matrix approach,'' \emph{IEEE Trans.
  Commun.}, vol.~58, no.~11, pp. 3219--3230, November 2010.

\bibitem{9.22}
------, ``Generalized space-time shift keying designed for flexible diversity-,
  multiplexing- and complexity-tradeoffs,'' \emph{IEEE Trans. Wireless
  Commun.}, vol.~10, no.~4, pp. 1144--1153, April 2011.

\bibitem{9.23}
Z.~{Zhang}, X.~{Chai}, K.~{Long}, A.~V. {Vasilakos}, and L.~{Hanzo}, ``Full
  duplex techniques for {5G} networks: self-interference cancellation, protocol
  design, and relay selection,'' \emph{IEEE Commun. Mag.}, vol.~53,
  no.~5, pp. 128--137, May 2015.

\bibitem{9.24}
Z.~{Zhang}, K.~{Long}, A.~V. {Vasilakos}, and L.~{Hanzo}, ``Full-duplex
  wireless communications: Challenges, solutions, and future research
  directions,'' \emph{Proceedings of the IEEE}, vol. 104, no.~7, pp.
  1369--1409, July 2016.

\end{thebibliography}
%

%
%
%




\end{document}